%% file: ExtentKinpaper.tex
\documentclass[useAMS,usenatbib]{mn2e}

\usepackage{graphicx,subfigure,color}
\usepackage[total={17.8cm,24.0cm},centering]{geometry}
\usepackage{times}

\def\hi{H{\small I}}
\def\sauron{{\tt SAURON}}

\def\atlas{{{ATLAS}}$^{\rm 3D}$}
\def\kms{km s$^{-1}$}
\def\co{CO}

\def\arcsec{$^{\prime \prime}$}
\newcommand{\farc}{\mbox{\ensuremath{.\!\!^{\prime\prime}}}}
\definecolor{Mygrey}{gray}{0.75}

\newcommand{\ltsimeq}{\raisebox{-0.6ex}{$\,\stackrel{\raisebox{-.2ex}{$\textstyle <$}}{\sim}\,$}}
\newcommand{\gtsimeq}{\raisebox{-0.6ex}{$\,\stackrel{\raisebox{-.2ex}{$\textstyle >$}}{\sim}\,$}}
\usepackage[compact]{titlesec}
\titlespacing{\section}{0pt}{*2}{*1}

\title[Extent and kinematics of molecular gas in ETGs]{The \atlas\ Project - XIV. The extent and kinematics of molecular gas in early-type galaxies} 
\author[Timothy A. Davis et al.]{\parbox{\textwidth}{Timothy A. Davis,$^{1,2}$\thanks{E-mail:\texttt{tdavis@eso.org}}
Katherine Alatalo,$^{3}$
Martin Bureau,$^{1}$
Michele Cappellari,$^{1}$
Nicholas Scott,$^{4}$
Lisa M. Young,$^{5}$\thanks{Adjunct Astronomer with NRAO}
Leo Blitz,$^{3}$
Alison Crocker,$^{6}$
Estelle Bayet,$^{1}$
Maxime Bois,$^{7}$
Fr\'ed\'eric Bournaud,$^{8}$
Roger L. Davies,$^{1}$
P. T. de Zeeuw,$^{2,9}$
Pierre-Alain Duc,$^{8}$
Eric Emsellem,$^{2,10}$
Sadegh Khochfar,$^{11}$
Davor Krajnovi\'c,$^{2}$
Harald Kuntschner,$^{12}$
Pierre-Yves Lablanche,$^{2,10}$
Richard M. McDermid,$^{13}$
Raffaella Morganti,$^{14,15}$
Thorsten Naab,$^{16}$
Tom Oosterloo,$^{14,15}$
Marc Sarzi,$^{17}$
Paolo Serra,$^{14}$
and Anne-Marie Weijmans,$^{18}$\thanks{Dunlap Fellow}}\vspace{0.4cm}\\ 
\parbox{\textwidth}{$^{1}$Sub-Dept. of Astrophysics, Dept. of Physics, University of Oxford, Denys Wilkinson Building, Keble Road, Oxford, OX1 3RH, UK\\
$^{2}$European Southern Observatory, Karl-Schwarzschild-Str. 2, 85748 Garching, Germany\\
$^{3}$Department of Astronomy, Campbell Hall, University of California, Berkeley, CA 94720, USA\\
$^{4}$Centre for Astrophysics \& Supercomputing, Swinburne University of Technology, PO Box 218, Hawthorn, VIC 3122, Australia\\
$^{5}$Physics Department, New Mexico Institute of Mining and Technology, Socorro, NM 87801, USA\\
$^{6}$University of Massachussets, Amherst, USA\\
$^{7}$Observatoire de Paris, LERMA and CNRS, 61 Av. de l`Observatoire, F-75014 Paris, France\\
$^{8}$Laboratoire AIM Paris-Saclay, CEA/IRFU/SAp -- CNRS -- Universit\'e Paris Diderot, 91191 Gif-sur-Yvette Cedex, France\\
$^{9}$Sterrewacht Leiden, Leiden University, Postbus 9513, 2300 RA Leiden, the Netherlands\\
$^{10}$Universit\'e Lyon 1, Observatoire de Lyon, Centre de Recherche Astrophysique de Lyon and Ecole Normale Sup\'erieure de Lyon, 9 avenue Charles Andr\'e, F-69230 Saint-Genis Laval, France\\
$^{11}$Max-Planck Institut f\"ur extraterrestrische Physik, PO Box 1312, D-85478 Garching, Germany\\
$^{12}$Space Telescope European Coordinating Facility, European Southern Observatory, Karl-Schwarzschild-Str. 2, 85748 Garching, Germany\\
$^{13}$Gemini Observatory, Northern Operations Centre, 670 N. A`ohoku Place, Hilo, HI 96720, USA\\
$^{14}$Netherlands Institute for Radio Astronomy (ASTRON), Postbus 2, 7990 AA Dwingeloo, The Netherlands\\
$^{15}$Kapteyn Astronomical Institute, University of Groningen, Postbus 800, 9700 AV Groningen, The Netherlands\\
$^{16}$Max-Planck-Institut f\"ur Astrophysik, Karl-Schwarzschild-Str. 1, 85741 Garching, Germany\\
$^{17}$Centre for Astrophysics Research, University of Hertfordshire, Hatfield, Herts AL1 9AB, UK\\
$^{18}$Dunlap Institute for Astronomy \& Astrophysics, University of Toronto, 50 St. George Street, Toronto, ON M5S 3H4, Canada
}
}
\begin{document}
\date{Accepted 2012 November 05. Received 2012 November 05; in original form 2012 January 31}

\pagerange{\pageref{firstpage}--\pageref{lastpage}} \pubyear{2011}

\maketitle

\label{firstpage}
\clearpage
\begin{abstract}
We use interferometric $^{12}$CO(1-0) observations to compare and contrast the extent, surface brightness profiles and kinematics of the molecular gas in CO-rich \atlas\ early-type galaxies (ETGs) and spiral galaxies.  We find that the molecular gas extent is smaller in absolute terms in ETGs than in late-type galaxies, but that the size distributions are similar once scaled by the galaxies optical/stellar characteristic scalelengths. 
Amongst ETGs, we find that the extent of the gas is independent of its kinematic misalignment (with respect to the stars), but does depend on environment, with Virgo cluster ETGs having less extended molecular gas reservoirs, further emphasizing that cluster ETGs follow different evolutionary pathways from those in the field. 
Approximately half of ETGs have molecular gas surface brightness profiles that follow the stellar light profile. These systems often have relaxed gas out to large radii, suggesting they are unlikely to have had recent merger/accretion events. 
A third of the sample galaxies show molecular gas surface brightness profiles that fall off slower than the light, and sometimes show a truncation. These galaxies often have a low mass, and either have disturbed molecular gas, or are in the Virgo cluster, suggesting that recent mergers, ram pressure stripping and/or the presence of hot gas can compress/truncate the gas. The remaining galaxies have rings, or composite profiles, that we argue can be caused by the effects of bars. 
We investigated the kinematics of the molecular gas using position-velocity diagrams, and compared the observed kinematics with dynamical model predictions, and the observed stellar and ionised gas velocities. 
We confirm that the molecular gas reaches beyond the turnover of the circular velocity curve in $\approx$70\% of our CO-rich \atlas\ ETGs, validating previous
work on the CO Tully-Fisher relation. In general we find that in most galaxies the molecular gas is dynamically cold, and the observed CO rotation matches well model predictions of the circular velocity. 
In the galaxies with the largest molecular masses, dust obscuration and/or population gradients can cause model predictions of the circular velocity to disagree with observations of the molecular gas rotation, however these effects are confined to the most star-forming systems. 
Bars and non-equilibrium conditions can also make the gas deviate from circular orbits. In both these cases one expects the models circular velocity to be higher than the observed CO velocity, in agreement with our observations.
Molecular gas is a better direct tracer of the circular velocity than the ionised gas, justifying its use as a kinematic tracer for Tully-Fisher and similar analyses. 

\end{abstract}


\section{Introduction}

The buildup of galaxies through hierarchical merging has emerged in recent years as the predominant theory of galaxy evolution. In this paradigm, the collapse of dark matter haloes proceeds in a "bottom-up" fashion, smaller structures forming first and merging to create larger systems \citep[e.g.][]{Springel2005}. It seems however that galaxy star-formation histories do not share the same "bottom-up" evolution. Massive early-type galaxies (ETGs) are the end-point of hierarchical galaxy evolution; they are normally considered to be `red and dead', without significant star-formation, and containing mainly old stellar populations \citep[e.g.][]{Bower:1992p3541}. Lower mass spiral and irregular galaxies, which generally exist in less dense environments and have undergone fewer mergers, are usually gas rich and actively star-forming \citep[e.g.][]{RobertsHayes1994}. 

Some ETGs do,  however, have molecular (and atomic; see \citealt{Morganti:2006p1934,Oosterloo:2010p3376}; \citealt{SerraHI}; Paper XIII) gas reservoirs. Molecular gas was first detected in ETGs by \cite{Wiklind:1986p3494} and \cite{Phillips:1987p3495}. Since then various surveys have detected molecular gas in ETGs that are infrared bright \citep[e.g.][]{Wiklind:1989p3496,Sage:1989p3497,Knapp:1996p1859}, or that have dust obscuration in optical images \citep[e.g.][]{Wang:1992p3498}. It was not until recently however that  unbiased samples of ETGs were searched for molecular gas \citep[e.g.][]{Welch:2003p2521,Sage:2007p3467,Combes:2007p231}. \citet[][hearafter Paper IV]{Young:2011p3480} conducted the largest of these studies to date, on the complete, volume-limited \atlas\ sample of ETGs \citep[][hereafter Paper I]{Cappellari:2011p3515}. This work found that 22\% of ETGs in the local volume (out to 42 Mpc) have a detectable molecular gas reservoir {(M$_{\rm H_2}$\gtsimeq 1$\times$10$^8$M$_{\odot}$)}. 
It is thus important to understand the origin, properties and fate of this gas, to understand downsizing and the evolution of massive galaxies.

To explain the tight correlations between the properties of ETGs (e.g the fundamental plane, \citealt{Dressler1987,Djorgovski:1987p3499}; the colour-magnitude relation, \citealt{Baldry:2004p3398}), it is thought that the cold inter-stellar medium (ISM) must be removed or destroyed (or potentially made stable; e.g. \citealt{Martig:2009p2923}), such that star-formation is abruptly quenched. How one reconciles the requirement for quenching with the observation of molecular and atomic gas reservoirs in some ETGs is an open question. 
In Paper X of this series \citep{Davis:2011b}, we showed that much of the molecular gas in field ETGs is kinematically misaligned and likely has an external origin. Thus, consistent with the quenching scenario, these galaxies could have been red and dead in the past, and have recently acquired more gas. Virgo cluster ETGs also have molecular gas however, without replenishing this externally. 

\cite{Khochfar:2011} (hereafter Paper VIII) suggest, from a modelling perspective, that star formation and hence molecular gas are important in the transformation of galaxies already on the red-sequence (e.g. in turning slow rotators into fast rotators). In order to test such a hypothesis one needs to understand the state of the molecular gas, and its sources. 

The molecular gas in ETGs exists in different conditions than those within the discs of spiral galaxies.
For example, ETGs have different gravitational potentials (affecting e.g.\ the \citealt{Toomre:1964p3272} stability parameter $Q$), different elemental abundances (high metallicity and $\alpha$-element over-abundance), higher prevalence of hot gas ($10^6$~K) and active galactic nuclei (AGN).  Old stellar populations also create a harder inter-stellar radiation field (e.g. the UV-upturn phenomenon; see \citealt{Yi:2008p3435,Bureau:2011}). Furthermore, as mentioned above, many galaxies have kinematically misaligned gas with an external origin, rotating along a different axis than the majority of the stars. Spiral galaxies have had their star-forming gas in situ for many gigayears (even if it is also being topped up externally; e.g. \citealt{Fraternali:2008p3502}).  All these differences may affect the properties of the gas, including its extent, ability to form stars, and kinematics.

To investigate these issues, it is vital to have spatially-resolved data on the distribution and kinematics of the molecular gas. Until recently, only a small number of ETGs had been mapped with mm-interferometers \citep[e.g.][]{Wrobel:1992p982,Young:2002p943,Young:2008p788,Schinnerer:2002p981,Crocker:2008p946,Crocker:2009p3262,Wei:2010p3501,Crocker:2011p3500}, hence drawing conclusions based on a few objects has been difficult. In this paper, we attempt to understand the similarities and differences between the extent and distribution of the molecular gas in spirals and ETGs. For this we will use a sample of 41 ETGs with molecular gas mapped as part of the \atlas\ project (Alatalo et al., in prep; hereafter A12). Furthermore, we continue a theme presented in Paper V of this series \citep{Davis:2011p3472}, evaluating the use of molecular gas as a circular velocity tracer in ETGs, this time in a spatially-resolved manner.

 In Section 2, we present our galaxy sample and outline the observations and models our work is based on. In Section 3 we present a study of the extent of the molecular gas in our ETGs, and compare this to the extent of the gas in spirals, and this extent varies with gas origin and environment. In Section 4 we examine the surface brightness profiles of the molecular gas in ETGs. In Section 5 we evaluate the use of CO as a tracer of the circular velocity in ETGs, and how the observed CO rotation compares with circular velocity predictions from mass models and observed stellar and ionised gas rotation velocities. We discuss these results in Section 6, before concluding in Section 7.

\section{Data and Methods}

\subsection{Sample}

\label{sample}

We present results based on a sample of 41 ETGs that have $^{12}$CO aperture synthesis imaging available, either from the \atlas\ project (A12) or the literature.
As part of the \atlas\ survey, all \co(1-0) detections from Paper IV with an integrated flux greater than 19 Jy \kms\ that do not have interferometric data available in the literature have been observed with the Combined Array for Research in mm-wave Astronomy \citep[CARMA;][]{Bock:2006p2806}. Full details of this interferometric survey can be found in A12, but we summarise the observations in Section \ref{carmadata}.

A total of 28 galaxies included in this work have been mapped with CARMA (see Table \ref{extenttable}).
We also include galaxies for which data are already available from the literature, mostly from the SAURON survey \citep{deZeeuw:2002p1496} follow-ups. 
These are NGC\,0524, NGC\,2685, NGC\,2768, NGC\,3032, NGC\,3489, NGC\,4150, NGC\,4459, NGC\,4476, NGC\,4477 NGC\,4526, and NGC\,4550. Full references for the observations used are listed in Table \ref{extenttable}. Furthermore we include NGC\,2697 and NGC\,4292, observed with CARMA as part of this project but later removed from the final \atlas\ sample (due to not making the distance and/or magnitude cut). These two objects are nevertheless classified as ETGs, {and by including them in the sample we are unlikely to introduce any significant bias.} In total, this results in a sample of 41 galaxies that have CO(1-0) interferometric data.

This sample of mapped galaxies includes the brightest $\approx$2/3 of the CO detections from Paper IV. The mapped galaxies are thus biased towards higher molecular gas masses, but are statistically indistinguishable (using a Kolmogorov-Smirnov test) from the full sample of CO-detected systems in terms of their {gas fractions (the ratio of molecular mass to stellar mass), group scale} environment, and host galaxy properties e.g. velocity dispersion, luminosity, effective radius  (Paper I) and specific angular momentum parameter ($\lambda_R$; \citealt{Emsellem:2011p3493}, hearafter Paper III). The mapped galaxies also follow the same CO Tully-Fisher relation (Paper V). Other than molecular gas mass, the only parameter in which our sample may have some bias is cluster membership. Because the Virgo cluster is nearby, its members are relatively bright in CO(1-0), and we have mapped a greater percentage of cluster members than field galaxies. However, we do not  expect this to affect our conclusions. Overall, we thus consider that our mapped CO-rich ETG sample is reasonably free of biases, although it is worth remembering that we do not fully sample the parameter space at small molecular gas masses.

\subsection{CARMA data} 
\label{carmadata}

Observations of the 28 sample galaxies were taken in the D-array configuration at CARMA between 2008 and 2010, providing a spatial resolution of 4-5\arcsec. Full details of this interferometric survey can be found in A12 but we summarize the observations here. 
\co(1-0) was observed using a narrow-band correlator configuration, providing at least 3 raw channels per 10 \kms\ binned channel whilst ensuring adequate velocity coverage for all galaxies.
Bright quasars were used to calibrate the antenna-based gains and for passband calibration. The data were calibrated and imaged using the `Multichannel Image Reconstruction, Image Analysis and Display' (MIRIAD) software package \citep{Sault:1995p2768}. {Comparison of the CARMA total fluxes with the IRAM 30m single-dish observations from Paper IV shows that within the measurement errors we always recover all the flux in our CARMA maps, and often have higher fluxes due to detecting gas outside the 30m telescope beam (A12). The sensitivity of our observations vary from 3-30 mJy/beam (dependant on the channel size used in the final cube; average 10 \kms). 
This gives us a 3$\sigma$ surface density limit of between 3 and 30 M$_{\odot}$ pc$^{-2}$ with a median of 14 M$_{\odot}$ pc$^{-2}$ (if one uses a standard galactic $X_{\rm CO}$ factor as in Paper IV).
}  

The CO integrated intensity maps we use in this paper are presented in A12. 
They were produced by the masking method. The cleaned image 
cube was smoothed spatially {with a gaussian 1.4 times larger in diameter than the beam, and in velocity with a hanning width of 3 channels}, and then this 
smoothed cube was clipped at 
$\approx$2.5 times the rms noise in a 
channel. The clipped cube was then used 
as a mask to define a three-dimensional volume in the original, 
unsmoothed cleaned cube, over which we took the zeroth moment \citep[see][]{Regan:2001p3275}. {For some of the analysis in this paper, unclipped moment maps were created from the reduced data cubes of A12, to avoid biasing against faint emission.}  

\subsection{SAURON IFS data and Jeans Models}

The SAURON observations of the \atlas\ galaxies and the extraction of the stellar kinematics are described in detail in Paper I. In brief, for each target, individual datacubes were merged and analysed as described in \cite{Emsellem:2004p1497}, ensuring a minimum signal-to-noise ratio of 40 per spatial and spectral pixel 
using the binning scheme developed by \cite{Cappellari:2003p3284}. 
The SAURON stellar kinematics were derived using a penalized 
pixel fitting routine \citep{Cappellari:2004p3283}, providing
parametric estimates of the line-of-sight velocity distribution for each bin. During the extraction of the stellar kinematics, the GANDALF code \citep{Sarzi:2006p1474} was used to simultaneously extract the ionised gas line fluxes and kinematics.

\label{jam}
In this paper we utilize axisymmetric Jeans anisotropic dynamical modeling  \citep[JAM;][]{Cappellari:2008p2773} of the \atlas galaxies. Some examples of using this modeling approach with SAURON integral-field kinematics are presented in \cite{Scott:2009p3218}. For the \atlas\ survey a multi-Gaussian expansion \citep[MGE;][]{Emsellem:1994p723} was fitted to the SDSS \citep{Abazajian:2009p3430} or Isaac Newton telescope photometry (Scott et al., in preparation). The MGEs were then used to construct JAM models for all 260 \atlas\ galaxies \citep[see][]{Cappellari:2010p3429}, fiting to the SAURON stellar kinematics (Paper I).
The models have three free parameters, the inclination ($i_{\mathrm{JAM}}$), the mass-to-light ratio ($M/L$) assumed to be spatially constant, and the anisotropy also assumed to be spatially constant. 

From each mass model, we have calculated the predicted circular velocity curve in the equatorial plane of the galaxy. It is worth bearing in mind that where the CO is misaligned from the plane of the galaxy, using the circular velocity curve from the equatorial plane 
would only be valid if the potential was spherically symmetric. As these models do not include dark matter, the circular velocity often declines at large radii, and one must be careful to compare circular velocities only where the models are constrained by the IFS data.  {Even if substantial amounts of dark matter are present in the central parts of these galaxies, where the models are constrained by the IFU data the circular velocity predicted will be correct \citep{Cappellari:2006p1498}. As in this paper we only consider the inner-parts of the galaxies, where the models are fully constrained, these considerations are unlikely to affect our results.}

\section{Molecular Gas Extent}
\label{extentsec}

In this section we attempt to understand if the radial extent of the molecular gas in ETGs is different from that in spirals. One can imagine mechanisms that would make the gas in ETGs less extended than the gas in spirals. External gas, for instance, could fall into the galaxy centre, and create inner discs and rings \citep{ElicheMoral:2009p2918}. Furthermore, the interaction of kinematically-misaligned gas with stellar mass loss from stars could cause the gas to loose angular momentum and become less extended over time. 
To look for signatures of these processes, we measured the extent of the gas in 41 CO-mapped \atlas\ ETGs, and a comparison sample of 44 spirals from the Berkeley Illinois Maryland Array Survey of Nearby Galaxies \citep[BIMA-SONG;][]{Helfer:2003p3285}.

\subsection{Redshifting BIMA-SONG}
\label{redshiftbima}

BIMA-SONG mapped CO(1-0) molecular emission in a volume (and magnitude) limited sample of 44 nearby spiral galaxies. 
The survey included every galaxy of Hubble type Sa -- Sd (except M33 and M31), with declinations $\delta$ $>$ -20$^{\circ}$, visual magnitudes $B$ $<$ 11.0, velocities v$_{\rm hel}$ $<$ 2000 \kms, and inclinations $i$ $<$ 70$^{\circ}$. {The surface brightnesses sensitivity of the BIMA-SONG survey is very similar to our \atlas\ survey (with a 3$\sigma$ limit of $\approx$20 M$_{\odot}$ pc$^{-2}$ when adjusted to our preferred X$_{CO}$; \citealt{Helfer:2003p3285})}.
As such it is an ideal survey to compare with our \atlas galaxies, which are also drawn from a complete volume limited survey. We chose not to use the HERACLES survey \citep{Leroy:2009p3484} for our comparison, because the sample was not selected in a volume limited manner, the angular resolution achieved is 2-3 times worse than our observations (13\arcsec) and the galaxy sample is small (18 objects). We do however compare qualitatively with the HERACLES results in Section \ref{etgandspiralextent}.

The molecular emission in these objects was mapped with a typical spatial resolution of 6\arcsec\ (360 pc at 12 Mpc, the average distance of the sample galaxies; \citealt{Helfer:2003p3285}). {The array configurations used had a minimum baseline of $\approx$8m, meaning the largest detectable scales are similar to our CARMA data (which was observed with a minimum baseline of 11m). We caution, however, that \cite{Helfer:2003p3285} showed that they often resolved out significant fractions of the flux in these spiral galaxies. In order to minimise the effect of this missing flux we used the data cubes which included zero-spacing information (from on-the-fly mapping with the NRAO 12m telescope), where available (24/44 objects). In the other galaxies we were unable to correct for the distribution of this missing flux, and thus caution that the sizes derived here should be strictly considered as lower limits. If we only consider galaxies with zero-spacing information available, the trends reported below do not change however.}
 
Another difficulty we faced when comparing the samples was that the \atlas\ sample galaxies are on average twice as distant (24.5 Mpc) as the BIMA-SONG galaxies. 
To overcome this problem we artificially `redshifted' the BIMA-SONG galaxies to the average distance of the \atlas\ galaxies, using the procedure outlined below.

Firstly, we obtained the clean cubes from the BIMA-SONG archive. The angular size of the beam when observing a distant galaxy appears larger relative to the galaxy size, so each BIMA-SONG cube was convolved with a Gaussian to create the correct relative beam size. The size of the major and minor axes of the convolving Gaussian required in order to reach the correct final beam were calculated by assuming that the beams add in quadrature:

\begin{equation}
B_{\mathrm{min,maj}}^{\mathrm{*}} = B_{\mathrm{min,maj}}^{\mathrm{old}} \sqrt{\left(\frac{d}{24.5\,\mathrm{Mpc}}\right)^{-2}-1},
\label{quadbeam}
\end{equation}
where $B_{\mathrm{min,maj}}^{\mathrm{old}}$ is the original beam size, $B_{\mathrm{min,maj}}^{\mathrm{*}}$ the convolving beam size, d is the distance to the BIMA-SONG galaxy and 24.5 Mpc is the average distance to the ATLAS-3D galaxies.

Secondly the clean BIMA-SONG cube was scaled to the flux expected at the larger distance.
{This part of the procedure is unlikely to effect our ability to detect these objects, as surface brightness sensitivity is independent of distance.}

Thirdly, a noise cube was created with the Miriad task \textsc{imgen}. It was then scaled to the correct level, so that when added to the BIMA-SONG clean cube the resultant cube had the same RMS surface brightness threshold as a typical CARMA observation of the \atlas\ galaxies (15 M$_{\odot}$ pc$^{-2}$ in a 10 km s$^{-1}$ chan). If the BIMA-SONG cube already has an RMS greater than this level, no additional noise was added. This only applies in some cases, however, and we do not expect this to bias our results.

Finally, a moment zero map of the "redshifted" cube was created, {using the same procedure outlined in Section \ref{carmadata}}. The pixel size was set so the map is Nyquist sampled (3 pixels across the geometric mean of the beam size). 

\subsection{Measuring gas extents}
\label{measureextent}
The molecular gas extent of each galaxy was estimated from its \atlas\ or redshifted BIMA-SONG moment zero map. The diameter was measured in the direction of the largest molecular gas extent, a level of approximately three times the RMS noise in the moment zero maps. {As the $3\sigma$ surface brightness limits are approximately the same in the two surveys we should be probing similar fractions of the gas reservoirs}. 
 This procedure is similar to that followed by \cite{Chapman:2004p2084} when measuring CO sizes in sub-millimeter galaxies. {Individual objects vary somewhat in the surface brightness sensitivity achieved. Interpolating from the surface brightness profiles in Section \ref{surfdens} suggests that measuring the extent at this range of surface-densities introduces an error of upto $\approx$15\% into the measured sizes. Given the small radial extent of the gas in many of our early-type objects, this error is likely insignificant when compared with the uncertainty due to beam smearing.} The use of a scale radius obtained from a parametric fit to the CO surface density profile would be preferable, but the limited CO spatial extent (compared to the beam) in many objects, and the absence of an obvious suitable parametric form (e.g. Section \ref{surfdens}) make this impractical at this point.

The beam size of interferometric observations smears the edges of molecular gas distributions, making them appear larger than they truly are. As a first-order attempt to remove this beam smearing, the beam size in the direction of maximum extent was quadratically subtracted from the measured diameter, to give a \textit{`de-convolved'} diameter ($d^{'}$) as follows:

\begin{eqnarray}
d^{'}=\sqrt{d^2 - (2r_{\mathrm{beam}})^2},\\
\mathrm{where\,\,\,\,\,\,\,\,\,\,\,\,\,\,\,\,\,\,\,\,\,\,\,\,\,\,\,\,\,\,\,\,\,\,\,\,\,\,\,\,\,\,\,\,\,\,\,\,\,\,\,\,\,\,\,\,\,\,\,\,\,\,\,\,\,\,\,\,\,\,\,\,\,\,\,\,\,\,\,\,\,\,\,\,\,\,\,\,\,\,\,\,\,\,\,\,\,\,\,\,\,\,\,\,\,\,\,\,\,\,\,\,} \nonumber \\
r_{\mathrm{beam}} = \frac{b_{\mathrm{max}}b_{\mathrm{min}}}{\sqrt{(b_{\mathrm{min}}\cos\theta_b)^2 + (b_{\mathrm{max}}\sin\theta_b)^2}}, \\
\mathrm{and\,\,\,\,\,\,\,\,\,\,\,\,\,\,\,\,\,\,\,\,\,\,\,\,\,\,\,\,\,\,\,\,\,\,\,\,\,\,\,\,\,\,\,\,\,\,\,\,\,\,\,\,\,\,\,\,\,\,\,\,\,\,\,\,\,\,\,\,\,\,\,\,\,\,\,\,\,\,\,\,\,\,\,\,\,\,\,\,\,\,\,\,\,\,\,\,\,\,\,\,\,\,\,\,\,\,\,\,\,\,\,\,\,\,\,\,\,\,} \nonumber \\
\theta_b = \phi_{\mathrm{GPA}} - \theta_{\mathrm{BPA}} 
\end{eqnarray}
\\Here $d$ is the measured CO diameter, $b_{\mathrm{max}}$ is the beam semi-major axis, $b_{\mathrm{min}}$ is the beam semi-minor axis, $\theta_{\mathrm{BPA}}$ is the beam position angle in degrees, $\phi_{\mathrm{GPA}}$ is the measured position angle along the maximal CO extent in degrees, $r_{\mathrm{ beam}}$ is the radius of the beam at angle $\phi_{\mathrm{GPA}}$, and $\theta_b$ is the measured angle from the beam axis, to the molecular gas position angle.

\begin{figure}
\begin{center}
  \subfigure{\includegraphics[width=4cm,angle=0,clip,trim=0.0cm 0cm 0cm 0.0cm]{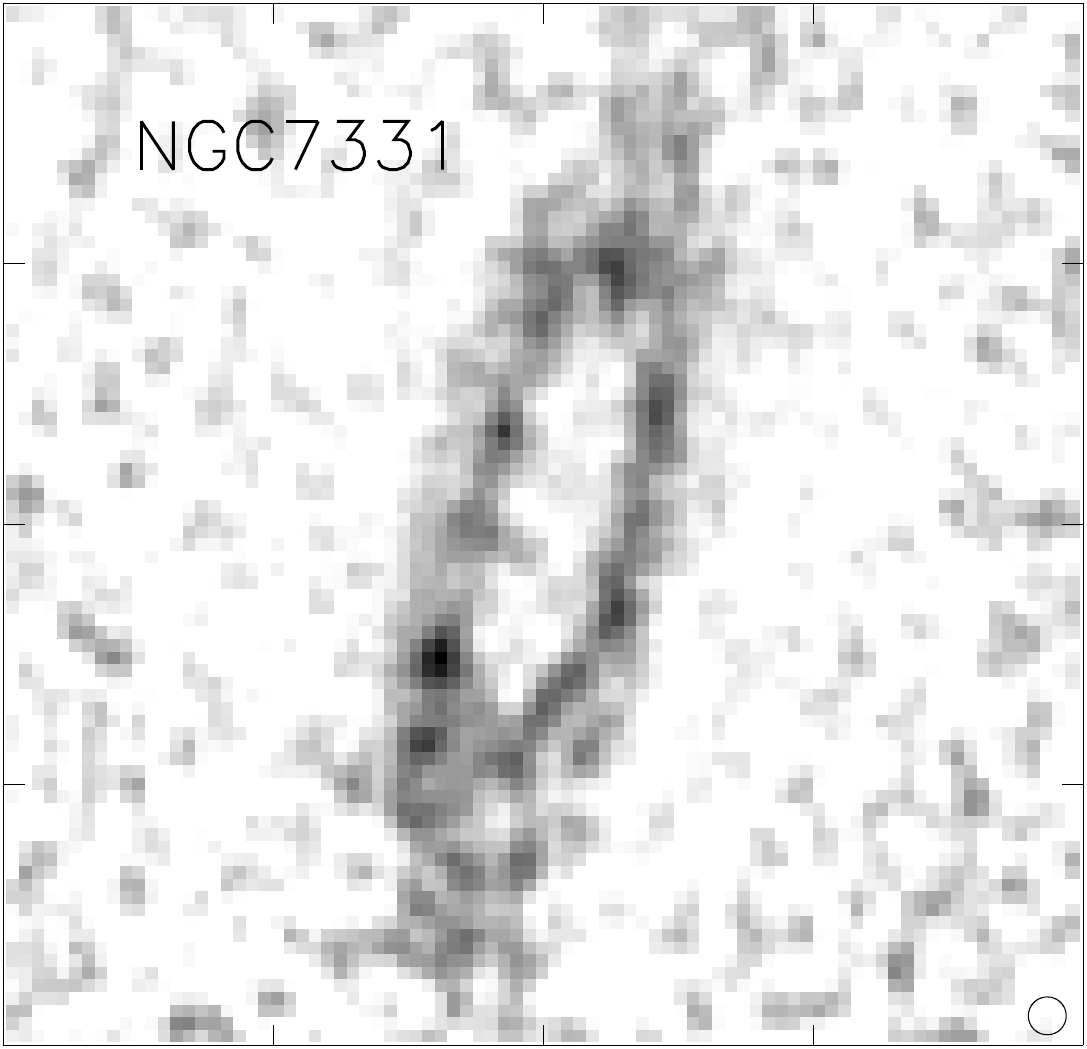}}
  \subfigure{\includegraphics[width=3.95cm,angle=0,clip,trim=0.0cm 0cm 0cm 0.0cm]{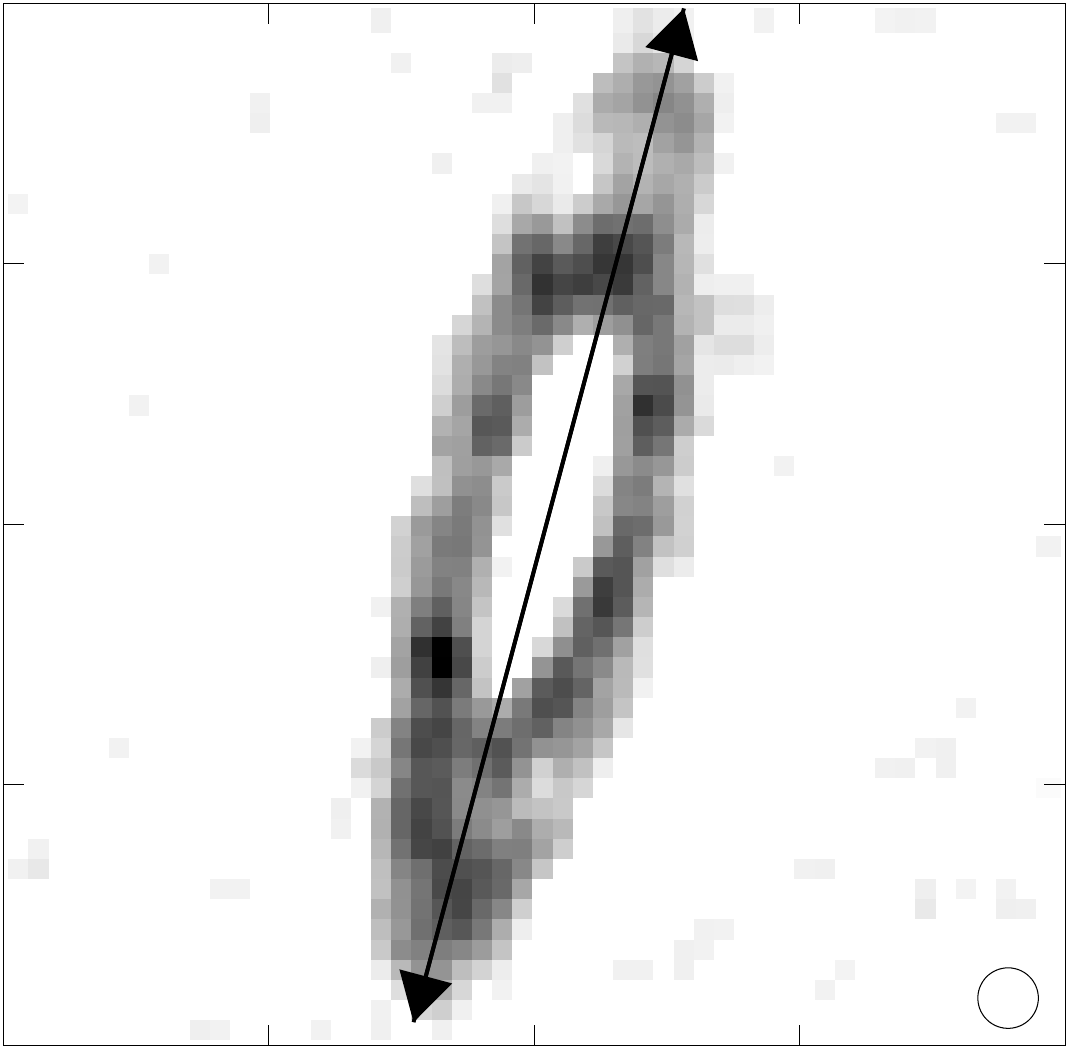}}
 \subfigure{\includegraphics[width=4cm,angle=0,clip,trim=0.0cm 0cm 0cm 0.0cm]{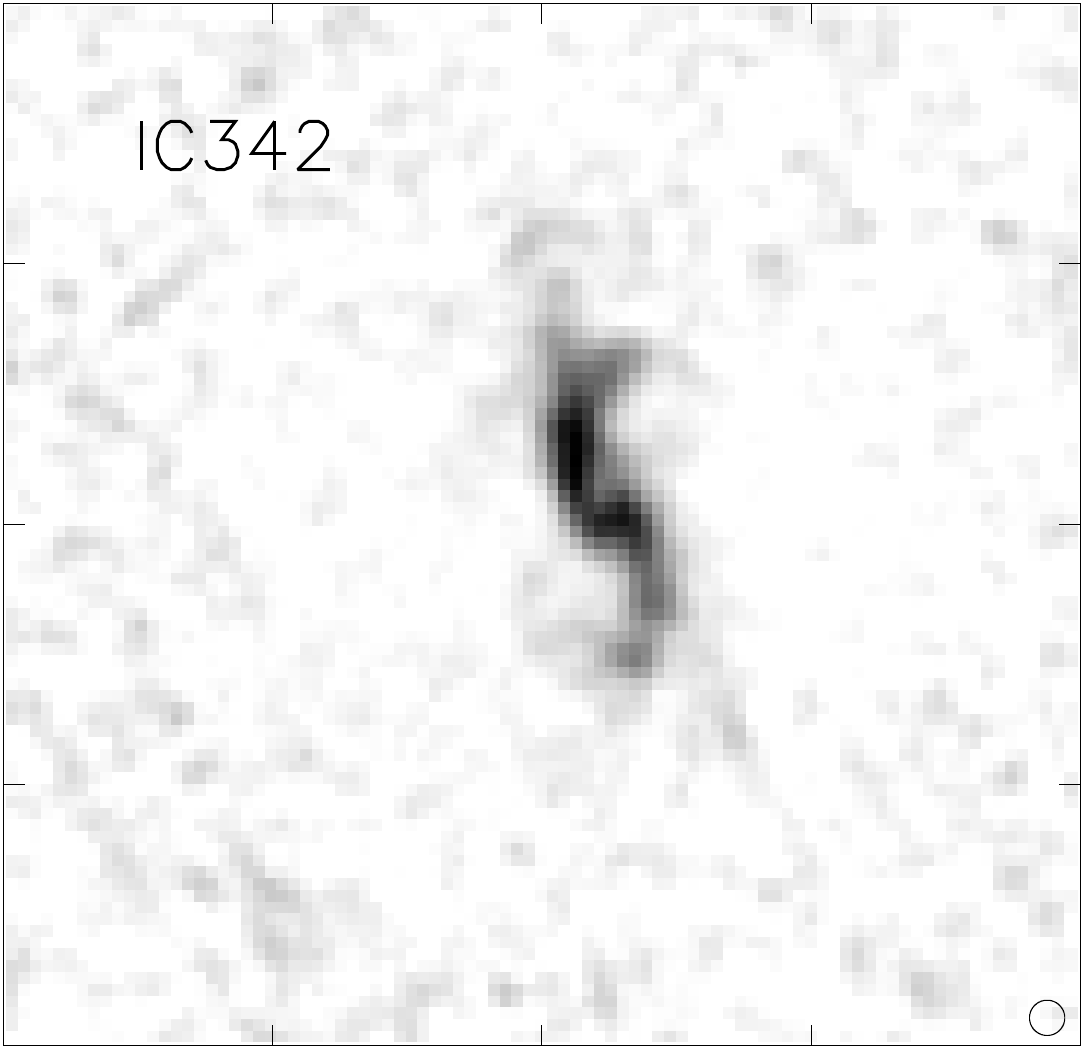}}
   \subfigure{\includegraphics[width=4cm,angle=0,clip,trim=0.0cm 0cm 0cm 0.0cm]{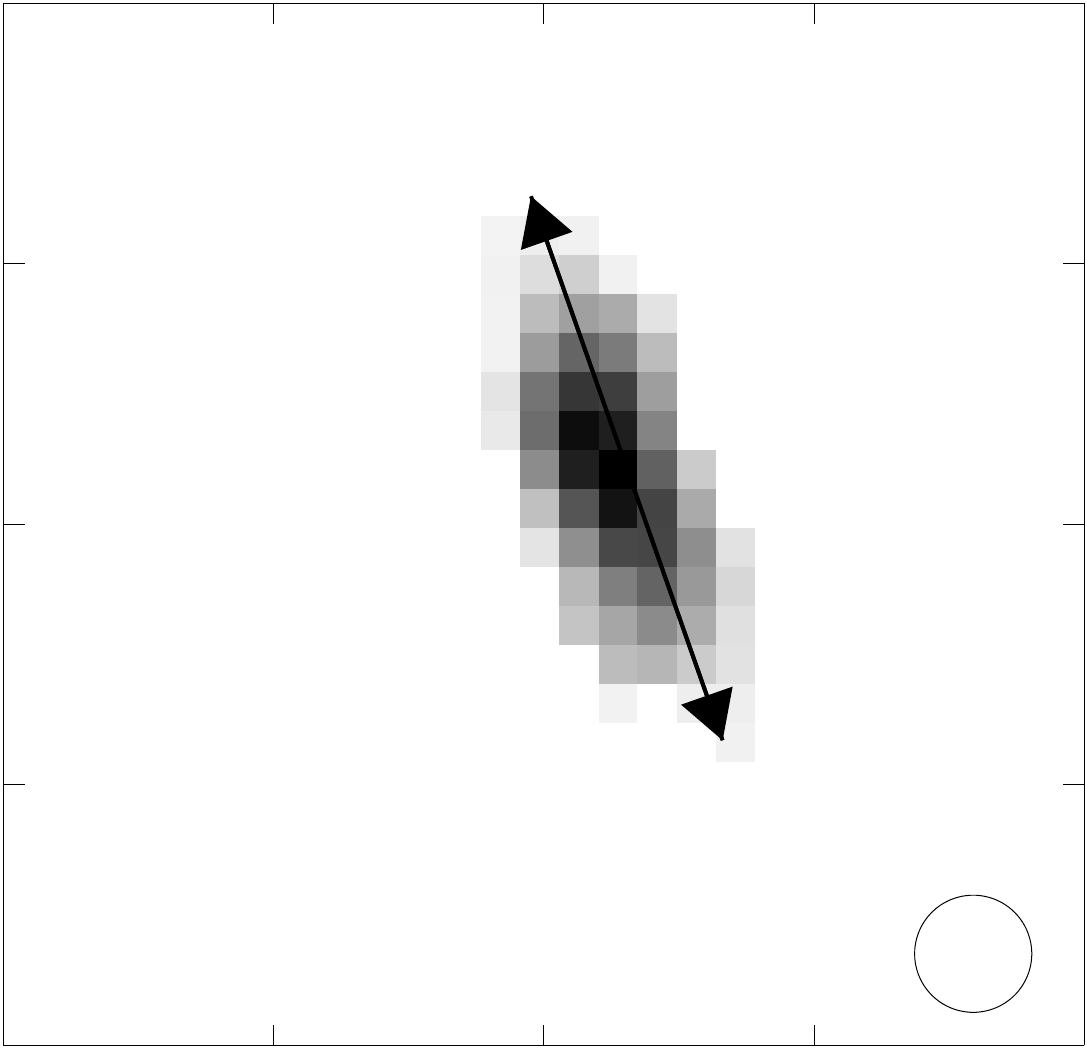}}
  \subfigure{\includegraphics[width=4cm,angle=0,clip,trim=0.0cm 0cm 0cm 0.0cm]{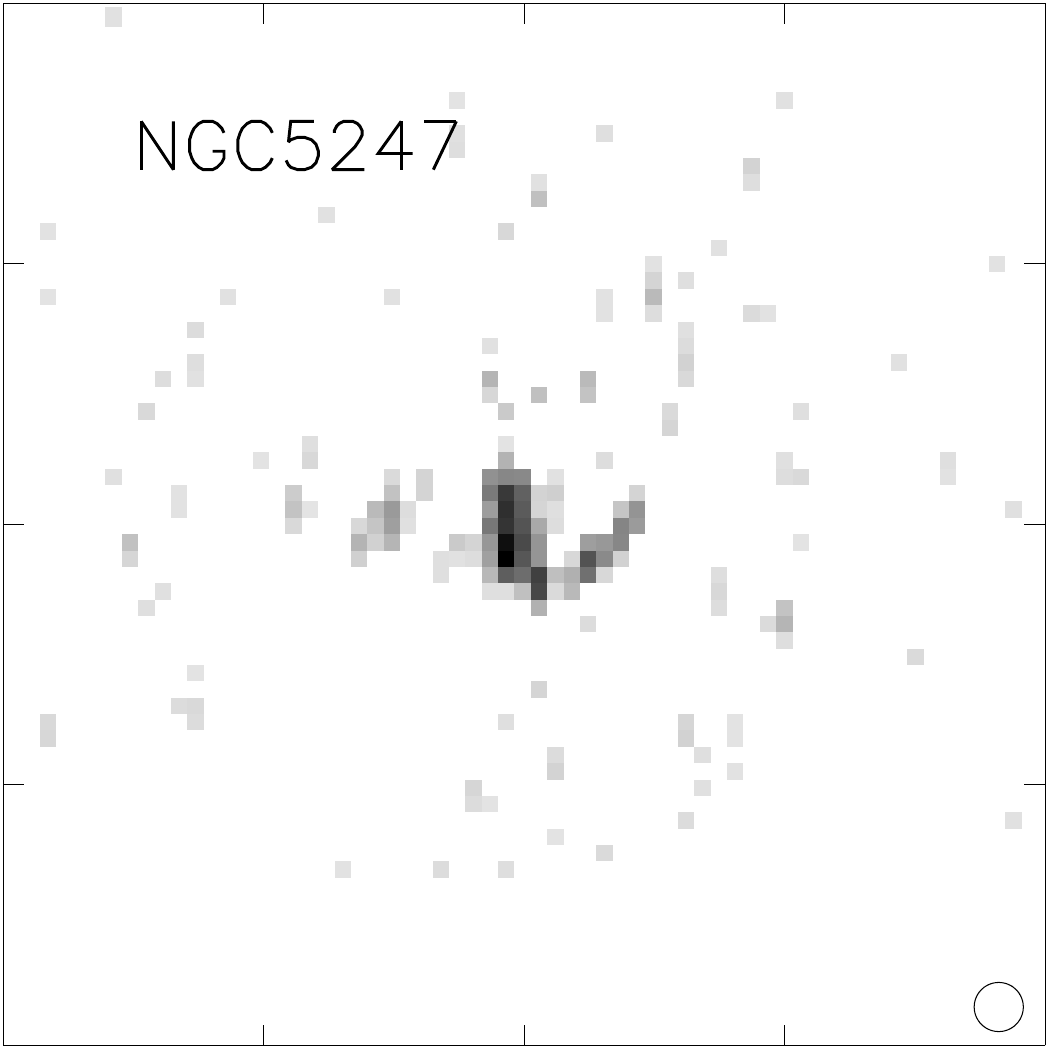}}
    \subfigure{\includegraphics[width=4cm,angle=0,clip,trim=0.0cm 0cm 0cm 0.0cm]{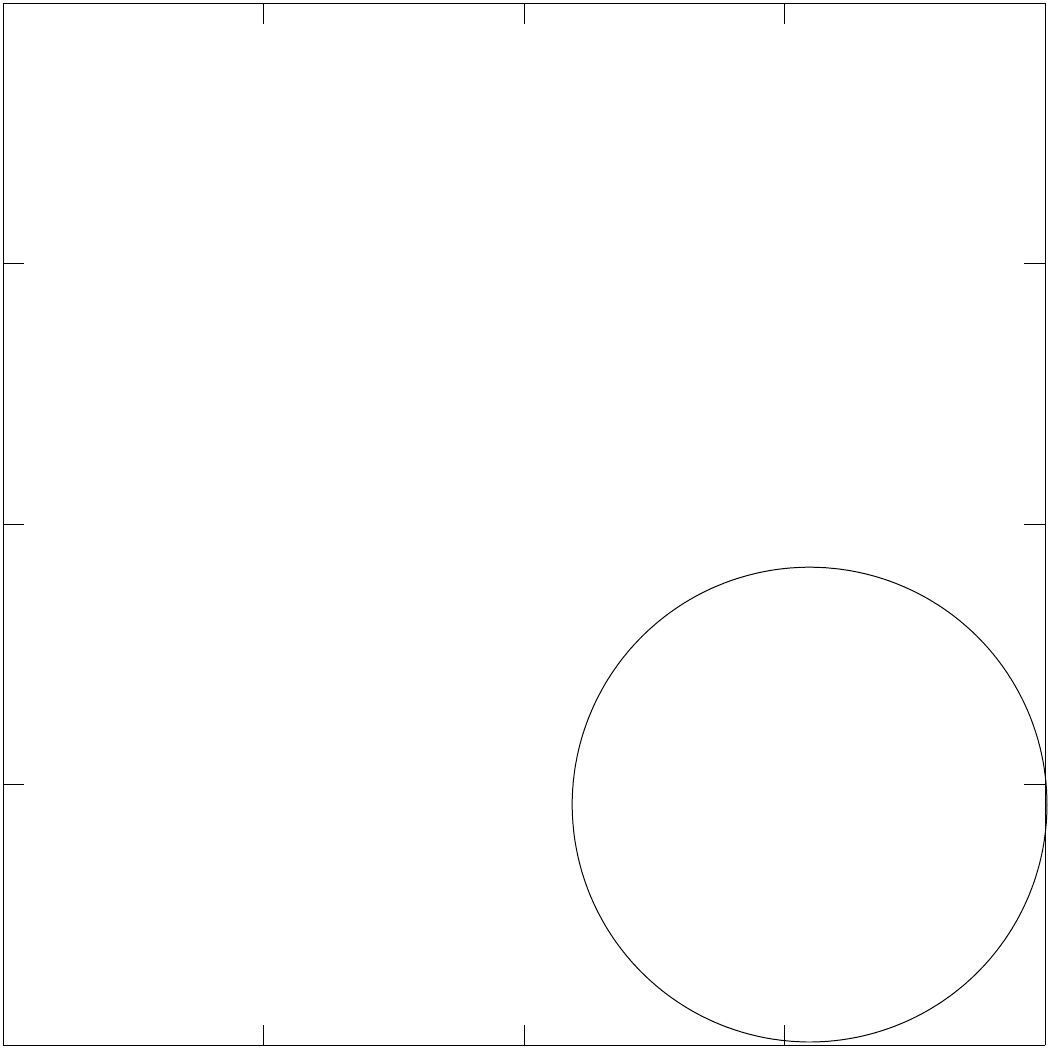}}
 \end{center}
 \caption{Three example CO(1-0) moment zero maps from the BIMA-SONG survey, presented as observed in the left column, and redshifted using the procedure outlined in Section \ref{redshiftbima} in the right column. The images in the left column are 85\arcsec\ square. The angular scale on the right hand images varies dependant on the distance to the object. The difference in beam size (displayed as a black oval in the lower left of each plot) gives an indication of the relative distance, and the new angular scale. The arrow in the upper and middle panels on the right indicate the measurement of maximum extent before beam subtraction. In the bottom right figure no emission is detected above 3$\sigma$.}
 \label{bimasongexamp}
 \end{figure}

Figure \ref{bimasongexamp} shows three examples of the redshifting and CO extent measurement process. In some cases, such as NGC7331 shown in the top panel, the structure and extent of the molecular gas in the original map is preserved. In other cases, like IC342 shown in the middle panels of  Figure \ref{bimasongexamp}, the extent of the molecular gas is similar in the redshifted map, but the structure observed originally has disappeared. Finally, in some cases the spiral galaxies from BIMA-SONG would not be detected at all at the \atlas\ distance and resolution. Such an example is NGC5247, shown in the bottom panels of Figure \ref{bimasongexamp}. 

In total, 31 of the 44 BIMA-SONG late-type galaxies had measurable CO extents after redshifting. {The majority of the galaxies which are undetectable after redshifting become so because of the greater noise introduced. Galaxies with zero-spacing information available all have measurable extents after the redshifting procedure. Only the brightest 24 spiral-galaxies were selected to have on-the-fly maps created, and hence it is not clear if these galaxies survive redshifting better simply because they are brighter, or because more of the large scale emission has been recovered (reducing the effect of beam-dilution). Removing the four objects without zero-spacing information does not change our results, suggesting any bias introduced is small.}

40 of our 41 ETGs are included here. One ETG, NGC\,2685 was removed from this analysis, as it has multiple components making measurement of the extent difficult. The field of view of the \atlas\ observations is often smaller than that achieved by BIMA-SONG. However, we do not expect this to bias our results, as very few of the ETGs have gas extended enough to reach the half-power beam point, and none reach to the edge of the field of view. The molecular gas extents of the \atlas\ and BIMA-SONG galaxies, and the parameters used to derive them are listed in Tables \ref{extenttable} and \ref{bimaextenttable}. For the \atlas\ galaxies, the beam position angle and the beam major and minor axes used in the beam correction procedure are taken from the original data source, as listed in Table \ref{extenttable}.

\subsection{Comparing spirals and ETGs}
\label{compspiraletg}

\begin{figure}
\begin{center} 
\includegraphics[scale=0.55,clip,trim=0cm 0cm 0cm 2cm]{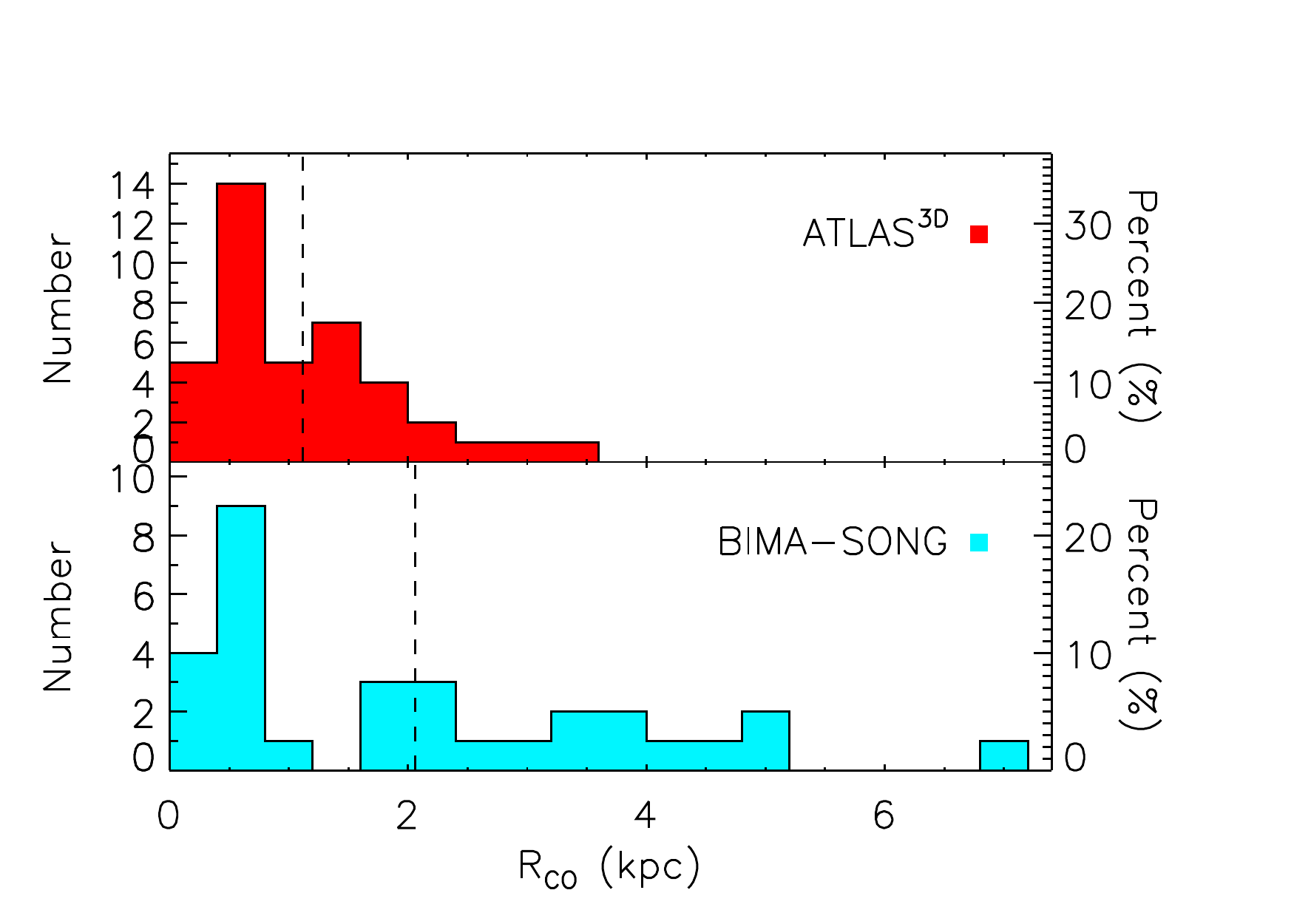}
\end{center}
\caption{\small Histogram showing the absolute molecular gas extent of the BIMA-SONG (blue) and \atlas\ (red) sample galaxies. The \atlas\ histogram contains 40 galaxies, and the BIMA-SONG histogram contains 31. The dashed lines show the mean extent of the galaxies in that histogram.}
\label{co}
\end{figure}

The distribution of absolute molecular gas extents in the redshifted BIMA-SONG and \atlas galaxies are plotted in Figure \ref{co}. The average BIMA-SONG galaxy CO radial extent is 2.1 kpc, while that for \atlas\ galaxies is 1.1 kpc. {The average surface density these are measured at is the same ($\Sigma_{H_2}$=15 M$_{\odot}$ pc$^{-2}$).}  A Mann--Whitney U (MW-U) test gives only a 3\% chance that they are drawn from the same parent distribution. We prefer the Mann--Whitney U test to a Kolmogorov-Smirnov (KS) test here, as it is less sensitive to differences in the shapes of the distributions in the two samples, which are uncertain due to the low number of mapped objects. In all cases presented here, using a KS test gives similar results.

We are also able to estimate how much more extended the \atlas\ galaxies would have to be, on average, for the distributions to be consistent. We do this by finding a multiplicative scaling factor that maximises the statistical similarity between the distributions (as calculated from a MW-U test).  
Performing this analysis, we find a scaling factor of 1.6. {The measured average surface density increases by a factor of between 3 and 1.5 when one decreases by a factor of 1.6 the radius at which the CO extent is measured, suggesting that the molecular gas really is much more extended (in absolute terms) in late-type galaxies than in ETGs.}

\begin{figure} 
\begin{center} 
\includegraphics[scale=0.55,clip,trim=0cm 0cm 0cm 2cm]{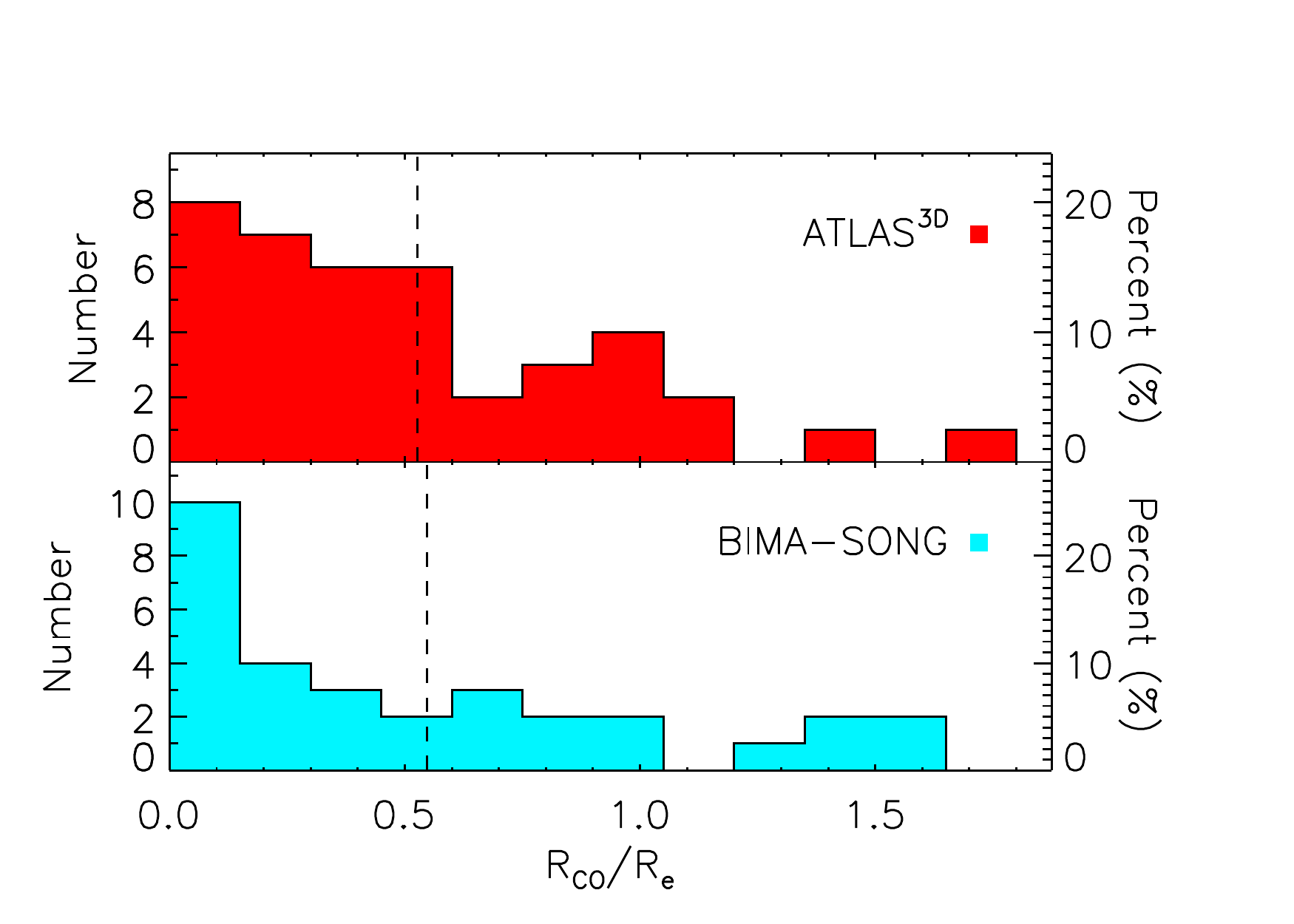}
\includegraphics[scale=0.55,clip,trim=0cm 0cm 0cm 2cm]{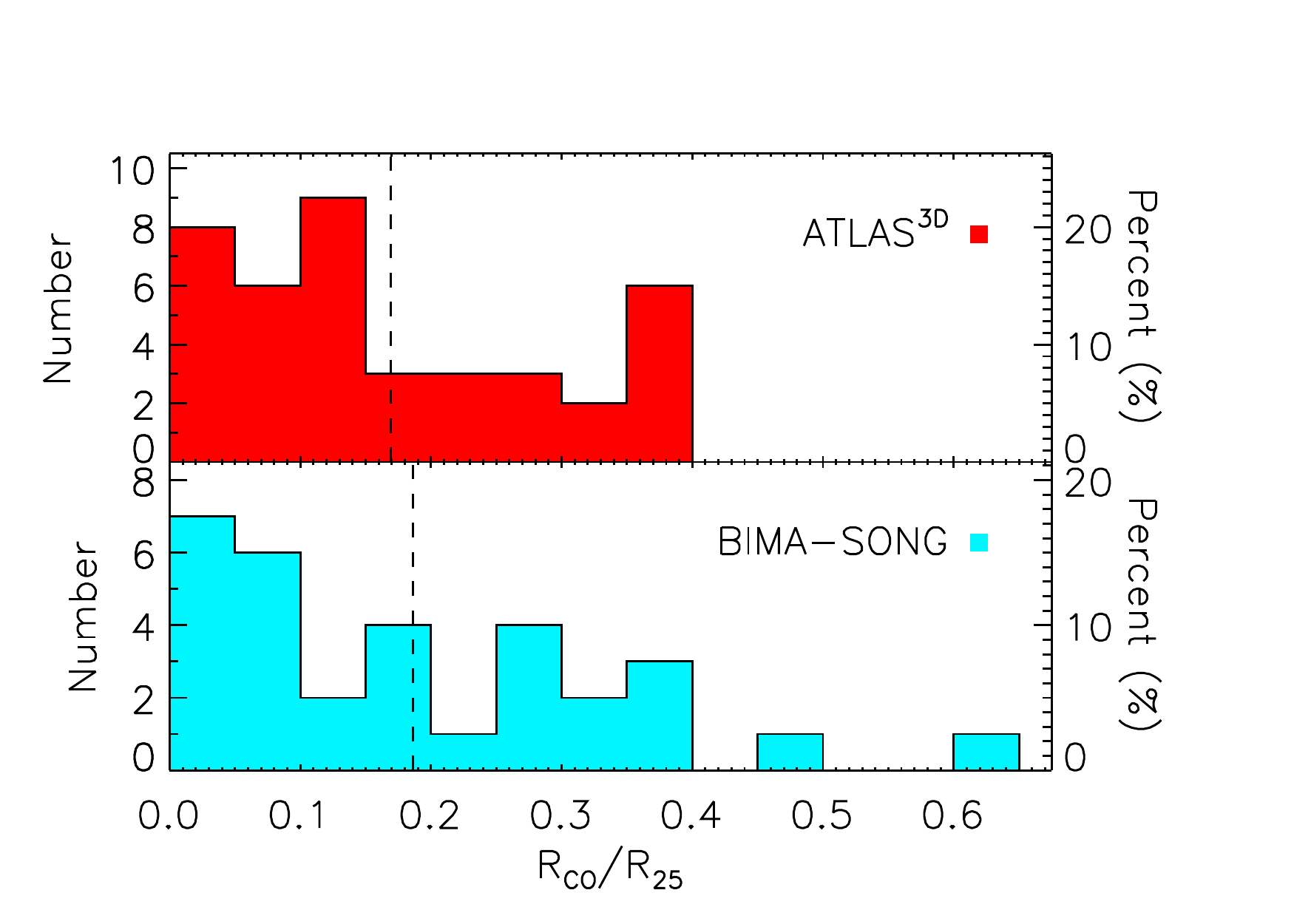}
\includegraphics[scale=0.55,clip,trim=0cm 0cm 0cm 2cm]{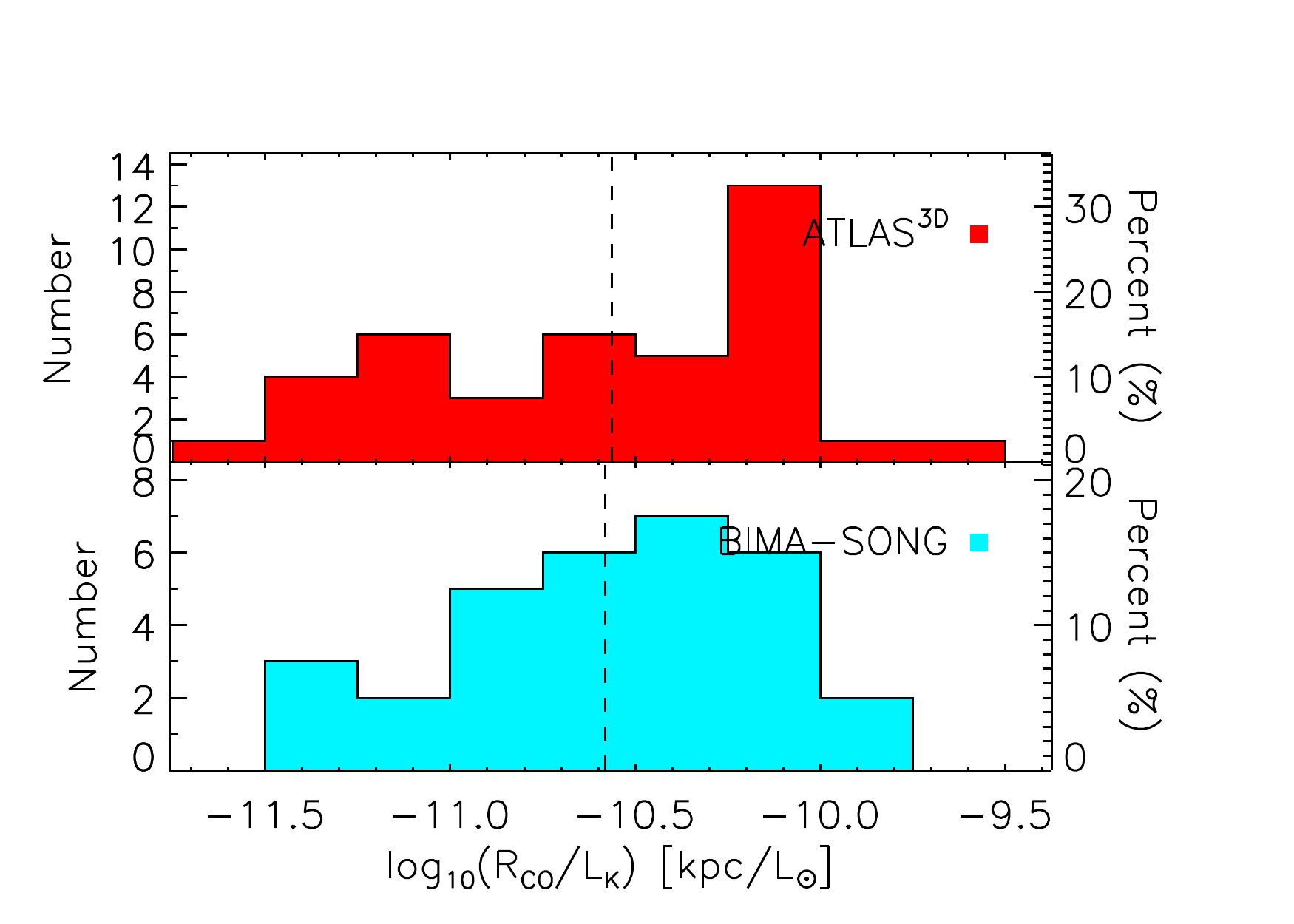}
\end{center}
\caption{\small As Figure \ref{co} but showing the relative molecular gas extent of the BIMA-SONG and \atlas\ sample galaxies. This has been normalized by the scaled mean of the effective radius of the galaxy at $B$, $H$, $J$ and $K_s$ band in the top panel, by the radius to the 25 mag arcsecond$^{-2}$ isophote in $B$-band in the middle panel, and by the galaxies $K_s$-band luminosity in the bottom panel. The dashed lines show the mean normalized extent of the galaxies in that histogram.}
\label{co_a3d_bima_norm}
\end{figure}

The average distribution of stars in an ETG and in a spiral are quite different, the former generally being more concentrated because of the contribution of a large bulge. Indeed we find here that the effective radius of the BIMA-SONG galaxies is 2 kpc larger than the \atlas\ galaxies, and thus it is important to investigate the gas extent relative to the extent of the stellar matter.
The top panel of Figure \ref{co_a3d_bima_norm} shows the molecular gas extent normalized by the effective radius ($R_{\mathrm{e}}$) of each galaxy. This is the standard scalelength for ellipticals, and represents the radius at which half the light is contained, assuming circular symmetry. The $R_{\mathrm{e}}$ values used here come from Paper I for \atlas\ galaxies, and for BIMA-SONG we use a combination of the RC3 \citep{deVaucouleurs:1991p2406} and 2MASS \citep{Jarrett:2000p2407} $R_{\mathrm{e}}$ measurements. In an identical way as was done in the \atlas\ sample paper, we use a scaled mean of the RC3 and 2MASS $R_{\mathrm{e}}$ values, where the scaling factor takes into account the wavelength dependance of the effective radius for a given stellar population (see Paper I for the full justification). 

\begin{equation}
R_{\mathrm{e}} = 0.5\,(R_{\mathrm{e,RC3}} + X_\mathrm{2MASS}\times \tilde{R}_{\mathrm{e, 2MASS}}),
\end{equation}
\noindent where $R_{\mathrm{e,RC3}}$ is the RC3 value at $B$-band, $\tilde{R}_{\mathrm{e, 2MASS}}$ is the median of the $J$,$H$ and $K_s$-band values in the 2MASS catalogue. The X$_\mathrm{2MASS}$ factor depends on the average stellar population of the galaxies. Paper I showed a X$_\mathrm{2MASS}$ of $\approx$1.7 is appropriate for ETGs. 
Using the 2MASS and RC3 values for the BIMA-SONG spirals {(in the same way as Paper I) we calculate that an X$_\mathrm{2MASS}$ factor of $\approx$1.26 is more appropriate in these late-type galaxies. We note however that if we used the same X$_\mathrm{2MASS}$ factor for both sets of galaxies this would not change our results.}

The top panel of Figure \ref{co_a3d_bima_norm} shows that the average BIMA-SONG galaxy normalized radial extent is 0.54, while the \atlas\ galaxy normalized mean is 0.47, however this difference is not statistically significant. A MW-U test shows we are unable to reject the hypothesis that they are drawn from the same parent distribution (with a probability of 46\%). A similar analysis to that conducted above suggests that the distributions are close to maximum similarity, with a scaling factor of 1.02.  Overall this suggests that on average the molecular gas in early- and late-type galaxies covers a similar fraction of the effective radius.

In an attempt to verify if the reduction in the average extent difference is due to the normalization radius adopted, we next attempted to normalize the CO diameter by $R_{25}$, the radius of the 25 mag arcsecond$^{-2}$ isophote in $B$-band. We use the average $R_{25}$ value from LEDA \citep{Paturel:1991p2408} for each galaxy. The results are shown in the middle panel of Figure \ref{co_a3d_bima_norm}. The average BIMA-SONG galaxy normalized radial extent is 0.18, and the \atlas\ normalized mean is 0.16. The small extent difference implied is not however statistically significant. A MW-U test gives a 28\% chance that the BIMA-SONG and \atlas\ normalized extents are drawn from the same parent distribution (a scaling factor of 1.14 is required to maximise this probability). On average, once again, the molecular gas in these systems seems to cover a reasonably similar fraction of the 25 mag arcsecond$^{-2}$ isophote. 

The result that the molecular gas distributions have similar distributions of CO extent to the scale radii suggests that on average the molecular gas in larger galaxies is more extended, and that there is little dependence on Hubble type. If this is true, one would expect a similar trend when normalizing by the $K_{s}$-band luminosity (bottom panel of Figure \ref{co_a3d_bima_norm}). The $L_{K_s}$ values used for BIMA-SONG galaxies are the average value from the LEDA catalogue \citep{Paturel:1991p2408}, and for \atlas\ galaxies we use the value tabulated in Paper I. These absolute magnitudes are converted into a luminosity assuming that the absolute magnitude of the sun at $K_s$-band is 3.28 mag \citep{Binney:1998p3454}. The average BIMA-SONG galaxy normalized radial extent is 10$^{-10.58}$ kpc L$_{\odot}^{-1}$, while the \atlas\ galaxies normalized mean is 10$^{-10.63}$ kpc L$_{\odot}^{-1}$. A MW-U test gives a 40\% chance they are drawn from the same parent distribution, and this distribution is close to maximal similarity (scaling factor of 0.997).

The significance of this similarity between the extent of the gas in ETGs and spirals in the context of their evolution will be discussed further in Section \ref{etgandspiralextent}.

\begin{figure}
 \begin{center} 
\includegraphics[scale=0.55,clip,trim=0cm 0cm 0cm 2cm]{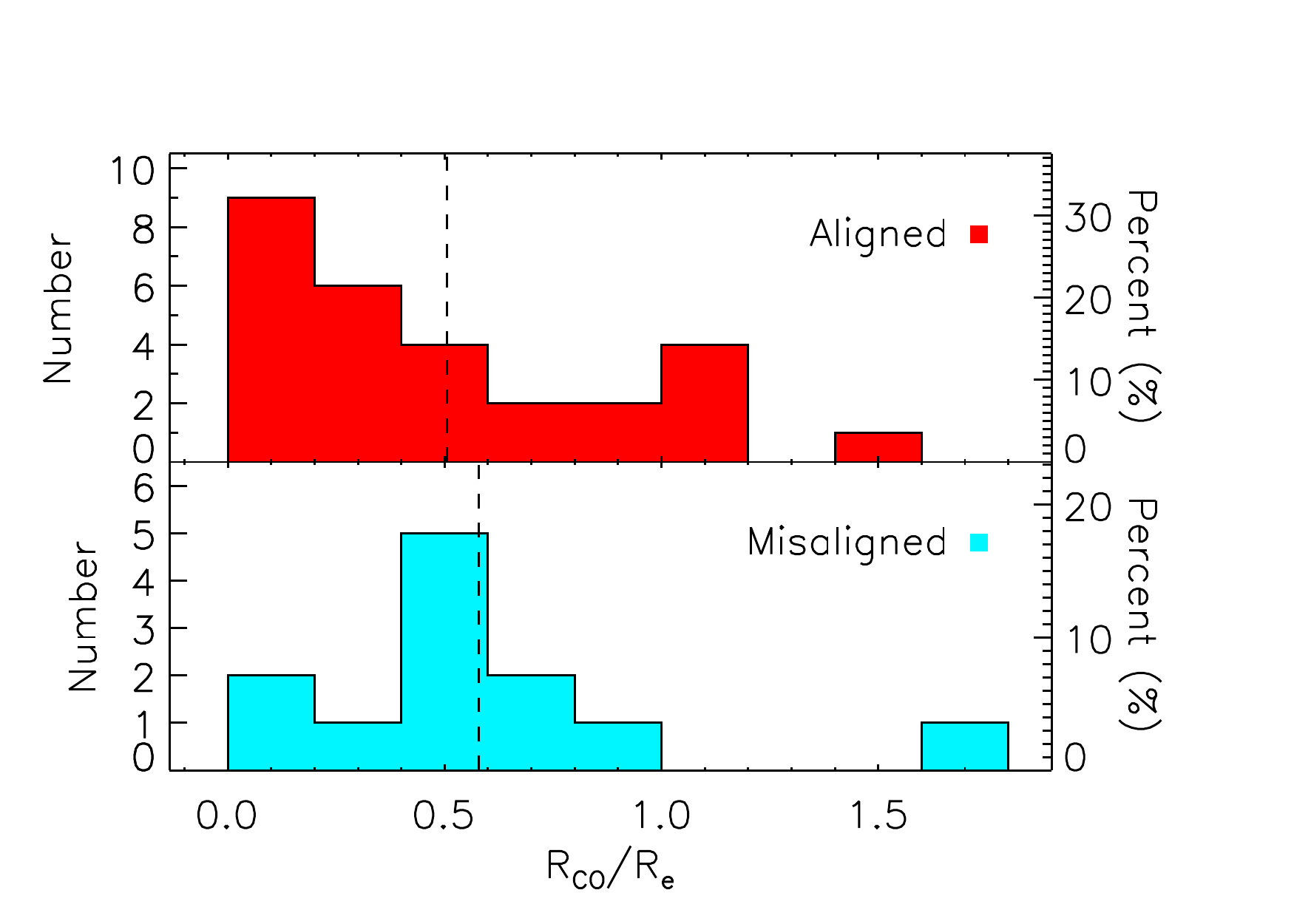}
\includegraphics[scale=0.55,clip,trim=0cm 0cm 0cm 2cm]{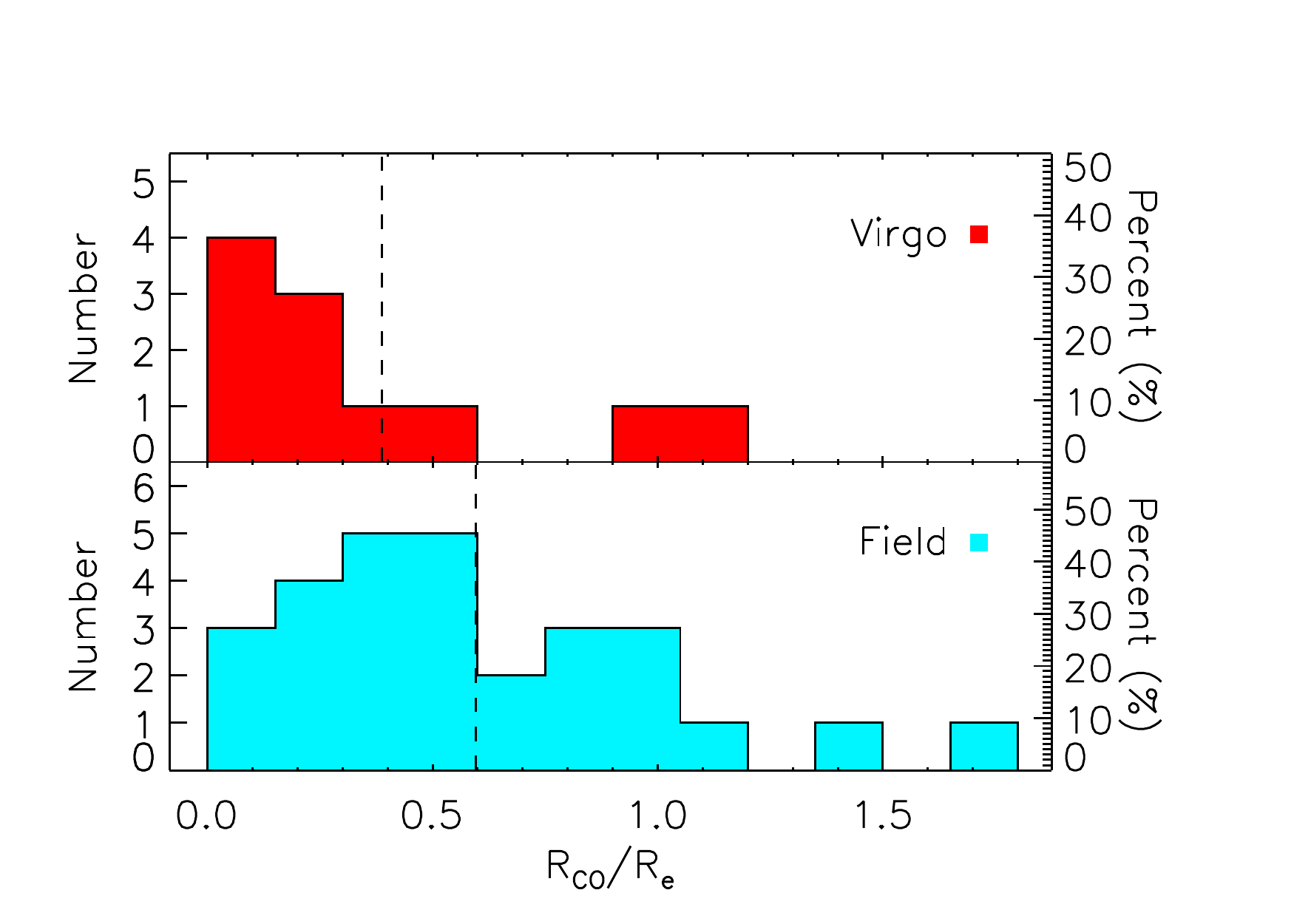}
\end{center}
\caption{\small The relative molecular gas extent of \atlas\ sample galaxies, normalized by the galaxies effective radius. In the top panel the \atlas\ galaxies are split into those with kinematically-aligned molecular gas (top), and misaligned (bottom) with respect to the stars. Kinematic misalignments are taken from Paper X. In the bottom panel the \atlas\ systems are split into Virgo cluster members (top) and field galaxies (bottom). Virgo cluster membership is taken from Paper I.}
\label{comis_virgo}
\end{figure}

\subsection{Effect of environment and gas origin}
\label{envgasorigin}
We can also break down the \atlas\ sample to investigate trends in the molecular gas extent with other parameters. The top panel of Figure \ref{comis_virgo} shows the extent of the molecular gas in the \atlas\ galaxies normalised by the effective radius, split into those galaxies where the kinematic misalignment between the molecular gas and the stars is greater than 30 degrees (see Paper X), and those with kinematically-aligned molecular gas.  The aligned galaxies have an average extent of 0.50$ R_{\mathrm{e}}$ and the misaligned galaxies 0.58 $R_{\mathrm{e}}$. The small difference between these figures is not statistically significant, as a MW-U test gives a 26\% chance that the kinematically-aligned and misaligned galaxies are drawn from the same parent population (a scaling factor of 1.23 maximizes this probability).

In Paper X, we also reported that the origin of the molecular gas displayed a strong environmental dependence. The bottom panel of Figure \ref{comis_virgo} shows a comparison of the extent of the molecular gas, normalised by the effective radius, in the Virgo cluster and field ETGs. The  Virgo cluster galaxies have molecular gas distributions that extend on average to 0.39 $R_{\mathrm{e}}$, while field galaxies have more extended molecular gas distributions, reaching out to 0.60 $R_{\mathrm{e}}$. A MW-U test gives only a 3\% probability that the Virgo and field ETGs are drawn from the same parent distribution, and a scale factor of 1.87 is required to maximize this probability.
We discuss possible reasons for this environmental difference in Section \ref{envgasorigin_discuss}.

{Given this environment dependance we revisited the correlations for spirals and ETGs discussed above, this time comparing only the field ETGs with the BIMA-SONG spirals (all of which are in the field). Removing the cluster galaxies leads us to estimate the mean absolute extent of the gas in field ETGs as 1.23 Kpc. Thus the average field spiral galaxy has gas which is $\approx$1.5 times as extended as the average field ETG. When considering the normalised trends however, the increase in the average normalised radius actually makes it statistically more likely that the field ETGs and spirals are drawn from the same population, strengthening our conclusions.}

\begin{table*}
\caption{Parameters used in calculating the maximal extent of the molecular gas in the \atlas\ early-type sample galaxies. }
\begin{tabular*}{0.9\textwidth}{@{\extracolsep{\fill}}l c c r r c c c c}
\hline
Galaxy & $R_{\mathrm{25}}$ & $D_{\mathrm{CO}}$ & $\theta_{\mathrm{GPA}}$ & $R_{\mathrm{CO}}$ & $\frac{R_{\mathrm{CO}}}{R_{\mathrm{e}}}$ & $\frac{R_{\mathrm{CO}}}{R_{\mathrm{25}}}$ & $\log\left(\frac{R_{\mathrm{CO}}}{\mathrm{L}_{\mathrm{Ks}}}\right)$ & Ref. \\
 & (kpc) & (\arcsec) & (deg) & (kpc) & & &   (kpc/L$_{\odot}$) & \\
 (1) & (2) & (3) & (4) & (5) & (6) & (7) & (8) & (9) \\
\hline
IC0676 & \hspace{4pt}8.81 & 22.0 & 164 & 1.22 & 0.46 & 0.14 & -10.13 & 1 \\
IC0719 & \hspace{4pt}5.39 & 26.8 & 50 & 1.85 & 0.48 & 0.34 & -10.13 & 1 \\
IC1024 & \hspace{4pt}3.27 & 25.0 & 23 & 1.21 & 1.08 & 0.37 & \hspace{4pt}-9.84 & 1 \\
NGC0524 & \hspace{4pt}9.36 & 19.4 & 47 & 1.05 & 0.21 & 0.11 & -11.17 & 5 \\
NGC1222 & \hspace{4pt}5.18 & 25.0 & 72 & 1.90 & 1.00 & 0.37 & -10.09 & 1 \\
NGC1266 & \hspace{4pt}5.92 & \hspace{4pt}8.9 & 145 & 0.62 & 0.22 & 0.10 & -10.65 & 1 \\
NGC2697 & \hspace{4pt}8.07 & 30.0 & 58 & 2.02 & 0.64 & 0.25 & -10.08 & 1 \\
NGC2764 & \hspace{4pt}8.94 & 36.5 & 21 & 3.42 & 1.45 & 0.38 & -10.05 & 1 \\
NGC2768 & 23.72 & 13.0 & 20 & 0.64 & 0.10 & 0.03 & -11.38 & 3 \\
NGC2824 & \hspace{4pt}4.61 & 16.0 & 142 & 1.36 & 1.03 & 0.29 & -10.31 & 1 \\
NGC3032 & \hspace{4pt}4.40 & 14.0 & 315 & 0.50 & 0.54 & 0.11 & -10.41 & 2 \\
NGC3182 & \hspace{4pt}9.52 & 16.5 & 120 & 1.16 & 0.34 & 0.12 & -10.52 & 1 \\
NGC3489 & \hspace{4pt}6.05 & 20.0 & 44 & 0.54 & 0.43 & 0.09 & -10.77 & 5 \\
NGC3607 & 14.84 & 30.0 & 122 & 1.49 & 0.36 & 0.10 & -11.03 & 1 \\
NGC3619 & 10.47 & 11.1 & 85 & 0.46 & 0.13 & 0.04 & -11.08 & 1 \\
NGC3626 & \hspace{4pt}7.65 & 27.0 & 148 & 1.22 & 0.50 & 0.16 & -10.55 & 1 \\
NGC3665 & 11.85 & 21.0 & 65 & 1.58 & 0.32 & 0.13 & -11.08 & 1 \\
NGC4119 & \hspace{4pt}4.90 & 17.6 & 107 & 0.63 & 0.20 & 0.13 & -10.55 & 1 \\
NGC4150 & \hspace{4pt}3.98 & 30.0 & 161 & 0.91 & 0.78 & 0.23 & -10.01 & 1 \\
NGC4292 & \hspace{4pt}3.72 & \hspace{4pt}9.2 & 44 & 0.31 & 0.15 & 0.08 & -10.44 & 1 \\
NGC4324 & \hspace{4pt}6.63 & 48.0 & 48 & 1.90 & 1.19 & 0.29 & -10.08 & 1 \\
NGC4429 & 13.53 & 14.0 & 90 & 0.48 & 0.14 & 0.04 & -11.36 & 1 \\
NGC4435 & \hspace{4pt}6.71 & 10.5 & 23 & 0.28 & 0.11 & 0.04 & -11.39 & 1 \\
NGC4459 & \hspace{4pt}9.59 & 20.0 & 113 & 0.51 & 0.19 & 0.05 & -11.16 & 2 \\
NGC4476 & \hspace{4pt}4.41 & 21.6 & 24 & 0.78 & 0.60 & 0.18 & -10.13 & 1 \\
NGC4477 & \hspace{4pt}9.15 & \hspace{4pt}7.4 & 47 & 0.21 & 0.07 & 0.02 & -11.50 & 5 \\
NGC4526 & 17.52 & 14.0 & 111 & 0.44 & 0.13 & 0.03 & -11.51 & 2 \\
NGC4550 & \hspace{4pt}7.48 & \hspace{4pt}8.0 & \hspace{4pt}1 & 0.30 & 0.26 & 0.04 & -10.74 & 4 \\
NGC4694 & \hspace{4pt}7.61 & 26.0 & 97 & 1.01 & 0.43 & 0.13 & -10.17 & 1 \\
NGC4710 & \hspace{4pt}6.25 & 60.0 & 30 & 2.41 & 0.96 & 0.39 & -10.34 & 1 \\
NGC4753 & 19.13 & 29.9 & 92 & 1.59 & 0.29 & 0.08 & -11.15 & 1 \\
NGC5379 & \hspace{4pt}9.57 & 41.3 & 57 & 2.95 & 0.97 & 0.31 & \hspace{4pt}-9.67 & 1 \\
NGC5866 & 10.16 & 63.0 & 144 & 2.26 & 0.86 & 0.22 & -10.56 & 1 \\
NGC6014 & \hspace{4pt}9.18 & \hspace{4pt}9.7 & 128 & 0.46 & 0.12 & 0.05 & -10.83 & 1 \\
NGC7465 & \hspace{4pt}4.36 & 25.6 & 71 & 1.67 & 1.75 & 0.38 & -10.17 & 1 \\
PGC029321 & \hspace{4pt}4.26 & 11.3 & 125 & 0.84 & 0.55 & 0.20 & -10.05 & 1 \\
PGC058114 & \hspace{4pt}2.07 & 16.1 & 91 & 0.77 & 0.72 & 0.37 & -10.05 & 1 \\
UGC05408 & \hspace{4pt}6.80 & \hspace{4pt}5.9 & 120 & 0.66 & 0.43 & 0.10 & -10.31 & 1 \\
UGC06176 & \hspace{4pt}7.88 & \hspace{4pt}8.0 & 36 & 0.32 & 0.16 & 0.04 & -10.87 & 1 \\
UGC09519 & \hspace{4pt}3.10 & 12.9 & 145 & 0.69 & 0.77 & 0.22 & -10.21 & 1 \\
\hline
\end{tabular*}
\parbox[t]{0.9 \textwidth}{ \textit{Notes:}  Column 1 lists the names of the galaxies in the \atlas\ sample considered here. Column 2 lists the radius of the 25 mag arcsec$^{-2}$ isophote at $B$-band, extracted from HyperLEDA and converted into kpc using the distance to the galaxy as tabulated in Paper I. Column 3 contains the maximum CO angular diameter, calculated as described in Section \ref{measureextent}. Column 4 contains the galaxy position angle. 
Column 5 contains the maximum CO extent, beam corrected and converted to a linear size using the distance to the galaxy. 
Column 6 is the ratio of the CO extent to the effective radius of the galaxy (tabulated in Paper I). Column 7 lists the ratio with respect to $R_{25}$. Column 8 lists the ratio to the $K_s$-band luminosity of the galaxy, calculated from the magnitudes listed in Paper I, and converted into a luminosity assuming that the absolute magnitude of the Sun at $K_s$-band is 3.28 mag \citep{Binney:1998p3454}. Column 9 lists the data reference for the CO interferometry. The beam position angle and the beam major and minor axes used in the beam correction are taken from these papers: (1) A12 (2) \cite{Young:2008p788}, (3) \cite{Crocker:2008p946}, (4) \cite{Crocker:2009p3262}, (5) \cite{Crocker:2010p3342}.}
\label{extenttable}
\end{table*}

\begin{table*}
\caption{Parameters used in calculating the extent of the molecular gas in the BIMA-SONG late-type galaxies. }
\begin{tabular*}{0.9\textwidth}{@{\extracolsep{\fill}}l c c c c r r r r r c c c c}
\hline
Galaxy & Dist. & M$_{K_s}$ & $R_{\mathrm{e}}$ & $R_{\mathrm{25}}$ & $D_{\mathrm{CO}}$ & $\theta_{\mathrm{GPA}}$ & $\theta_{\mathrm{BPA}}$ & B$^*_{\mathrm{maj}}$ & B$^*_{\mathrm{maj}}$ & $R_{\mathrm{CO}}$ & $\frac{R_{\mathrm{CO}}}{R_{\mathrm{e}}}$ & $\frac{R_{\mathrm{CO}}}{R_{\mathrm{25}}}$ & $\log\left(\frac{R_{\mathrm{CO}}}{\mathrm{L}_{\mathrm{Ks}}}\right)$ \\
 & (Mpc) & (mag) & (kpc) & (kpc) & (\arcsec) & (deg) & (deg) & (\arcsec) & (\arcsec) & (kpc) & & &   (kpc/L$_{\odot}$) \\
 (1) & (2) & (3) & (4) & (5) & (6) & (7) & (8) & (9) & (10) & (11) & (12) & (13) & (14)\\
\hline
IC342 & \hspace{4pt}3.9 & -23.40 & \hspace{4pt}9.80 & \hspace{4pt}9.14 & 10.3 & 10 & 33 & 18.9 & 17.2 & 0.51 & 0.05 & 0.06 & -10.96 \\
NGC0628 & \hspace{4pt}7.3 & -23.14 & \hspace{4pt}4.42 & 11.13 & 34.3 & 338 & \hspace{4pt}8 & 7.2 & 5.3 & 2.02 & 0.46 & 0.18 & -10.26 \\
NGC1068 & 14.4 & -24.96 & \hspace{4pt}3.10 & 14.85 & 49.7 & 12 & 16 & 8.9 & 5.6 & 2.91 & 0.94 & 0.20 & -10.83 \\
NGC2903 & \hspace{4pt}6.3 & -22.96 & \hspace{4pt}3.11 & \hspace{4pt}7.11 & 39.9 & 30 & 21 & 36.5 & 34.1 & 2.12 & 0.68 & 0.30 & -10.17 \\
NGC3184 & \hspace{4pt}3.6 & -20.57 & \hspace{4pt}2.32 & \hspace{4pt}3.74 & \hspace{4pt}3.3 & 10 & 11 & 16.7 & 15.2 & 0.20 & 0.08 & 0.05 & -10.25 \\
NGC3351 & 10.1 & -23.36 & \hspace{4pt}3.68 & \hspace{4pt}7.89 & 11.5 & \hspace{4pt}5 & \hspace{4pt}6 & 17.9 & 12.6 & 0.23 & 0.06 & 0.03 & -11.29 \\
NGC3368 & 10.1 & -23.70 & \hspace{4pt}3.03 & \hspace{4pt}9.93 & \hspace{4pt}9.2 & 40 & \hspace{4pt}4 & 16.9 & 12.5 & 0.55 & 0.18 & 0.05 & -11.06 \\
NGC3521 & \hspace{4pt}7.2 & -23.50 & \hspace{4pt}2.61 & \hspace{4pt}5.75 & 39.7 & 340 & \hspace{4pt}0 & 30.0 & 19.3 & 2.11 & 0.81 & 0.37 & -10.39 \\
NGC3627 & 11.1 & -24.35 & \hspace{4pt}5.01 & 11.17 & 116.4 & 10 & 80 & 16.0 & 12.7 & 6.87 & 1.37 & 0.61 & -10.21 \\
NGC3726 & 11.1 & -22.40 & \hspace{4pt}5.39 & \hspace{4pt}9.25 & \hspace{4pt}9.1 & 10 & 10 & 8.7 & 7.3 & 0.35 & 0.06 & 0.04 & -10.73 \\
NGC3938 & 17.0 & -23.34 & \hspace{4pt}6.13 & 10.79 & 66.1 & 90 & 327 & 8.4 & 7.7 & 3.86 & 0.63 & 0.36 & -10.06 \\
NGC4051 & 17.0 & -23.48 & \hspace{4pt}7.65 & 12.68 & 10.4 & \hspace{4pt}0 & 358 & 10.1 & 7.1 & 0.62 & 0.08 & 0.05 & -10.91 \\
NGC4258 & \hspace{4pt}8.1 & -24.08 & \hspace{4pt}6.09 & 17.03 & 29.6 & 340 & \hspace{4pt}4 & 18.5 & 16.2 & 1.60 & 0.26 & 0.09 & -10.74 \\
NGC4303 & \hspace{4pt}8.1 & -22.70 & \hspace{4pt}3.28 & \hspace{4pt}7.22 & 31.6 & 20 & 356 & 11.8 & 8.8 & 1.81 & 0.55 & 0.25 & -10.13 \\
NGC4321 & 16.1 & -24.45 & \hspace{4pt}9.81 & 15.83 & 82.1 & 350 & \hspace{4pt}9 & 10.9 & 7.5 & 4.83 & 0.49 & 0.31 & -10.41 \\
NGC4414 & 19.1 & -24.46 & \hspace{4pt}2.53 & \hspace{4pt}8.20 & 67.9 & 340 & \hspace{4pt}6 & 8.2 & 6.4 & 3.97 & 1.57 & 0.48 & -10.50 \\
NGC4535 & \hspace{4pt}7.8 & -22.08 & \hspace{4pt}4.26 & \hspace{4pt}7.67 & 28.2 & \hspace{4pt}0 & \hspace{4pt}2 & 11.2 & 8.7 & 1.63 & 0.38 & 0.21 & \hspace{4pt}-9.93 \\
NGC4559 & \hspace{4pt}9.7 & -22.37 & 13.65 & 15.07 & \hspace{4pt}8.4 & 100 & \hspace{4pt}-16 & 5.9 & 5.6 & 0.42 & 0.03 & 0.03 & -10.64 \\
NGC4569 & \hspace{4pt}9.7 & -23.35 & \hspace{4pt}5.33 & 12.46 & 57.0 & 15 & 358 & 9.7 & 8.0 & 3.35 & 0.63 & 0.27 & -10.13 \\
NGC4579 & 16.8 & -24.64 & \hspace{4pt}5.83 & 13.18 & 11.7 & 90 & 348 & 11.8 & 10.4 & 0.69 & 0.12 & 0.05 & -11.33 \\
NGC4736 & \hspace{4pt}4.3 & -23.06 & \hspace{4pt}1.02 & \hspace{4pt}4.43 & \hspace{4pt}2.9 & 90 & 62 & 39.3 & 28.6 & 0.17 & 0.17 & 0.04 & -11.30 \\
NGC4826 & \hspace{4pt}4.1 & -22.73 & \hspace{4pt}1.94 & \hspace{4pt}4.52 & 15.7 & 120 & \hspace{4pt}2 & 44.6 & 31.2 & 0.68 & 0.35 & 0.15 & -10.58 \\
NGC5005 & \hspace{4pt}4.1 & -21.62 & \hspace{4pt}0.70 & \hspace{4pt}3.10 & 15.9 & 80 & 22 & 7.1 & 6.9 & 0.93 & 1.33 & 0.30 & \hspace{4pt}-9.99 \\
NGC5033 & 21.3 & -24.69 & \hspace{4pt}5.61 & 21.88 & 68.7 & 110 & 85 & 8.0 & 7.1 & 4.00 & 0.71 & 0.18 & -10.58 \\
NGC5055 & \hspace{4pt}7.2 & -23.68 & \hspace{4pt}4.22 & 10.72 & 13.8 & 110 & 17 & 19.9 & 18.6 & 0.49 & 0.12 & 0.05 & -11.09 \\
NGC5194 & \hspace{4pt}8.4 & -24.13 & \hspace{4pt}5.89 & 12.79 & 86.1 & 10 & \hspace{4pt}3 & 18.5 & 16.2 & 5.05 & 0.86 & 0.39 & -10.26 \\
NGC5247 & 22.2 & -24.20 & 12.54 & 18.05 & 59.8 & 68 & \hspace{4pt}2 & 12.2 & 4.7 & 3.30 & 0.26 & 0.18 & -10.47 \\
NGC5248 & 22.7 & -24.55 & 11.72 & 20.44 & 42.6 & 70 & \hspace{4pt}1 & 5.8 & 6.9 & 2.44 & 0.21 & 0.12 & -10.74 \\
NGC5457 & \hspace{4pt}7.4 & -23.50 & 12.34 & 30.95 & 13.9 & 90 & 59 & 5.7 & 5.4 & 0.80 & 0.06 & 0.03 & -10.81 \\
NGC6946 & \hspace{4pt}5.5 & -23.33 & \hspace{4pt}5.51 & \hspace{4pt}8.38 & 14.8 & 10 & 14 & 26.6 & 22.0 & 0.63 & 0.11 & 0.08 & -10.84 \\
NGC7331 & 12.5 & -24.46 & \hspace{4pt}4.46 & 17.81 & 79.3 & 340 & \hspace{4pt}5 & 9.8 & 8.0 & 4.67 & 1.05 & 0.26 & -10.43 \\
\hline
\end{tabular*}
\parbox[t]{0.9 \textwidth}{ \textit{Notes:}  Column 1 lists the names of the galaxies in the BIMA-SONG  survey that had measurable extents after undergoing the redshifting procedure outlined in Section \ref{redshiftbima}. Column 2 contains the distance to the galaxy. Column 3 is the absolute magnitude of the galaxy at $K_s$-band, extracted from HyperLEDA. Columns 4 and 5 list the effective radius and the radius of the 25 mag arcsec$^{-2}$ isophote at $B$-band, extracted from HyperLEDA and converted to kpc using the (real) distance to the galaxy. Column 6 contains the maximum CO angular diameter (at the new redshifted distance), calculated as described in Section \ref{measureextent}. Columns 7 to 10 contain the galaxy position angle, the beam position angle and transformed beam major and minor axes, respectively. 
Column 11 contains the maximum CO extent, beam corrected, and converted to a linear size using the new distance to the galaxy.
Column 12 is the ratio of the CO extent to the effective radius of the galaxy (tabulated in Paper I). Column 13 lists the ratio of the CO extent with respect to $R_{25}$. Column 14 contains the ratio of the CO extent with respect to the $K_s$-band luminosity of the galaxy, calculated from the absolute magnitudes, which have been converted into a luminosity assuming that the absolute magnitude of the sun at $K_s$-band is 3.28 mag \citep{Binney:1998p3454}.}
\label{bimaextenttable}
\end{table*}

\section{Gas Surface Brightness Profiles}
\label{surfdens}

Various authors \citep[e.g.][]{Young:1982p3483,Regan:2001p3275,Leroy:2009p3484} have studied the correspondence between the radial profile of the molecular gas and that of the stars in spiral galaxies.  On small scales  ($\sim$100 pc), the clumpy nature of the ISM causes the molecular gas to deviate from a simple exponential profile, but at larger scales (such as those probed by our CARMA observations; 4\farc5 at 24.5 Mpc corresponds to $\approx$ 500pc) the molecular gas in the disk of many spiral galaxies is found to have a reasonably similar profile to that of the stellar surface density, and has a similar scale-length. Although molecular rings, and excesses and deficits of central molecular emission are found in spiral galaxies, outside the bulge region the surface-brightness profile is still often comparable to the stellar surface brightness. As the overall distribution of extents is similar in spirals and ETGs, one might expect a similar result, with the molecular gas density profiles following those of the stars. 
However, the observed molecular gas structures in our galaxies vary widely, and include central discs, rings, bars, spirals and disturbed gas. These galaxies are also much more bulge dominated than spirals. Molecular gas in ETGs could therefore have very different surface brightness profiles than the relaxed discs seen in the majority of spiral galaxies.

In order to investigate this issue, we extracted azimuthally averaged radial surface brightness profiles from the {unclipped CO integrated intensity maps (to avoid any biases against faint emission)} for all the sample galaxies. We used the CO kinematic position angle (from Paper X) and the 'best' inclination for the molecular gas (from Table 1, column 8 in Paper V) to estimate the average CO brightness in concentric elliptical annuli of one beam width (centred at the optical nucleus of the galaxy as tabulated in Paper I). For the galaxies close to edge-on (inclination $>$80$^{\circ}$), as we do not resolve the molecular gas discs thickness we estimate the surface brightness in rectangular regions one beam width thick along the major axis of the disc.
 These CO surface brightness profiles were then compared with the $r$-band stellar luminosity surface density profile convolved to the same resolution. The stellar surface brightness profiles was extracted from the MGE \citep{Emsellem:1994p723} model of each \atlas\ galaxy (Scott et al., in prep) with the software of \cite{Cappellari:2002p3482}. We used the MGE values rather than those extracted directly from the $r$-band images, as the MGE models have been carefully fitted to remove the effects of foreground stars and other observational effects. MGE models were not available for NGC\,4292 and PGC\,058114 (due to a lack of suitable $r$-band images), so in these two cases we used the method described above directly on $K_s$-band images. Using $K_s$-band images for all sources would not affect our results.

\begin{figure*}
\begin{center}
\subfigure{\includegraphics[scale=0.5,clip,trim=0.75cm 0.5cm 0.9cm 1cm]{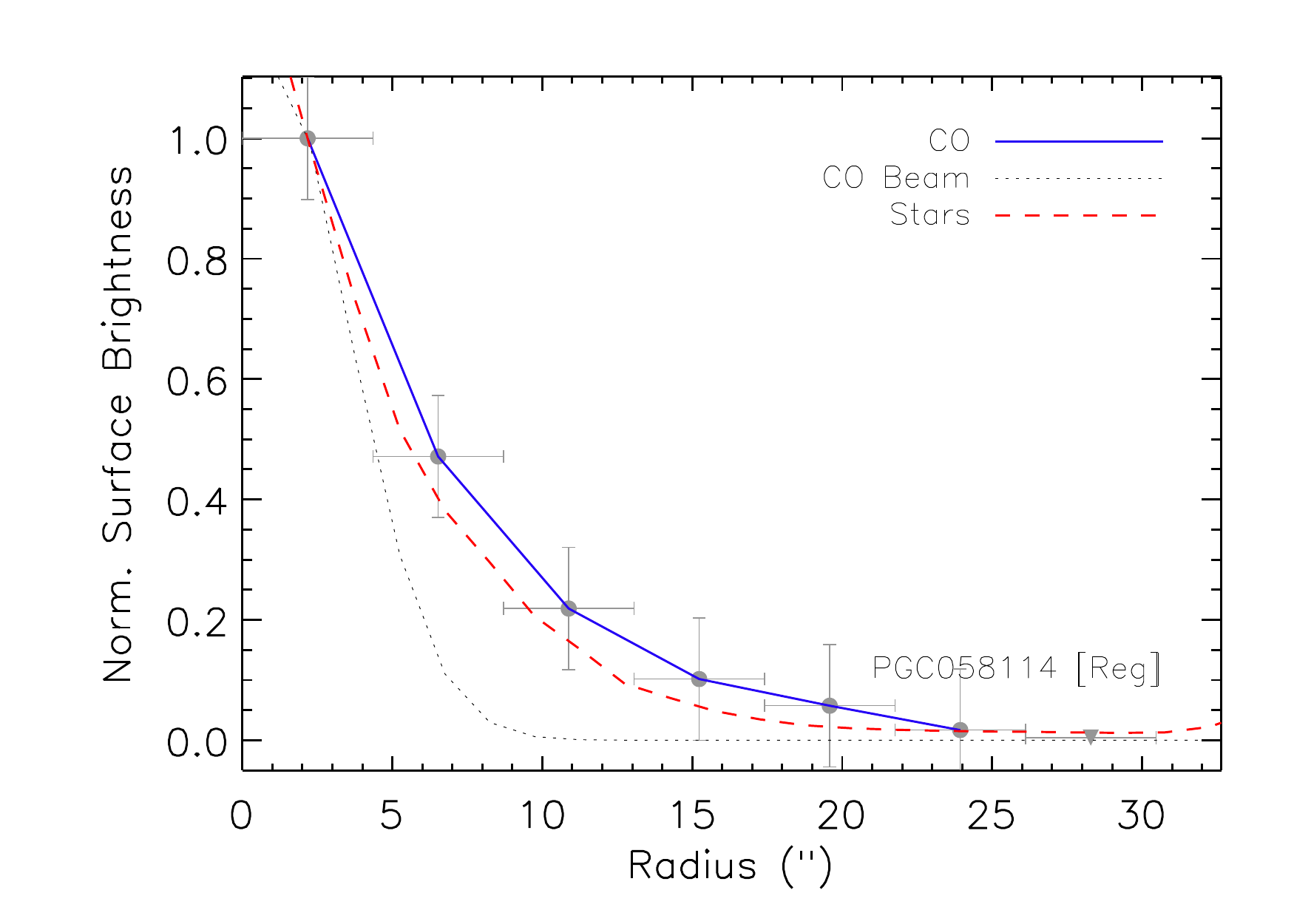}}
\subfigure{\includegraphics[scale=0.5,clip,trim=0.75cm 0.5cm 0.9cm 1cm]{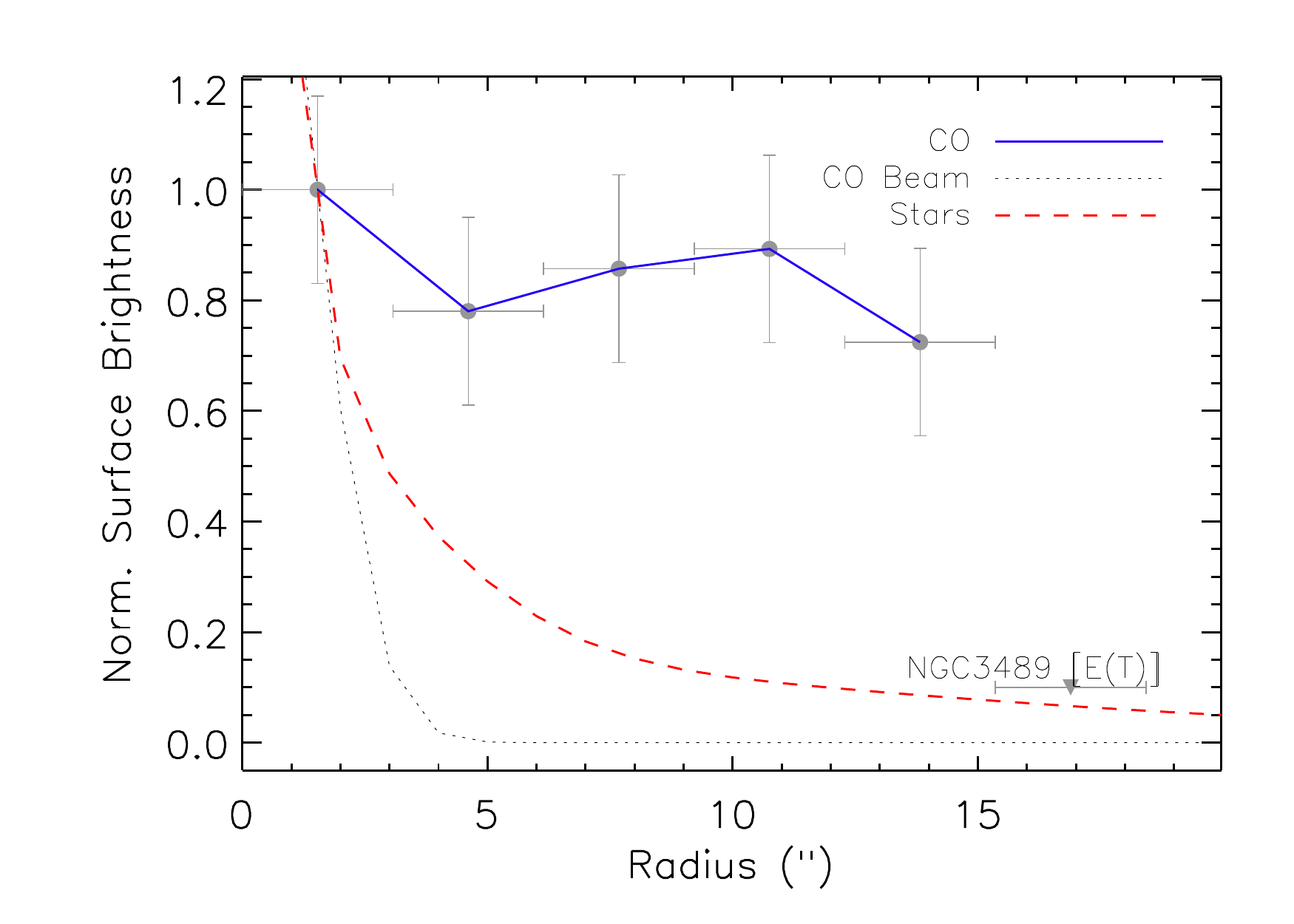}}\\
\subfigure{\includegraphics[scale=0.5,clip,trim=0.75cm 0.5cm 0.9cm 1cm]{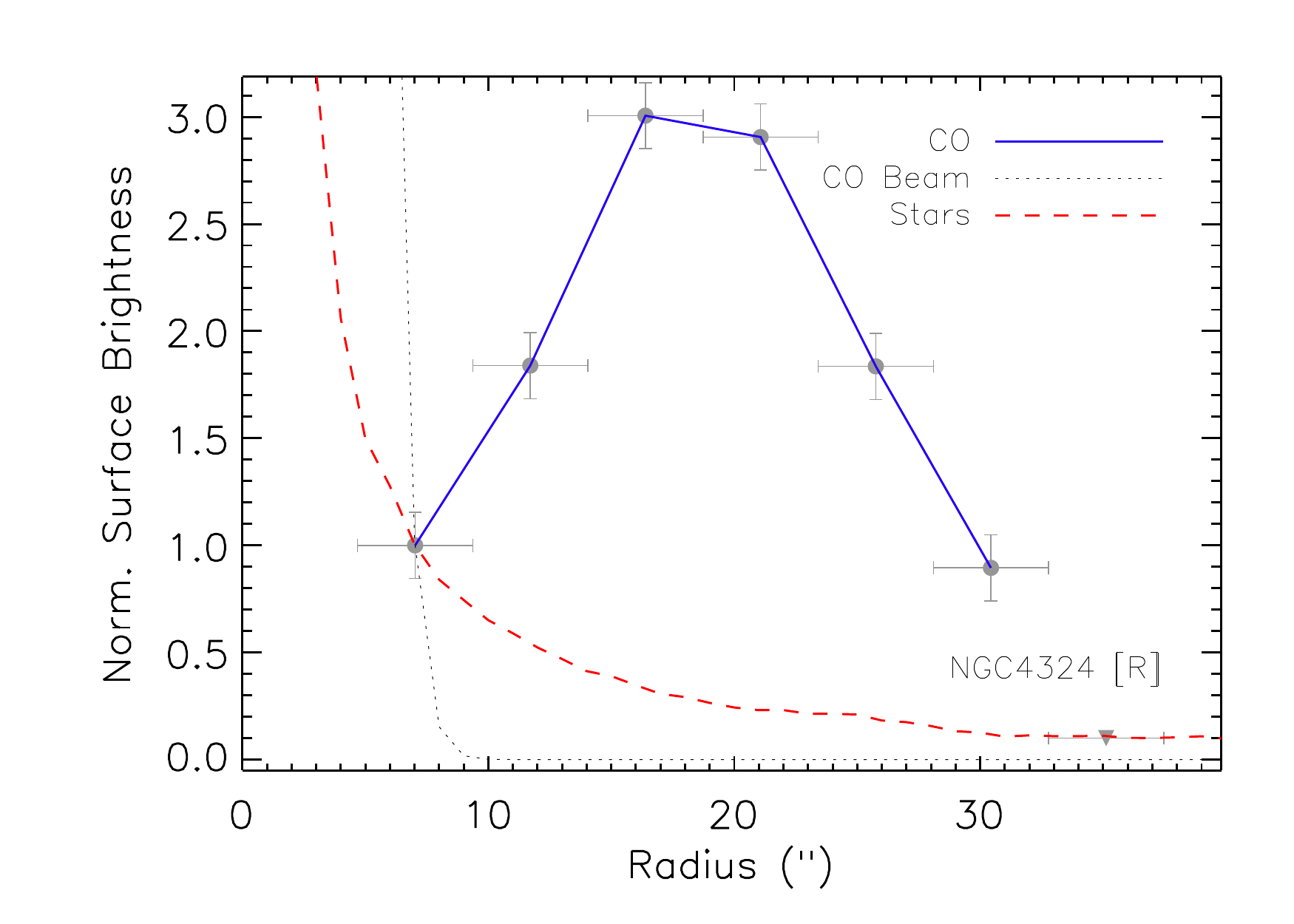}}
\subfigure{\includegraphics[scale=0.5,clip,trim=0.75cm 0.5cm 0.9cm 1cm]{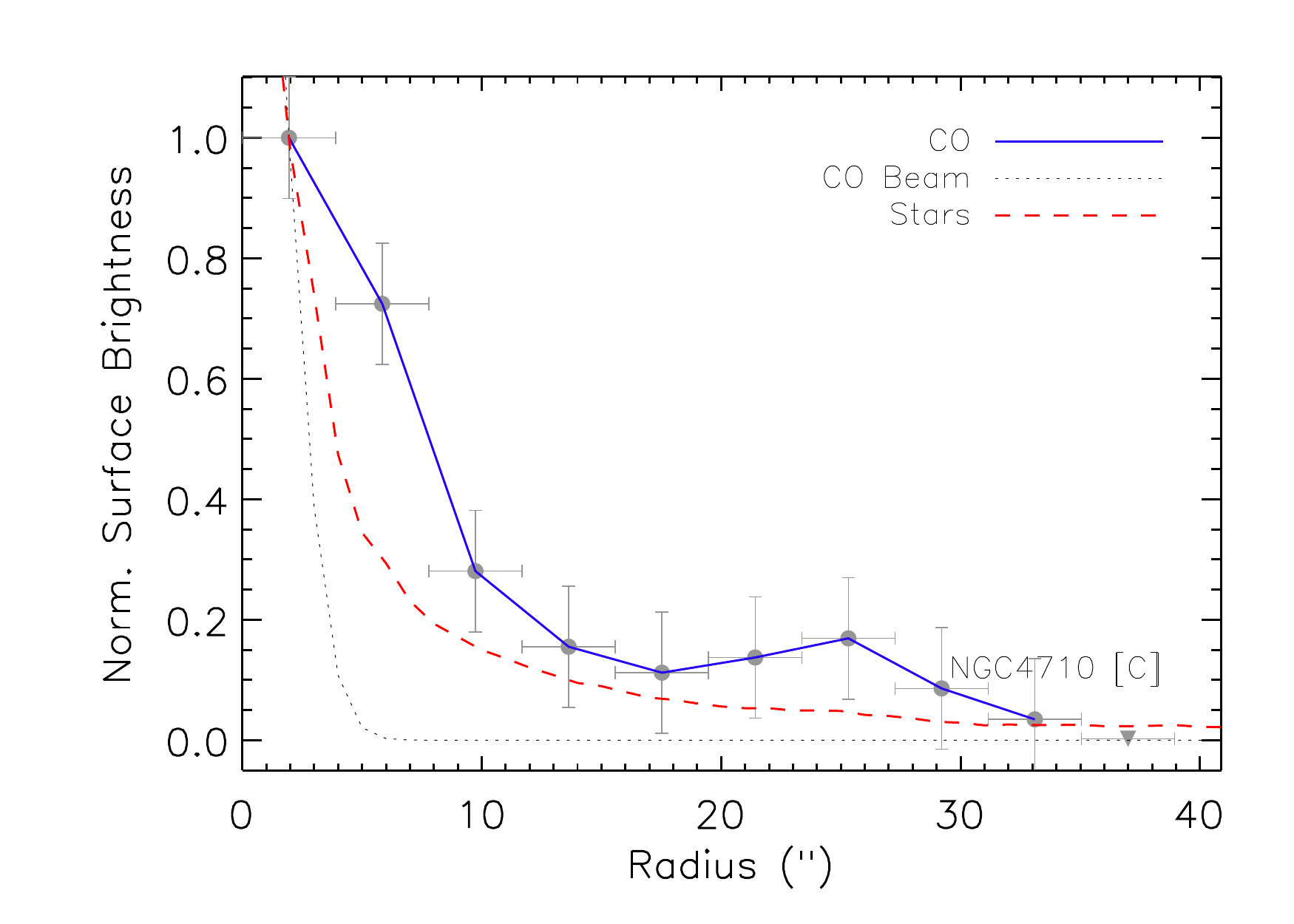}}
\caption{\small  Elliptically averaged radial surface brightness profiles of the stars (red dashed line; $r$-band) and molecular gas (blue solid line with errors) for four of our \atlas\ ETGs (normalized at the first CO datapoint). The galaxies shown are PGC058114 (top-left; classed as a regular galaxy), NGC\,3489 (top-right; classed as an excess galaxy), NGC\,4324 (bottom-left, classed as a ring) and NGC\,4710 (bottom-right; classed as a composite system).The CO beamsize is shown as a black dotted gaussian. The error bars on the CO measurements denote the width of the elliptical annuli in the x-direction and the RMS noise in the elliptical annulus in the y-direction. The letter(s) after the galaxy name denotes the profile class. "Reg" denotes a regular profile, "E" an extended profile, "E(T)" an extended profile with truncation, "R" a ring, "C" a composite, and "U" an unresolved profile.}
\label{SDexample}
\end{center}
\end{figure*}

In Figure \ref{SDexample}, we show four examples of profiles chosen to be representative of the types of radial molecular gas surface density profiles we see. Profiles for all of the CO mapped \atlas\ galaxies are shown in Appendix A (in the online material). Some galaxies (such as PGC\,058114, shown in the top-left panel of Figure \ref{SDexample}) follow the stellar luminosity surface density closely, falling off with a similar scale length as the stars. We denote such systems as \textit{regular} in this work. We find that unlike in spirals where they make up $\approx$80\% of the population \citep{Regan:2001p3275}, regular systems make up only half (21/41; 51\%) of our ETG sample.

Around a third (12/41; 29\%) of the galaxies in this work have molecular gas radial profiles that fall off slower than those of the stars, and 7/12 of these systems appear to be truncated at large radii (the CO surface brightness does not smoothly drop to our detection limit, but ends abruptly). {In each plot we include the first undetected radial bin with a 3$\sigma$ upper limit, to show this.} The galaxy (NGC\,3489) shown in the upper-right panel of Figure \ref{SDexample} is an example of such a profile. We classify such systems as having \textit{excess} emission (with or without truncation) from now on.

Approximately 7\% of systems (3/41) have profiles that show a lack of emission in the central regions, with a peak further out before the profile falls off (and is sometimes truncated). We have included the Helix galaxy (NGC\,2685) in this category, based on the information available in \cite{Schinnerer:2002p981}. The width of such features is normally about that of the synthesised beam, suggesting that they are not resolved. NGC\,4324 (Figure \ref{SDexample}, bottom left panel) is an example of such a profile (which we denote as \textit{rings}).

The last major category of profiles we see are like NGC\,4710 (bottom-right panel of Figure \ref{SDexample}). The gas surface density falls quickly from a central maximum before peaking again at larger radii. Such systems make up $\approx$7\% (3/41) of the galaxies in our sample. {These objects are all highly inclined, and such profiles are thus not present in the BIMA-SONG sample, which was selected to avoid edge on galaxies.} We denote these systems as \textit{composites}, and discuss their nature (and the nature of the other classes defined above) in Section \ref{surfden_discuss}. 

Two galaxies (NGC\,1266 and UGC\,05408) were not resolved at the resolution of our D-array CARMA data (but see \citealt{Alatalo:2011p3489} for a full discussion of the gas in the molecular outflow of NGC\,1266). We simply denote these surface-brightness profile as \textit{unresolved}. Table \ref{pvsdtable} lists the surface brightness profile class assigned to each galaxy. For a discussion of how these profile types arise, {and any dependance on galaxy properties}, see Section \ref{surfden_discuss}.

\section{Gas kinematics}
\label{gaskinem}
\subsection{Comparing CO velocities to circular velocities}
\label{pvJAM}
In Paper V of this series, we compared the CO single-dish and interferometric line-widths with JAM models of the circular velocity of each galaxy. We showed that most of the galaxies with double-peaked or boxy spectra are likely to have relaxed gas reaching beyond the peak of the rotation curve. This allowed us to use these systems to investigate the CO Tully-Fisher relation of ETGs. It is however possible to go beyond this analysis, and for each galaxy compare the position-velocity diagram (PVD) of the mapped molecular gas with the circular velocity curve derived from the JAM models, as well as the observed stellar and ionised gas velocities \citep[as done for four of these galaxies in][]{Young:2008p788}. 

We extracted the CO PVDs from our data cubes along the kinematic position angle determined in Paper X, as described in A12. Two example PVDs are shown in Figure \ref{pvexample}, while those for the whole of the sample are shown in Appendix B (in the online material). We do not include the PVDs for the two non-\atlas\ galaxies (no corresponding IFU data), NGC\, 2685 (kinematics discussed in \citealt{Schinnerer:2002p981}), NGC\,1266 (molecular outflow discussed in \citealt{Alatalo:2011p3489}), PGC058114 (insufficient photometry for a JAM model) and UGC05408 (which is unresolved at the resolution of our data).

The observed PVDs are overlaid with the projected JAM circular velocity curves in black (see Section \ref{jam} for details). If the molecular gas is counter-rotating, we also overlay a mirrored version of the circular velocity curve, to allow a direct comparison with the CO, and to investigate if they are in agreement. If the CO tightly follows the circular velocity curve, this suggests that CO is dynamically cold and relaxed, even in ETGs. Substantial disagreement would indicate that the gas is dynamically disturbed, not on circular orbits, or indicate problems with the JAM models \citep[e.g. invalid assumption of a constant anisotropy or mass-to-light ratio ($M/L$); see][]{Young:2008p788}. 

By inspection of the PVDs and circular velocity curves, we establish that the molecular gas reaches beyond the turnover radius in 68$\pm$6\% of our ETGs (as indicated in Table \ref{pvsdtable}). This compares well to the 71$\pm$7\% of galaxies which were classified as having double-peaked or boxy single-dish spectra in Paper V (and hence were assumed to reach beyond this turnover).

 \subsection{Simulated observations}
 \label{pvsims}

{To investigate if the observed PVDs are consistent with dynamically cold gas rotating at the predicted circular velocities, while including observational effects, we have developed a new \texttt{IDL} observation simulation tool. This routine, the KINematic Molecular Simulation (KinMS) tool, allows one to create a mock interferometric data cube based on arbitrary surface-brightness and velocity profiles, with a realistic treatment of projection, disk-thickness and gas velocity dispersion, as well as simulating observational effects such as beam-smearing, velocity binning, etc. We make this tool freely available to the community, along with full documentation and examples, at http://www.eso.org/$\sim$tdavis.}

{In this work we set up the KinMS tool to simulate our CARMA observations, imposing a fixed velocity resolution (10 \kms\ channels), bandwidth ($\approx$900 \kms), pixel size (1\arcsec) and the angular resolution of our CARMA data (typically 4\farc5, A12). We make the assumption that the gas is dynamically cold, with a velocity dispersion $\sigma_{\mathrm{gas}}$ $\approx8$ \kms, and that it rotates at the predicted JAM circular velocity. We give the molecular gas disk/rings a fixed vertical scale height of 100pc, similar to that observed in the Milky Way \citep[e.g.][using any sensible value of this parameter does not change our results]{Nakanishi2006}. We project the model to the observed inclination, as derived from the JAM models (see Section \ref{jam}). We then extract a PVD in the same way as for the real data, as described in Section \ref{pvJAM}.}

Given the above setup, each model is only left with two free parameters, controlling the scaling and shape of the surface brightness profile. For the majority of galaxies, we assume that the CO is distributed in an exponential disc, with a scale length varied to match our observations. In the systems with molecular rings, we characterised the surface brightness profile with a Gaussian of FWHM $\approx$100 pc centred on the observed ring radius. The parameters and profile type for each galaxy are listed in Table \ref{pvsdtable}.
{It would be possible to remove this modelling step by using the observed surface-brightness profiles presented in Section \ref{surfdens}, or even the observed moment zero maps.
However these observed profiles would have to be de-projected, and de-convolved in some manner to remove the effect of the beam, both processes which introduce significant uncertainties, especially given the limited CO spatial extent (compared to the beam) in many objects. In this paper we only aim to assess the agreement between the JAM circular velocities and
our data, not the internal details of gas distribution, and using simplified surface brightness profiles suffices for this purpose. Tests on well resolved galaxies from this sample showed that using the observed profiles would not change our conclusions.} We do however note any major discrepancies between the model and the observed PVD in Section \ref{pvdiagsec}.  Two example models are shown in the right column of Figure \ref{pvexample}, and models for the whole sample are shown in Appendix B (in the online material).

 \begin{figure*}
\begin{center}
\includegraphics[scale=0.525,clip,trim=0cm 0cm 0cm 0cm]{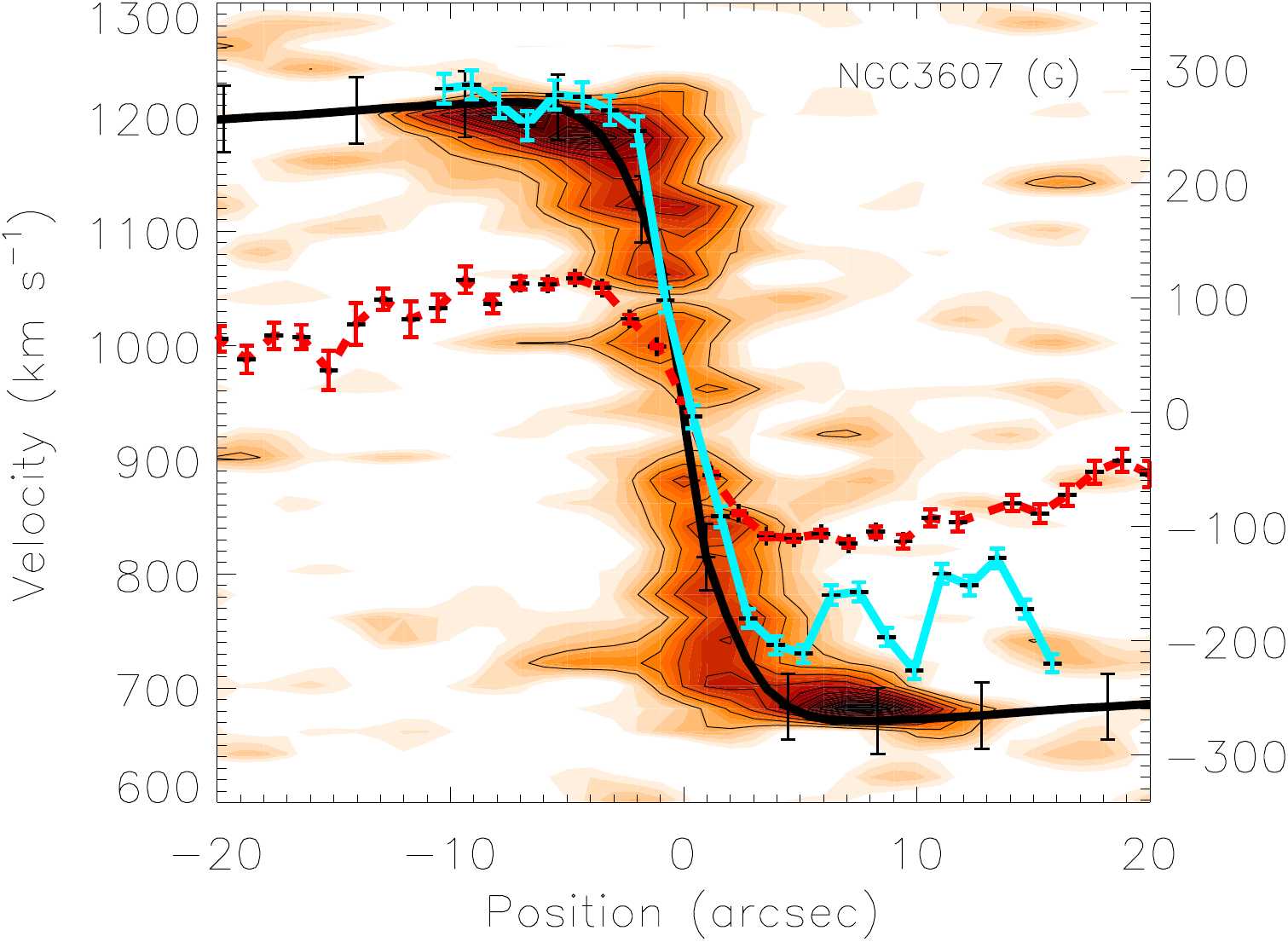}
\includegraphics[scale=0.525,clip,trim=0cm 0cm 0cm 0cm]{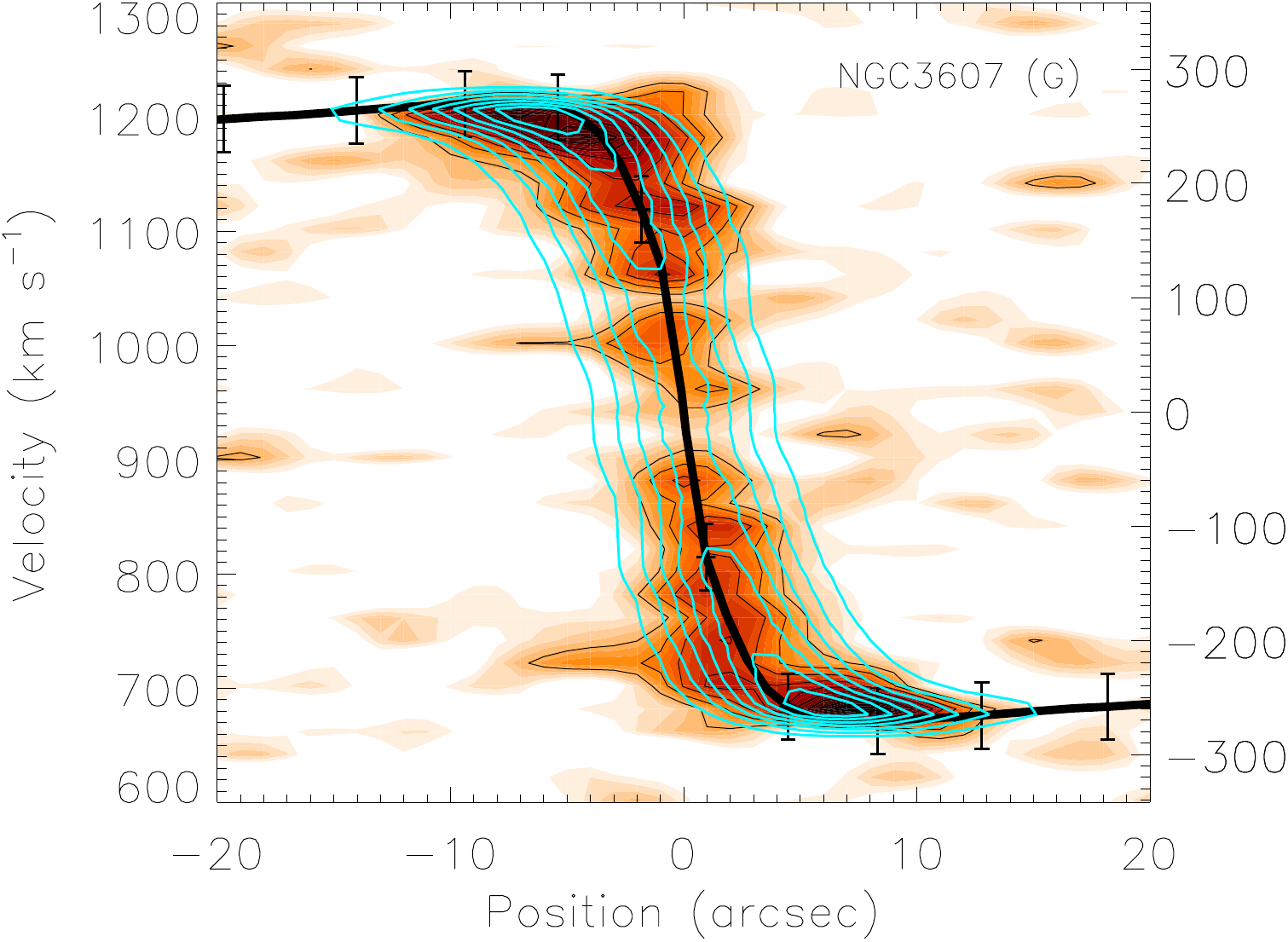}\\
\includegraphics[scale=0.525,clip,trim=0cm 0cm 0cm 0cm]{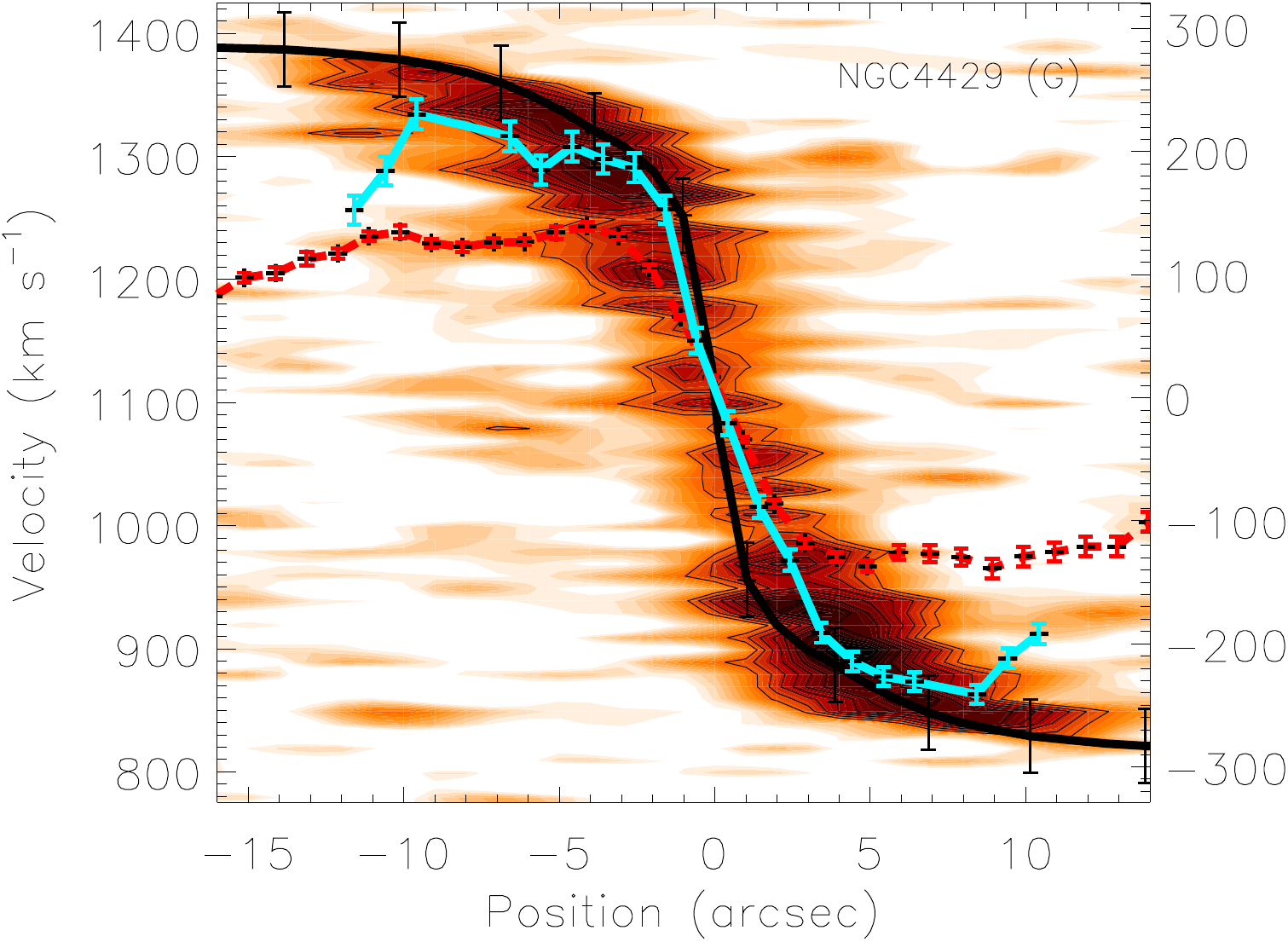}
\includegraphics[scale=0.525,clip,trim=0cm 0cm 0cm 0cm]{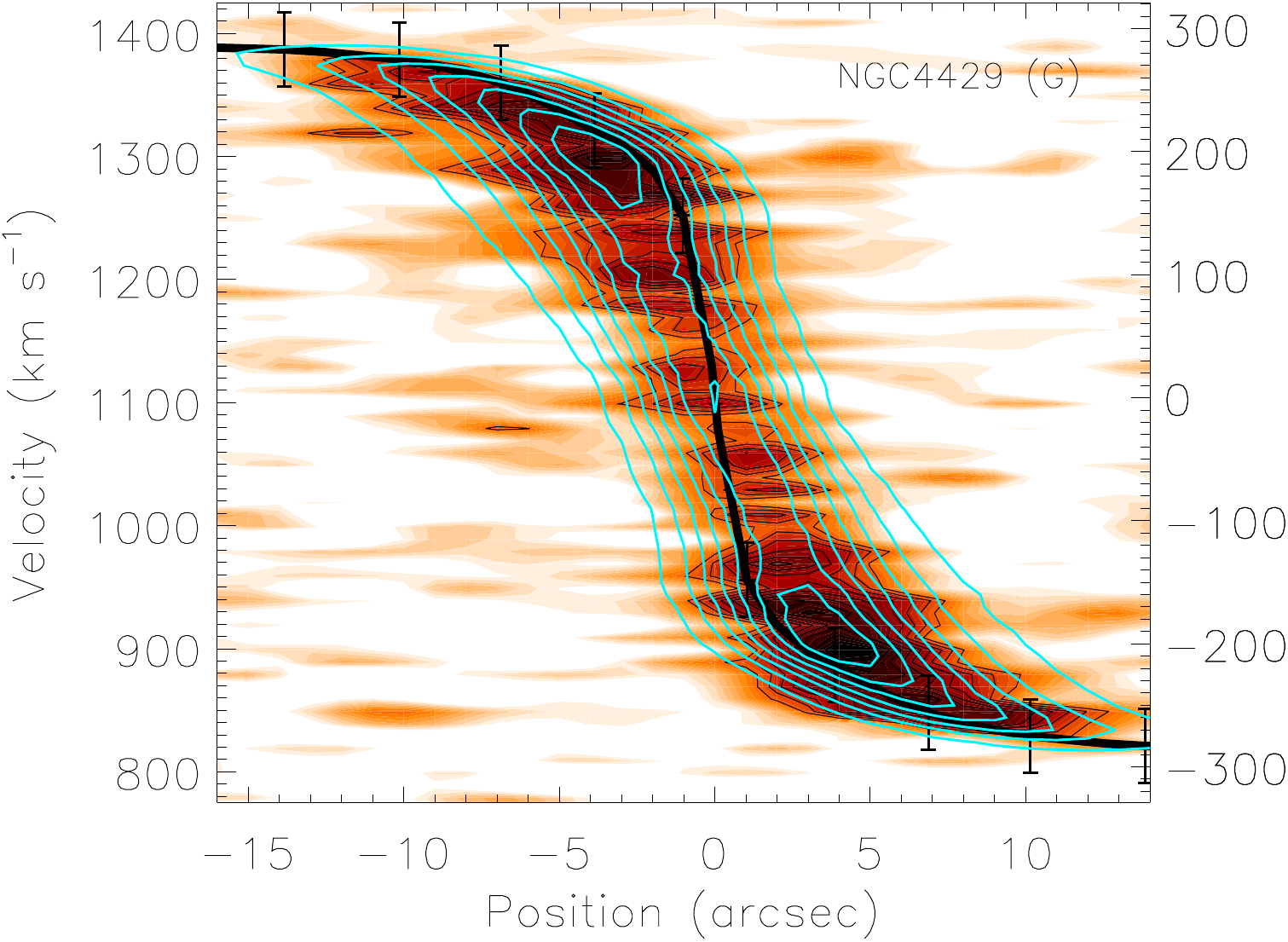}
\caption{\small CO Position-Velocity diagrams (orange with black contours) of NGC\,3607 (top) and NGC\,4429 (bottom), overlaid with the JAM circular velocity curve (thick black line). On the left, these have also been overlaid with the observed stellar (red dashed line) and ionised gas (blue line) rotation curve. On the right we overlay the modeled observations of each galaxy (blue contours), created assuming that the gas is dynamically cold and rotating at the predicted circular velocity, as described in Section \ref{pvsims}. These systems have good JAM models predictions of the circular velocity, as denoted by the profile class "G" displayed after the galaxy name.}
\label{pvexample}
\end{center}
\end{figure*}

Comparing the CO PVDs to the JAM model circular velocities and modelled PVDs (by eye) allows us to identify different classes of galaxies. These are described in detail below:

\begin{itemize}
\item ``G" (good) -- In 51\% (18/35) of our sample the models match well both the shape and position of the CO PVDs.
\item ``O" (over) -- In a further 20\% (7/35) of the galaxies the shape of the JAM circular velocity curve and modeled PVD is similar to that of the observed CO PVD, but the models predict higher velocities than those seen in CO.
\item ``X" (disturbed) --  In 11\% (4/35) of the galaxies where the CO is disturbed (as identified in A12), the CO appears to have little coherent rotation and its kinematics are, as expected, unrelated to the JAM model predictions.
\item ``P" (polar) -- In 9\% (3/35) of the galaxies the molecular gas rotates around the polar plane of the galaxy (Paper X). The model PVDs are systematically above the observed rotation, as expected if the potential of the galaxy is flattened. 
\item ``?" (misc) --  Of the remaining 9\% (3/35) of galaxies, two (NGC\,3032, NGC\,2824) show signs that the assumption of a constant $M/L$ as a function of radius is invalid (likely due to strong star-formation in the central parts), and one (NGC\,0524) is so face-on that a large degeneracy exists between the model inclination and $M/L$ (as discussed in \citealt{Cappellari:2006p1498}).
\end{itemize}

Membership of these classes was determined by comparing the model and observed PVDs, and noting significant discrepancies by eye. It is thus important to note that these classifications rely on the \textit{projected} velocity differences between the model and observed PVD. Where the gas PVD is X shaped we only compare the model to the arm of the X-shape with larger extent, as in the inner arm is likely to come from gas in the X2 orbits within bars, which can be highly elliptical in shape \citep[e.g.][]{Bureau:1999p3649,Athanassoula:1999p3488}. {These galaxies almost all also show evidence for bars in the stellar light profile/kinematics (see Section \ref{surfden_discuss})}.

In order to more quantitatively compare the velocity of the CO at a given radius with other measures of rotation we extracted a `trace' from the observed and model PVDs. This trace was constructed by finding the velocity of the pixel with the highest CO flux at each radius in the PVD (as long as this pixel has a flux greater than three times the RMS noise in the cube). This trace will likely lie below the true circular velocity of the gas (as discussed in Section \ref{compwithjam}), so some offset between this measure and the other velocity tracers is expected. Given this caveat however, it is possible to use this trace as a measure of relative variations between sample members. {Edge on galaxies, where the line-of-sight passes through material at many different radii, could be expected to show systematic offsets from less edge on systems. This effect is not observed in this sample however (the edge on systems do not lie show any systematic offset from the other systems in any analysis we have conducted), likely due to the proper treatment of the 3D structure of these systems within the KinMS code.}

\begin{figure}
\begin{center}
\includegraphics[scale=0.525,clip,trim=0.0cm 0cm 0cm 0cm]{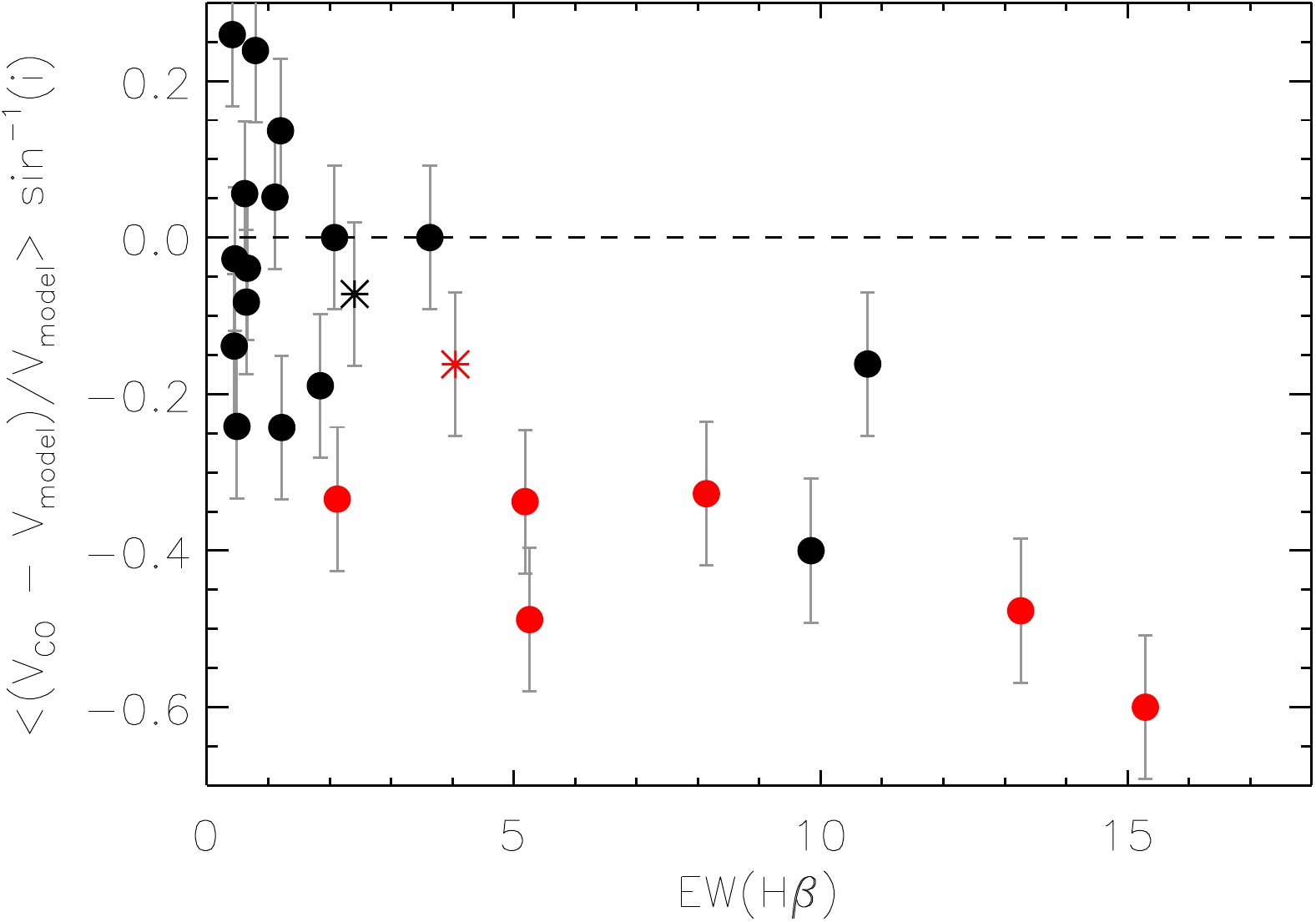}
\caption{\small Normalized radially averaged velocity differences between the trace of the CO and model PVD, plotted against the luminosity average equivalent width of the H${\beta}$ emission within an elliptical aperture (with the same extent, PA and axis ratio as the molecular gas). Black points denote galaxies in which the JAM model prediction of the circular velocity matches well the observed molecular gas rotation (black and red stars indicate galaxies where the gas is counter-rotating; see Paper X). Red points are those galaxies in which the JAM model predicts higher velocities than those observed in CO. All other JAM profile classes are not plotted on this diagram. The dashed line indicates a velocity percentage offset of zero. NGC5866 is missing from this figure because it lacks a measurement of the equivalent width of H${\beta}$.}
\label{JAMveldif}
\end{center}
\end{figure}

In Figure \ref{JAMveldif} we plot the average equivalent width (EW) of the H${\beta}$ emission against the normalized average difference between the CO trace and the model trace. This measure allows us to estimate the fractional difference between the observed CO velocity and the model. The average EW is the H${\beta}$ luminosity weighted mean calculated of all \sauron\ bins within an elliptical aperture of the same extent,  axis ratio and position angle (Table \ref{extenttable}) as the molecular gas. Black points identify galaxies where the JAM model prediction of the circular velocity is \textit{good} (black stars indicate galaxies where the gas is counter-rotating; see Paper X). Red points identify galaxies in which the JAM model prediction is \textit{over} the CO velocities.  
The red points, where the JAM model predicts higher velocities than observed in CO, although not selected via this diagram, have as expected larger offsets than the black points, with (de-projected) offsets of $<$-0.2 dex (equivalent to a higher prediction of the velocity by $\gtsimeq$30 \kms).
Importantly, however, almost all the galaxies with EW(H${\beta}$) $\gtsimeq$4 \AA\ have significant ($<$-0.3 dex) offsets, and most are classified as having JAM models that predict higher velocities than those observed in CO. We discuss these findings, and why these classes arise in Section \ref{compwithjam}.

It is clear that our discrete classification by eye of the projected PVDs is not completely clean, and a few galaxies could be put in either the good or over category. The exact fraction of galaxies in the good and over classes are uncertain at the $\approx$5-10\% level, however this is not enough to change the results presented here.

\subsection{Stellar velocities}
\label{copvvsSIGMA}
In the left column of Figure \ref{pvexample} and Appendix B (in the online material) we also overplot the observed stellar rotation curves onto the CO PVDs with red dashed lines. These were extracted from the \sauron\ IFU data, using pseudo-slits along the same PAs as those used for the CO (tabulated in Paper X). The observed velocity field was fitted with a three dimensional (minimum curvature spline) surface, and the required values along the slit were interpolated. This process allows us to minimize the effect of bad bins which happen to lie within the slit.  

Stellar systems with large velocity dispersions rotate slower, at a given radius, than dynamically cold components like the molecular gas. The difference between the velocities is because the stars do not move on circular orbits (known as asymmetric drift in a disk system). One would thus expect the velocity difference between the stars and the molecular gas to be larger in higher dispersion (and thus likely more massive) galaxies. Comparing the stellar rotation to a cold gas rotation tracer in principle allows one to estimate the magnitude of the asymmetric drift in each galaxy and disentangle these from the effect of turbulence, using e.g. the formalism of \cite{Weijmans:2008}. Such a detailed comparison is outside the scope of this work, but will be attempted in a future work in this series. Here we derive a simple zeroth order estimate of the velocity dispersion in these galaxies by assuming that the second moments of the molecular gas and stars will be equal, as they trace the same underlying potential: 

\begin{eqnarray}
\sqrt{V_*^2 + \sigma_*^2} = \sqrt{V_{\rm gas}^2 + \sigma_{\rm gas}^2},
\end{eqnarray}
and thus
\begin{eqnarray}
 \sigma_{*\rm ,pred} = \sqrt{(V_{\rm gas}^2-V_*^2) + \sigma_{\rm gas}^2},
\end{eqnarray}
where $V_*$ and $V_{\rm gas}$ are the measured stellar and molecular gas velocities at a given point, $\sigma_{\rm gas}$ is the molecular gas velocity dispersion (here we assume 8 \kms, the dispersion of \hi\ in the solar neighborhood; \citealt{vanderKruit:1982p3503,Dickey:1990p3504}), and $ \sigma_{*\rm ,pred}$ is thus the predicted stellar velocity dispersion required to produce the observed velocity differences between the CO and stars.

\begin{figure}
\begin{center}
\includegraphics[scale=0.525,clip,trim=0.0cm 0cm 0cm 0cm]{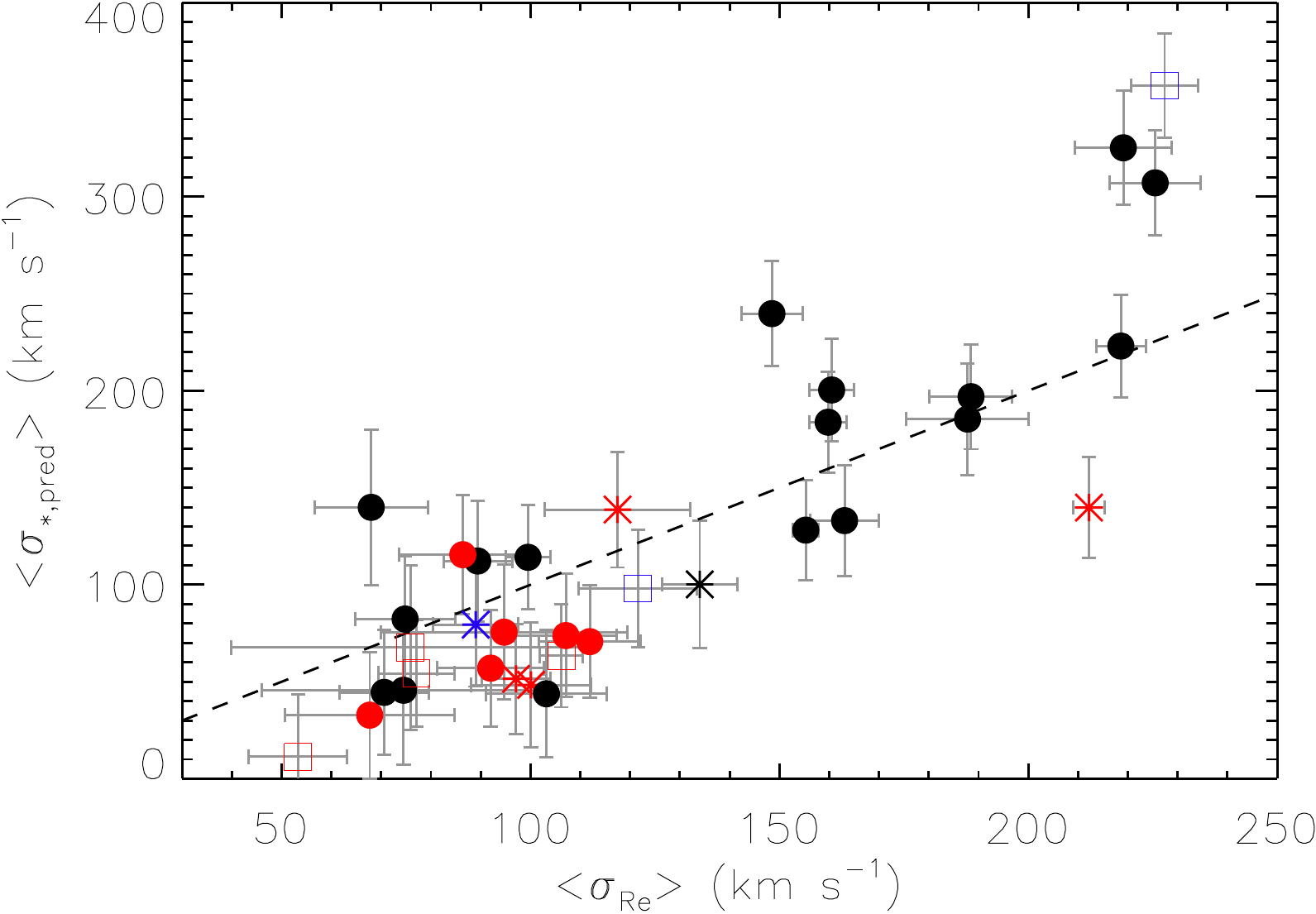}
\caption{\small 
Radially averaged stellar velocity dispersions predicted using the simple approximations in Equation 9, plotted against  the observed stellar velocity dispersion (Paper I) within one effective radius. The error bars on $\sigma_{Re}$ are not the formal error, but show the variation in observed velocity dispersion along the slit.
The points in this figure are colour coded to denote the JAM model profile class. Red and black points have the same meaning as in Figure \ref{JAMveldif}, and in addition red stars demote galaxies with polar molecular gas, open red squares represent disturbed galaxies and open blue squares are the remaining systems. Stars of these colours indicate galaxies which, in addition to the above criteria, also have misaligned molecular gas. The dashed line shows the one-to-one relation as a guide to the eye.}
\label{dvSIGMA}
\end{center}
\end{figure}

 Figure \ref{dvSIGMA} shows a comparison of the radially averaged predicted velocity dispersions derived using the zeroth order approximation described above,  plotted against  the observed stellar velocity dispersion of each galaxy (Paper I) measured within one effective radius. The error bars on $\sigma_{Re}$ are not the formal error, but show the variation in observed velocity dispersion along the slit. 
The points in this figure are colour coded to denote the JAM model profile class; red and black points have the same meaning as in Figure \ref{JAMveldif}, and in addition red stars denote galaxies with polar molecular gas, open red squares represent disturbed galaxies and open blue squares are the remaining systems (discussed in Section \ref{pvJAM}). The dashed line shows the one-to-one relation as a guide to the eye.

Figure \ref{dvSIGMA} shows that the velocity difference between the molecular gas and stellar velocities does appear to scale with the degree of pressure support in the system, as expected. 
{The highest velocity dispersion systems ($\sigma\gtsimeq$200\kms) show some sign of preferentially lying above the relation, which may indicate that the zeroth order approximations we use to derive the predicted velocity dispersions break down as the velocity dispersion becomes similar to the circular velocity. Given the low number of objects, however, the statistical significance of any such trend is low. }
Figure \ref{dvSIGMA} also highlights that the systems which show differences between the CO rotation and the JAM prediction (red solid points) tend to have low ($>$120\kms) velocity dispersions. 


\subsection{Ionized gas velocities}

In the left column of Figure \ref{pvexample} and Appendix B (in the online material), we also overplot the observed ionised gas rotation curves onto the CO PVDs with a blue line.
These were extracted from the \sauron\ IFU data, using a pseudo-slit along the same PA as used for the stars and CO (tabulated in Paper X). The observed velocity field was fitted with a surface, and the required values along the slit interpolated, as was done for the stars.

Ionized gas discs would normally be expected to rotate slower than the dynamically cold molecular gas, due to their larger dispersion, but faster than the dispersion-dominated stars. 
The relative velocities of the ionized gas and the CO can provide clues on the source and ionization mechanism of the gas emission. {For instance strong H$\beta$ emission in these molecular gas rich galaxies is often associated with H{\small II} regions embedded in the dynamically cold star-forming disk, while regions where [OIII] emission dominates seems to have a significant contribution from a dynamically hotter component of the ionized gas which is not related to star formation \citep[e.g.][]{Sarzi:2006p1474,Young:2008p788}. }

\begin{figure}
\begin{center}
\includegraphics[scale=0.525,clip,trim=0.0cm 0cm 0cm 0cm]{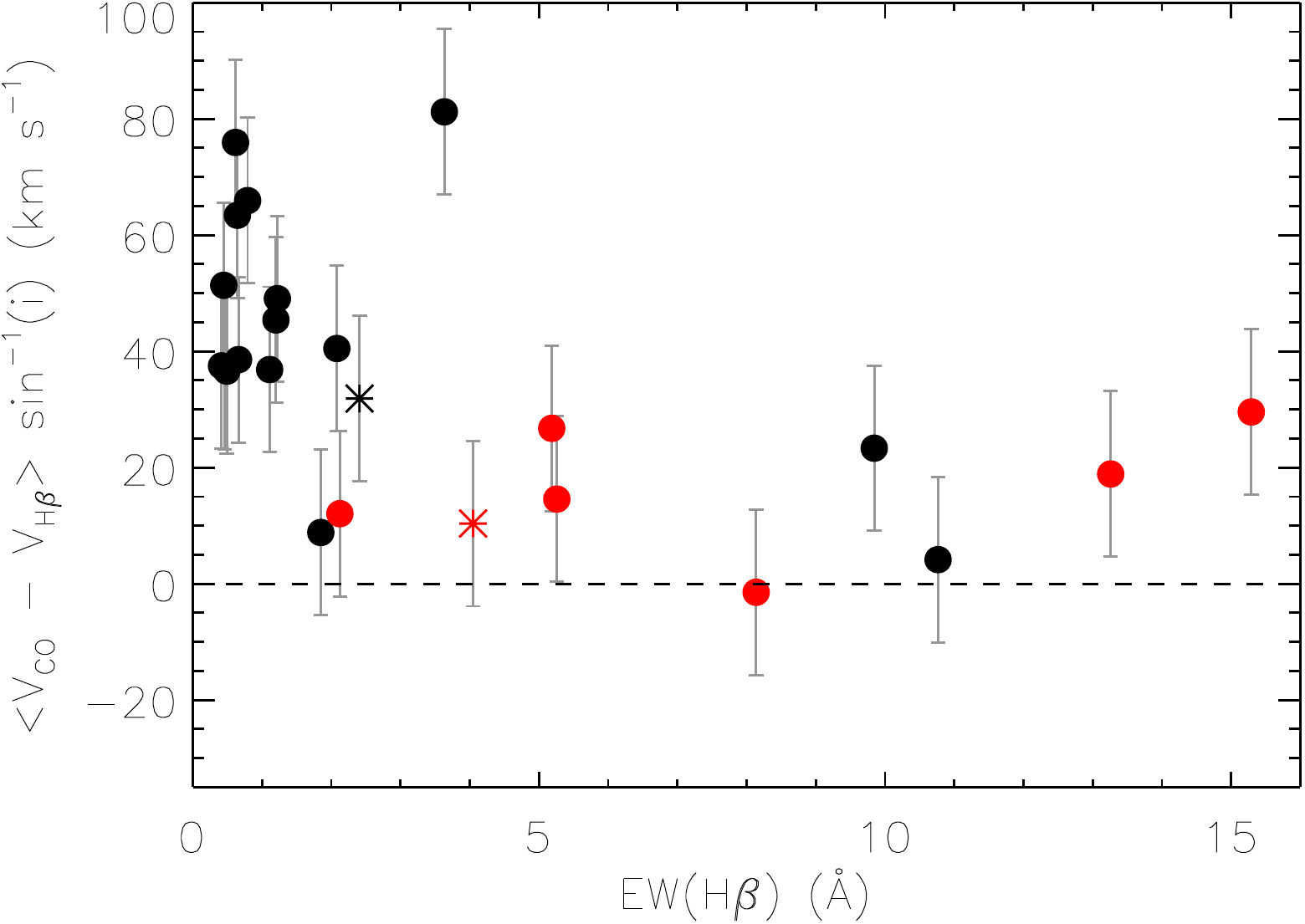}
\includegraphics[scale=0.525,clip,trim=0.0cm 0cm 0cm 0cm]{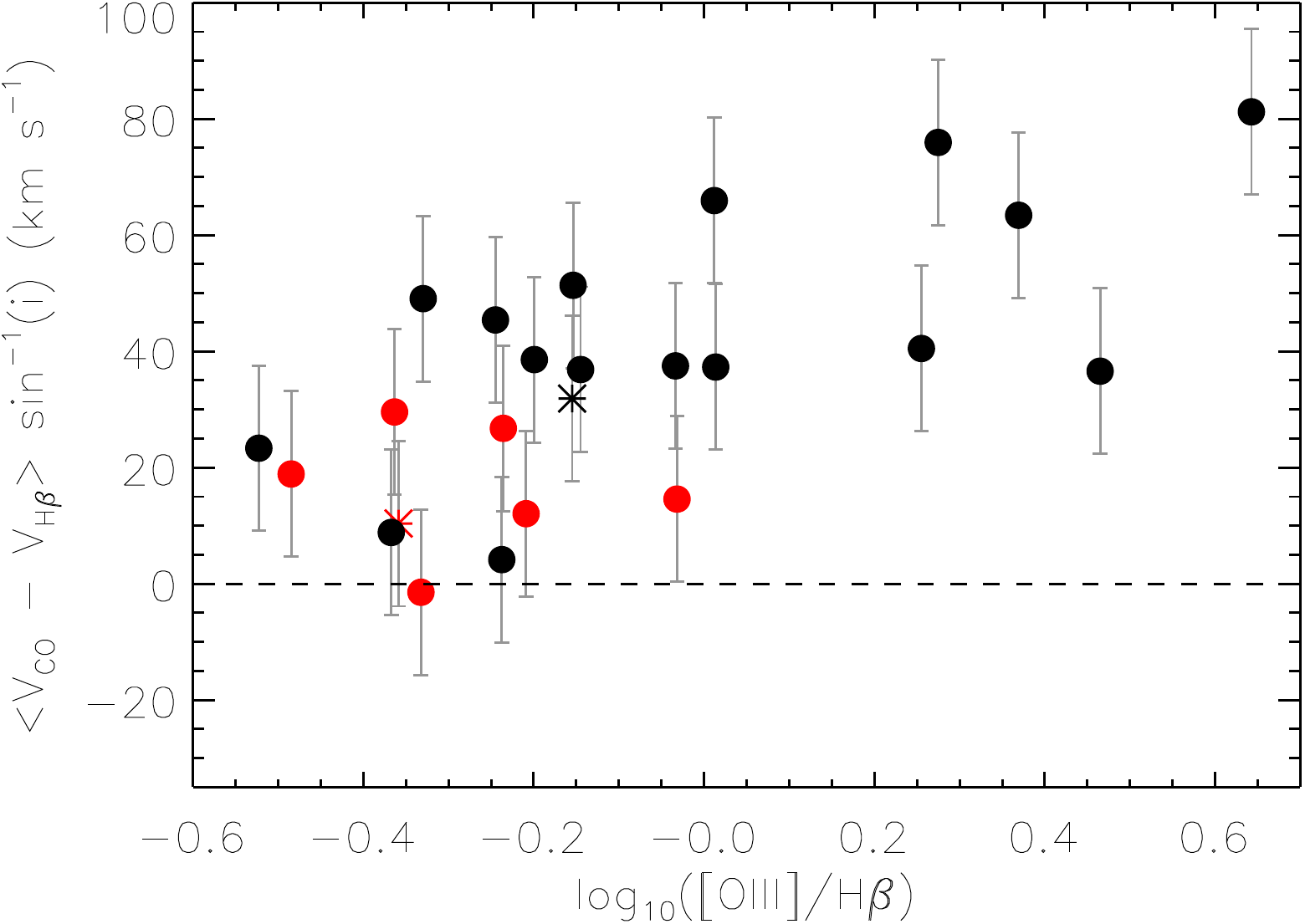}
\caption{\small Radially averaged velocity difference between the molecular and ionised gas plotted against the luminosity weighted EW(H${\beta}$) (top panel; as in Figure \ref{JAMveldif}) and the [OIII]/H$\beta$ ratio (bottom panel). Symbols, and their colour coding have the same meaning as in Figure \protect \ref{JAMveldif}. The dashed line denotes zero velocity offset}
\label{ionveldif}
\end{center}
\end{figure}

The top panel of Figure \ref{ionveldif} shows the radially averaged velocity difference between the molecular and ionised gas plotted against the average EW(H${\beta}$) (as in Figure \ref{JAMveldif}). The bottom panel shows the same measure plotted against the mean [OIII]/H$\beta$ ratio, measured over the same area as used for the EW(H${\beta}$) measurements. Galaxies with an EW(H${\beta}$) $>$ 3\AA\, and those with log([OIII]/H$\beta$) $<$ 0.0 have ionised gas velocities close to those of the molecular gas. Systems with a lower EW(H${\beta}$), and higher [OIII]/H$\beta$ ratios have a wide spread of velocity differences, and many have CO profiles which agree well with the JAM model prediction. 
These findings are discussed in more detail in Section \ref{compwithgas}.

\begin{table}
\centering
\caption{Surface brightness profiles and position-velocity diagram classes and model parameters for the \atlas\ ETG sample. }
\begin{tabular}{l r r r r r}
\hline
Galaxy & SB class & PVD class & Scale rad  & Type & Turnover \\
 & & &(arcsec)  & & \\
 (1) & (2) & (3) & (4) & (5) & (6)\\
\hline
IC0676 & E & O & 2.0 & d & x\\
IC0719 & E(T) & O & 4.0 & d & x\\
IC1024 & E & G & 2.0 & d & -\\
NGC0524 & Reg & ? & 2.0 & d & x\\
NGC1222 & E(T) & X & 5.0 & d & -\\
NGC1266 & U & -- & -- & -- & -\\
NGC2697 & R & -- & -- & -- & -\\
NGC2764 & Reg & O & 5.0 & d & x\\
NGC2768 & Reg & P & 2.0 & d & x\\
NGC2824 & Reg & ? & 5.0 & d & x\\
NGC3032 & Reg & ? & 3.0 & d & x\\
NGC3182 & E(T) & O & 1.5 & d & -\\
NGC3489 & E(T) & G & 4.0 & d & x\\
NGC3607 & Reg & G & 5.0 & d & x\\
NGC3619 & Reg & O & 4.0 & d & x\\
NGC3626 & E(T) & G & 13.0 & r & x\\
NGC3665 & Reg & G & 2.2 & d & x\\
NGC4119 & E & O & 3.0 & r & -\\
NGC4150 & E(T) & X & 1.0 & d & x\\
NGC4292 & Reg & -- & -- & -- & -\\
NGC4324 & R & G & 20.0 & r & x\\
NGC4429 & Reg & G & 3.0 & d & x\\
NGC4435 & Reg & G & 2.0 & r & -\\
NGC4459 & Reg & G & 2.3 & d & x\\
NGC4476 & Reg & G & 3.5 & d & x\\
NGC4477 & Reg & G & 1.0 & r & -\\
NGC4526 & E & G & 4.0 & d & x\\
NGC4550 & E(T) & X & 1.5 & d & -\\
NGC4694 & C & X & 0.5 & d & -\\
NGC4710 & C & O & 25.0 & r & x\\
NGC4753 & E & G & 5.0 & d & x\\
NGC5379 & Reg & O & 20.0 & r & x\\
NGC5866 & C & G & 25.0 & r & x\\
NGC6014 & Reg & O & 1.5 & d & -\\
NGC7465 & Reg & P & 3.0 & d & x\\
PGC029321 & Reg & ? & 0.5 & d & x\\
PGC058114 & Reg & -- & -- & -- & o\\
UGC05408 & U & -- & -- & -- & o\\
UGC06176 & Reg & O & 1.0 & d & -\\
UGC09519 & Reg & P & 2.5 & d & -\\

\hline
\end{tabular}
\parbox[t]{0.5 \textwidth}{ \textit{Notes:} Column 1 contains the galaxy name. Column 2 lists the surface brightness profile class described in Section \ref{surfdens}. "R$_{\rm eg}$" denotes a regular profile, "E" an excess profile, "E(T)" an excess profile with truncation, "R" a ring and "C" a composite profile. Column 3 lists the position-velocity class described in Section \ref{pvsims}. "G" denotes a good profile, "O" a profile where the JAM model predicts a higher rotation velocity than observed in CO, "P" denotes a polar molecular gas disk, "X" denotes a galaxy with disturbed gas, and "?" denotes systems with known problems. A dashes shows that the galaxy was not included in this analysis, for reasons discussed in Section \ref{pvJAM}. Column 4 shows the scale radius used to create the model PVDs described in Section \ref{pvsims}. Column 5 shows the type of surface brightness profile used to create the model PVD: "e" denotes an exponential disc and "r" a Gaussian ring.  Column 6 contains an "x" if the molecular gas extends beyond the turnover of the JAM circular velocity curve, and a dash if the gas is insufficiently extended (in two cases we do not have the data to determine this, these galaxies are denoted with a "o" in this column).  }
\label{pvsdtable}
\end{table}

\section{Discussion}

\subsection{Molecular gas extent}

\subsubsection{ETGs and spirals}
\label{etgandspiralextent}

Earlier in this paper we identified scenarios that would result in the gas in ETGs being less extended than that in spiral galaxies.
In Section \ref{compspiraletg} (Figure \ref{co}), we indeed found convincing evidence that the absolute extent of the molecular gas in ETGs is smaller than that in spiral galaxies. However, spiral galaxies are generally more extended (due to their exponential light profiles; the average effective radius of the BIMA-SONG galaxies is 2 kpc larger than that for \atlas\ galaxies).  
Figure \ref{co_a3d_bima_norm} shows that the distribution of molecular gas extents seem to scale with the size and luminosity of the galaxy. That is, the distributions of the relative CO-extent (with respect to $R_{\mathrm{e}}$, $R_{25}$ and $K_s$-band luminosity) of ETGs and spirals are similar.

A similar correlation between the size scale of spiral discs and their large-scale CO distribution has been discussed before by several authors. \cite{Young:1982p3483} found that the major- and minor-axis profiles of the molecular gas in NGC\,6946 and IC\,342 were in good agreement with optical B-band radial profiles. \cite{Regan:2001p3275} found a similar result for a subsample of 15 BIMA-SONG spirals, for which the mean ratio of the CO to stellar (exponential) scale lengths was 0.88$\pm$0.14. 
\cite{Leroy:2009p3484} also find an average CO scale-length of 0.2$\pm$0.05 $R_{25}$ for their HERACLES sample spiral galaxies. 
{Although these parametric scale radii are hard to compare with the (isophotal) radii used in this paper, it appears that some link exists between the average size of the molecular reservoirs in galaxies and their stellar distributions. }

Despite the large intrinsic variation in scale-length between galaxies, \cite{Regan:2001p3275} suggest that in a time-averaged sense, the stellar and molecular discs must be closely coupled.  {They posit} that this coupling could arise from star formation associated with the molecular gas, that would over time cause the scale lengths to become equal. {Alternatively they suggest} that gravitational effects could exert some feedback on the molecular disc, affecting angular momentum transfer, damping out variations and causing the molecular disc to conform more closely to the stellar potential.  Here we suggest that mass loss from previous generations of stars could also play a role. The mass loss material would interact, exchange angular momentum, and provide a means of coupling the stellar and molecular discs.

The fact that we find a similar relative molecular gas extent in both the \atlas\ ETGs and BIMA-SONG spirals leads to the suggestion that some combination of these processes that \cite{Regan:2001p3275} suggested were active in spirals could also be at work in ETGs. When one considers the 
low molecular gas mass fractions {and star-formation rates} present in our ETGs (typically molecular to stellar mass ratios for our ETGs are $\approx$0.5\%, as compared to $>$10\% for spirals; e.g. \citealt{McGaugh:1997p3670}), gravitational processes or stellar mass loss coupling seem more likely to be the dominant driver(s). {Both of these processes require the gas to have been present and relaxed in the potential of the galaxy for some time, however. While the majority of the ETGs in our sample have dynamically relaxed gas (See Section \ref{gaskinem}), often the gas is misaligned from the stellar body, and/or not distributed in a `disky' way (Section \ref{surfdens}). As we find a similar extent for most subsamples of our ETGs (e.g. misaligned and aligned gas reservoirs; Section \ref{envgasorigin}), this suggests that the dominant process which controls the extent of the gas has to work on short timescales, and must be independent of the orientation of the gas within the galaxy potential.}

\subsubsection{Gas kinematic misalignments and extent}
\label{envgasorigin_discuss}
In Section \ref{envgasorigin} we investigated the extent of the molecular gas in subsamples of the \atlas\ ETGs. We found in the top panel of Figure \ref{comis_virgo} that kinematically-misaligned galaxies (where the gas has an external origin) have a similar distribution of extents as aligned galaxies (where the gas could be from internal sources). This result is initially hard to understand, as gas with an external origin can have any amount of angular momentum, aligned in any direction, and thus the radii at which incoming gas will settle depends strongly on the initial parameters of the source {\citep[e.g.][]{1996ApJ...471..115B,2010MNRAS.406.2405B}}. 
In Paper X, however, we discussed evidence that the majority of the gas in (field) ETGs could be of external origin, even when it is aligned with the stars. If all the gas was from external sources, regardless of misalignment, this could provide a natural explanation for the similarity between the two distributions. Indeed, when one removes cluster galaxies (which we showed are more likely to have internally generated gas) the kinematically-aligned and misaligned field galaxies extent distributions are still statistically indistinguishable {(A MW-U test gives a 30\% chance that the BIMA-SONG and field galaxy \atlas\ normalized extents are drawn from the same parent distribution)}. 

Molecular gas (and hence likely the star formation) in spirals, however, cover a similar proportion of the galaxy as discussed above. In a fast-rotator with a spiral progenitor, gas created from stellar mass loss (which comes predominantly from stars $<$1 Gyr old) would thus occupy a similar area. 
 
\subsubsection{Gas extent and environment}
 
In Section \ref{envgasorigin}, we showed that the molecular gas in Virgo cluster ETGs covers only around half as much of the effective radius as the gas in an average field galaxy. This result is statistically significant at the 97\% level, even given the low number of mapped galaxies in the Virgo cluster.
This suggests that entering the cluster has significantly affected these molecular reservoirs. \hi\ in cluster environments is known to be substantially affected by ram-pressure stripping, but it is thought that denser molecular gas can remain bound until much deeper in the inter-cluster medium \citep[e.g.][]{2005A&A...441..473V}. It is however possible that ram pressure could have compressed the molecular gas reservoirs in these objects.

Paper X reports that all but one of these fast-rotating Virgo cluster galaxies have kinematically-aligned gas, and thus the molecular reservoir has either been substantially affected by the cluster environment or new gas has been generated internally. 
{The misalignment of the gas in field ETGs was not found to be radius dependent at the resolution limit of Paper X. Removal of the outer gas, as implied here, would thus not naturally lead to the gas being measured as more aligned. }

Over time it is expected that stellar mass loss from old stars and any kinematically-misaligned molecular gas would collide, reducing the velocity and angular momentum of {the gas, and causing the molecular reservoir} to become smaller. If such a process is important one would expect misaligned galaxies to show a negative correlation between the $\lambda_{\rm R}$ parameter (a proxy for the amount of rotation present within the stellar component of the galaxy; see Paper III) and the extent of the misaligned gas reservoir. Indeed, in our sample we do see a weak correlation in this sense, with faster rotating ETGs having less extended misaligned gas disks, but as we only have a total of twelve misaligned objects further work will be required to determine if this trend is confirmed in a statistically significant sample. 

Given the above mechanism, the observation that the kinematically-aligned gas in Virgo galaxies is less extended is consistent with a scenario where some proportion of cluster ETGs at some stage had misaligned gas, which has been forced to align (due to the stellar potential and interaction with stellar mass loss). {Interactions between misaligned molecular material and the ICM could also have a similar effect, although to first order one would expect the pressure exerted by the ICM to not correlate with the stellar rotation axis, and thus to create both co- and counter-rotating gas reservoirs. Any of these processes would have to force the gas into an aligned state quickly}, however, as:
\begin{enumerate}
\item we observe very few systems in the cluster that still have kinematically-misaligned gas.
\item the molecular gas fractions observed inside and outside the cluster are similar. Assuming the galaxies entered the cluster looking like field ETGs, {and that they form stars with the same efficiency as field ETGs}, then star-formation would use up significant fractions of the molecular material if the timescale for the gas to relax into the equatorial plane were long.
\end{enumerate}
If the molecular gas can be held stable against star formation (or equivalently have its star formation efficiency reduced) then the problem of gas depletion through star-formation may be bypassed. However the fact that we do not observe many misaligned systems remains. {The dynamical timescale (over which gas is expected to relax into the plane of the galaxy) is shorter for smaller gas reservoirs, and thus the observation that Virgo cluster galaxies have smaller molecular reservoirs can perhaps somewhat help to explain this discrepancy.}

Alternatively, if the gas in Virgo galaxies were regenerated through stellar mass loss (and thus created kinematically-aligned with the stars), then this mass loss would have to come preferentially from stars at the very centre of these galaxies (perhaps because the youngest stars, which lose the most mass, are formed in a central starburst as the galaxies enter the cluster). Alternatively, the mass loss from the outer parts has to be swept away by ram pressure or transported inwards (e.g. by gravitational torques from interactions, or by a bar).

\subsection{Molecular gas surface brightness profiles}
\label{surfden_discuss}
In Section \ref{surfdens} we presented the molecular gas radial surface brightness profiles of the \atlas\ ETGs, and compared them with the stellar profiles. We showed that, broadly speaking, they can be separated into four classes, discs: excess (with or without truncation), rings and composite systems. These four classes are very different, and hence it is meaningful to ask how these differences arise. 

Molecular reservoirs with regular profiles are almost all found in galaxies where the molecular gas is co-rotating (or exactly counter-rotating) with respect to the stars (the kinematic misalignments are tabulated in Paper X). The exceptions are  NGC\,2768, NGC\,7465 and UGC\,09519, where the molecular gas is in a polar disc (see \citealt{Crocker:2008p946}, A12).   
 \hi\ is detected in 66\% of the (observed) field galaxies with regular profiles- a higher detection rate than observed in the parent population (Paper XIII). This \hi\ is often also in relaxed discs (class $D$ or $d$ in Paper XIII), suggesting many of these galaxies are relaxed systems even out to larger radii. This also suggests galaxies with regular profiles have not had a large merger or accretion event recently, which would disturb the molecular (and atomic) gas disc.

A third (12/41; 29\%) of the galaxies in our sample show a CO profile that does not fall off as fast as the stellar profile. These systems appear to be divided into two types, those that show truncations in their molecular gas distribution (7/12; 58\%), and those that fall off regularly but with a different scale-length than that of the stars (5/14; 42\%). The truncated galaxies either have messy/counter-rotating gas with signs of ongoing interaction (IC\,719, NGC\,1222;  NGC4150, NGC4550; A12, \citealt{Beck:2007p2941,Crocker:2009p3262}) or they have evidence of spiral structure or rings in the molecular gas (NGC\,3182, NGC\,3626; A12, NGC\,3489, \citealt{Crocker:2010p3342}).
Similarly, the galaxies which have excess profiles without a clear truncation are often messy (IC\,676, IC1024; A12, NGC\,4753, \citealt{SteimanCameron:1992p2507}), or may have unresolved rings (NGC\,4119; A12, NGC\,4526; Davis et al., 2012, in prep).

The excess galaxies (including those with and without truncation) are generally found to be low-luminosity ETGs (83\% have luminosities $>$-23.5 $M_{\textrm{Ks}}$). Lower luminosity (hence presumably lower mass) galaxies have shallower potentials, are less able to retain their gas, and may require longer to force accreted molecular gas into discs with similar scalelengths as the stars.  All of the excess class of galaxies in the field which were observed in \hi\ were detected, and many have dynamically and morphologically unsettled gas (as defined in Paper XIII) at large radii. {\hi\ at these radii takes longer to relax than the centrally concentrated molecular gas, and (as argued in Paper XIII and \citealt{Oosterloo:2010p3376}) the presence of unsettled \hi\ gas suggests recent gas accretion or mergers. Such events may thus be the cause of the extended/truncated profiles in field galaxies.}

Galaxies with a ring may betray the presence of resonance systems, potentially caused by a bar. 
The molecular gas will then accumulate onto stable, non-intersecting closed orbits, as often seen in spiral discs \citep[e.g.][]{Combes:1991p3648}. {All of these objects show signatures of bars in their optical morphology or stellar kinematics (Paper II).} By inspection of the CO PVDs (Section \ref{pvdiagsec}), we suspect that the galaxies that we have classified as having rings based on the surface brightness profile generally have gas around the turnover radius (often coincident with co-rotation, i.e. inner rings; \citealt{Mazzuca:2011p3505}) or further out. Gas within the inner Lindblad resonance (i.e. nuclear rings or spirals) is unlikely to be resolved enough to show a central dip at our surveys spatial resolution (e.g. the gas in NGC\,4119 is likely to be distributed in an nuclear ring given its morphology, but this is not apparent in its surface brightness profile). It has been shown that minor mergers can also potentially cause gas to accumulate in rings without the formation of a bar \citep[e.g.][]{ElicheMoral:2009p2918}.
 
The systems with composite profiles are NGC\,4710, NGC\,4694 and NGC\,5866. NGC\,4694 is a disturbed system with the majority of its molecular gas outside the body of the galaxy (A12), and this leads to the derived profile shape. The other galaxies in this class have X-shaped PVDs, that usually arise from gas caught in two rings within barred galaxy discs (with little gas inbetween), the first being gas caught in $X_2$ orbits, and the second gas beyond co-rotation \citep[e.g.][]{Bureau:1999p3649,Athanassoula:1999p3488}. 
As discussed in Section \ref{surfdens} such profiles are not present in the BIMA-SONG sample, which was selected to avoid edge on systems. These galaxies would have two clearly separated peaks visible in the surface-brightness profiles if viewed at lower inclinations. Thus this types of surface brightness profile is not intrinsic to these galaxy, but arises due to projection effects.
Both systems are edge on, discy fast-rotators, and NGC\,4710 has a peanut-shaped bulge \citep[typically thought of as being caused by the buckling of a bar; see][]{Bureau:1999p3487}, while NGC\,5866 is thought to be unbarred with a prototypical classical bulge. If the X-shaped PVD of NGC\,5866 is indeed due to material caught in barred orbits, then this interpretation would have to be revisited. 
It is possible that we are seeing material in inner and outer rings, not associated with a bar, although the creation of this double-ring structure may then be difficult to explain self-consistently.

\subsection{Velocity profiles}
\label{pvdiagsec}

In Figure \ref{pvexample} (and Appendix B in the online material) we presented the CO PVDs of our sample galaxies, overlaid with the modelled circular velocity curve, and the stellar and ionised gas rotation curves extracted from the \sauron\ data. We also presented modeled PVDs, where we use the JAM model circular velocities and inclinations, and attempted to fit the observed PVDs with models of a relaxed molecular disk or ring (see Section \ref{pvsims}). 

These simple gas surface brightness profiles perform well in general, and allow us to test the JAM model circular velocity predictions. As revealed in Section \ref{surfdens} however, our real galaxies can have quite complicated surface brightness profiles, and thus we do not expect to fully reproduce the structures seen in the PVDs. In some objects we note that the central parts of the PVD show a gap, which is not present in the model data. It seems likely that the CO does not extend into the very centre of some of these galaxies (e.g NGC\,4435, NGC\,4459, NGC\,6014). Such central molecular deficiencies have been observed in spiral galaxies \citep[e.g. $\approx$ 50\% of the BIMA-SONG galaxies][]{Regan:2001p3275}.
In barred galaxies these depressions could be due to gravitational torques forcing gas outwards to the inner-lindblad resonance, or  inward onto a central black-hole ({for the gas within the dynamical sphere of influence}; \citealt{Combes2001}).

\subsubsection{Circular velocities}
\label{compwithjam}

In Sections \ref{pvJAM} and \ref{pvsims} we described the behaviour of the observed CO PVDs, and how they compare to the predicted JAM circular velocities.  The molecular gas reaches beyond the turnover of the circular velocity curve in 68$\pm$6\% of the sample ETGs, as predicted from the single-dish spectra in Paper V. This confirms that using the shape of the single-dish line profile is a reliable method to select galaxies with have molecular gas beyond the turnover radius.

We separated the galaxies into five rough classes based on the relation between the observed CO kinematics and the JAM models: those in which the model agreed well with the observations, those in which the shape of the model was similar to the observed profile but predicted higher velocities, those which have polar rings, or disturbed gas (and thus we don't expect agreement between the model and the observed kinematics), and lastly those in which we do not consider due to lack of resolution or known problems with the models.

By comparison with the results of Paper X we find that all the galaxies where the models agree well with observed CO profiles have their gas kinematics aligned with the stellar kinematics (15/17 galaxies), or exactly counter-rotating (2/14 galaxies). The CO thus rotates in the plane in which the model predictions are valid. Over half of these systems have regular CO surface brightness distributions (Section \ref{surfdens}). The JAM circular velocity curves trace the outer envelope of the CO PVD in many cases. This behaviour is reproduced by our models which include the effects of beam smearing, and is often seen in \hi\ PVDs (e.g. Fig 6 in \citealt{Oosterloo:2007}). 
 In these galaxies with good models the average percentage offset between the peak of the observed CO flux at each position (the PVD `trace' introduced in Section \ref{pvJAM}) and the model PVD trace is only -3\% ($\approx$ -5\kms; smaller than the observational channel width), with an RMS scatter of 15\%. 
Thus we consider the observed CO kinematics and the predicted circular velocity curves of these systems to be consistent. The CO in these systems is thus likely to be dynamically cold and an excellent tracer of the circular velocity.

In 20\% (7/35) of CO-mapped galaxies the shape of the model PVD is similar to that of the observed CO rotation curve but is at systematically higher velocities. The average offset between the CO trace and the model trace is -0.55 dex, larger than would be expected from our models which include the effect of beam smearing. The CO PVDs show that the gas is regularly rotating, and half of these systems have regular surface brightness profiles (which we argued in Section \ref{surfden_discuss} are present in relaxed systems). Such large differences are thus unlikely to be caused by a dynamical disturbance (e.g. a merger). These systems all have co-rotating or exactly counter-rotating molecular gas, so the model predictions are for the correct plane. {These galaxies are also not offset from the CO Tully-Fisher relation for ETGs (presented in Paper V) suggesting the gas is relaxed and traces the underlying potential.} The remaining disagreement between the model and observations must be due to other factors. Below we discuss possible causes for these remaining offsets, either due to physical effects within the gas, or potential problems with the modeling technique.

A simple explanation of the offsets between the observations and models would be if the molecular gas was not dynamically cold. Using the zeroth order approximation shown in equation 9 reveals you would require very large molecular gas velocity dispersions ($\sigma_{\mathrm{CO}}\approx 80$ \kms) to explain an offset from the JAM circular velocity of the magnitude observed in these systems. We used the KinMS tool to simulate galaxies with such large velocity dispersions, and although the outer-envelopes of the PVD distributions can be made to agree, the PVD shape is inconsistent with that observed.  
We also consider such large velocity dispersions to be unlikely considering the average value for cold atomic gas in nearby spirals is only $\sigma_{\mathrm{\hi}}\approx8$ \kms \citep{vanderKruit:1982p3503,Dickey:1990p3504}. 

Other potential causes of the offset are beam smearing, the finite thickness of the gas disc, non-circular motions in the gas (e.g. caused by bars) and changes in the density profile (e.g. spiral arms), which could all act to lower the gas (apparent) rotation velocity. The first two effects are included in our models, and are thus unlikely to be the cause of the offsets, and thus we concentrate on the other possible causes here.

Around half of the systems in which the JAM model circular velocities are higher than the gas velocity have central CO rings (NGC\,3182), or have (single or multiple) solid-body components in their observed PVDs (NGC\,2764, NGC\,4710, NGC\,5379).  Such morphological and kinematic features often betray the presence of resonance systems, potentially caused by a bar \citep[e.g.][]{Combes:1991p3648}. The majority of the galaxies (6/7) also have a bar visible in optical imaging (Paper II). In barred systems material follows elliptical rather than circular orbits, and hence some deviation from the circular velocity is expected. Non-circular motions of this type can provide a possible explanation for the offset between the observations and models in galaxies with bars.
One must also be cautious, as material caught in a circular ring at a resonance point such as co-rotation does rotate at the circular velocity. Without understanding the resonance structure of each galaxy it is thus difficult to say with certainty which galaxies have lower CO velocities due to non-circular motions in the gas.   In principle one can detect non-circular motions by the effect they have on the observed molecular gas velocity fields \citep[see e.g.][]{Schoenmakers1997,Krajnovic:2006p2929}, however our modest spatial resolution limits our ability to do this with the current data set.

An example of a galaxy which may have gas in non-circular orbits is IC\,0676. This galaxy has a strong optical bar, and the CO rotation is aligned with the bar, not the main disk. As the PVD of IC\,676 does not show solid body rotation it is unlikely to have a molecular ring, and hence the gas velocity could be affected by non-circular motions, and produce the observed discrepancy. 

Lablanche et al., 2012 (hearafter Paper XII) have also shown that the presence of a bar can induce an additional uncertainty (of the order of $\approx$15\%) in the $M/L$ predicted by the JAM model.
As the assumed $M/L$ sets the normalisation of the rotation curve ($v \propto \sqrt{M/L}$), this could explain some (but likely not all; the average effect would be $\approx$15 -- 20 \kms) of the offsets of the JAM predictions. As we have 7 galaxies with JAM profiles that predict higher velocities than the CO, and few where the predicted circular velocity is low, statistically it is unlikely that this is the dominant cause of the difference between the observed CO rotation and JAM predictions.

The models we show here are based on the JAM circular velocity curve for an edge on system. These were derived from mass models of the galaxies, and the observed SAURON kinematics. The output circular velocity curves have been projected back into the line of sight with the best fitting JAM inclination, in order to allow comparison with the observed CO PVD. If the JAM models systematically underestimated the inclination of the system then this could result in the model predictions being too high. Careful comparison with dust distributions in these systems, and other inclination indicators (e.g. those presented in Paper V) suggests that the derived JAM inclinations are unlikely to be biased in this manner however (Cappellari et al., in prep). 

Although the early-type galaxies in this sample were selected to show no evidence of spiral arms in optical imaging, it is possible for them to have spiral structures in the molecular gas disk \citep[c.f. NGC\,3489;][]{Crocker:2011p3500}. Such structures may remain unnoticed if the beam size of our observations is too large to resolve them, or if the gas disk is observed close to edge on (5/7 of the galaxies where the observations and models do not agree have inclinations $>$60$^{\circ}$; Paper V).
{Both spiral structure itself, and associated streaming motions along spiral arms could have a noticeable effect on the observed kinematics of the gas, and the PVD.
Streaming motions can have a large amplitude \citep[e.g. $\approx$50 \kms in M51;][]{2007ApJ...665.1138S}, however these systems would have weaker spiral structure than M51, and at the scales probed here the measured velocity should change by $<$10\kms, not enough to explain the offsets we see.} 

\begin{figure*}
\begin{center}
\includegraphics[width=\textwidth]{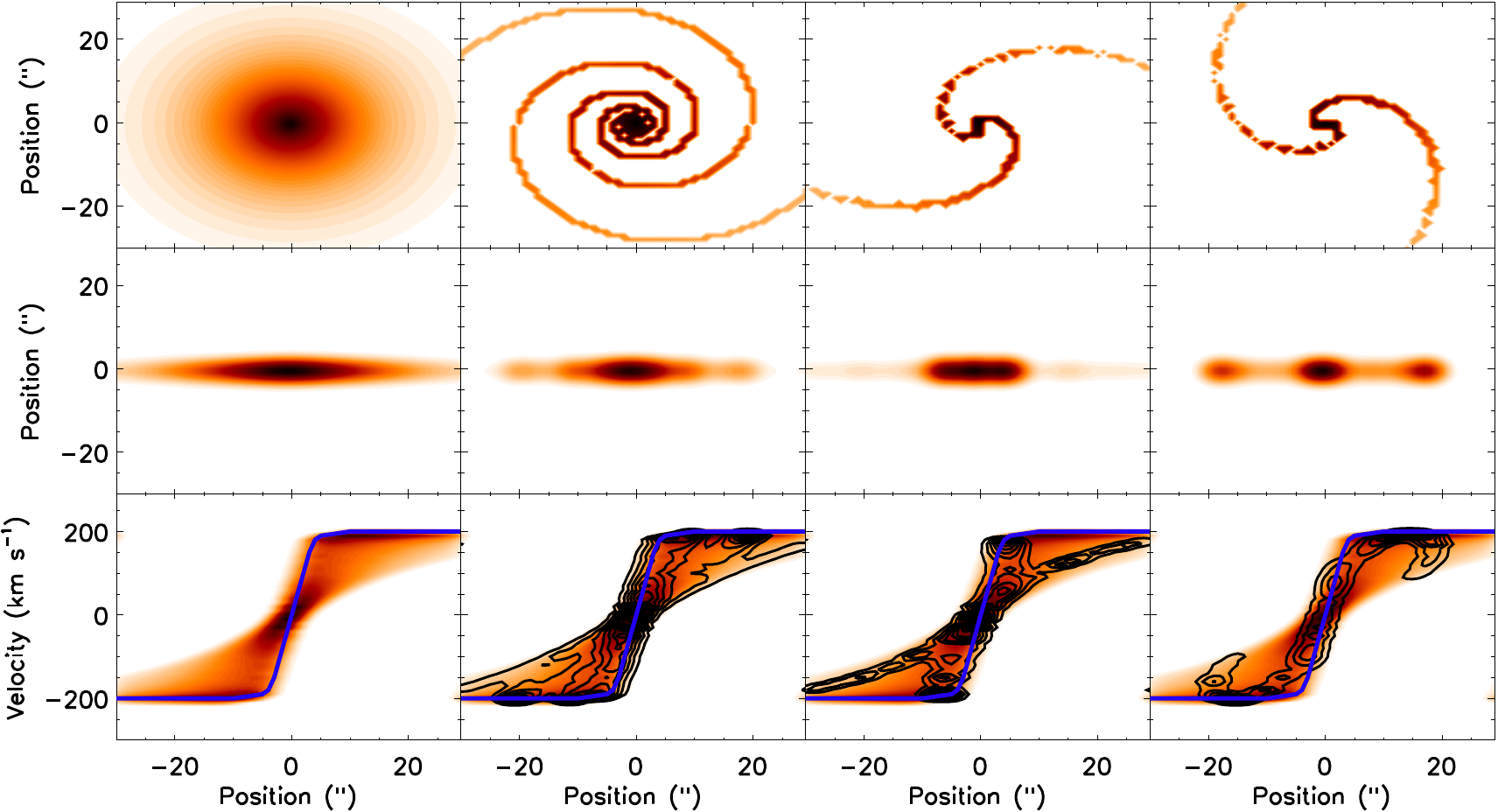}
\caption{\small Simulated observations showing the effect of spiral structure on the derived position-velocity diagrams. The left column shows a disk simulation, and its PVD for reference. The second column shows a tightly wrapped spiral like the Milky Way \citep[with a pitch angle of 12$^\circ$;][]{Vallee1995}, and the third and fourth show a loosely wrapped spiral patterns (pitch angle 35$^\circ$) observed with different angles to the line of sight. The top row shows the input surface density distribution in each simulation, viewed face on. The same radial surface brightness profile is used in each case. The middle panel shows the zeroth moment of the simulated observation, viewed edge on. The bottom panel shows the disk PVD (colours), and in the last three panels this is overlaid in black with contours of the PVD for the spiral density distribution. The blue line shows the input circular velocity curve, which is identical in each case. The simulations all were performed with a resolution of 4\farc5, a pixel size of 1\arcsec x 1\arcsec and a gas velocity dispersion of 8 \kms.}
\label{spiralsim}
\end{center}
\end{figure*}

The effect of spiral structure itself (e.g. not having gas at all radii along the PVD) depends on how tightly wound the spiral arms are, and the angle they make to the line of sight. 
Figure \ref{spiralsim} shows a simulation (made in the same fashion as described in Section \ref{pvsims}) of the PVD observed for three different spiral gas distributions. The first column shows a simple disk simulation as a reference, while the second column shows a tightly wound spiral, and the third and fourth panels show a loose spiral, viewed at different angles. For each simulation we show the face on view of the input surface brightness profile in the top row, and how it would look when projected to an edge on inclination in the second row. The bottom row shows the observed PVD for the edge on system in black, overlaid on the PVD of the disk system. A tightly wound spiral pattern displays a very similar PVD to a complete disk, and hence it seems unlikely that a tightly wound spiral could cause the offset between simulations and observations that we observe. A loose spiral pattern however displays a PVD that changes in form depending upon the angle at which it is viewed. It is important to note however that the observed PVD in these cases covers some fraction of the total area of the PVD of the disk simulation, but does not go outside it. One can thus potentially identify systems with such spirals by comparison with the PVD of a disk (as presented in Figure \ref{pvexample} and Appendix B in the online material). None of the cases in which the JAM and CO rotation disagree are clear candidates for having such spiral structure, based on the PVD, but we are unable to rule out such an explanation completely.  {Even if some objects do have such a structure, the amplitude of this effect would not be enough to explain all the offsets seen.}

A potential cause for the models circular velocities to not agree with the observed CO rotation is the effect of dust on the MGE. Many of the systems where the JAM models overpredict the CO velocity have central dust obscuration in the optical images used to create the MGEs (even after a dust correction based on $g$-$r$ colour is applied). Such system include IC\,676, NGC\,2764 and NGC\,4710. The models of these galaxies are likely to badly capture the light distribution in the very central parts, that are vital to set the circular velocity normalisation. The measured light profiles may become too shallow in the inner parts and the models will require a high $M/L$ to best fit the (steeper) rise in the stellar velocities.

Assuming a fixed dust-to-molecular gas ratio, it is clear that we would expect these problems to be worse at high molecular gas masses.
Systems with profiles that are at higher velocities than the observed CO rotation do tend to have larger molecular gas reservoirs (Paper IV). These galaxies have a mean H$_2$ mass of 6$\times$10$^8$M$_{\odot}$, nearly half an order of magnitude higher than that of galaxies with  good JAM model predictions ($<{\rm M_{H_2}}$$> =$2$\times$10$^7$M$_{\odot}$). Such problems with dust could be overcome to some degree by utilising near-infrared imaging (which is less affected by dust than optical images) in the centres of these objects when creating the MGE,  but creating such models is beyond the scope of this paper. Comparison of the rotation curve of NGC\,4710 derived by \cite{Williams:2009p2681} using infrared imaging shows that an offset of $\approx$30 km/s can be induced by this effect, enough to explain the offset between the JAM velocity and CO in this object.

One other potential cause of the offset between the models and observations would be if the assumption of a constant $M/L$ as a function of radius is invalid. The systems with model PVDs that lie above the CO profiles tend to have large H${\beta}$ emission equivalent widths, and low [OIII]/H$\beta$ ratios (see Figs. \ref{JAMveldif} and \ref{ionveldif}).  These systems also have larger molecular masses, on average, as stated above.
The EW of H${\beta}$ is often used as a star-formation tracer in spiral galaxies (where an EW(H${\beta}$) $\gtsimeq$2\AA\ is usually considered indicative of ongoing star-formation). One must be cautious when using EW(H${\beta}$) to trace star-formation in ETGs, as other sources of ionisation can cause H${\beta}$ emission \citep[e.g. old stellar populations,][]{Sarzi:2010p3476}.  However, in all but one of the galaxies here that have EW(H${\beta}$) $\gtsimeq$2\AA\ the ionisation is likely to be from star-formation (as Figure \ref{ionveldif} shows they have log([OIII]/H${\beta}$) ratios $\ltsimeq$ -0.2; \citealt{Sarzi:2010p3476}). The one galaxy where this is not the case is NGC\,7465, which is a composite source, with ionisation both from star-formation, and a X-ray bright AGN. 

Active star-formation would create a population gradient, with many young stars in the inner parts of the galaxy where the CO is concentrated, and older stars dominating in the gas-poor regions further out. The galaxy average $M/L$ (and thus circular velocity) predicted by the JAM model would thus be too high in the inner parts of the galaxy where the CO is, leading to an overestimation of the circular velocity (with a corresponding underestimation in the outer parts). \cite{Young:2008p788} show that the magnitude of this effect can be as high as a factor of 3 in velocity in the inner parts of NGC\,3032, a galaxy with a strong population gradient. Even in galaxies where the population gradient is shallower this effect could be large enough to explain the difference between the observed and predicted circular velocities. Cappellari et al., in prep identify galaxies with H$\beta$ absorption linestrengths $>2.3$ \AA\ (within an effective radius) as likely to have population gradients strong enough to affect the JAM models.  Of the systems classified as having JAM model circular velocity profiles that predict higher velocities than the observed CO rotation, 5/7 have such strong indicators of young stars (IC\,0676, IC\,0719, NGC\,2764, NGC\,5379, NGC\,6014).
The two galaxies with JAM models which match well the molecular gas PVD but which have large H${\beta}$ emission equivalent widths are IC1024 and UGC\,06719. In these galaxies the CO extends over the entire disc of the galaxy, to beyond 1 R$_{\mathrm{e}}$, and hence the variation of $M/L$ as a function of radius is likely to be lower. 

Figure \ref{dvSIGMA} shows that many of the systems in which the JAM model and CO rotation do not agree are at low velocity dispersions.  If the difference between the JAM model velocities and the CO rotation is due to ongoing active star-formation then this could be a consequence of `downsizing', where lower mass galaxies have more star-formation activity at low redshifts  \citep[e.g.][]{Cowie:1996p3650}. Even if the star-formation rate of all \atlas\ galaxies were the same, the luminosity weighted $M/L$ in a low-mass galaxy would be more affected than in a high-mass galaxy. As we touched on above, however, the molecular gas reservoirs of these low-mass systems are much larger on average, so the situation is even more extreme.

We thus conclude that in the galaxies with the largest molecular masses, and highest star-formation rates, dust obscuration, non circular motions in the gas and/or population gradients could be affecting the observed CO, or derived JAM circular velocities. 
The magnitude of these effects can be up to 0.6 dex in velocity. 
In the context of the entire \atlas\ sample however, $\ltsimeq$3\% of galaxies (8/260) have sufficient molecular gas, dust and star formation to affect the JAM model predictions, so these effects should not strongly bias results for the whole sample.
Indeed, all cases without the problems highlighted above we found excellent agreement between the CO observations and JAM models predictions.

In the 11\% (4/35) of CO-mapped galaxies where the CO is disturbed (as identified in A12), its rotation (or lack thereof) is unrelated to the JAM prediction of the circular velocity. {In one of these systems (NGC1222) the JAM model is also likely to be bad due to large scale stellar kinematic disturbances, however in the other cases the stars seem unaffected by the gas disturbance.} All of these galaxies have extended surface brightness profiles, with truncations, supporting our conclusion that these profiles indicate mergers and/or recent accretion. 

Three of the CO-mapped galaxies (9\%) have their molecular gas rotating in 
the polar plane (Paper X), hence the disagreement with the model 
predictions (for rotation along the photometric major axis of the 
galaxy) probably implies that the total mass distribution in these 
three systems is flattened {(as a spherically symmetric mass distribution would have the same rotation curve in any plane)}. The luminous matter in these three 
galaxies has ellipticities between 0.45 and 0.65, supporting this assumption. If one has a suitable tracer (such as \hi) that extends to large enough radii it may be possible to constrain the shape of the dark matter halo in these systems \citep[e.g. as done for IC2006 in][]{Franx1994}, however with this data such an analysis is impossible.

\subsubsection{The dark matter halo of NGC4753}
\cite{SteimanCameron:1992p2507} have discussed in detail the complex, twisted dust lanes of NGC\,4753, and showed that they are consistent with a disc that is strongly twisted by differential precession. They used this twisted disc to set constraints on the shape of the matter distribution, finding that the total dark and luminous mass is nearly spherically-distributed, with an ellipticity no higher than 0.16 and no smaller than 0.01. Due to a lack of kinematic information for the warped disc however, they were unable to determine if the dark matter halo is oblate or prolate. If the material in the warped disc rotates in the same direction as the stars, then the dark matter must have an oblate shape, while if it counter-rotates the halo is likely to be prolate. 
Assuming that the molecular gas we detect in this galaxy is associated with the warped dust disc, it is possible to break this degeneracy. The molecular gas is kinematically aligned with the stars and thus it follows, according to \cite{SteimanCameron:1992p2507}, that the dark halo in this galaxy is oblate, with a flattening between 0.16 and 0.01.

\subsubsection{Ionised gas velocities}
\label{compwithgas}

The ionised gas in the CO-mapped \atlas\ galaxies always (within the errors) rotates at the same speed, or slower, than the molecular gas. This confirms that CO is the best tracer of the circular velocity in the inner parts of ETGs. Due to the large mismatch in angular resolutions between the CO interferometry and the \hi\ data we possess (\citealt{Morganti:2006p1934,Oosterloo:2010p3376}; Paper XIII) we are unable to directly compare CO and \hi\ velocity profiles in this work, but we would expect \hi\ to be a comparable tracer of the circular velocity, and the best available probe at large radii. {Comparisons of the integrated profile shapes will be presented in a future work, discussing the \hi\ Tully-Fisher relation of ETGs.}

The differences between the velocities of the CO and ionised gas relates to the ionisation mechanism powering the optical emission. H{\small II} regions embedded in a dynamically cold star-forming molecular disc are likely to rotate much like the disc itself, and be dynamically cold. On the other hand, gas clouds ionised by other (most likely diffuse and stellar; \citealt{Sarzi:2010p3476}), sources do not have to necessarily display such a relaxed kinematic behaviour. In the specific case of emission powered by shocks between gas clouds,  for instance, a larger velocity dispersion for the ionised-gas is expected. In Figure \ref{ionveldif}, it is clear that the most star-forming galaxies (with the largest H$\beta$ equivalent widths and star-forming [OIII]/H$\beta$ ratios; Figure \ref{ionveldif}) have the smallest differences between the ionised gas and the molecular gas velocities. 

A simple first order estimate based on equating the second moments of molecular and ionised gas (as in Equations 8 \& 9) shows that the small systematic offset from the {dashed zero-velocity offset line} of Figure \ref{ionveldif} can be explained by the observed velocity dispersion providing dynamical pressure-support to the ionised gas (A $\Delta$V $\approx$21 \kms is expected given a $<\sigma_{gas}>$ $\approx$ 62 \kms). Galaxies with a smaller amount of star formation have a wide range of offset values between the ionised and molecular gas velocities, reflecting a variety of ionisation mechanisms and the presence of few, if any, gas clouds in the immediate vicinities of OB-stars and their giant molecular nurseries.

\section{Conclusions}

We have investigated the extent and surface brightness profiles of the molecular gas in the CO-rich \atlas\ ETGs, and compared them to spiral galaxies. We find that the molecular gas extent is smaller in absolute terms in early-type than in late-type galaxies, but that the size distributions appear to scale similary with optical/stellar characteristic scalelengths. 
Amongst ETGs, we find that the extent of the molecular gas is independent of the kinematic misalignment, despite the many reasons why misaligned gas might have a smaller extent. The extent of the molecular gas does depend on environment, with Virgo cluster ETGs having less extended molecular gas reservoirs. This may be due to ram pressure stripping, gravitational interactions within the cluster and/or a greater prevalence of bars driving gas towards the galaxy centres. 

We have identified four classes of surface brightness profiles in the CO-mapped \atlas\ ETGs. Around half have molecular gas that follows the stellar light profile, similarly to molecular gas in spirals. These galaxies appear to have relaxed gas out to large radii, suggesting that these galaxies have been left undisturbed long enough for the molecular gas to relax into a profile that follows the stellar mass. 
A third of the sample galaxies show molecular gas surface brightness profiles that fall off slower than the light, and sometimes show a truncation. These galaxies often have a low mass, and either have disturbed molecular gas, suggesting that a recent merger or gas accretion may cause the excess, or are in cluster environments where ram pressure stripping and the presence of hot gas may compress and/or truncate the gas. The remaining galaxies have rings, or composite profiles, that we argue can be caused by the effects of bars (or potentially minor mergers). 

We also investigated the kinematics of the molecular gas using position-velocity diagrams. On to these we overlaid the JAM model circular velocity predictions and the stellar and ionised gas velocities. 
We find that the molecular gas reaches beyond the turnover of the circular velocity curve in 75$\pm$6\% of our \atlas\ ETGs, validating previous work. The CO is also closer to the predicted circular velocity than the stars or ionised gas, validating its use as a kinematic tracer for Tully-Fisher and similar analyses. 
{This study suggests that the dust lanes in NGC\,4753 are likely to co-rotate with the stars. From this it follows, according to \cite{SteimanCameron:1992p2507}, that the shape of the dark-matter halo in NGC\,4753 is oblate, with a flattening between 0.16 and 0.01.}

We found that in $\approx$50\% of our galaxies the predicted circular velocity profiles and the observed CO rotation are consistent, the CO appears to be dynamically cold, and CO is an excellent tracer of the circular velocity of the system. In 20\% of CO-rich galaxies we found the JAM model prediction of the circular velocity is systematically above the observed CO velocity.  These galaxies have strong dust features, that make constructing a mass model from optical imaging challenging. These systems also often have bars, and higher molecular gas masses and H${\beta}$ equivalent widths than the rest of the sample. 
This suggests that dust obscuration and/or population gradients can affect model predictions of the circular velocity. At the same time bars, or non axisymmetric structures could cause the gas to rotate slower than the circular velocity {(however the amplitude of this effect is likely insufficient to explain the observed discrepancies).}
As systems with CO that show the problems highlighted above make up a small percentage ($\approx$3\%; 8/260) of the whole \atlas\ sample this will not strongly bias results based on JAM models of this type when applied to all galaxies. 
In all cases where the CO appears to be regularly rotating, where the MGE models well reproduce the observed photometry, and where star formation activity is low we found excellent agreement between these two completely independent measures of the circular velocity (CO observations and JAM models of the \sauron\ stellar kinematics), in agreement with an earlier study on a smaller number of objects by \cite{Young:2008p788}.

We also note that the ionised gas in the most star-forming galaxies appears to be dynamically colder, likely because it is associated with star formation in the molecular disc.

In conclusion, we find that the molecular gas reservoirs in many ETGs are very similar in terms of their relaxed kinematics and relative extent to those in spiral galaxies. However rich environments and merger activity can lead to some systems having different morphological and kinematic properties. Understanding how these differences affect the physical conditions within the gas, and thus star-formation, will be required to correctly understand the ongoing evolution of the most massive galaxies in our universe.

\vspace{10pt}
\noindent \textbf{Acknowledgments}

The research leading to these results has received funding from the European
Community's Seventh Framework Programme (/FP7/2007-2013/) under grant agreement
No 229517.

MC acknowledges support from a Royal Society University Research Fellowship.
This work was supported by the rolling grants `Astrophysics at Oxford' PP/E001114/1 and ST/H002456/1 and visitors grants PPA/V/S/2002/00553, PP/E001564/1 and ST/H504862/1 from the UK Research Councils. RLD acknowledges travel and computer grants from Christ Church, Oxford and support from the Royal Society in the form of a Wolfson Merit Award 502011.K502/jd. RLD also acknowledges the support of the ESO Visitor Programme which funded a 3 month stay in 2010.
SK acknowledges support from the the Royal Society Joint Projects Grant JP0869822.
RMcD is supported by the Gemini Observatory, which is operated by the Association of Universities for Research in Astronomy, Inc., on behalf of the international Gemini partnership of Argentina, Australia, Brazil, Canada, Chile, the United Kingdom, and the United States of America.
TN and MBois acknowledge support from the DFG Cluster of Excellence `Origin and Structure of the Universe'.
MS acknowledges support from a STFC Advanced Fellowship ST/F009186/1.
MBois has received, during this research, funding from the European Research Council under the Advanced Grant Program Num 267399-Momentum.
The authors acknowledge financial support from ESO.

This paper is partly based on observations carried out with the IRAM Thirty Meter Telescope. IRAM is supported by INSU/CNRS (France), MPG (Germany) and IGN (Spain).
Support for CARMA construction was derived from the states of California, Illinois, and Maryland, the James S. McDonnell Foundation, the Gordon and Betty Moore Foundation, the Kenneth T. and Eileen L. Norris Foundation, the University of Chicago, the Associates of the California Institute of Technology, and the National Science Foundation. Ongoing CARMA development and operations are supported by the National Science Foundation under a cooperative agreement, and by the CARMA partner universities. 
We acknowledge use of
the HYPERLEDA database ({\tt http://leda.univ-lyon1.fr}) and
the NASA/IPAC Extragalactic Database (NED) which is operated
by the Jet Propulsion Laboratory, California Institute of Technology,
under contract with the National Aeronautics and Space Administration.

\bsp
\bibliographystyle{mn2e}
\bibliography{bibMorphoKin}
\bibdata{bibMorphoKin}
\bibstyle{mn2e}
\label{lastpage}

\appendix
\section{Surface brightness profiles}
\label{sbprofap}
\input{Appendix_A}

\section{Position-velocity diagrams}
\label{pvdiagap}
\input{Appendix_B}

\end{document}

%% file: Appendix_A.tex

\begin{figure*}
\begin{center}
\subfigure{\includegraphics[scale=0.45]{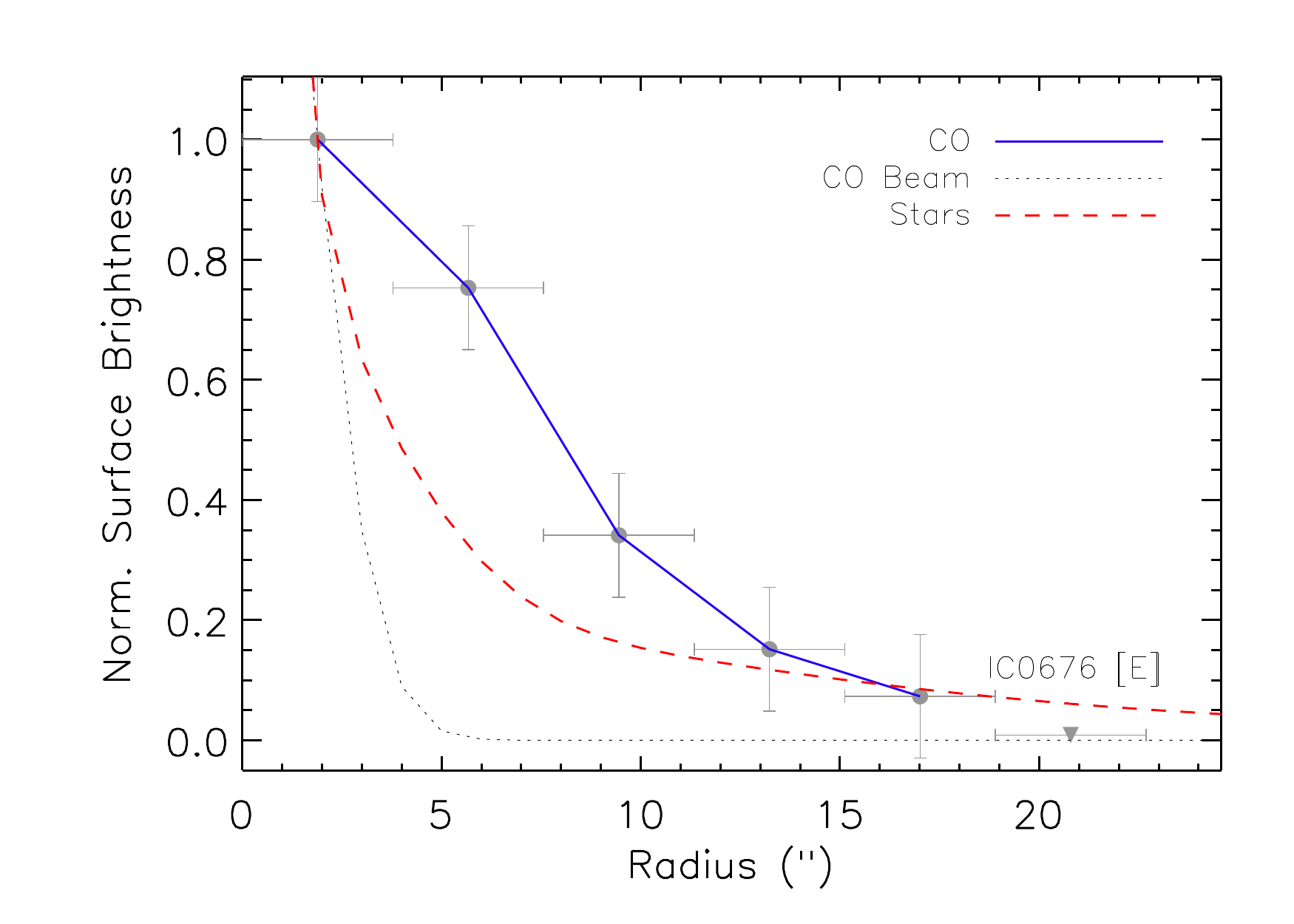}}
\subfigure{\includegraphics[scale=0.45]{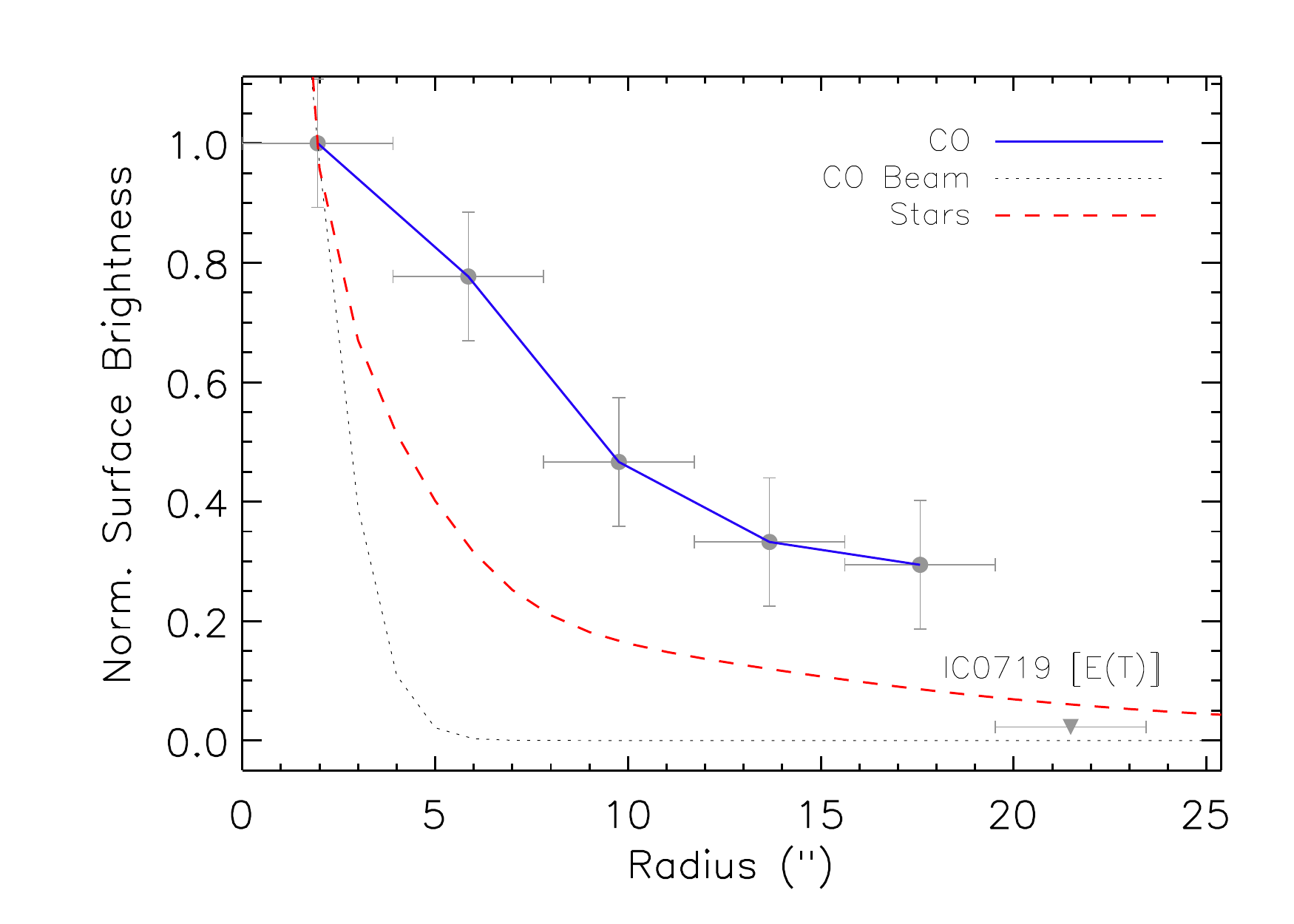}}
\subfigure{\includegraphics[scale=0.45]{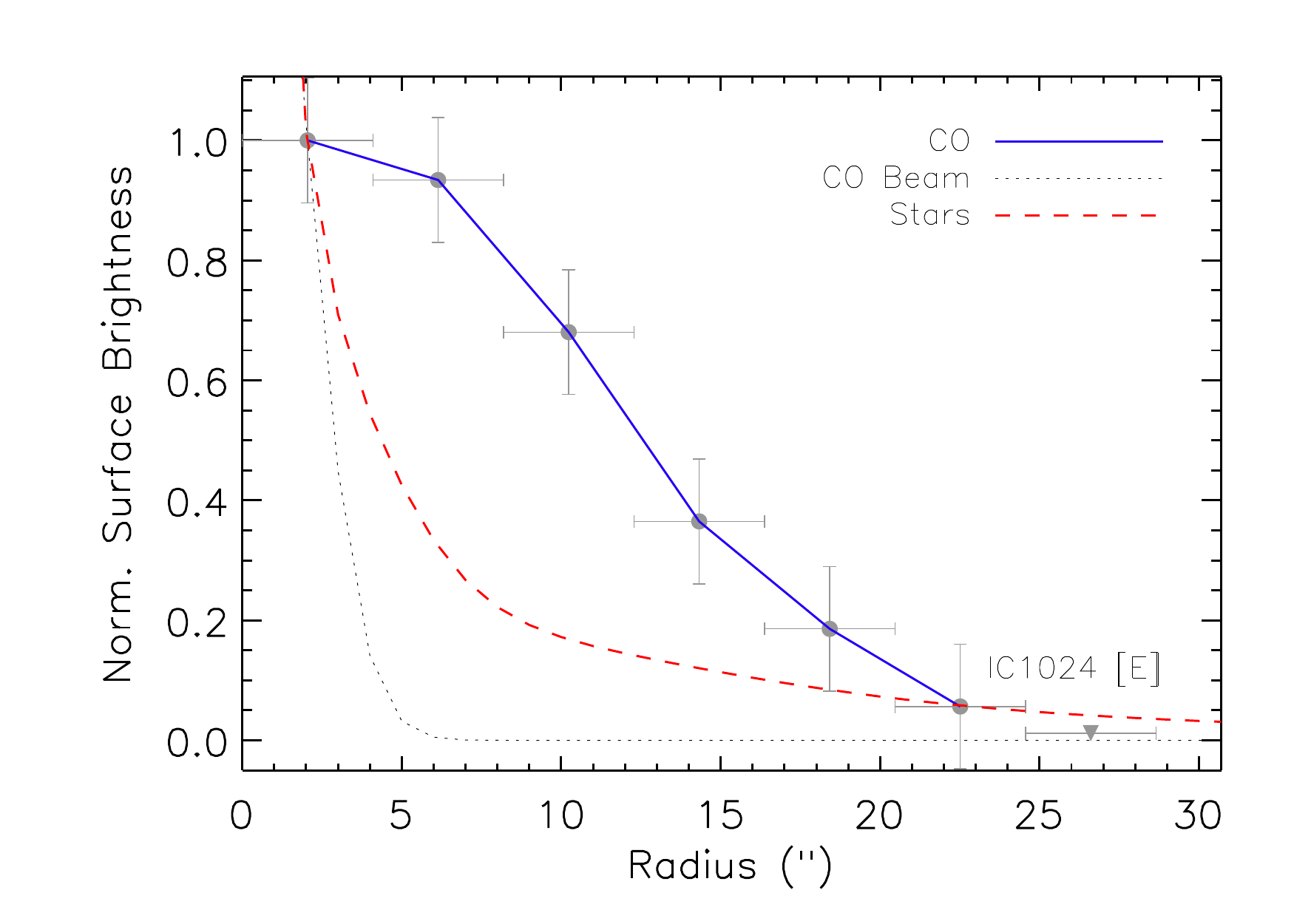}}
\subfigure{\includegraphics[scale=0.45]{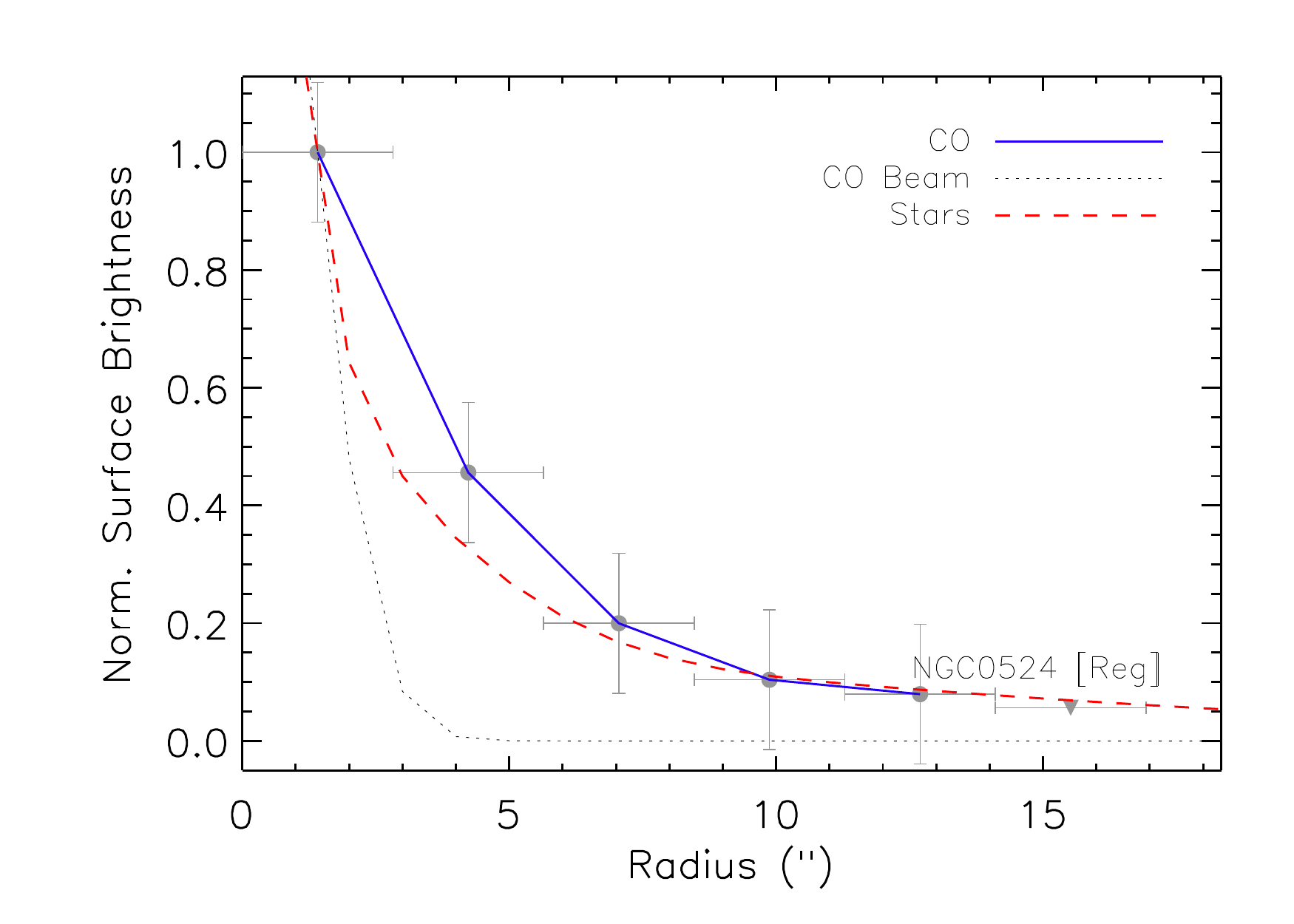}}
\subfigure{\includegraphics[scale=0.45]{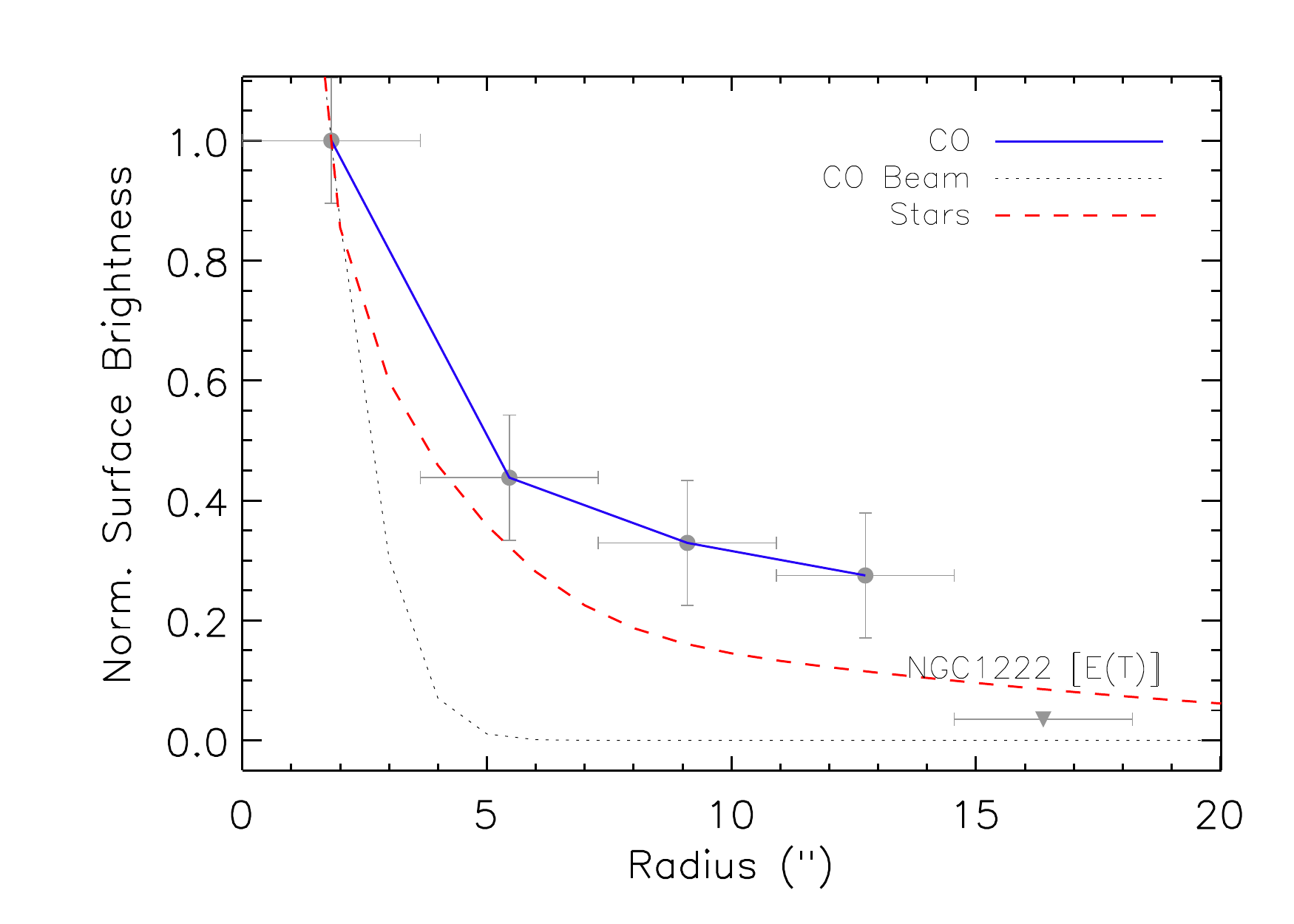}}
\subfigure{\includegraphics[scale=0.45]{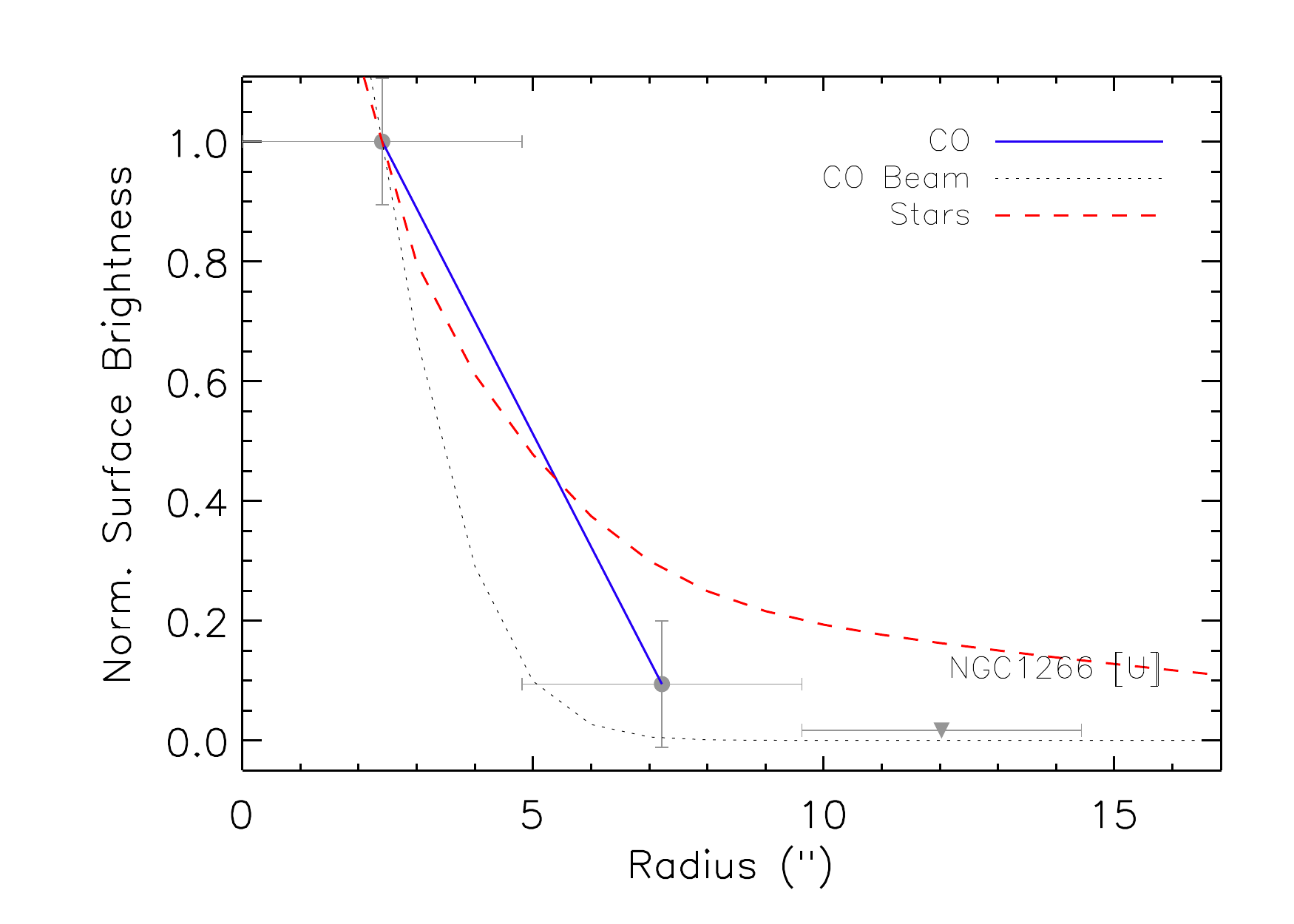}}
\parbox[t]{0.9 \textwidth}{ \caption{\small Radial surface brightness profiles of the stars (red dashed line; $r$-band) and molecular gas (blue solid line with errors) for our ATLAS$^{\rm 3D}$ ETGs (normalized at the first CO datapoint). The CO beamsize is shown as a black dotted line. The error bars on the CO measurements denote the width of the elliptical annuli in the x-direction and the RMS noise in the elliptical annulus in the y-direction. The letters at the bottom-right corner of each plot denote the profile class. "Reg" denoting a regular profile, "E" an extended profile, "E(T)" an extended profile with truncation, "R" a ring, "C" a composite, and "U" an unresolved profile. }}
\end{center}
\end{figure*}
\begin{figure*}
\begin{center}
\subfigure{\includegraphics[scale=0.45]{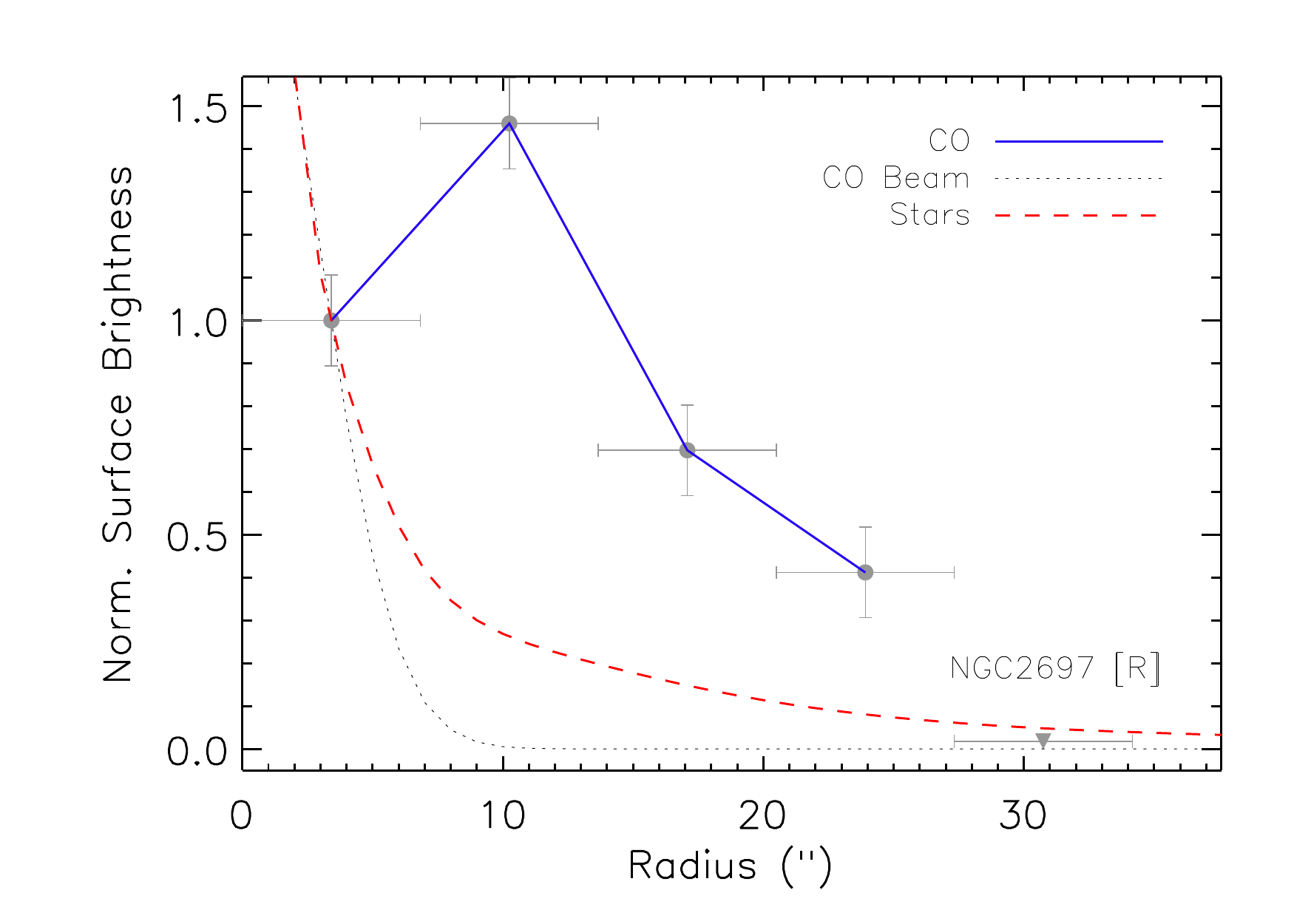}}
\subfigure{\includegraphics[scale=0.45]{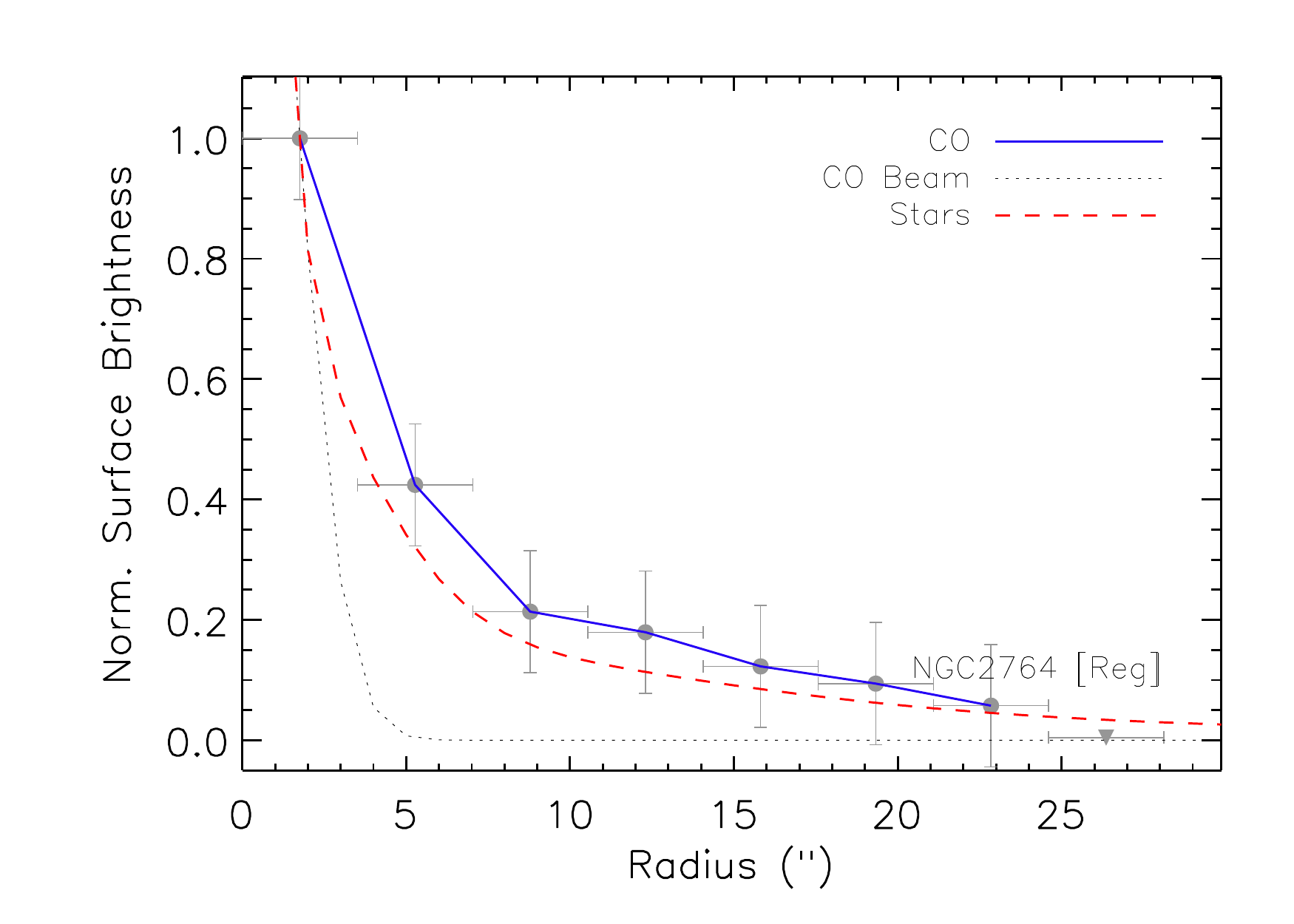}}
\subfigure{\includegraphics[scale=0.45]{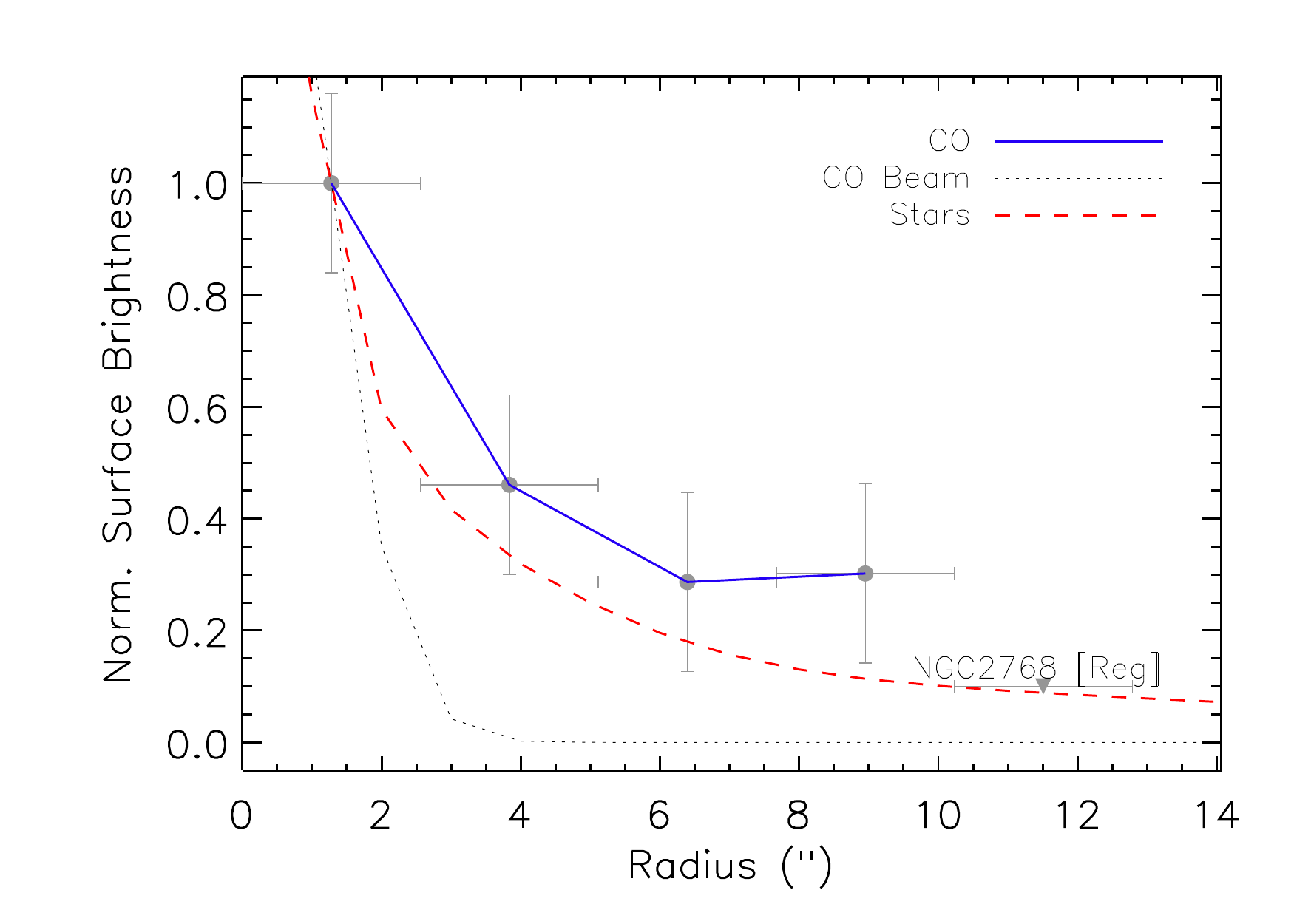}}
\subfigure{\includegraphics[scale=0.45]{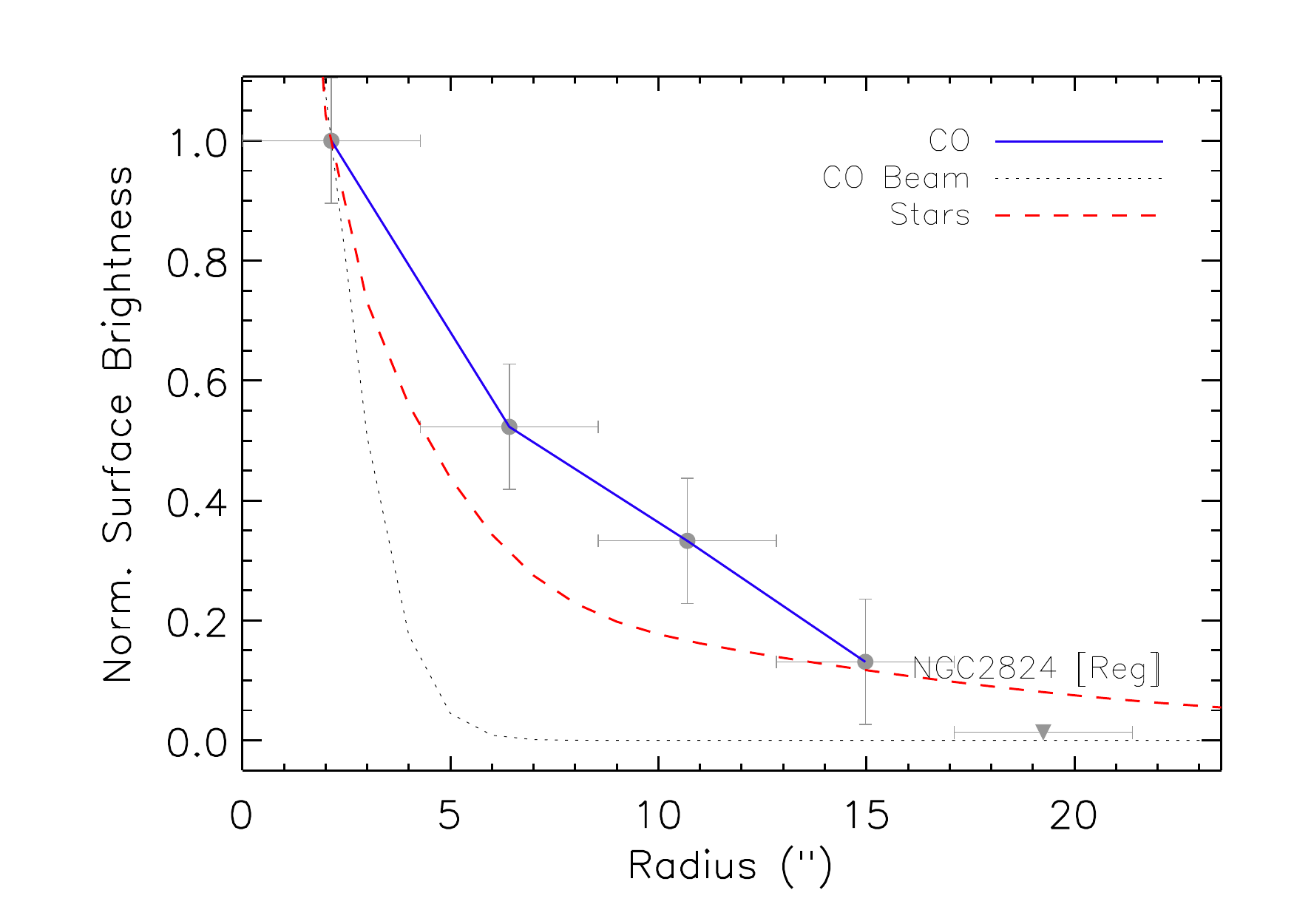}}
\subfigure{\includegraphics[scale=0.45]{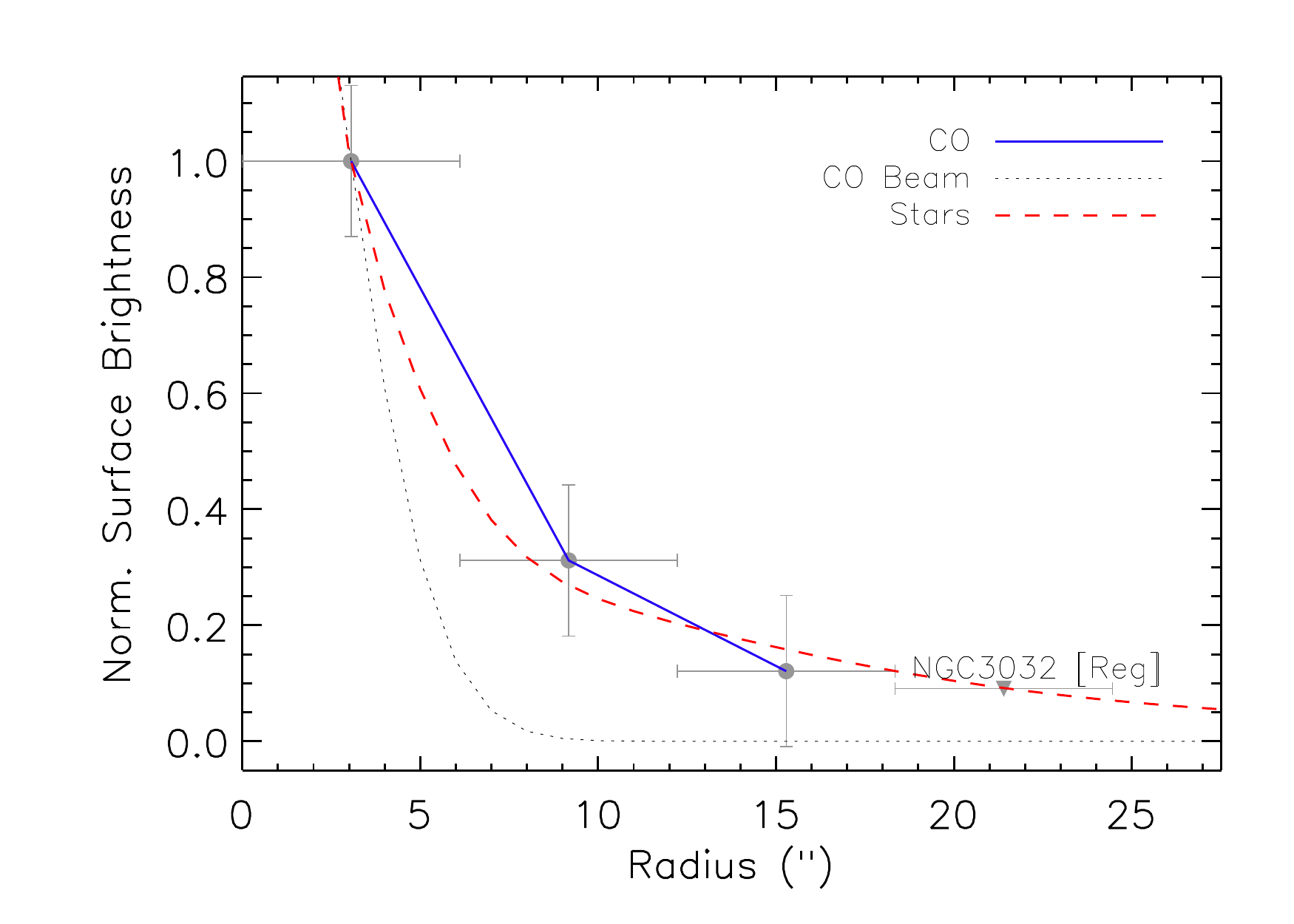}}
\subfigure{\includegraphics[scale=0.45]{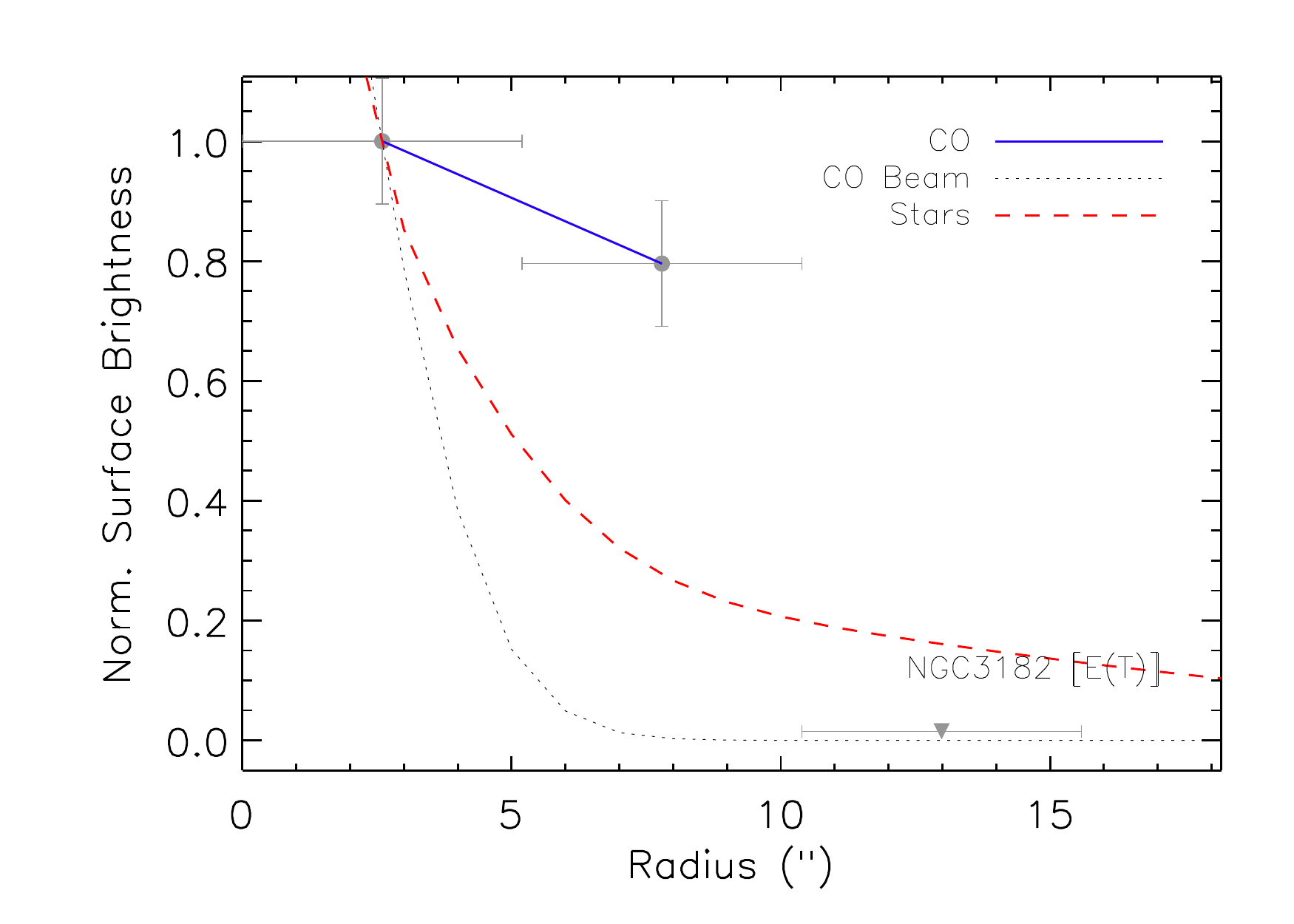}}
\subfigure{\includegraphics[scale=0.45]{sdplots/radial_sd_NGC3489.pdf}}
\subfigure{\includegraphics[scale=0.45]{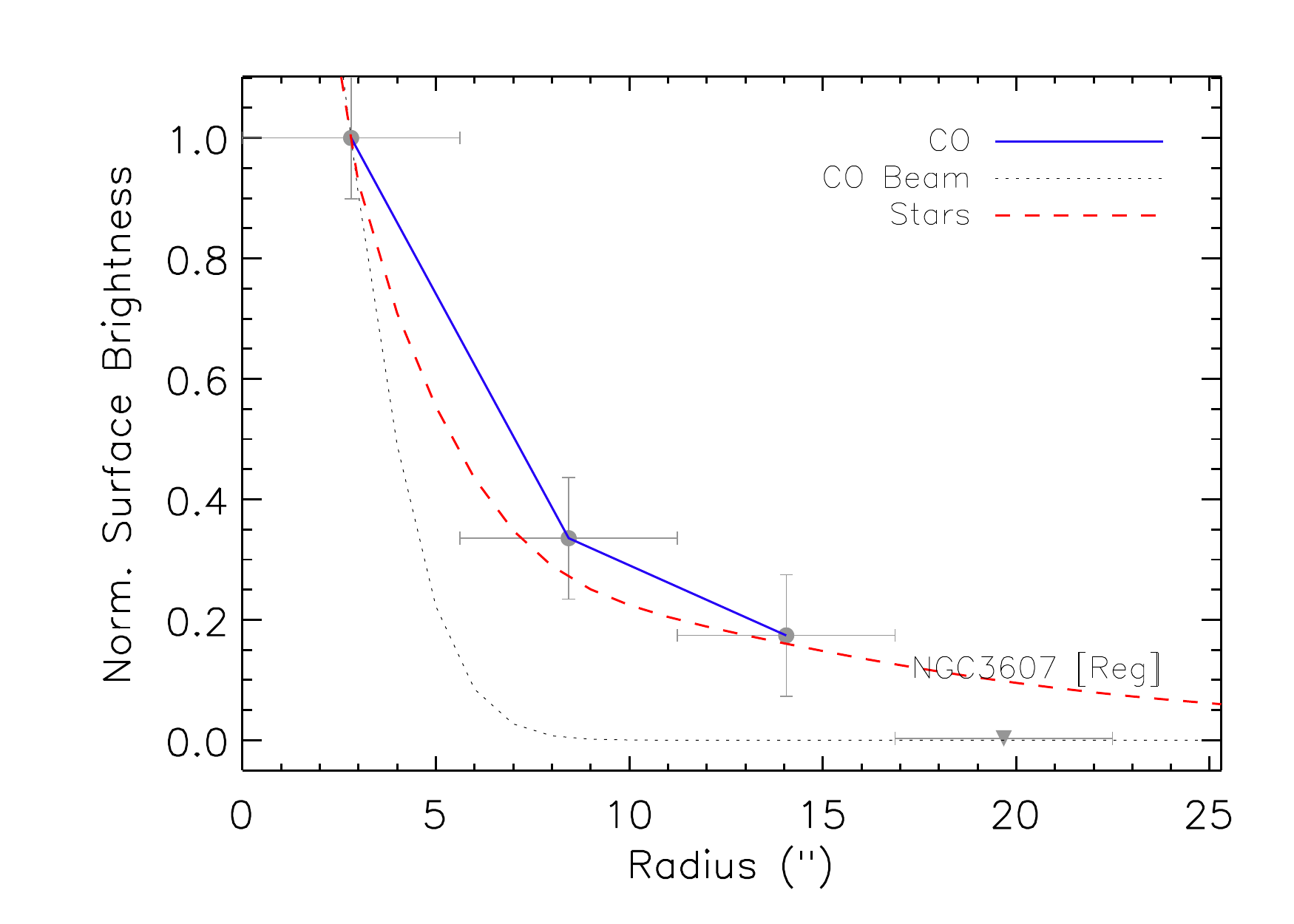}}
\parbox[t]{0.9 \textwidth}{\textbf{Figure A1.} continued}
\end{center}
\end{figure*}
\begin{figure*}
\begin{center}
\subfigure{\includegraphics[scale=0.45]{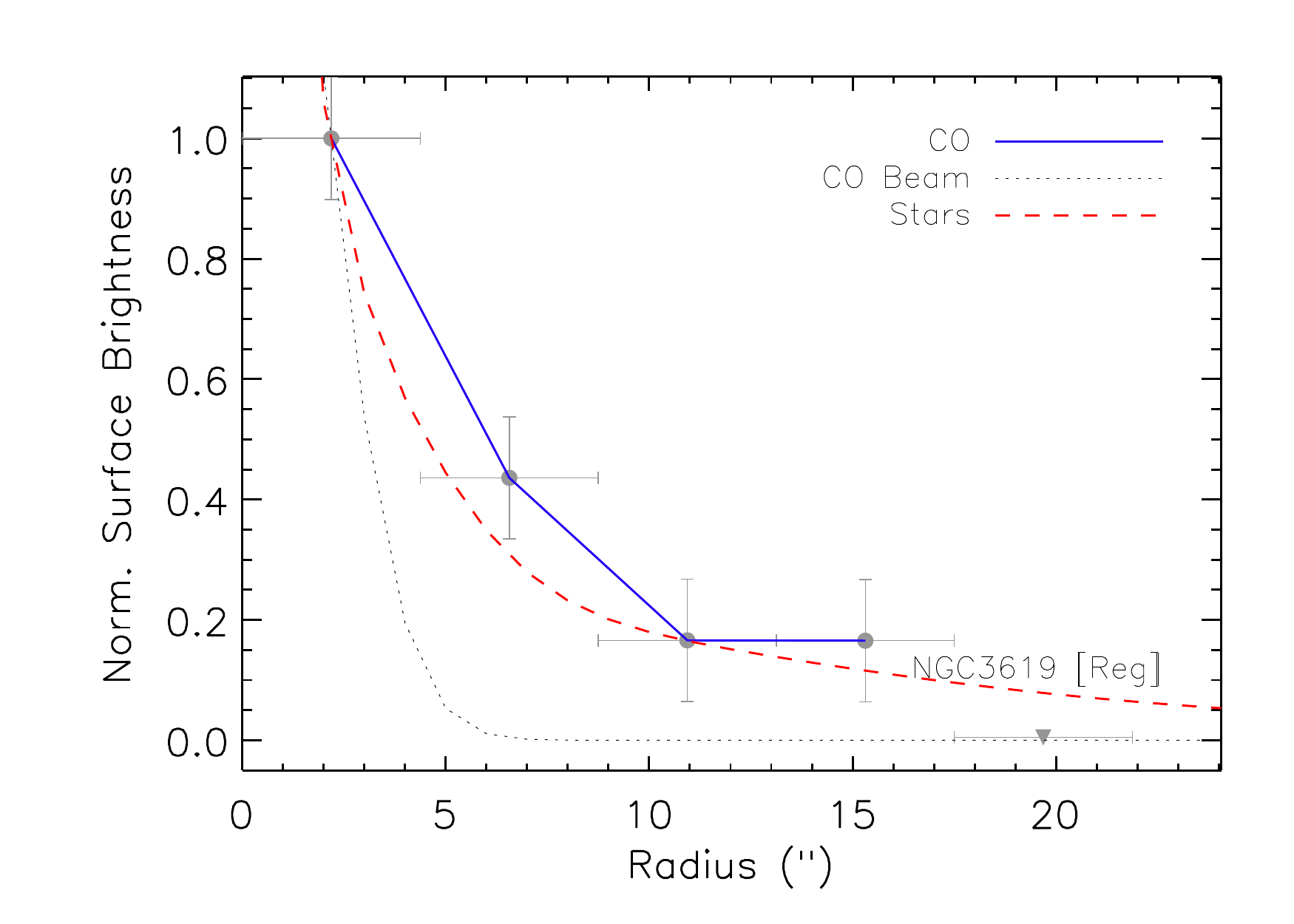}}
\subfigure{\includegraphics[scale=0.45]{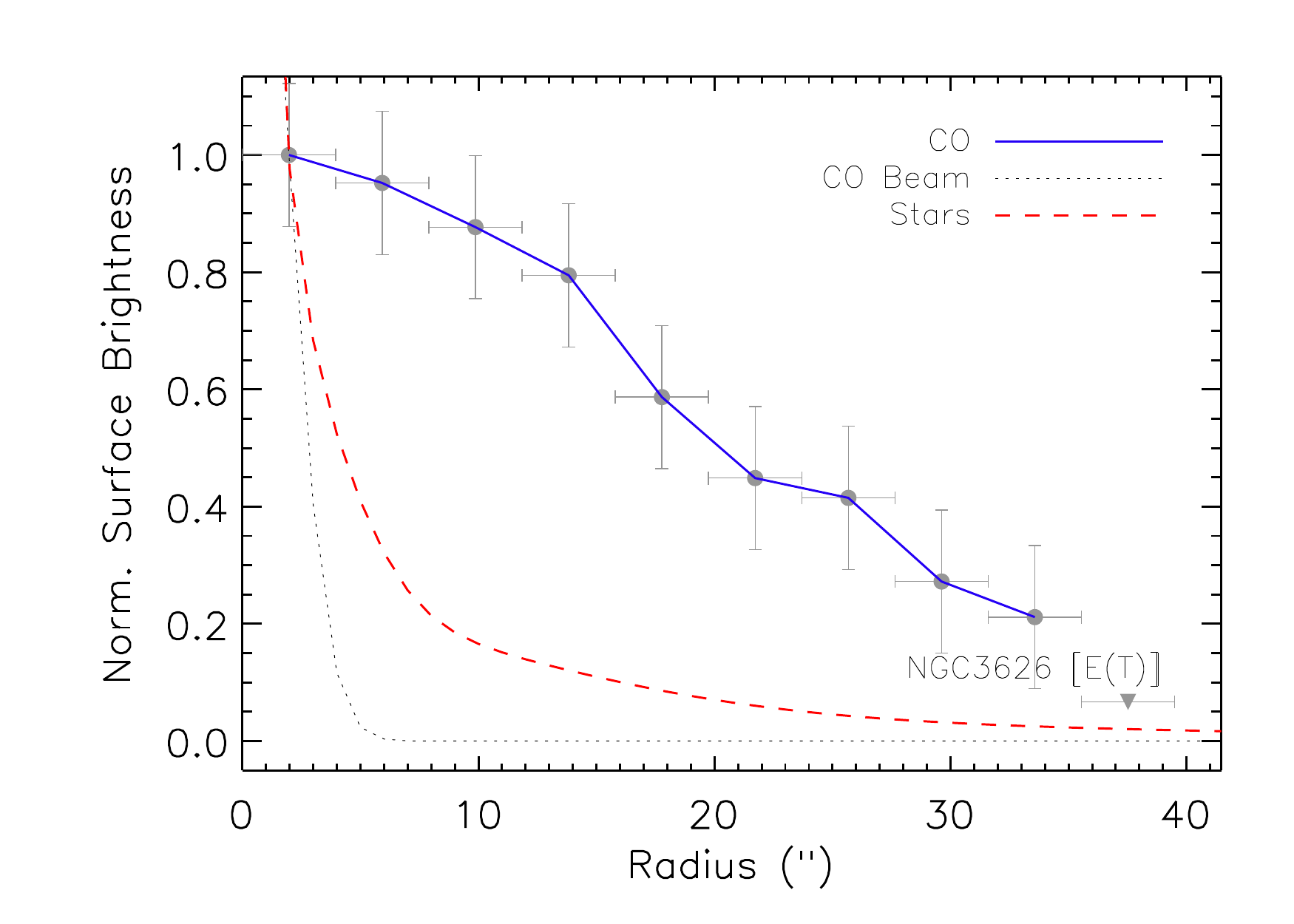}}
\subfigure{\includegraphics[scale=0.45]{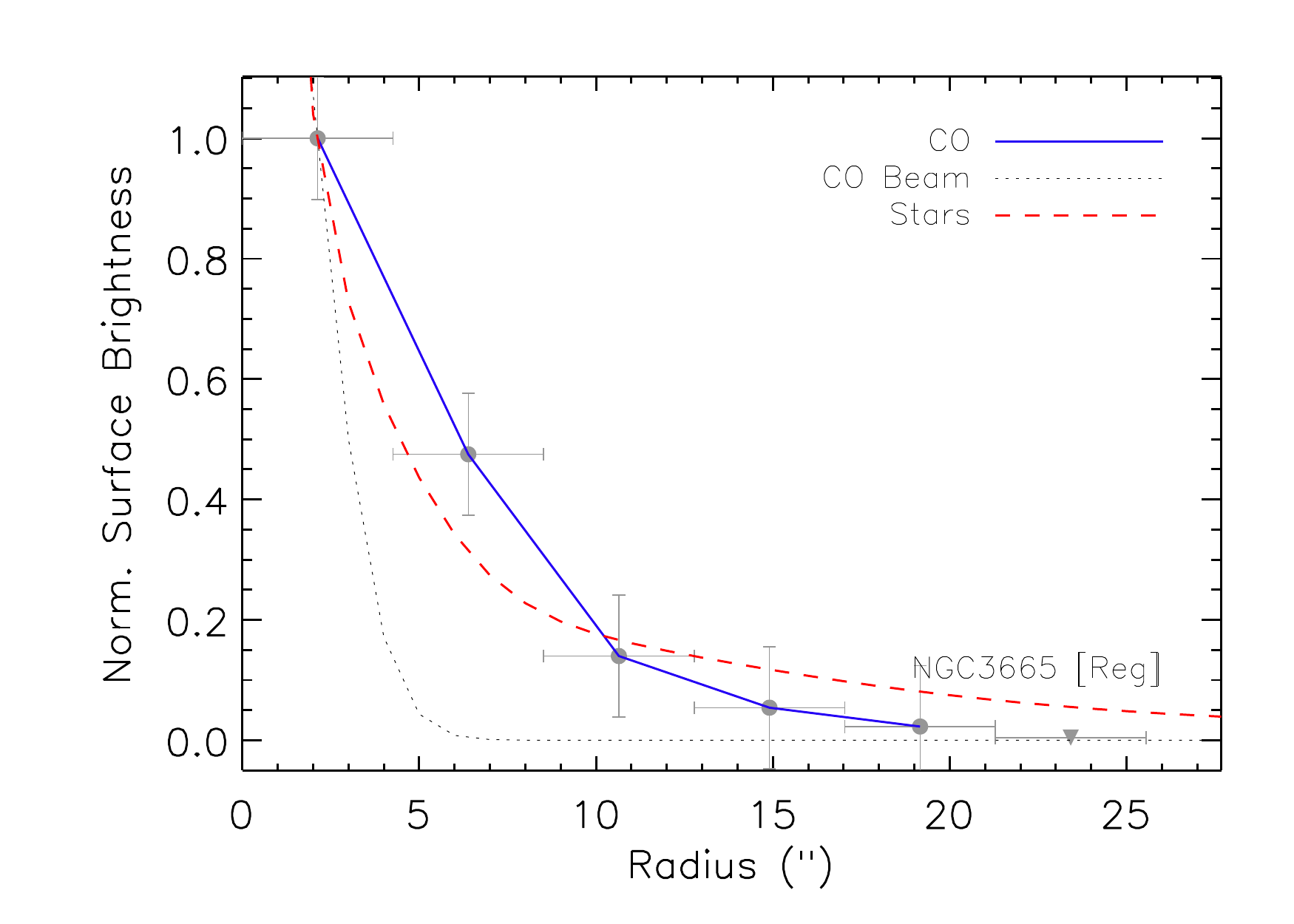}}
\subfigure{\includegraphics[scale=0.45]{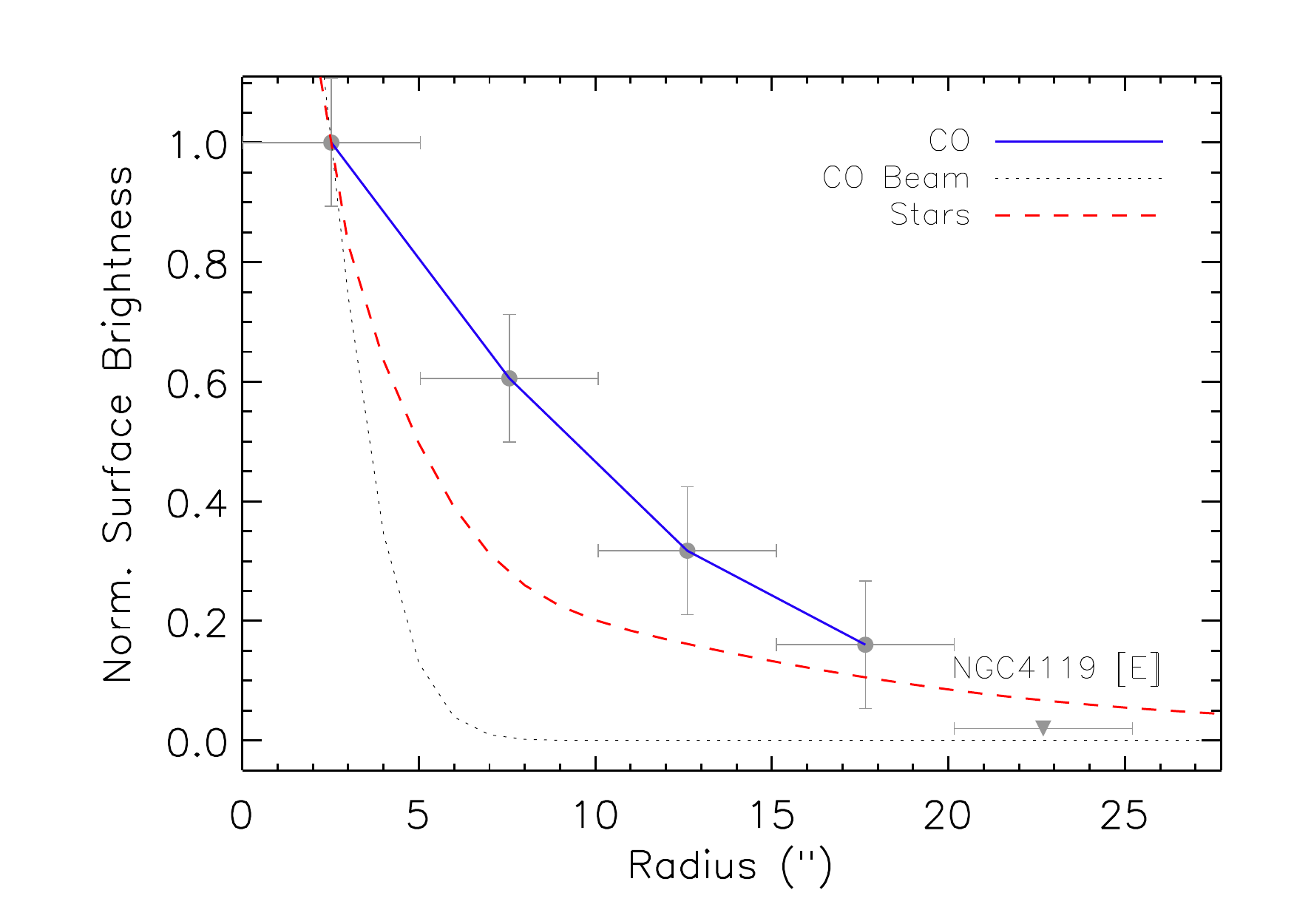}}
\subfigure{\includegraphics[scale=0.45]{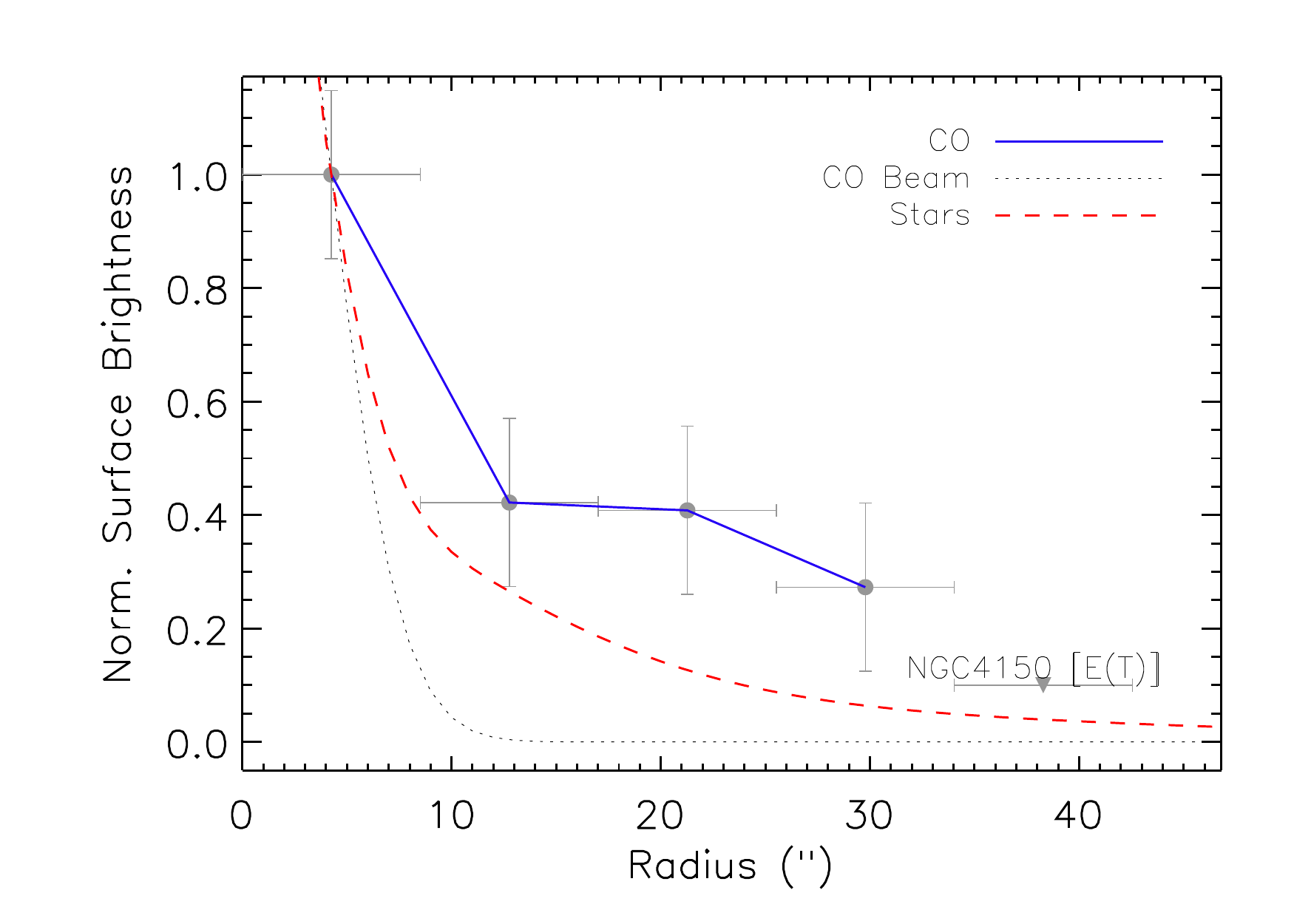}}
\subfigure{\includegraphics[scale=0.45]{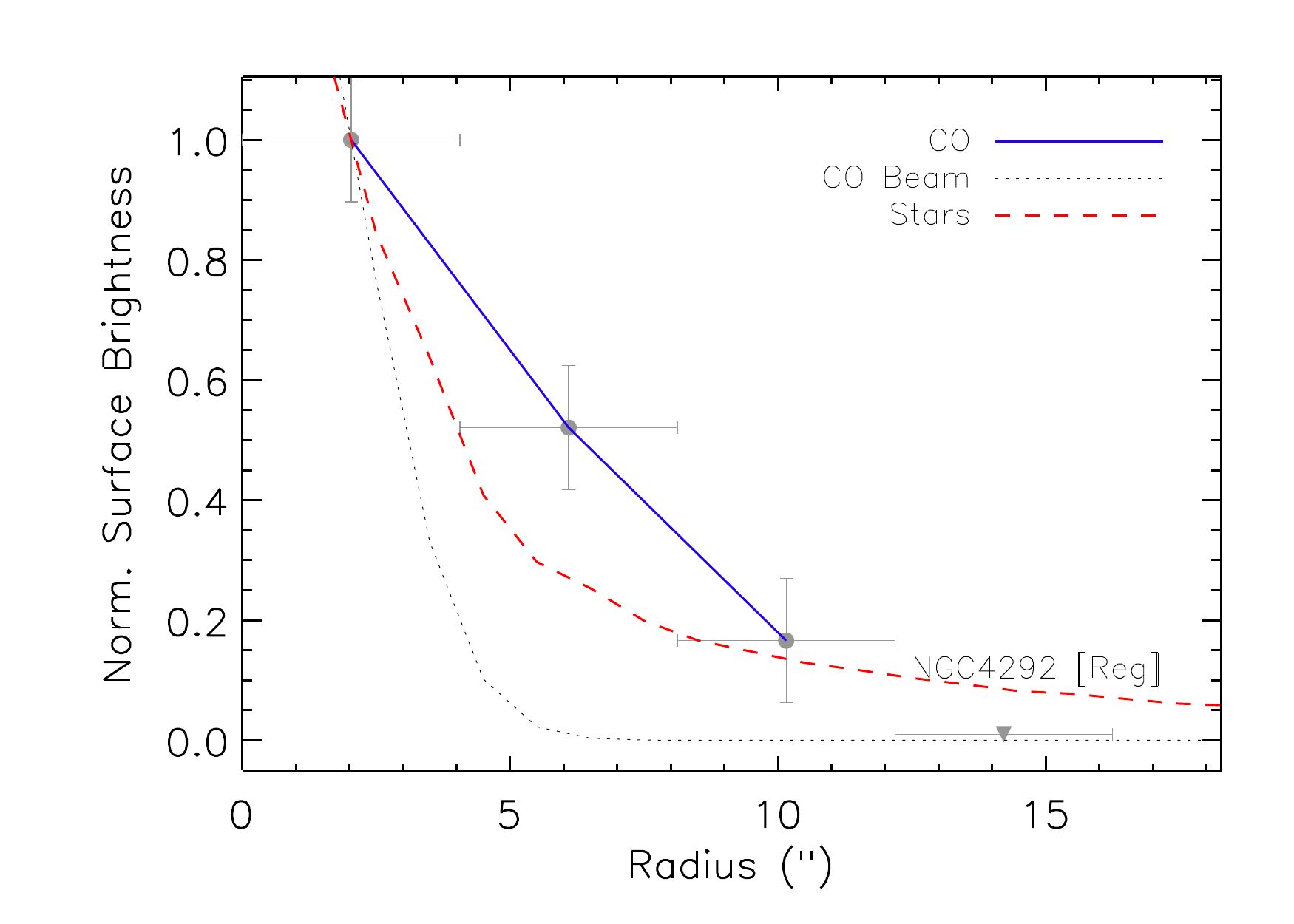}}
\subfigure{\includegraphics[scale=0.45]{sdplots/radial_sd_NGC4324.pdf}}
\subfigure{\includegraphics[scale=0.45]{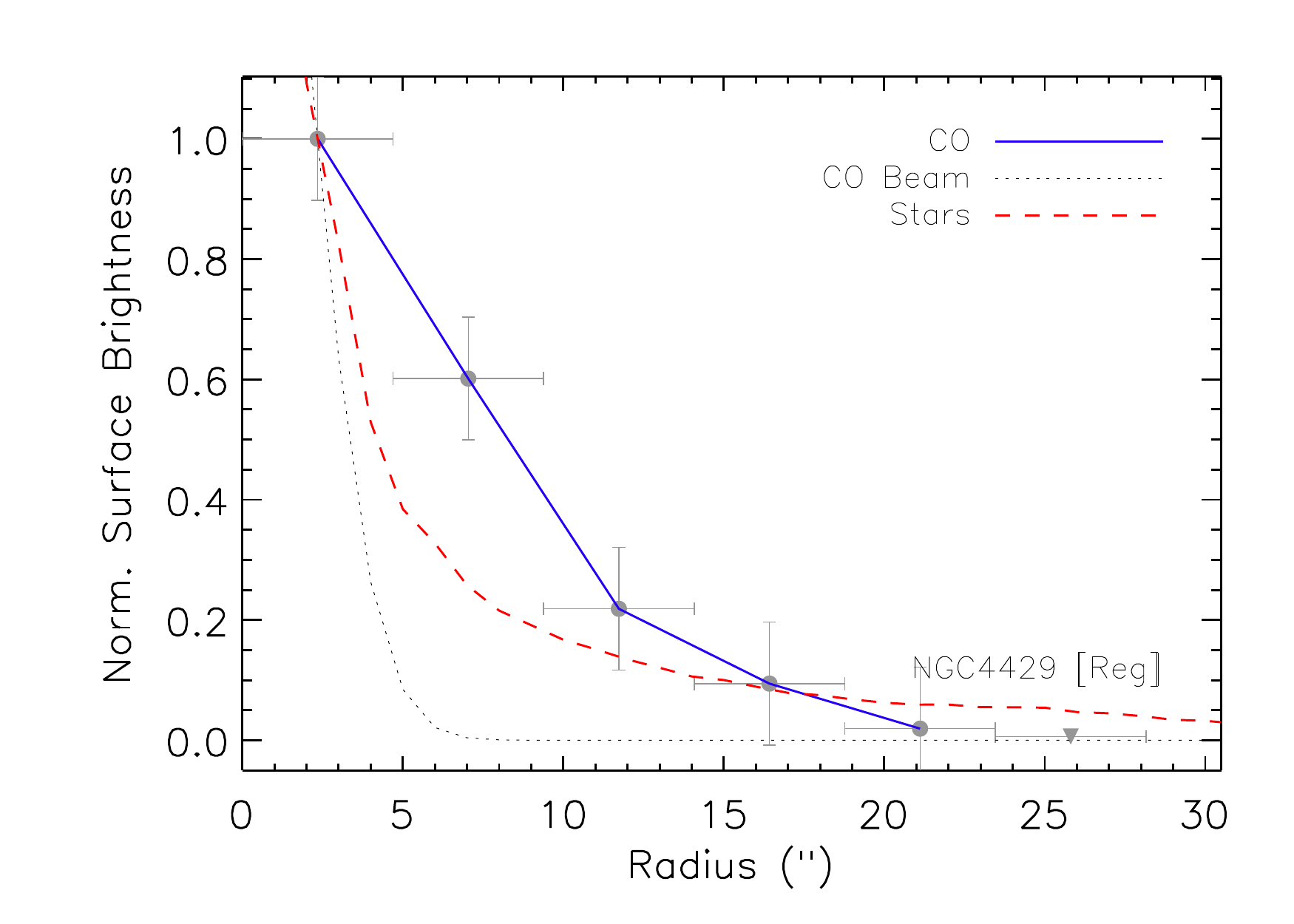}}
\parbox[t]{0.9 \textwidth}{\textbf{Figure A1.} continued}
\end{center}
\end{figure*}
\begin{figure*}
\begin{center}
\subfigure{\includegraphics[scale=0.45]{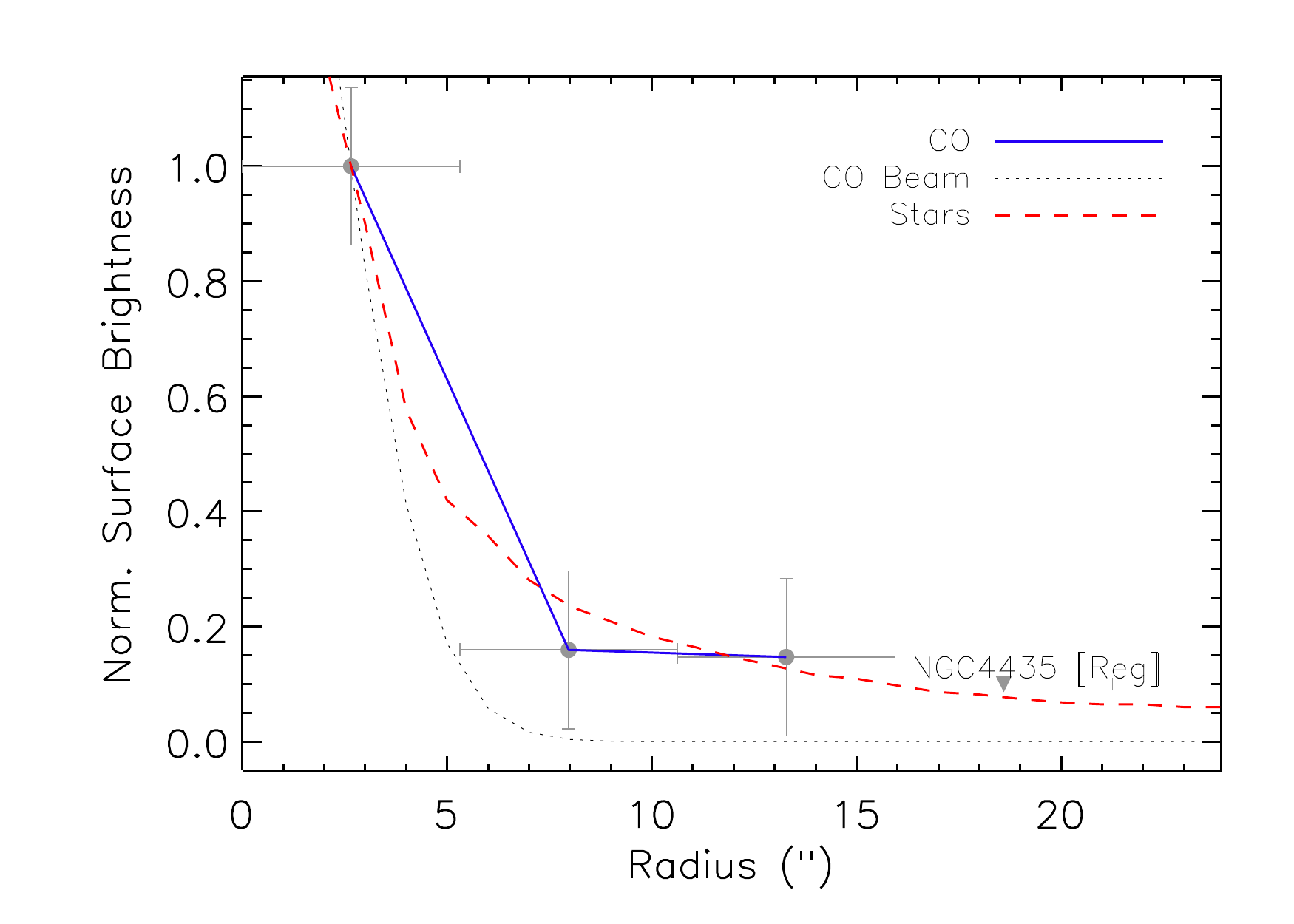}}
\subfigure{\includegraphics[scale=0.45]{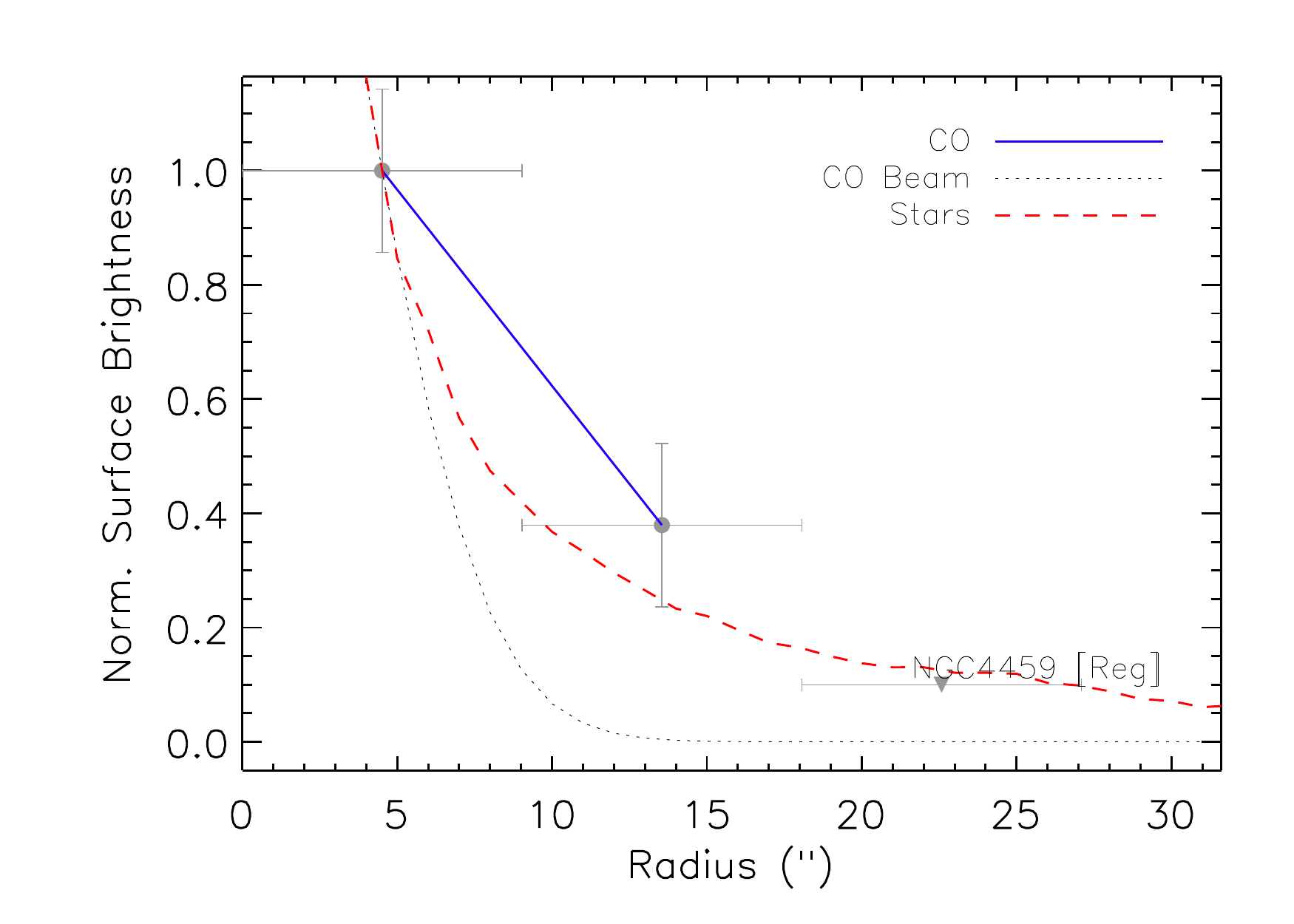}}
\subfigure{\includegraphics[scale=0.45]{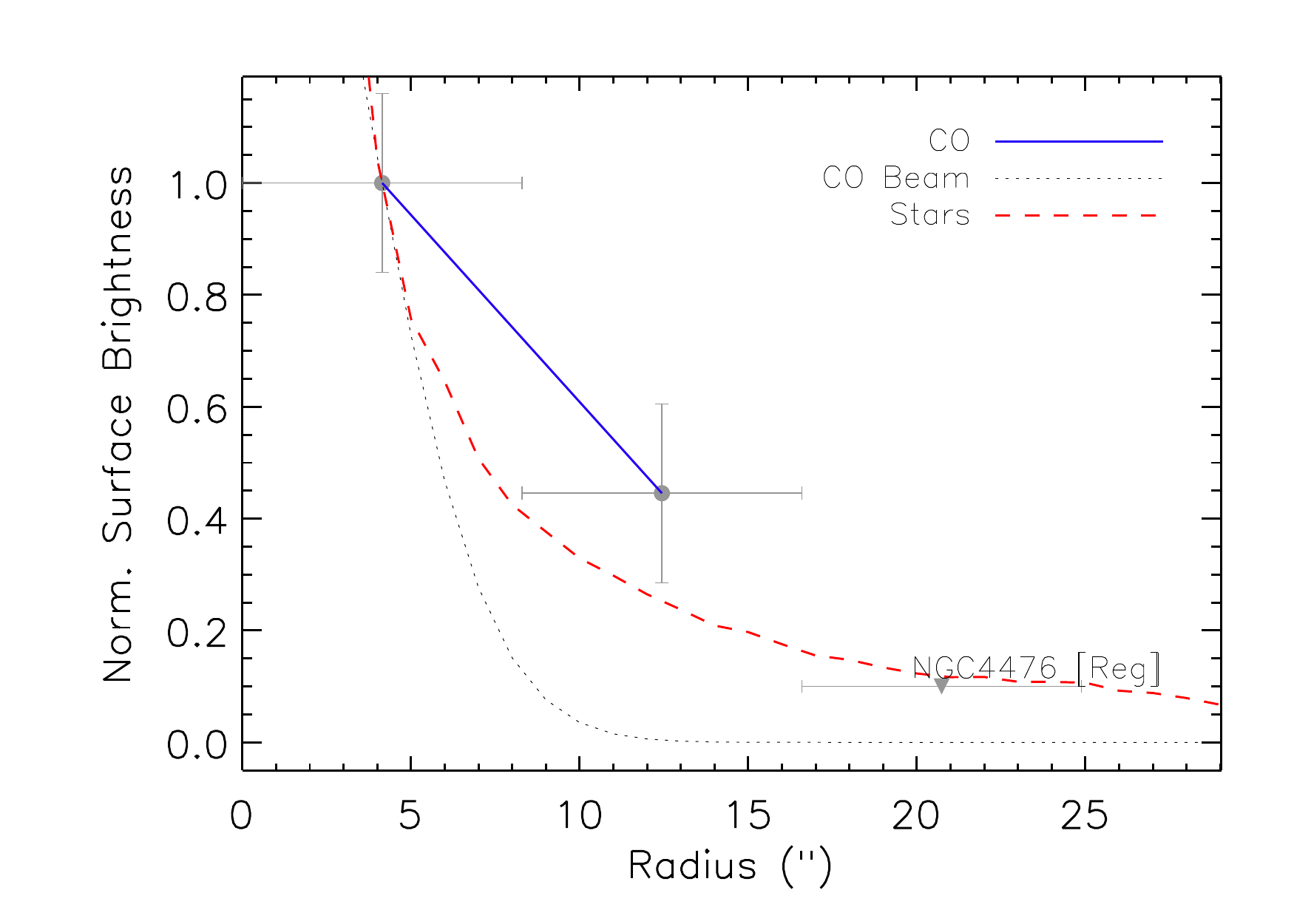}}
\subfigure{\includegraphics[scale=0.45]{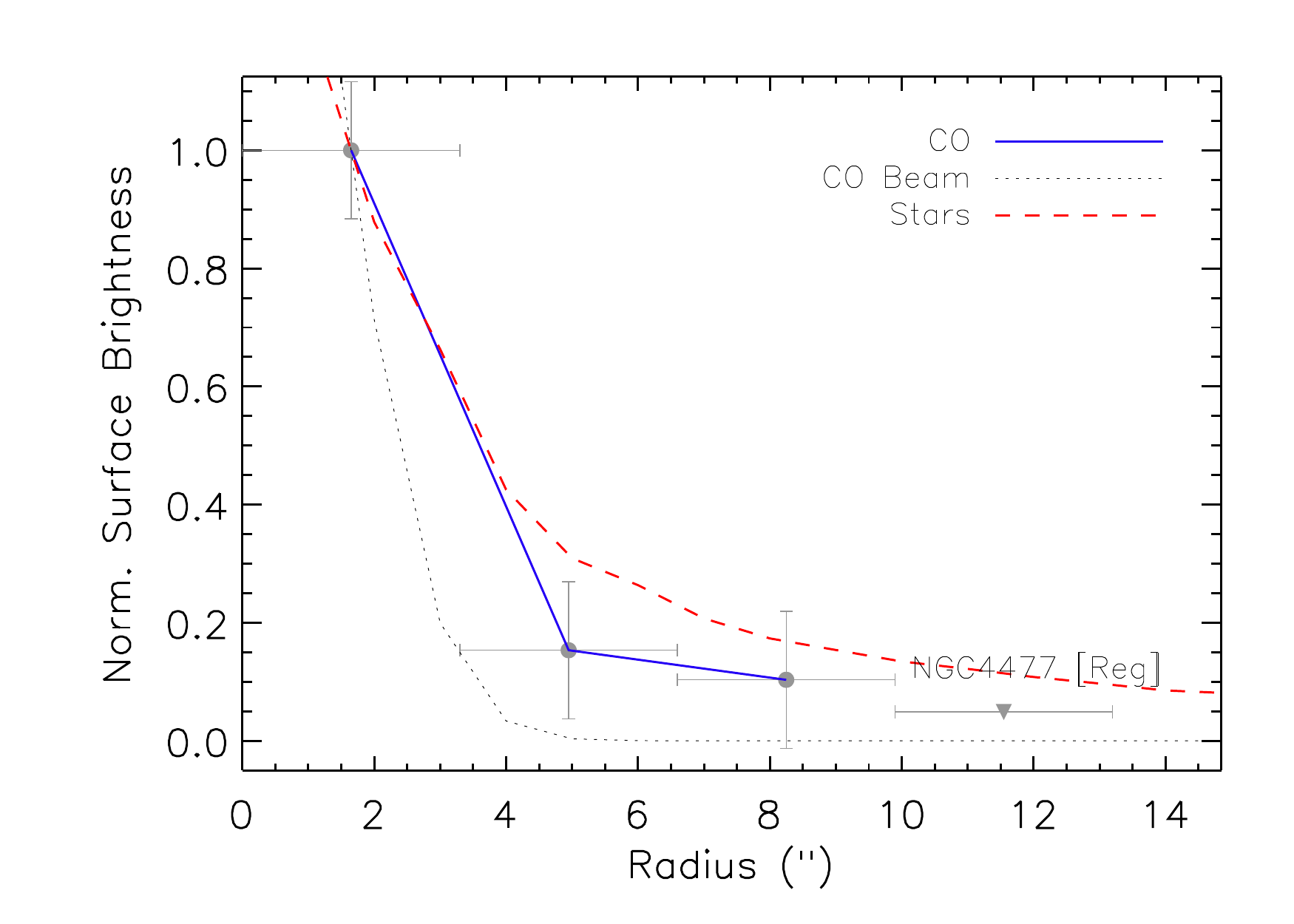}}
\subfigure{\includegraphics[scale=0.45]{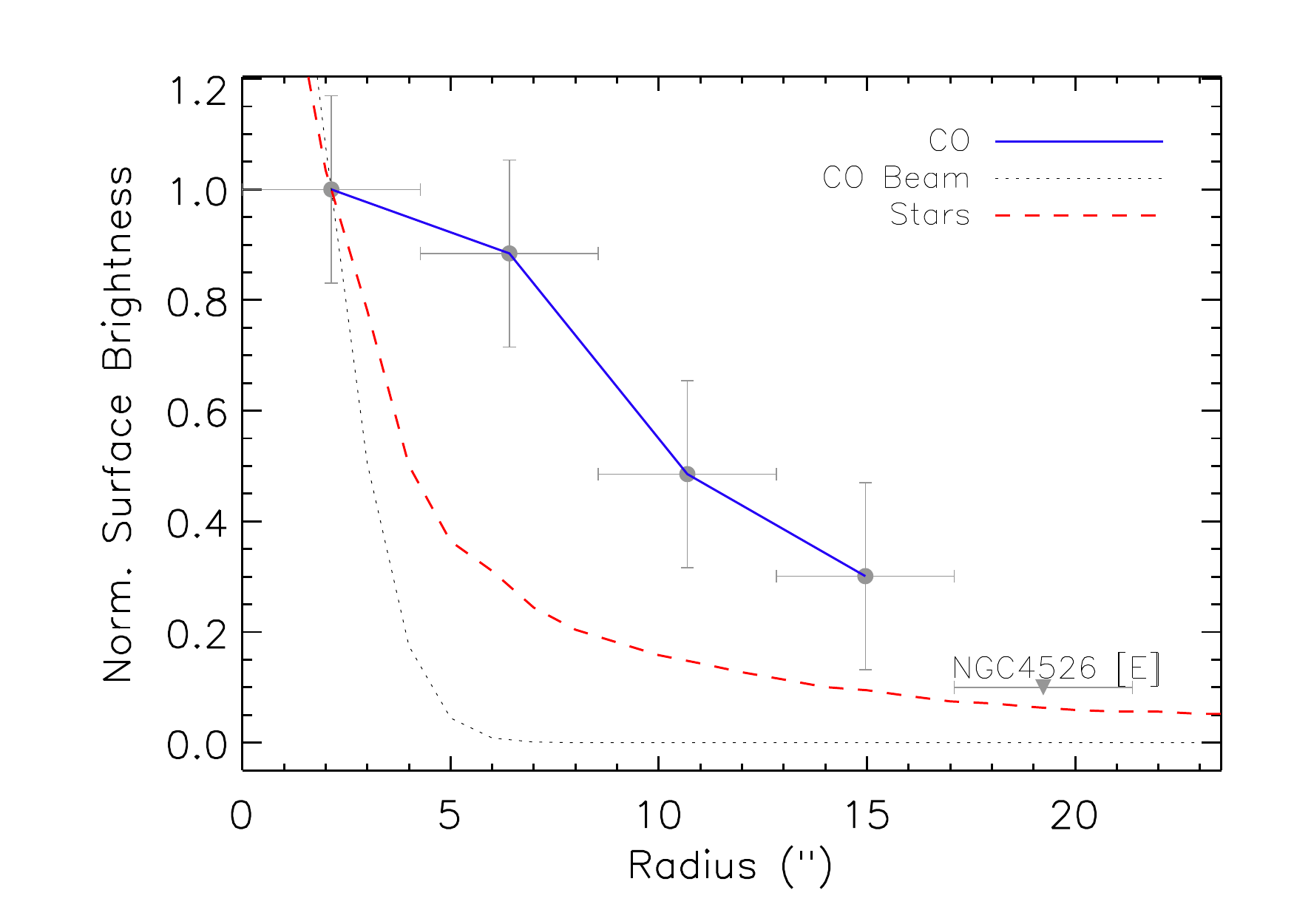}}
\subfigure{\includegraphics[scale=0.45]{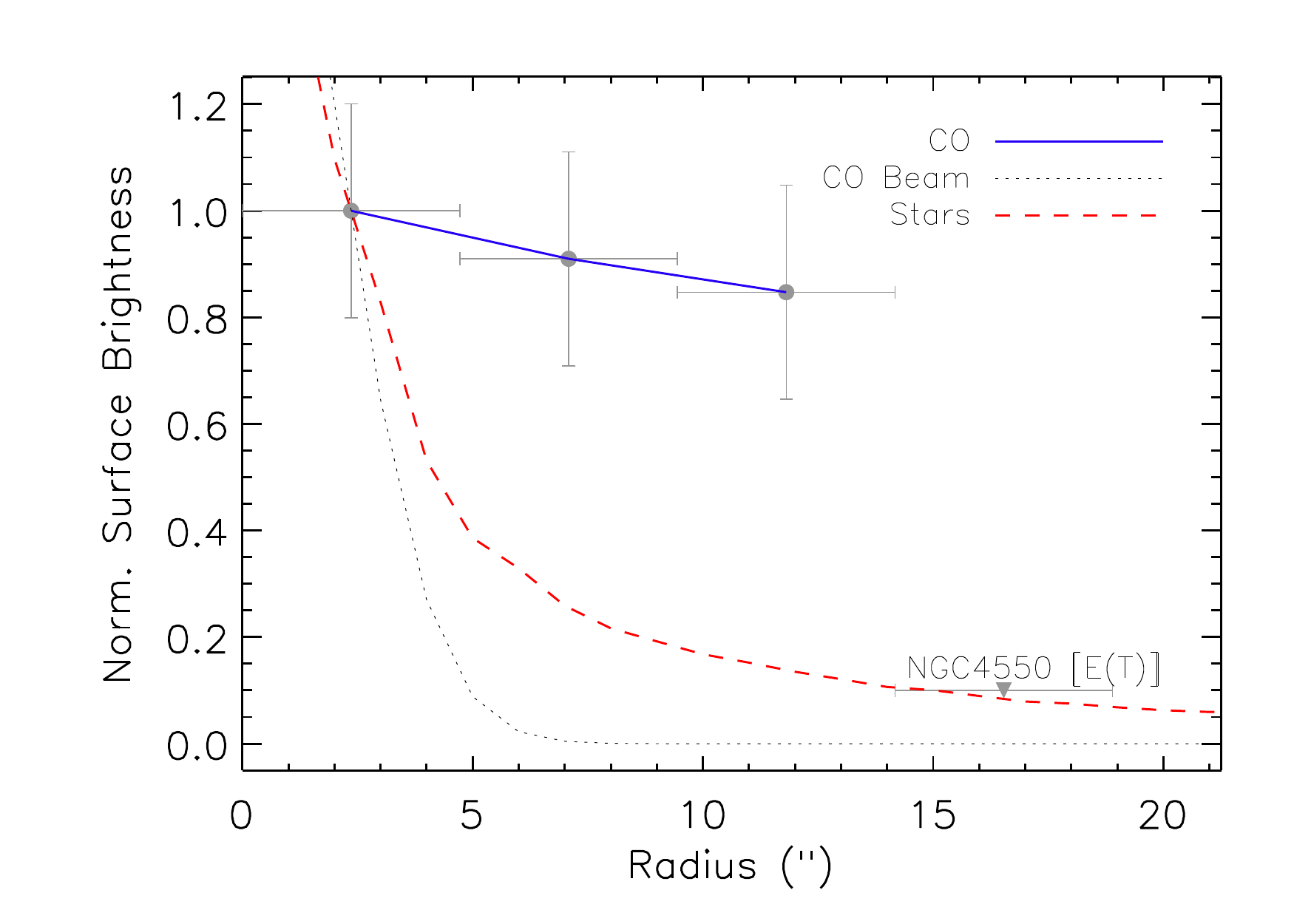}}
\subfigure{\includegraphics[scale=0.45]{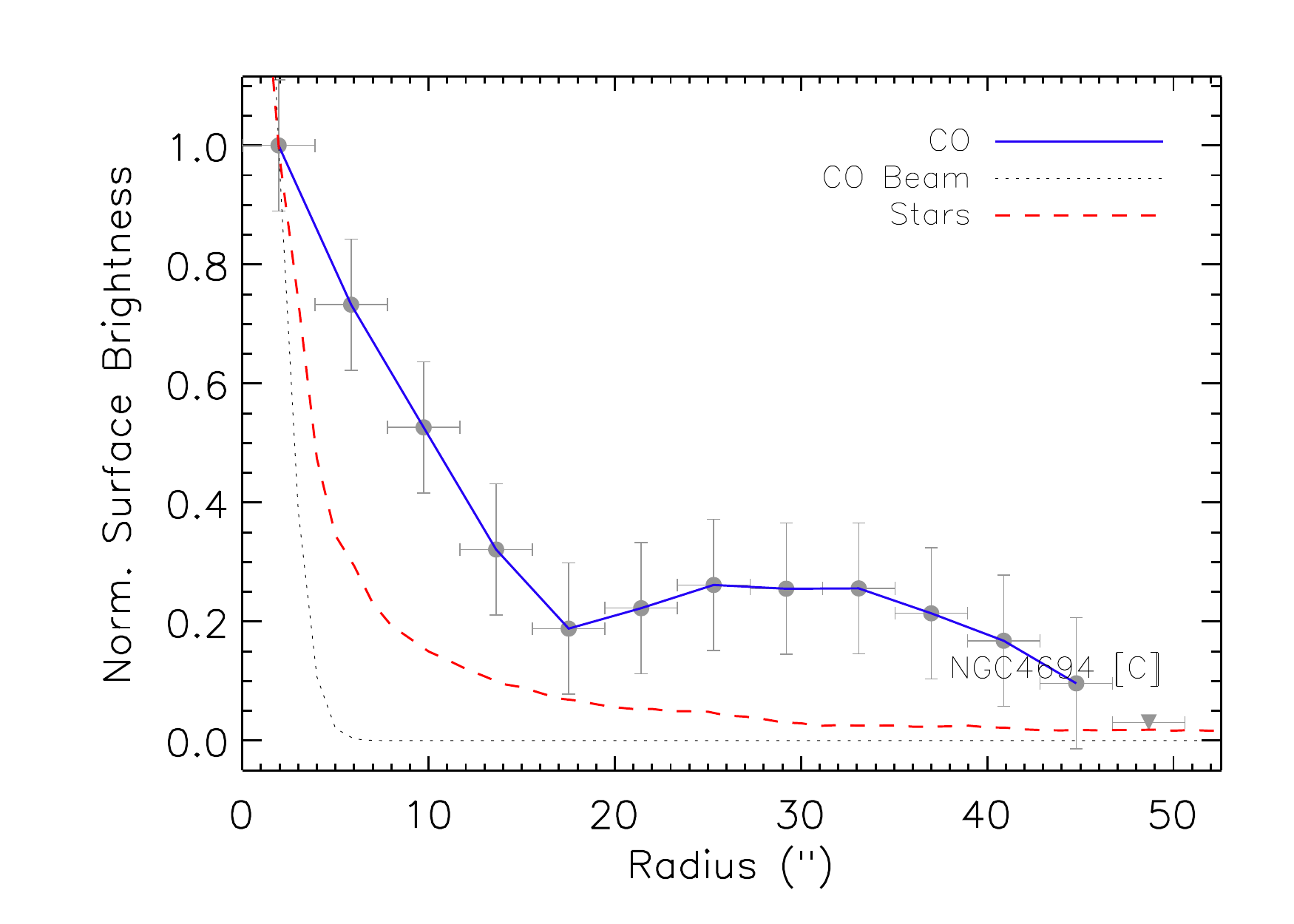}}
\subfigure{\includegraphics[scale=0.45]{sdplots/radial_sd_NGC4710.pdf}}
\parbox[t]{0.9 \textwidth}{\textbf{Figure A1.} continued}
\end{center}
\end{figure*}
\begin{figure*}
\begin{center}
\subfigure{\includegraphics[scale=0.45]{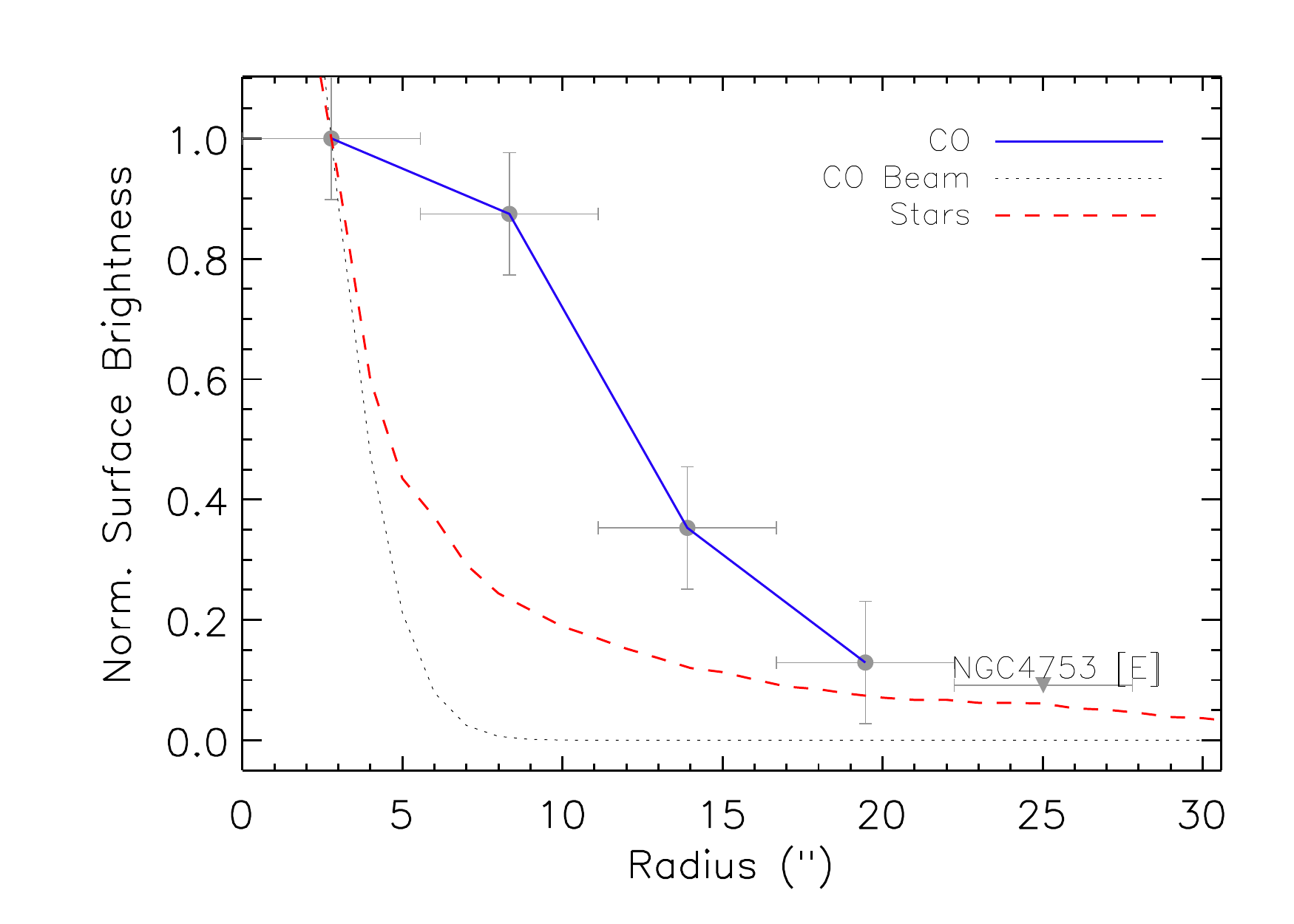}}
\subfigure{\includegraphics[scale=0.45]{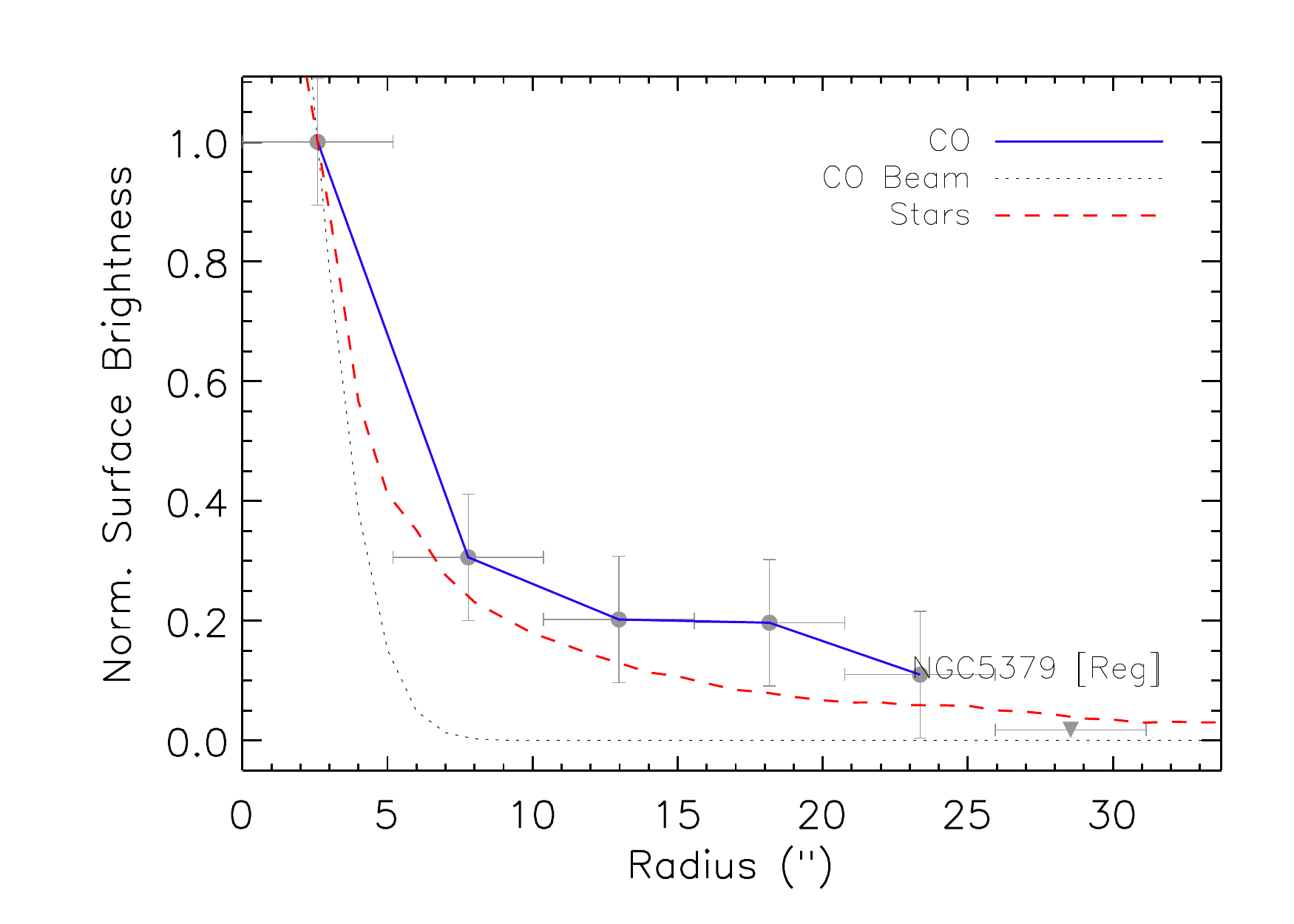}}
\subfigure{\includegraphics[scale=0.45]{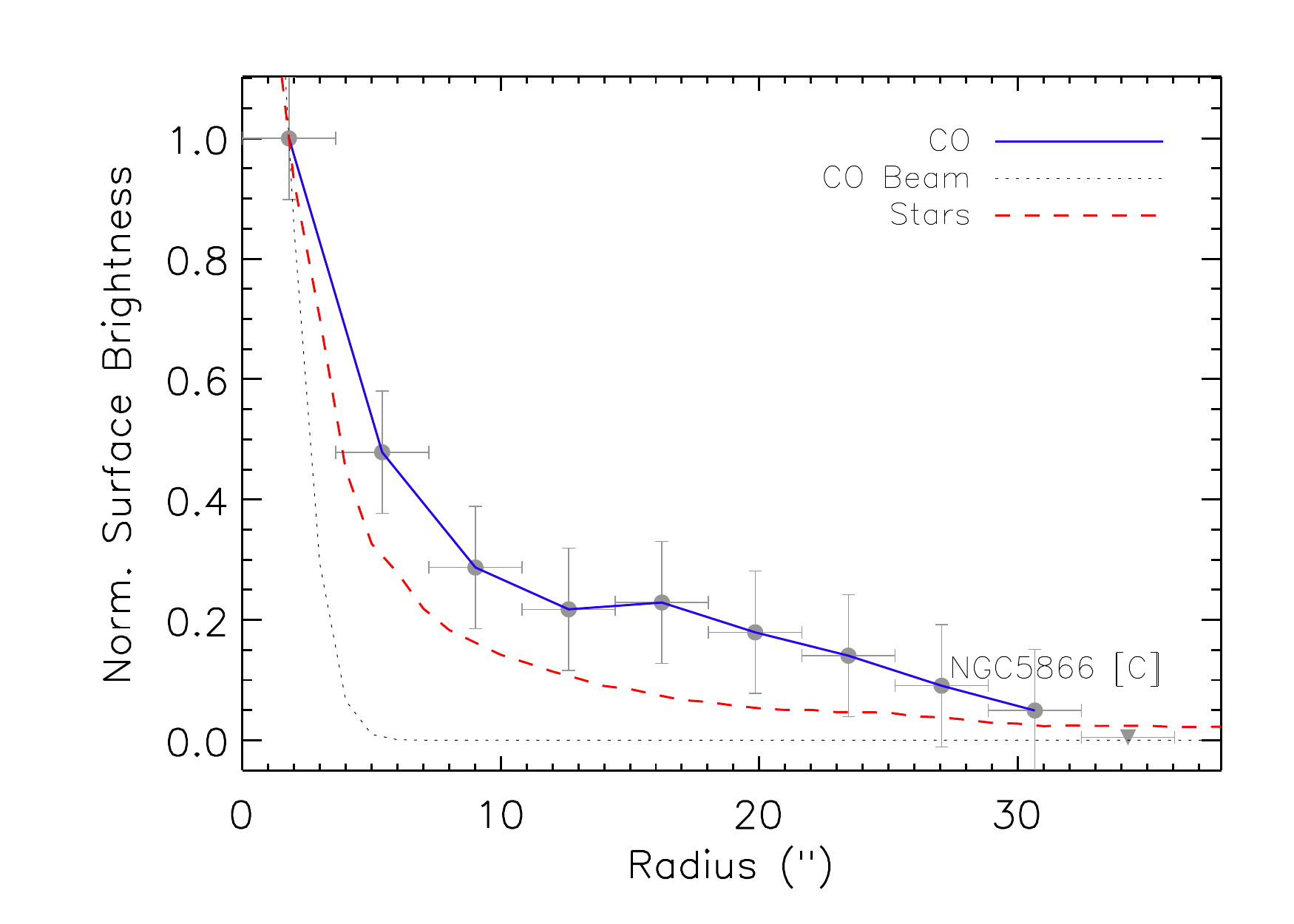}}
\subfigure{\includegraphics[scale=0.45]{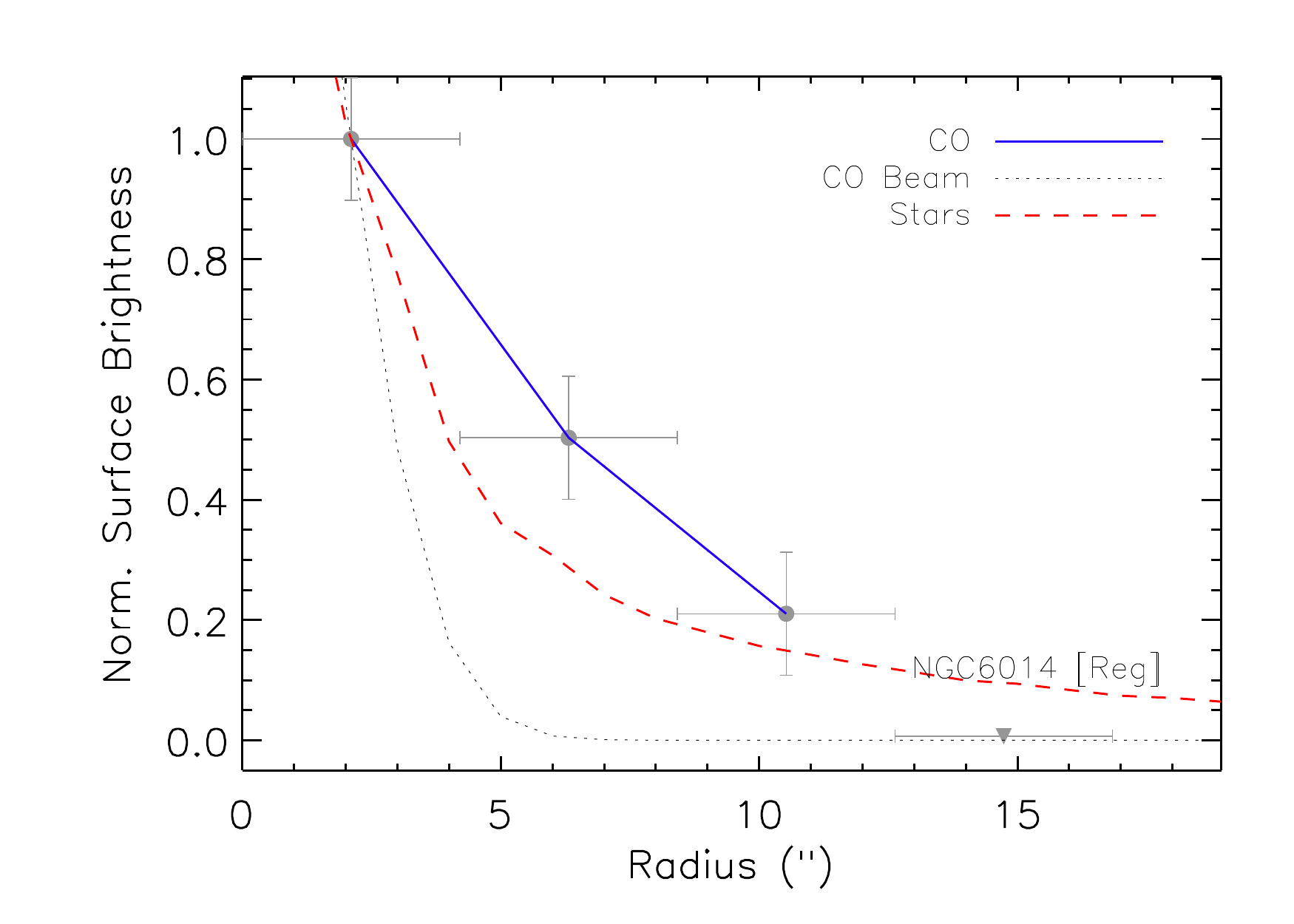}}
\subfigure{\includegraphics[scale=0.45]{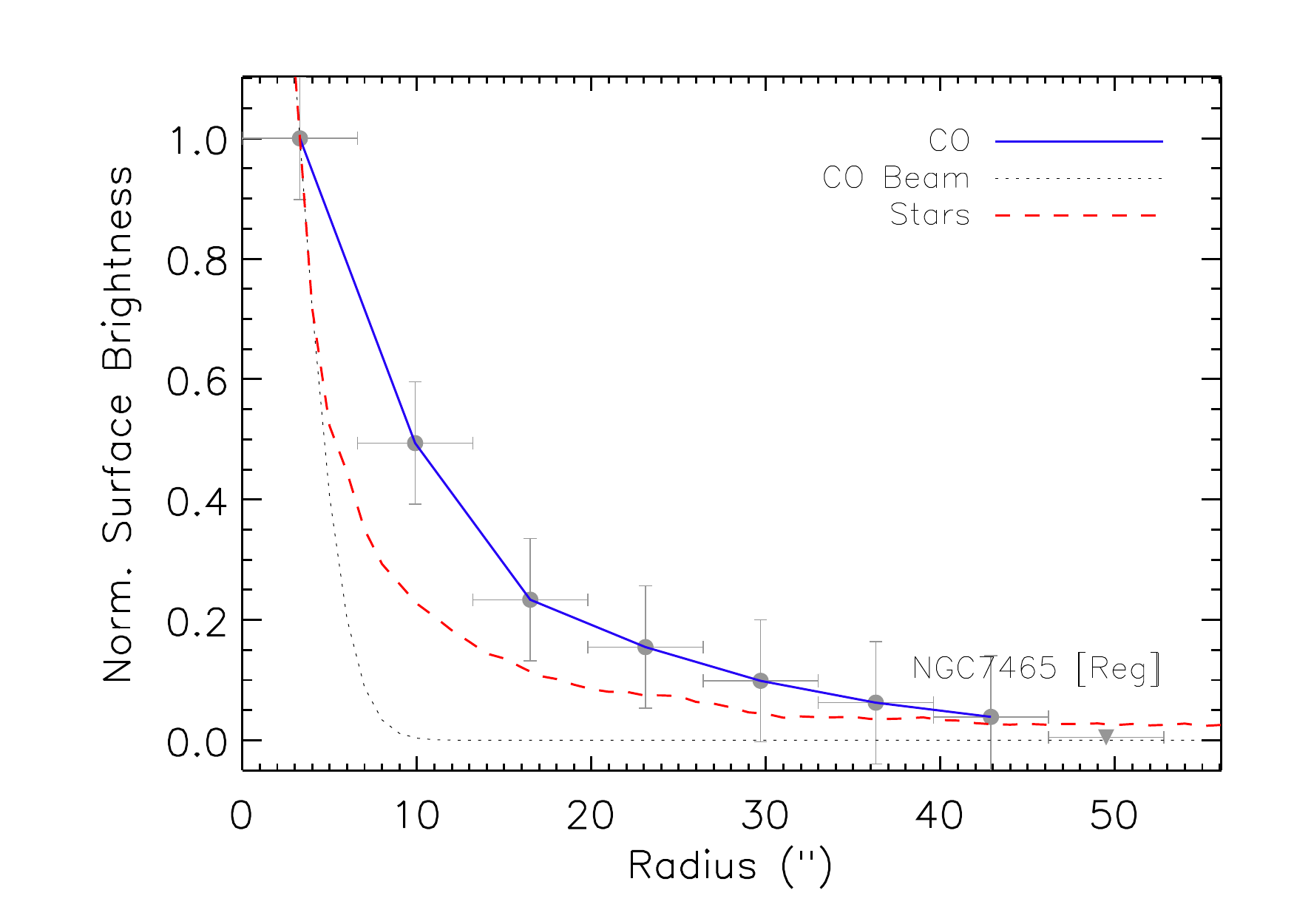}}
\subfigure{\includegraphics[scale=0.45]{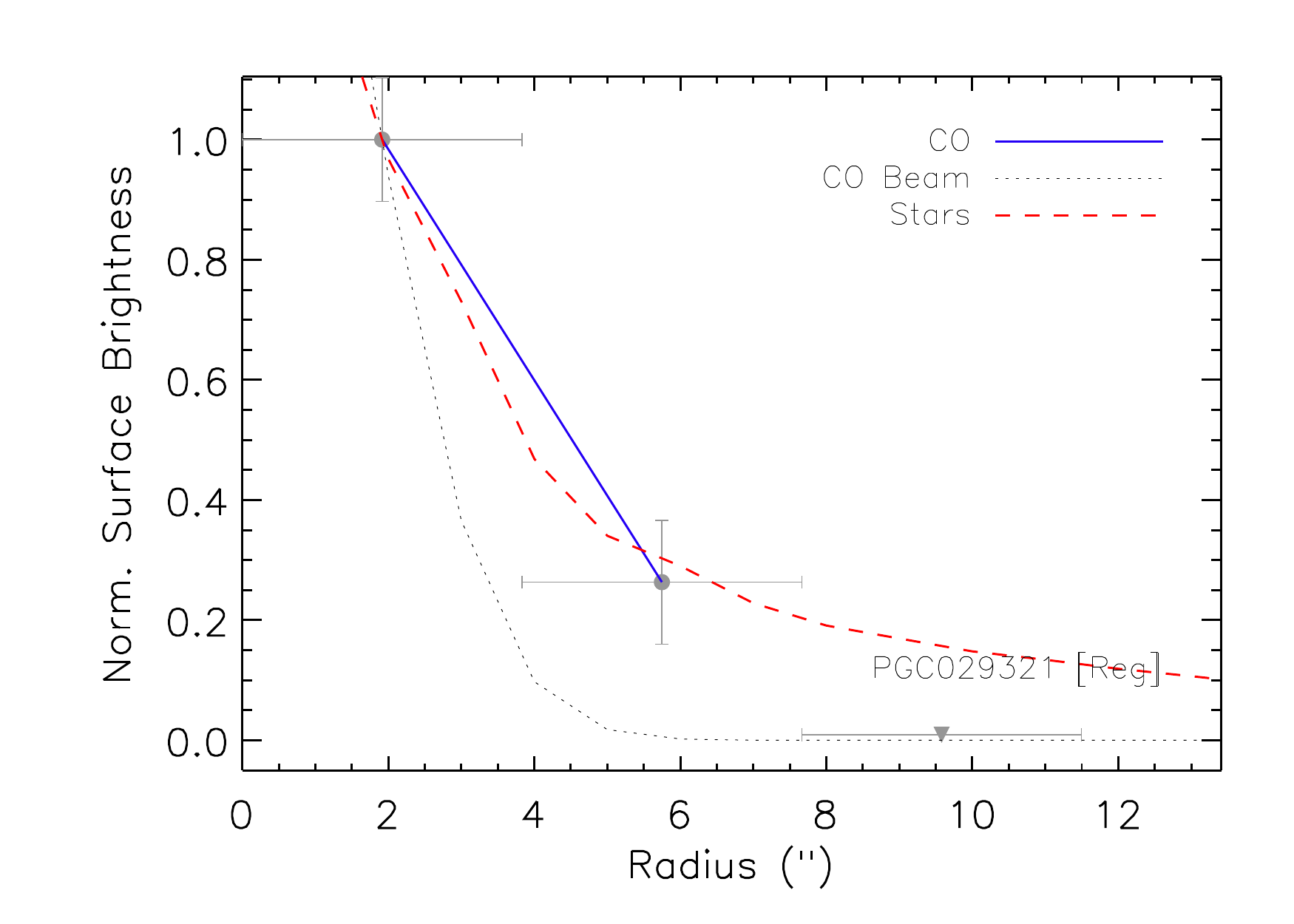}}
\subfigure{\includegraphics[scale=0.45]{sdplots/radial_sd_PGC058114.pdf}}
\subfigure{\includegraphics[scale=0.45]{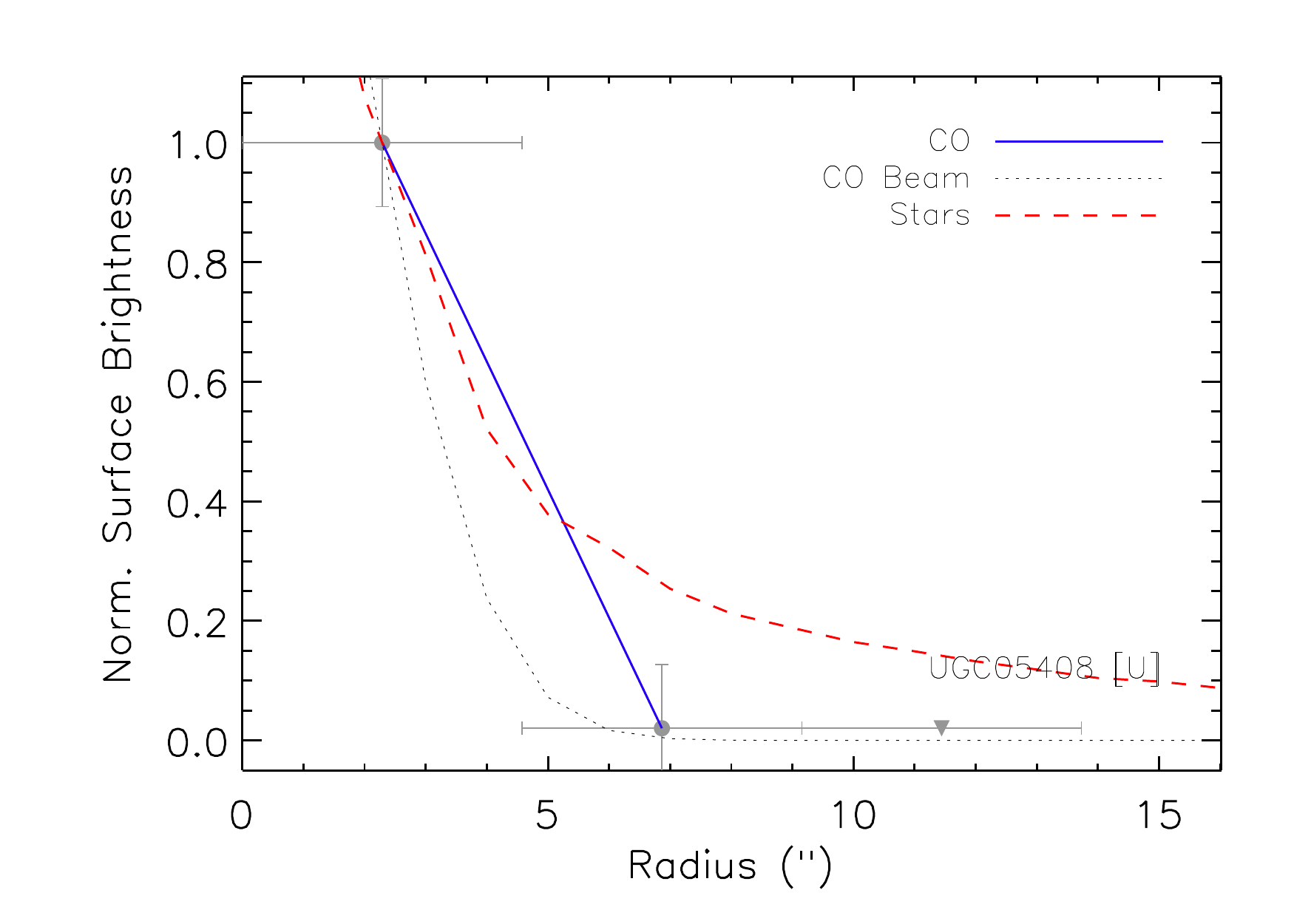}}
\parbox[t]{0.9 \textwidth}{\textbf{Figure A1.} continued}
\end{center}
\end{figure*}
\begin{figure*}
\begin{center}
\subfigure{\includegraphics[scale=0.45]{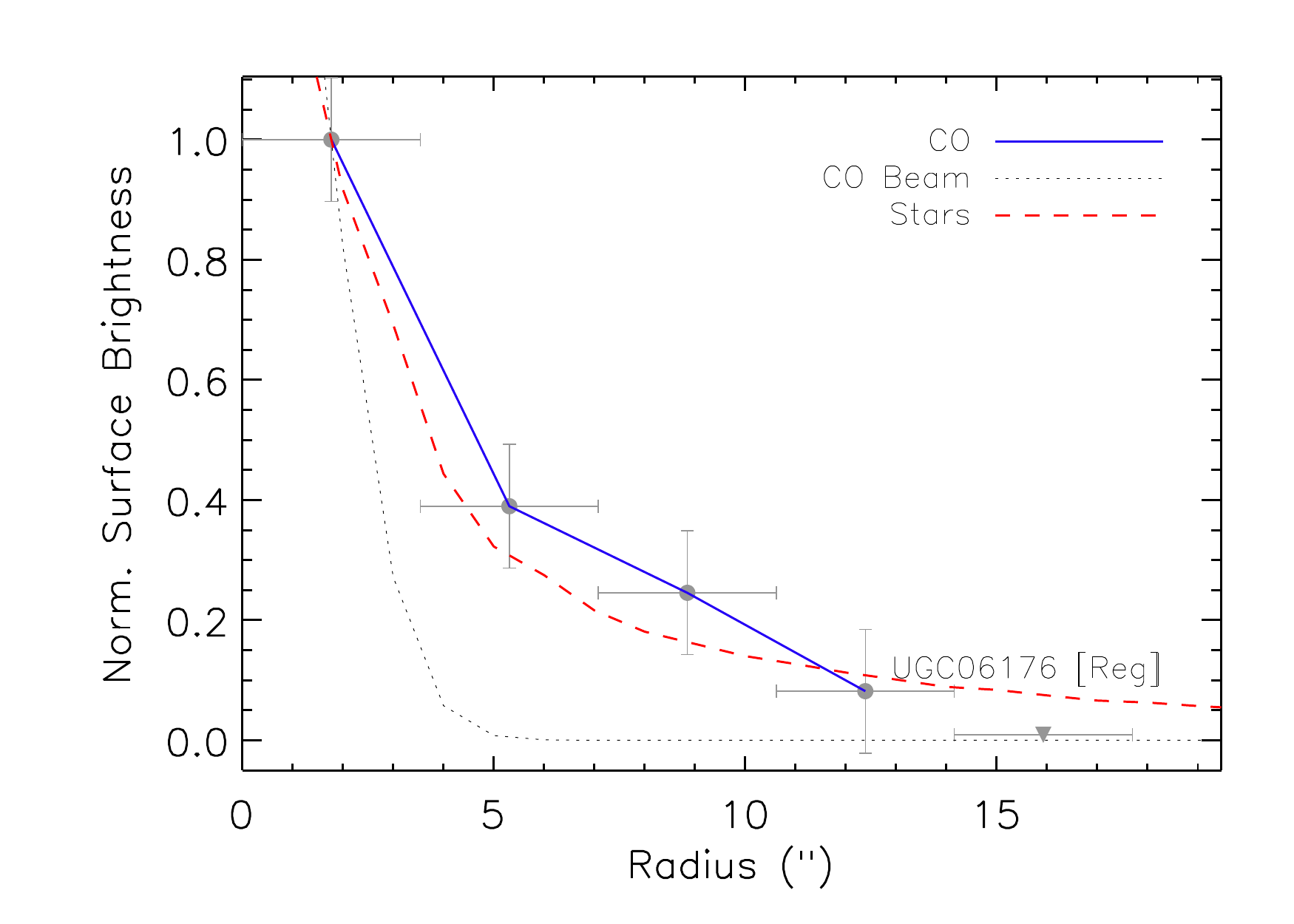}}
\subfigure{\includegraphics[scale=0.45]{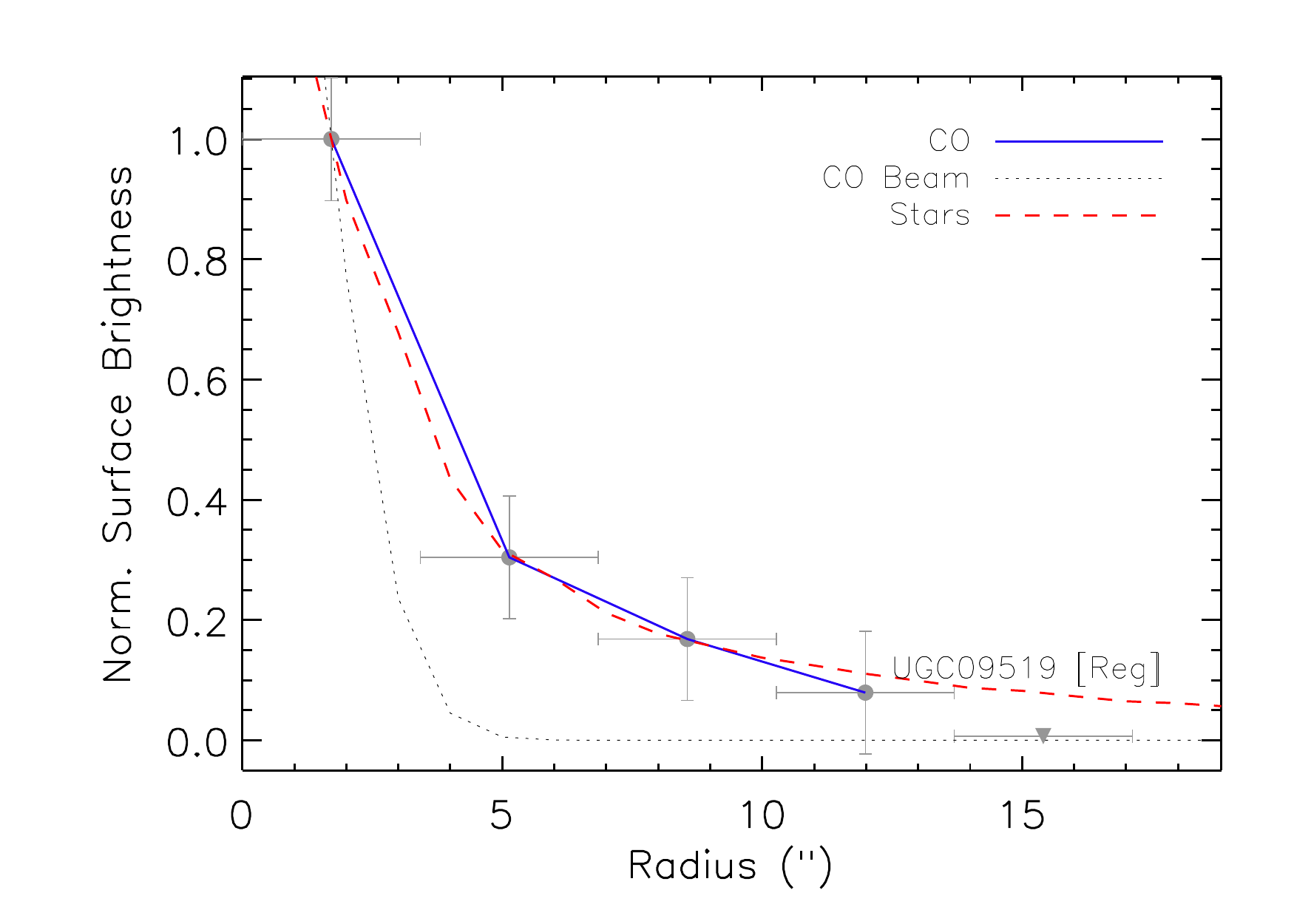}}
\parbox[t]{0.9 \textwidth}{\textbf{Figure A1.} continued}
\end{center}
\end{figure*}

%% file: Appendix_B.tex

\begin{figure*}
\centering
\subfigure{\includegraphics[scale=0.45]{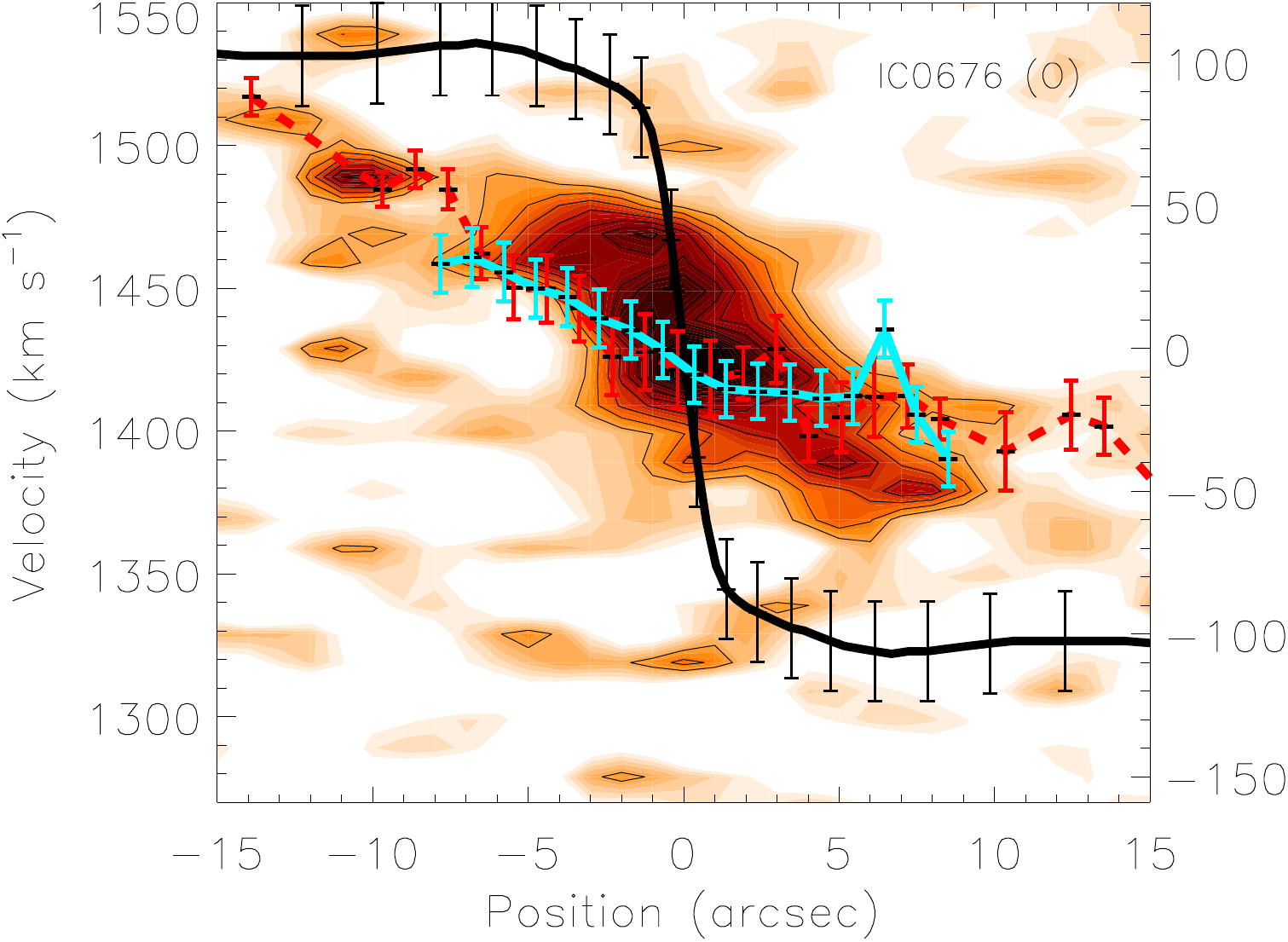}}
\subfigure{\includegraphics[scale=0.45]{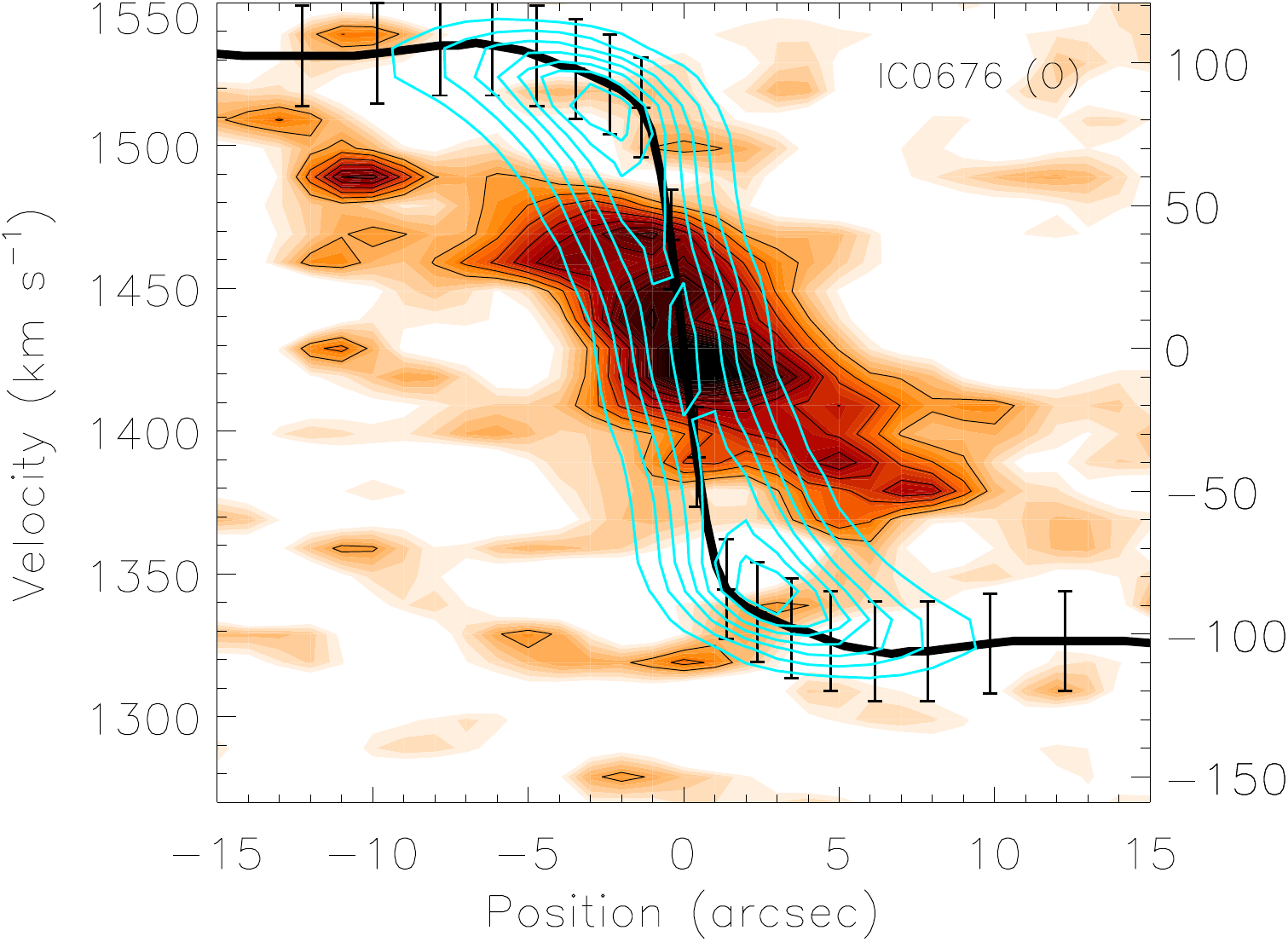}}
\subfigure{\includegraphics[scale=0.45]{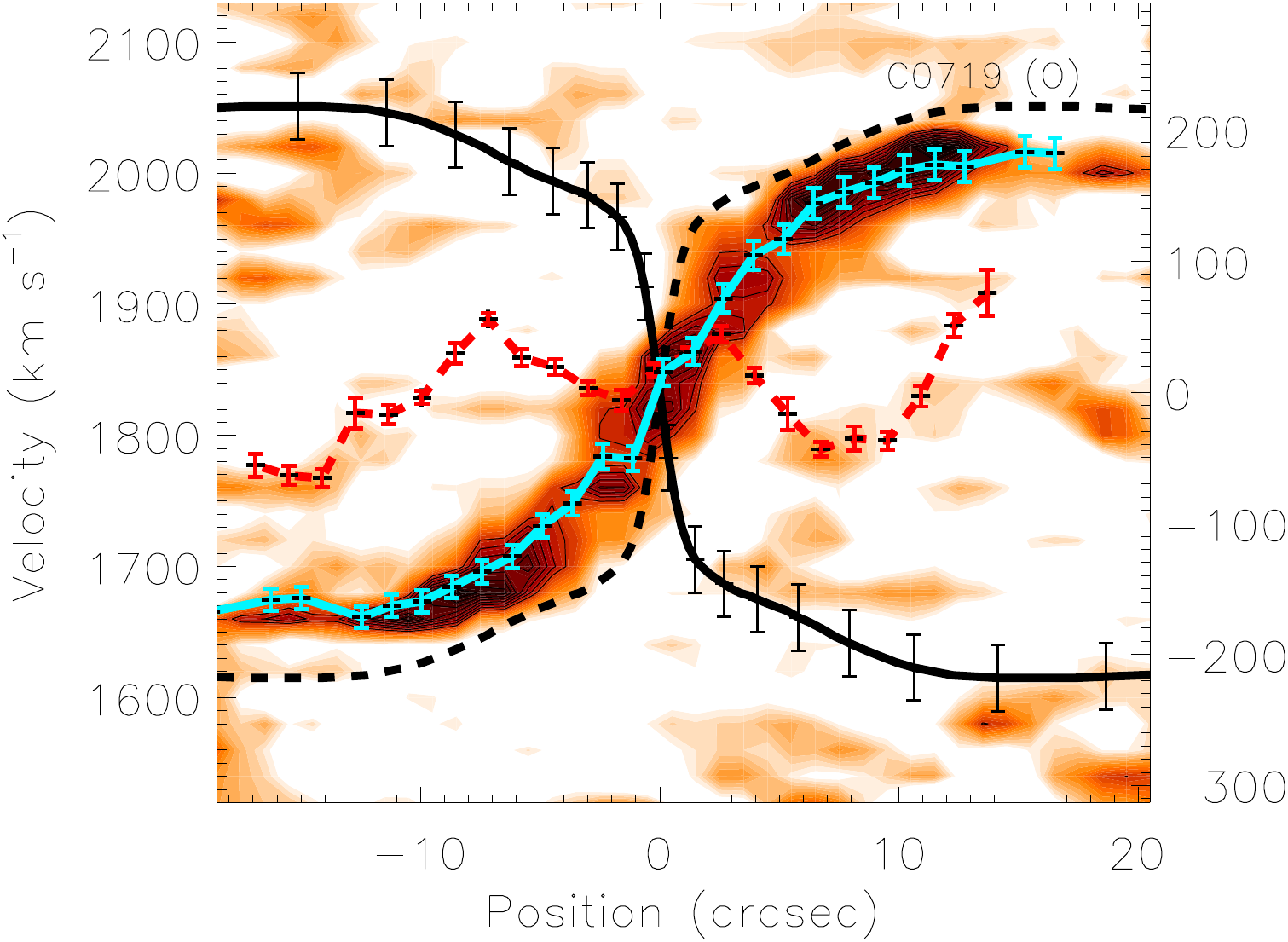}}
\subfigure{\includegraphics[scale=0.45]{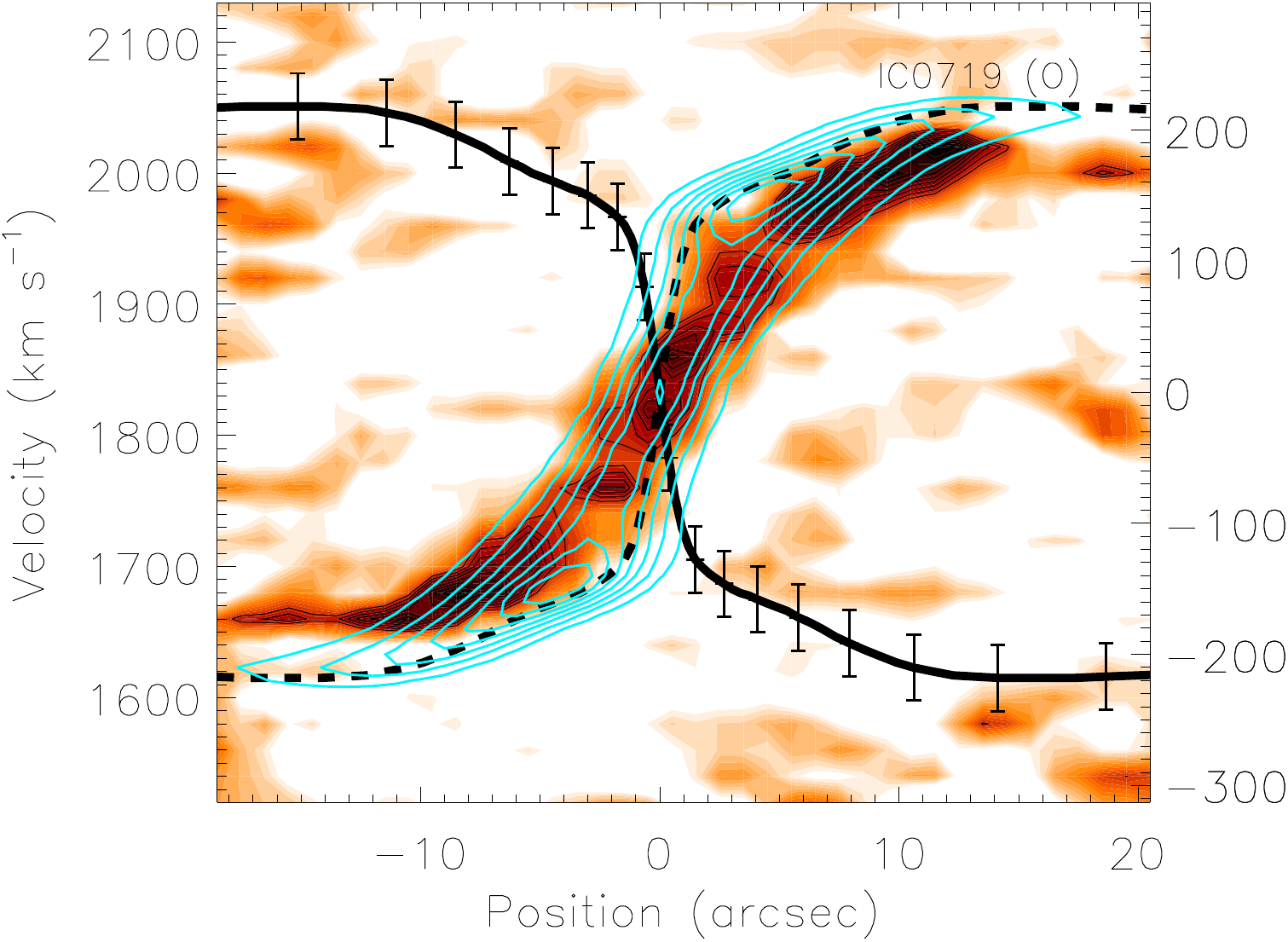}}
\subfigure{\includegraphics[scale=0.45]{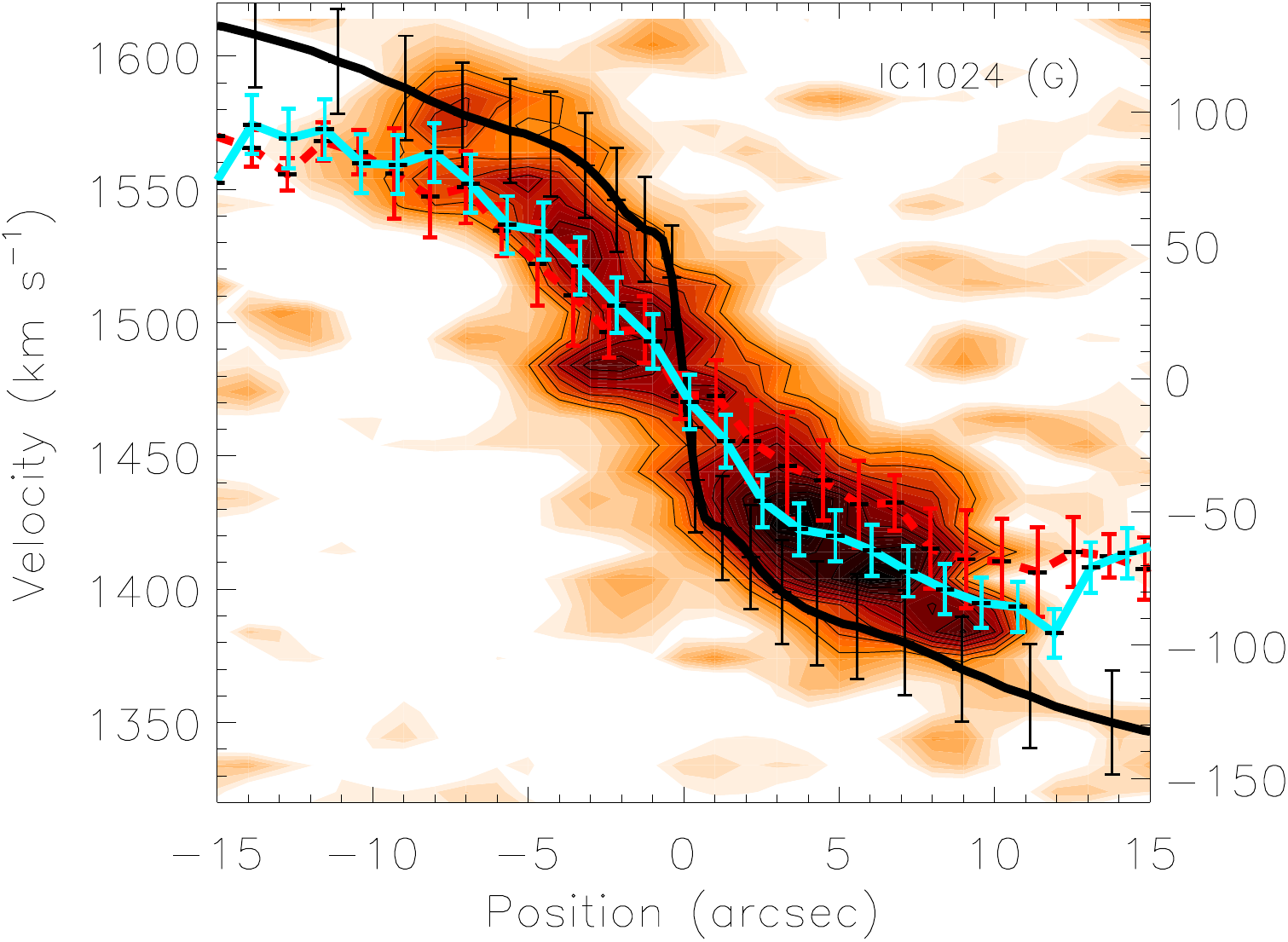}}
\subfigure{\includegraphics[scale=0.45]{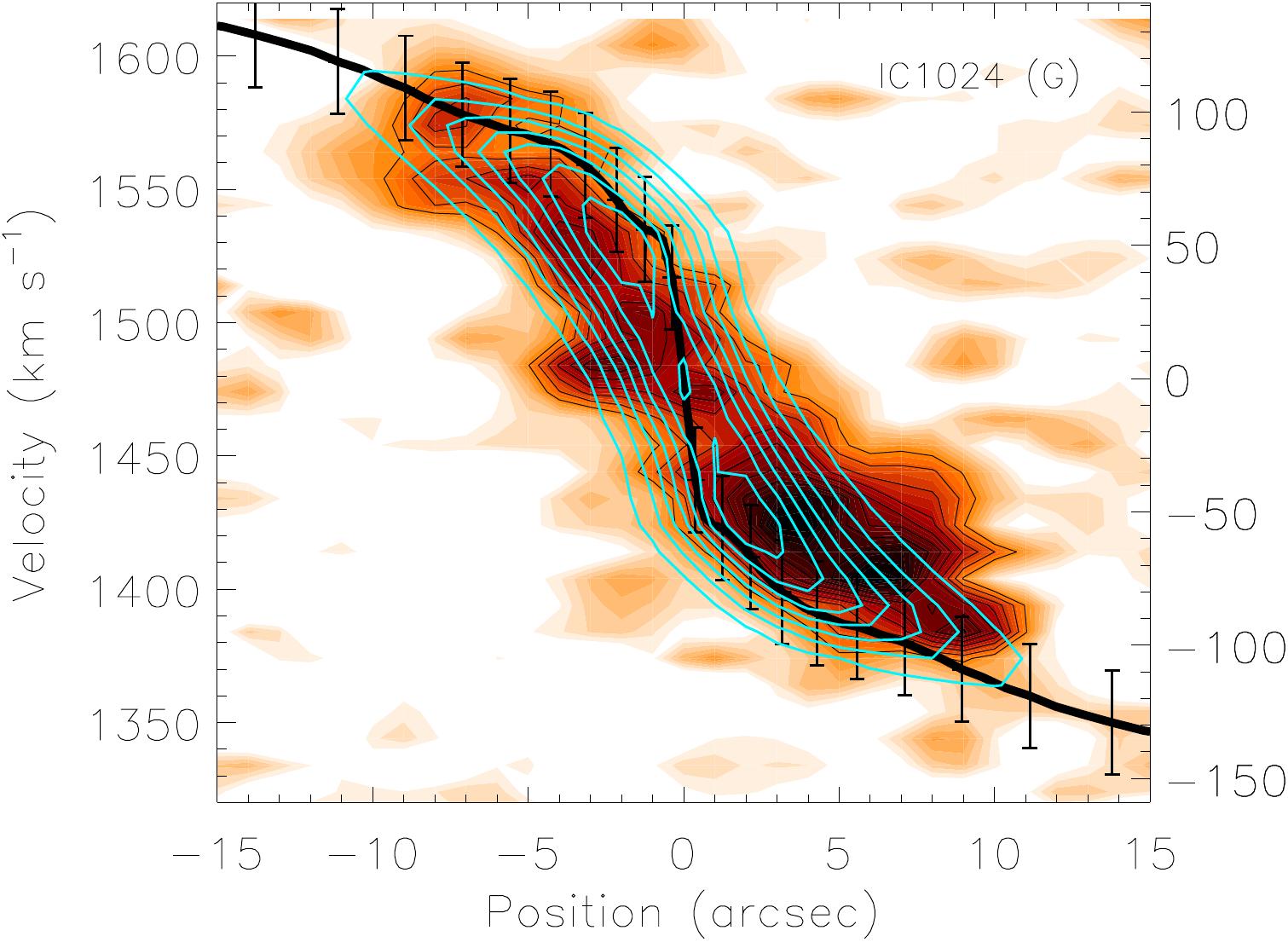}}
\subfigure{\includegraphics[scale=0.45]{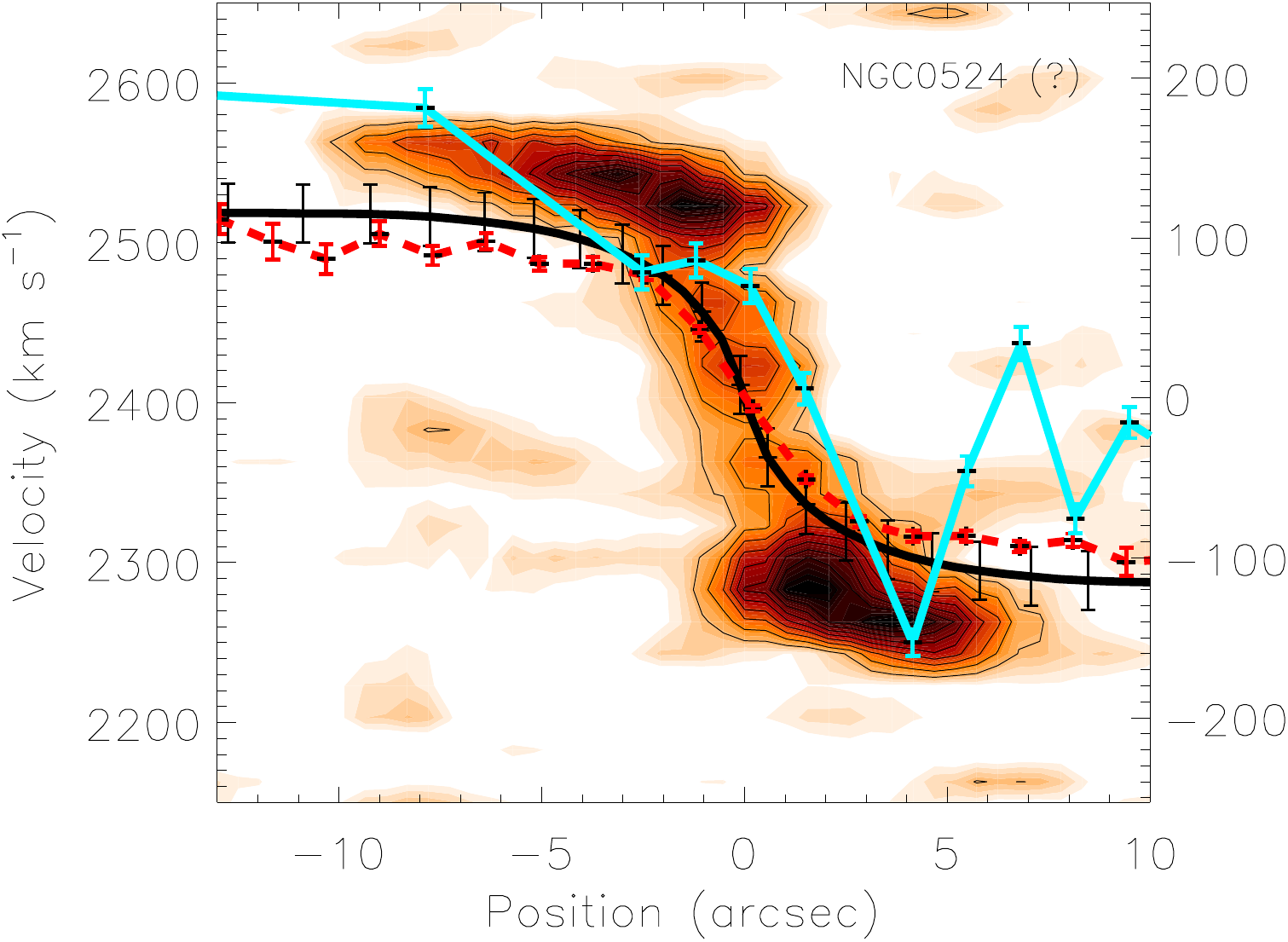}}
\subfigure{\includegraphics[scale=0.45]{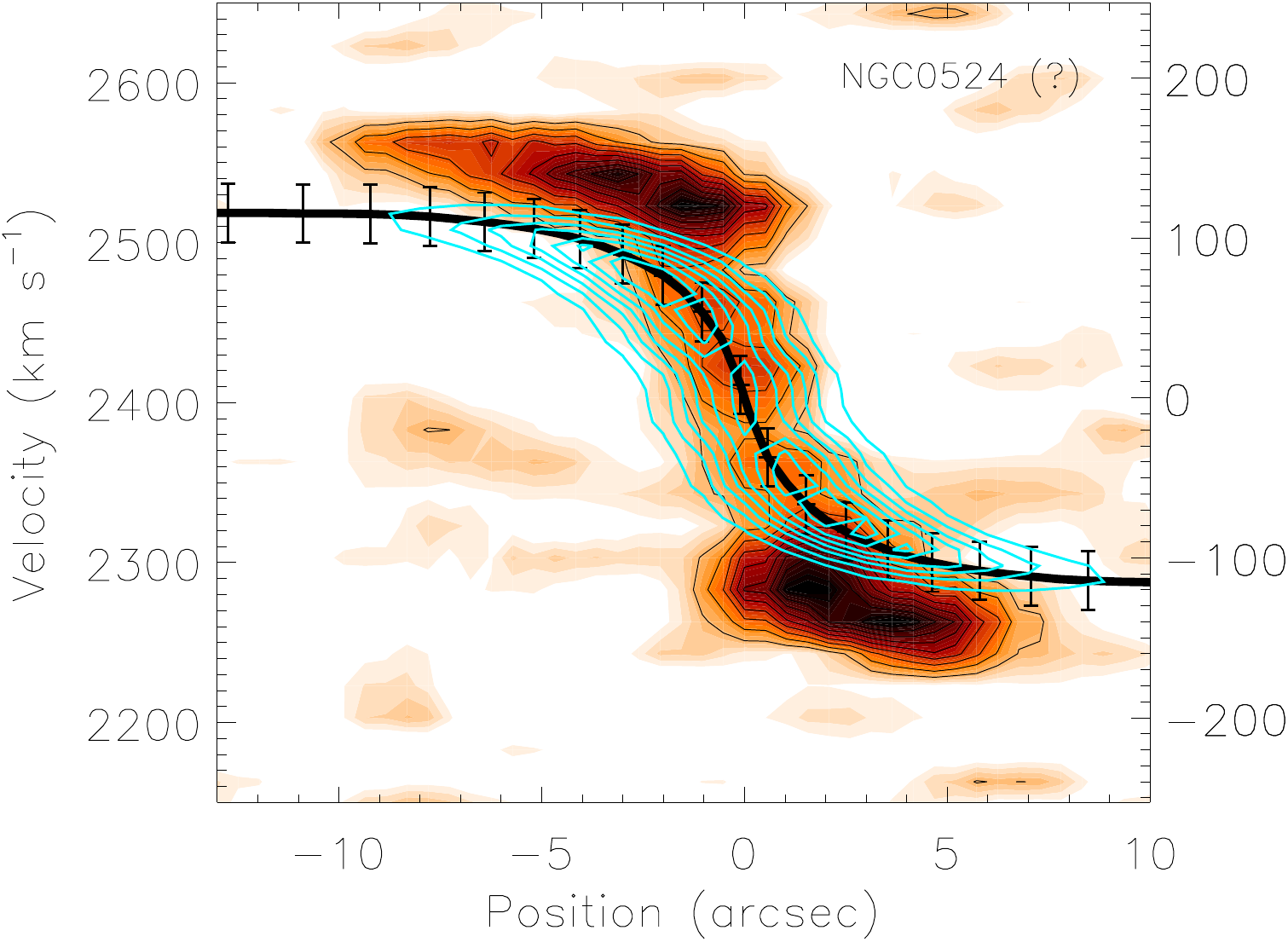}}
\parbox[t]{0.9 \textwidth}{ \caption{\small CO Position-Velocity diagrams (orange with black contours) for the ATLAS$^{\rm 3D}$ galaxies, overlaid with the JAM circular velocity curve (thick black line). On the left, these have also been overlaid with the observed stellar (red dashed line) and ionised gas (blue line) rotation curve. On the right we overlay the modeled observations of each galaxy projected into the line of sight (blue contours), created assuming that the gas has an intrinsic velocity dispersion of 8 km s$^{-1}$, is rotating at the predicted circular velocity, and has an exponential or gaussian surface brightness profile, as described in Section 5.2. The MGE model of IC\,0676 is effected by dust (Scott et al., in prep), and as discussed in the text the gas may have significant non circular motions due to a bar. These two effects likely cause the discrepancy. NGC\,0524 has a degenerate inclination solution which causes the disagreement between model and observations (see figure A1 in Cappellari et al., 2006)}}
\end{figure*}
\begin{figure*}
\centering
\subfigure{\includegraphics[scale=0.45]{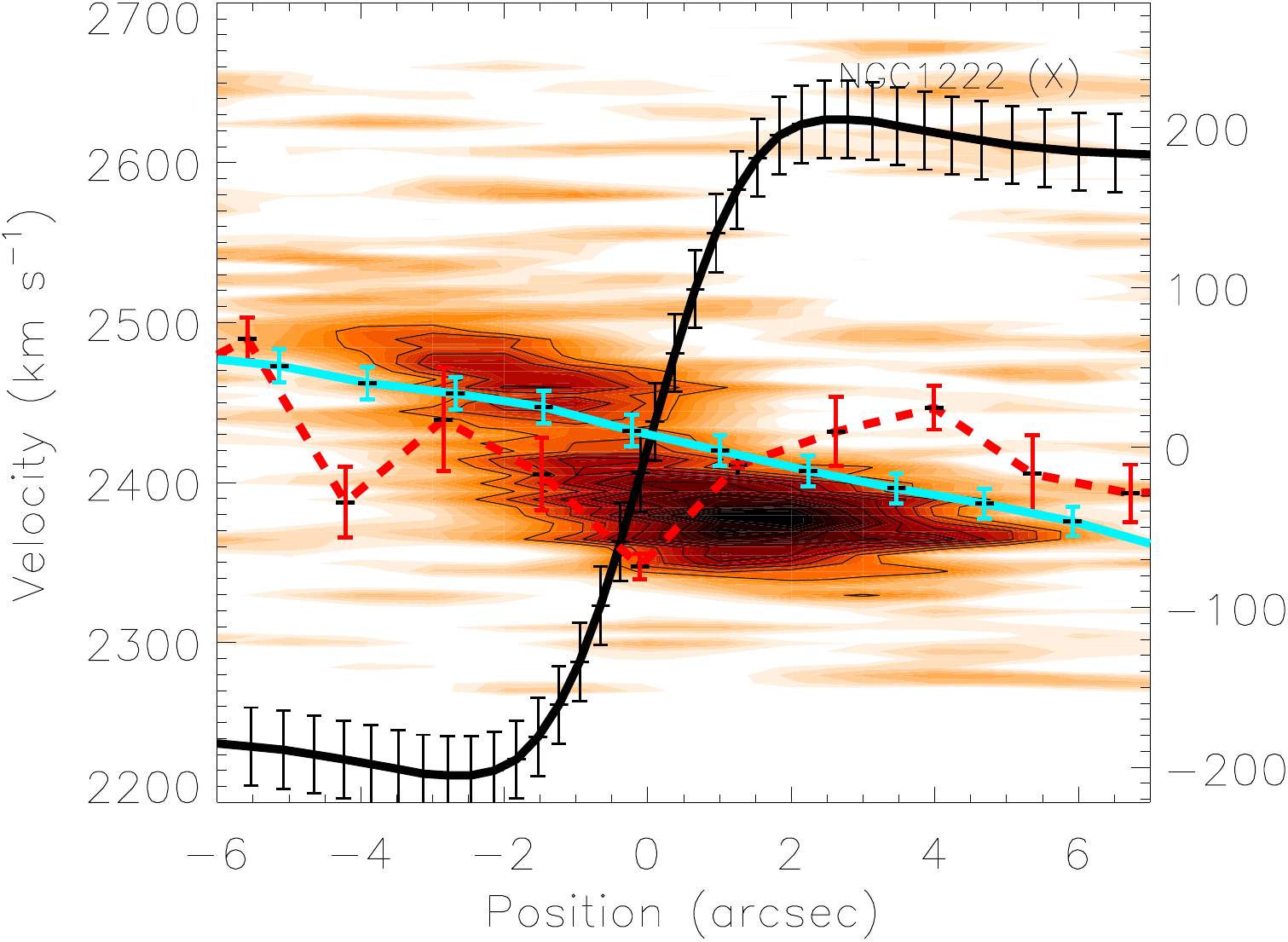}}
\subfigure{\includegraphics[scale=0.45]{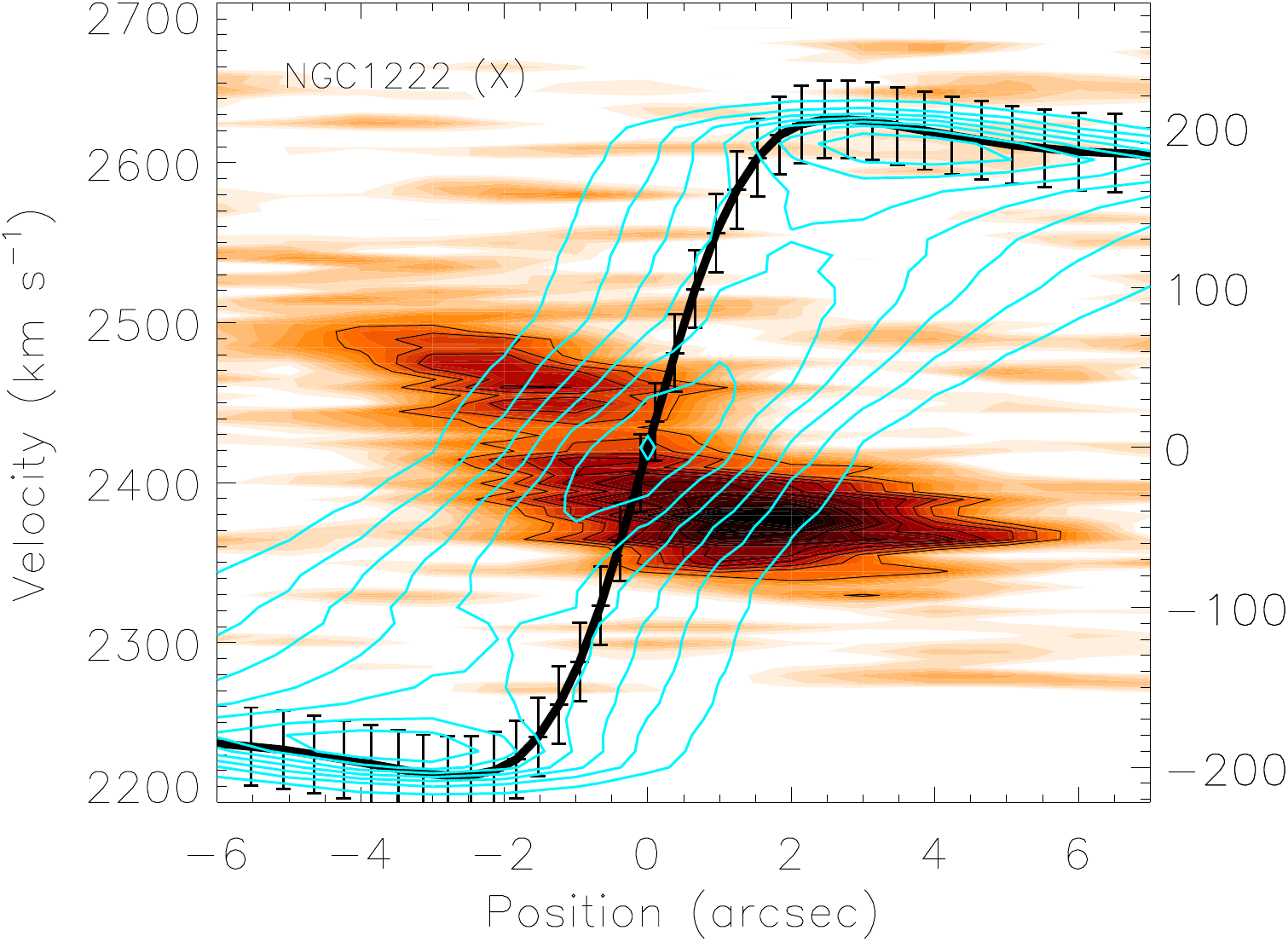}}
\subfigure{\includegraphics[scale=0.45]{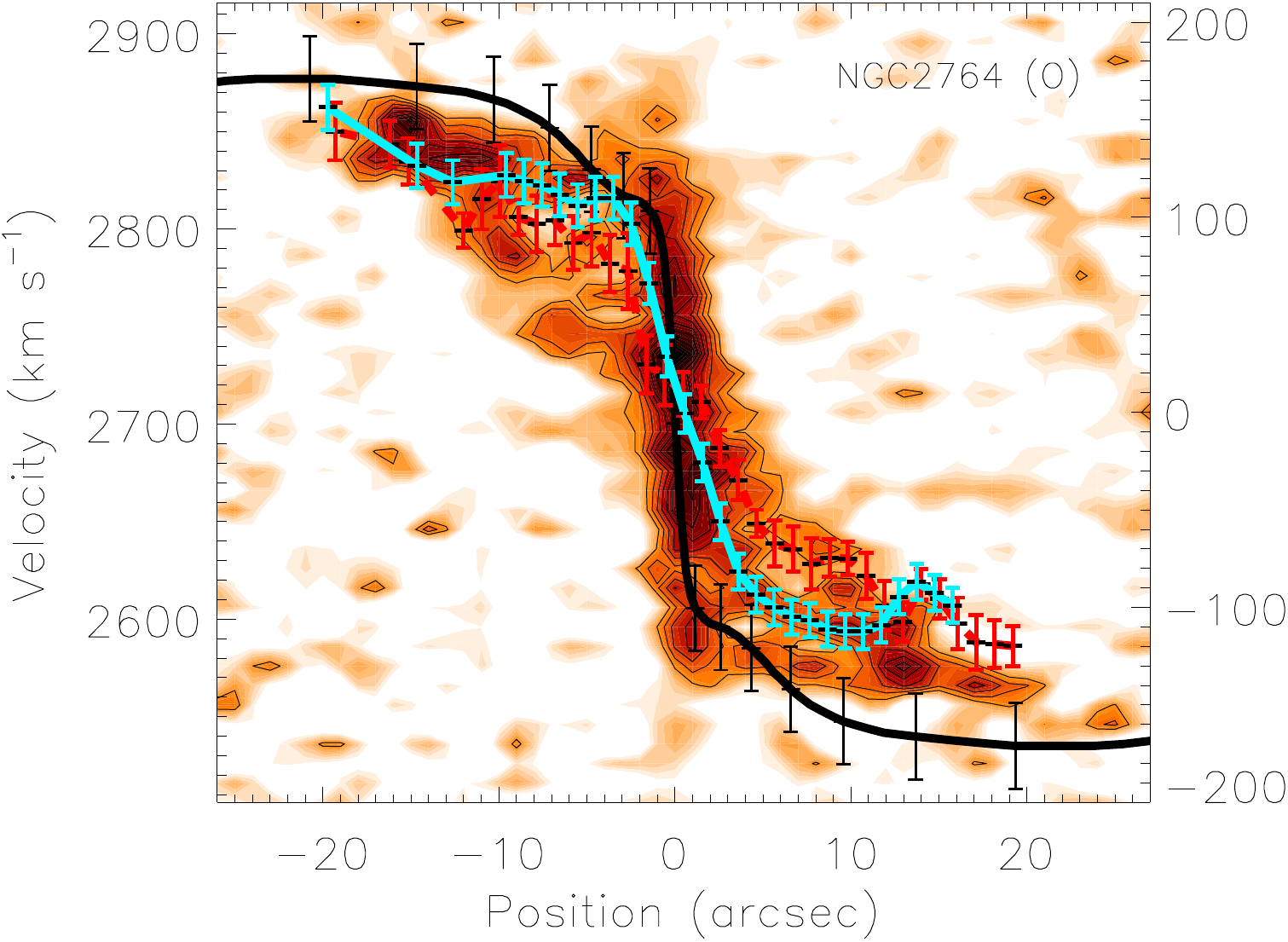}}
\subfigure{\includegraphics[scale=0.45]{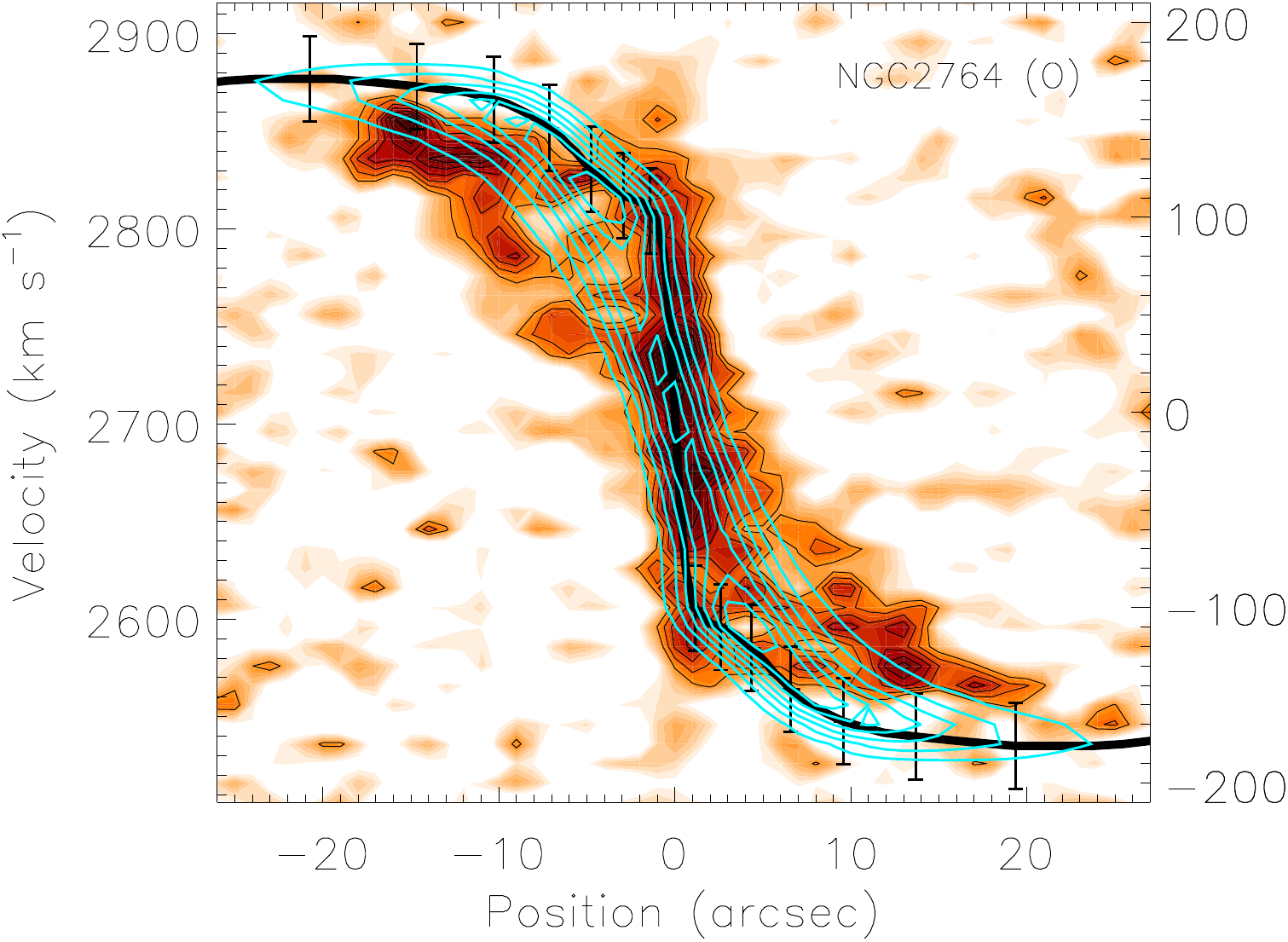}}
\subfigure{\includegraphics[scale=0.45]{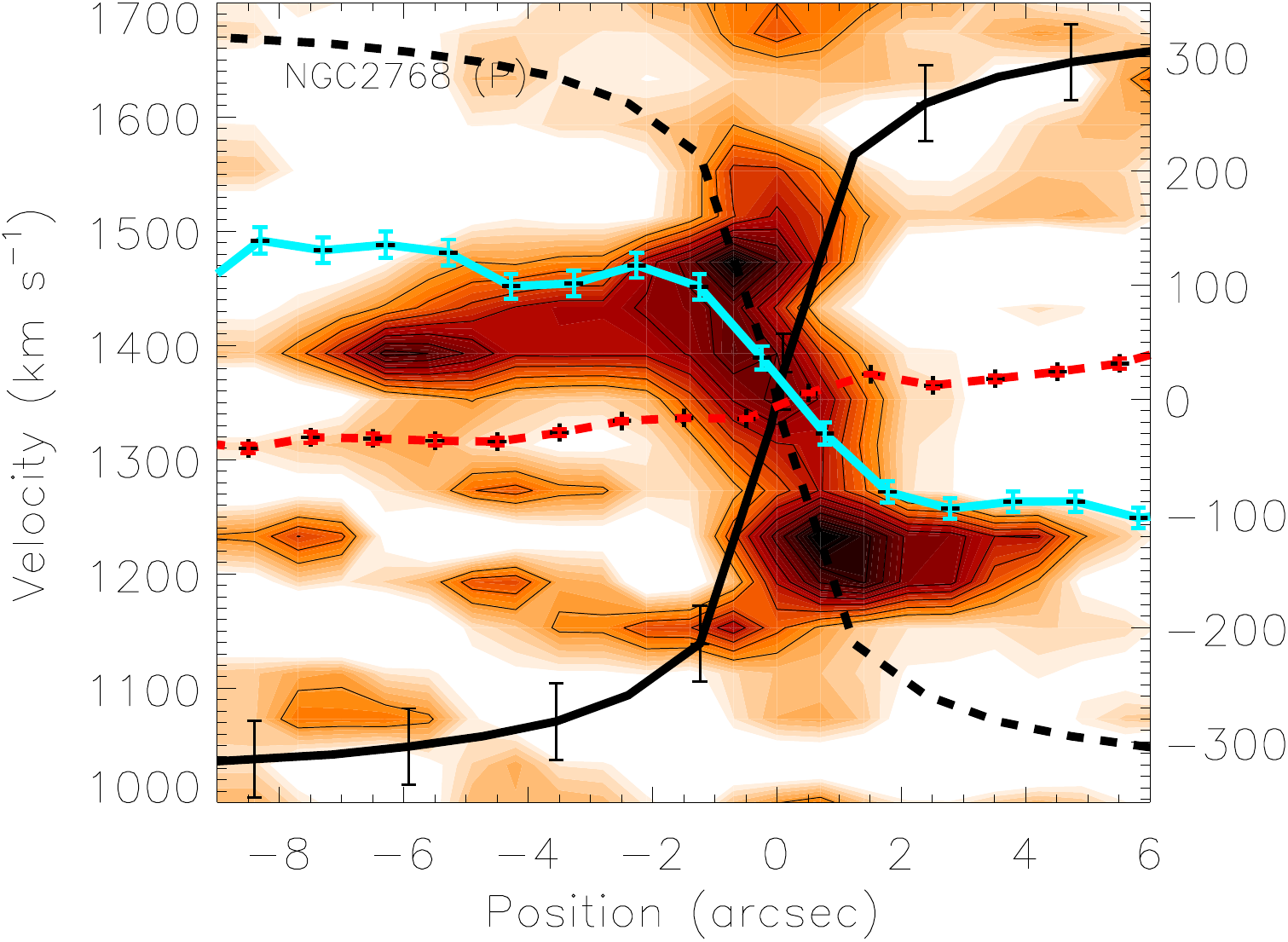}}
\subfigure{\includegraphics[scale=0.45]{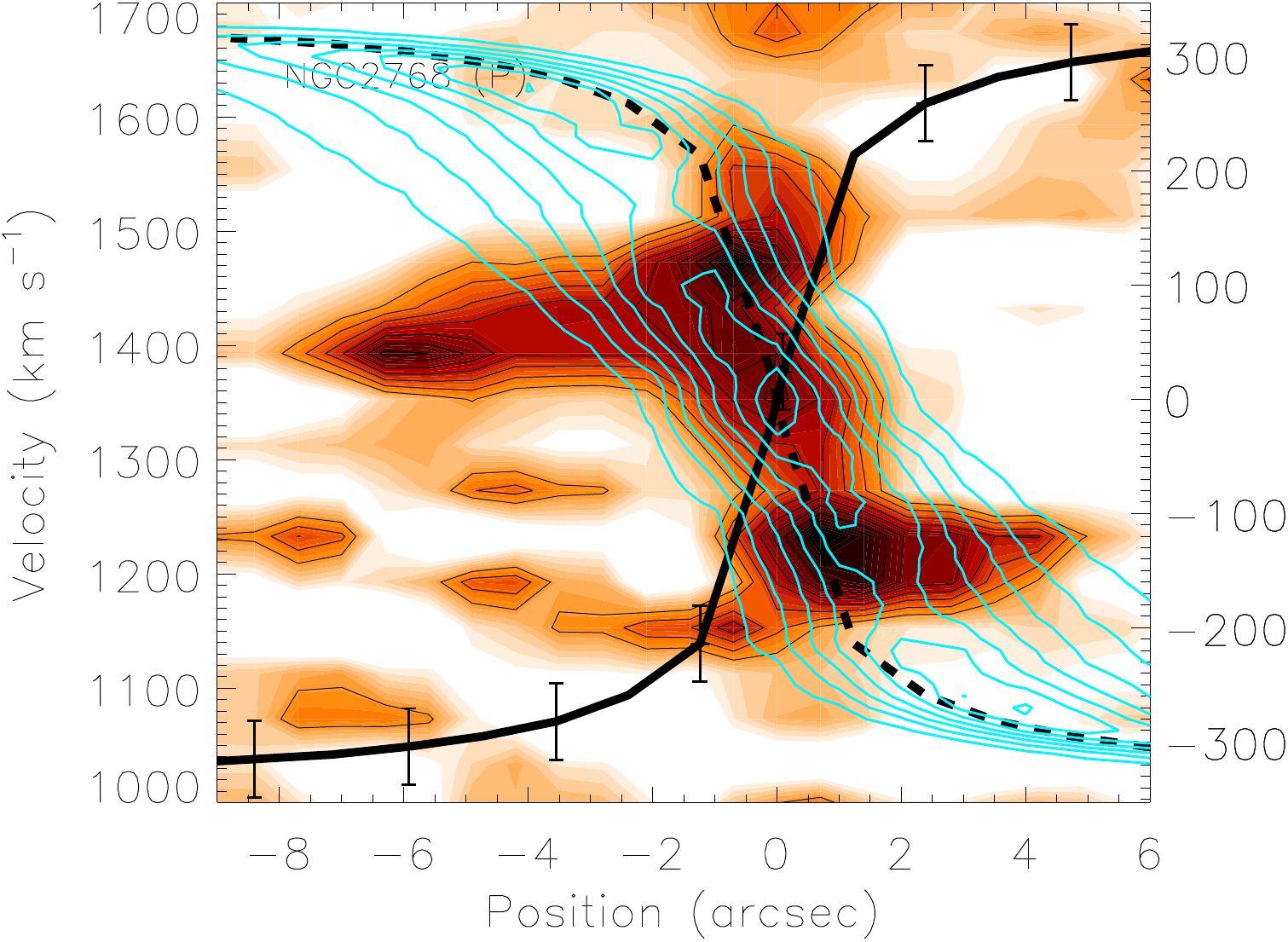}}
\subfigure{\includegraphics[scale=0.45]{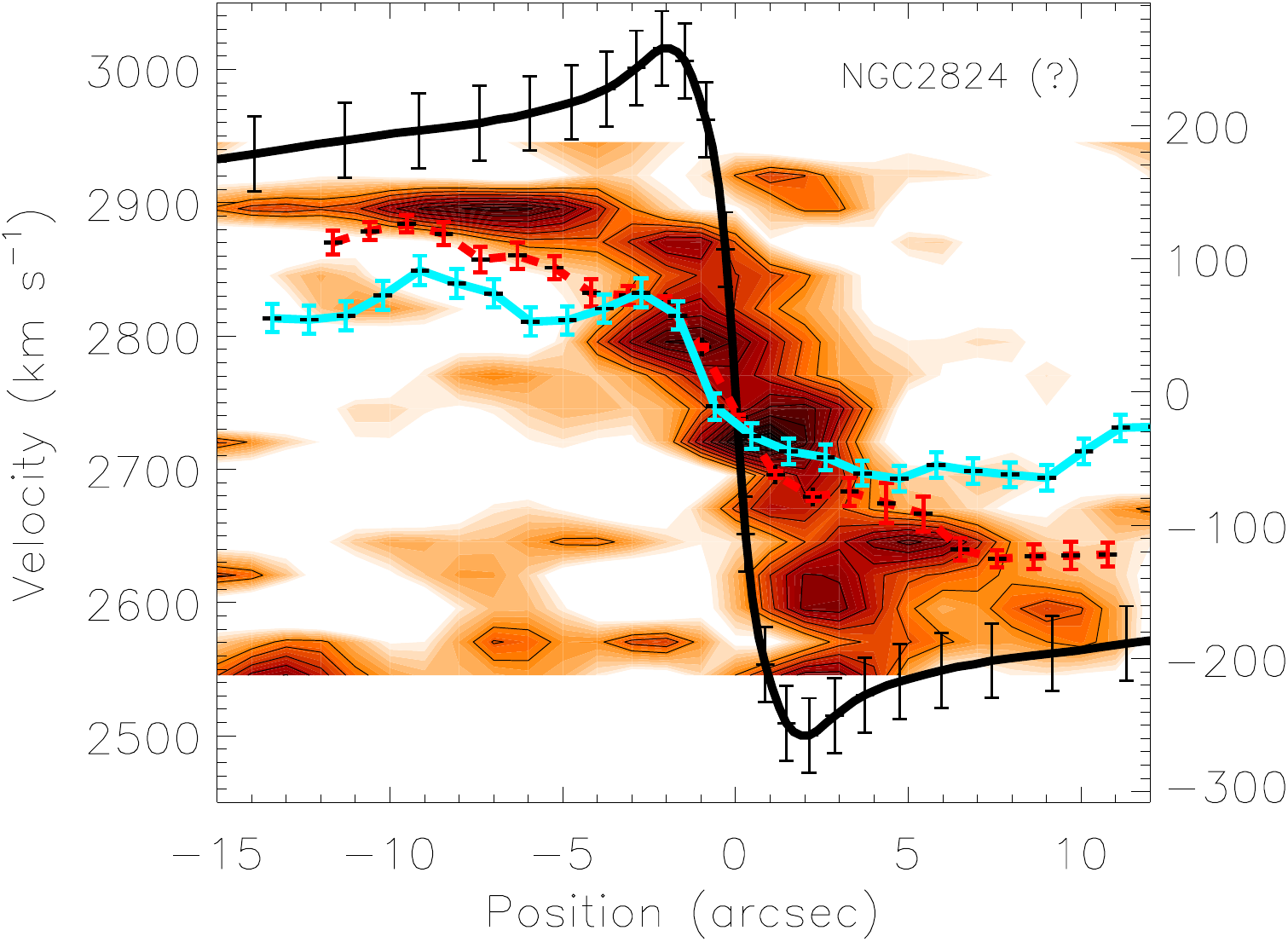}}
\subfigure{\includegraphics[scale=0.45]{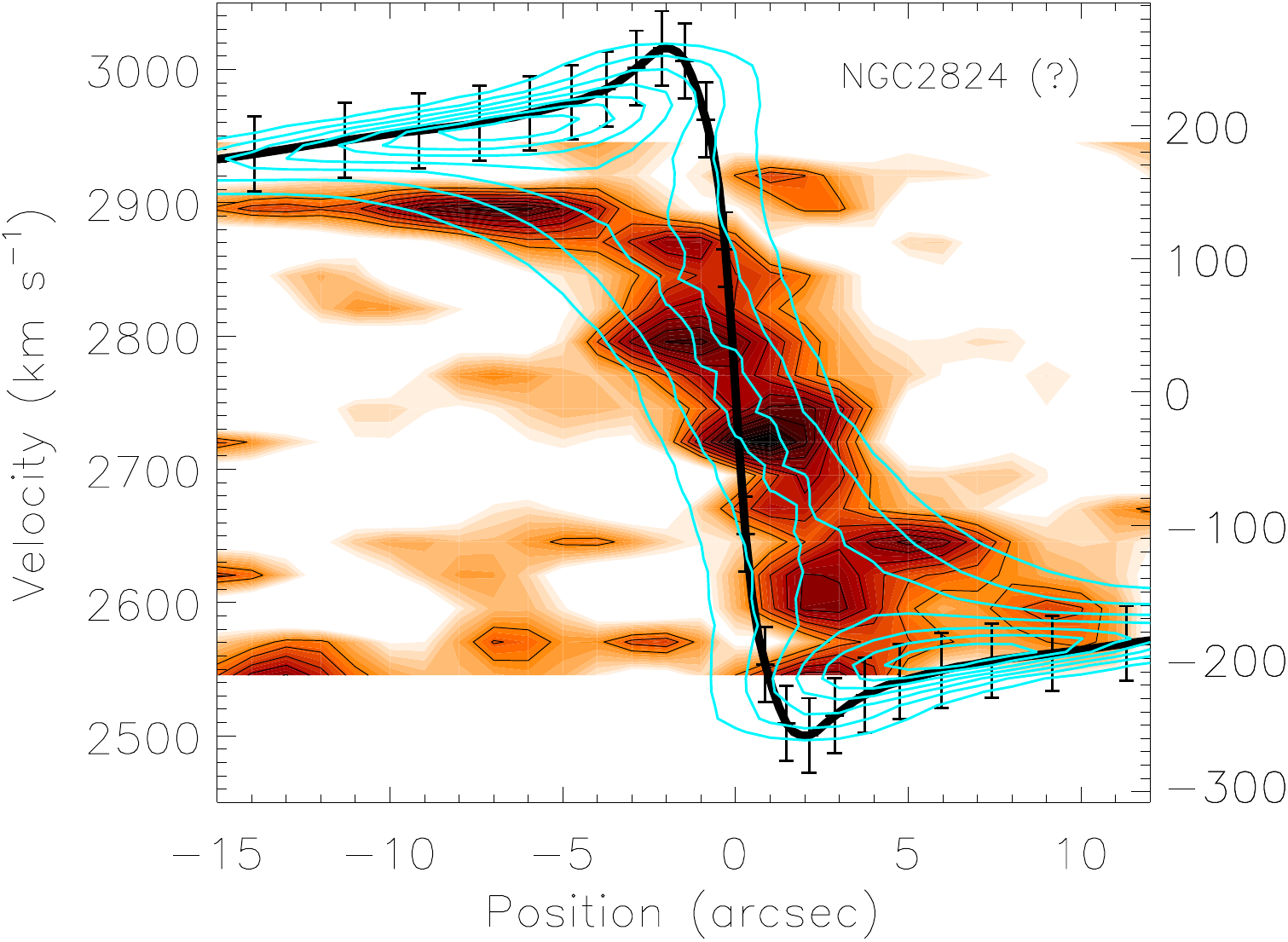}}
\parbox[t]{0.9 \textwidth}{ \textbf{Figure B1.} continued. NGC2764 has an MGE model which is heavily affected by dust (Scott et al., in prep).}
\end{figure*}
\begin{figure*}
\centering
\subfigure{\includegraphics[scale=0.45]{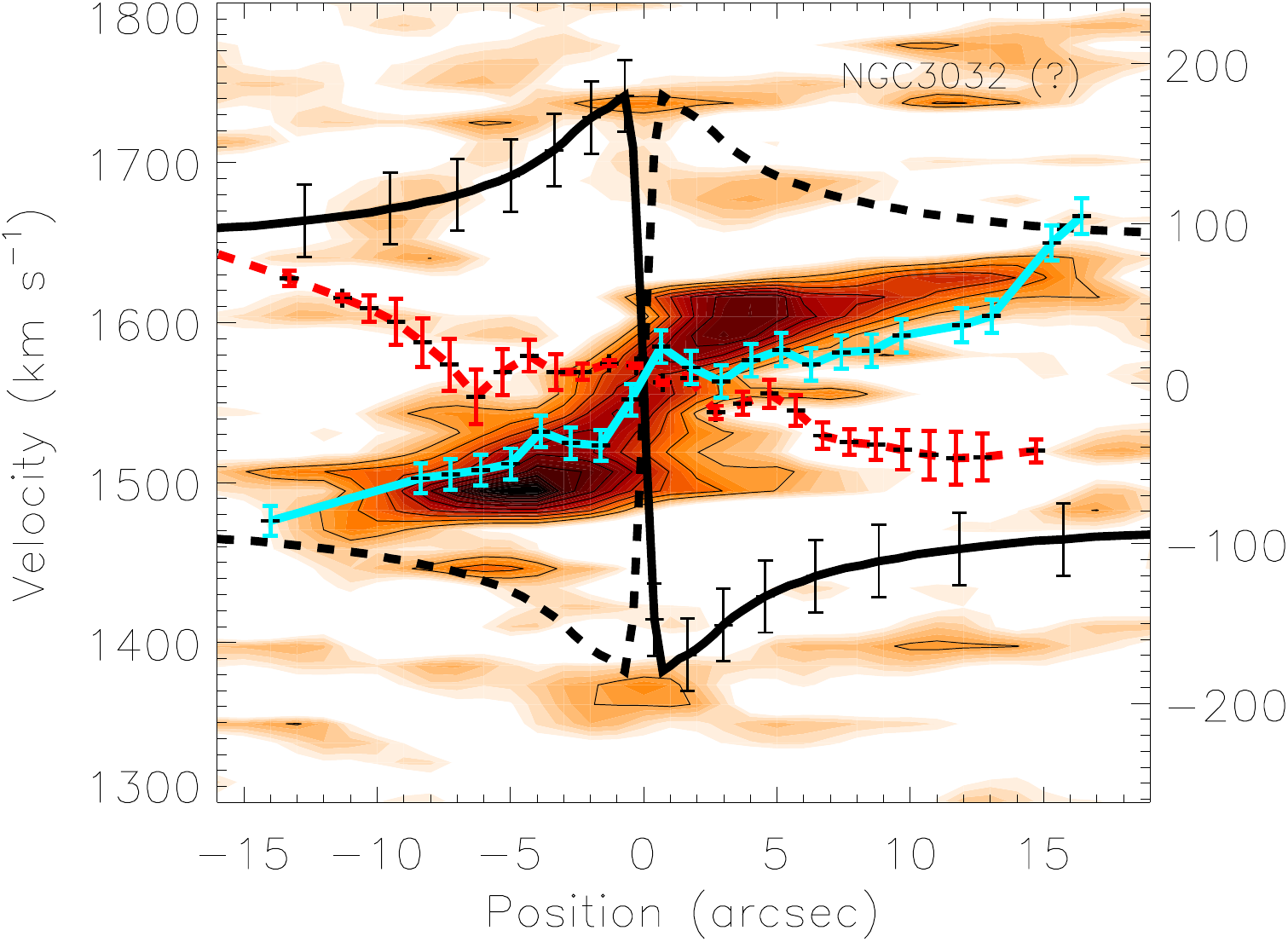}}
\subfigure{\includegraphics[scale=0.45]{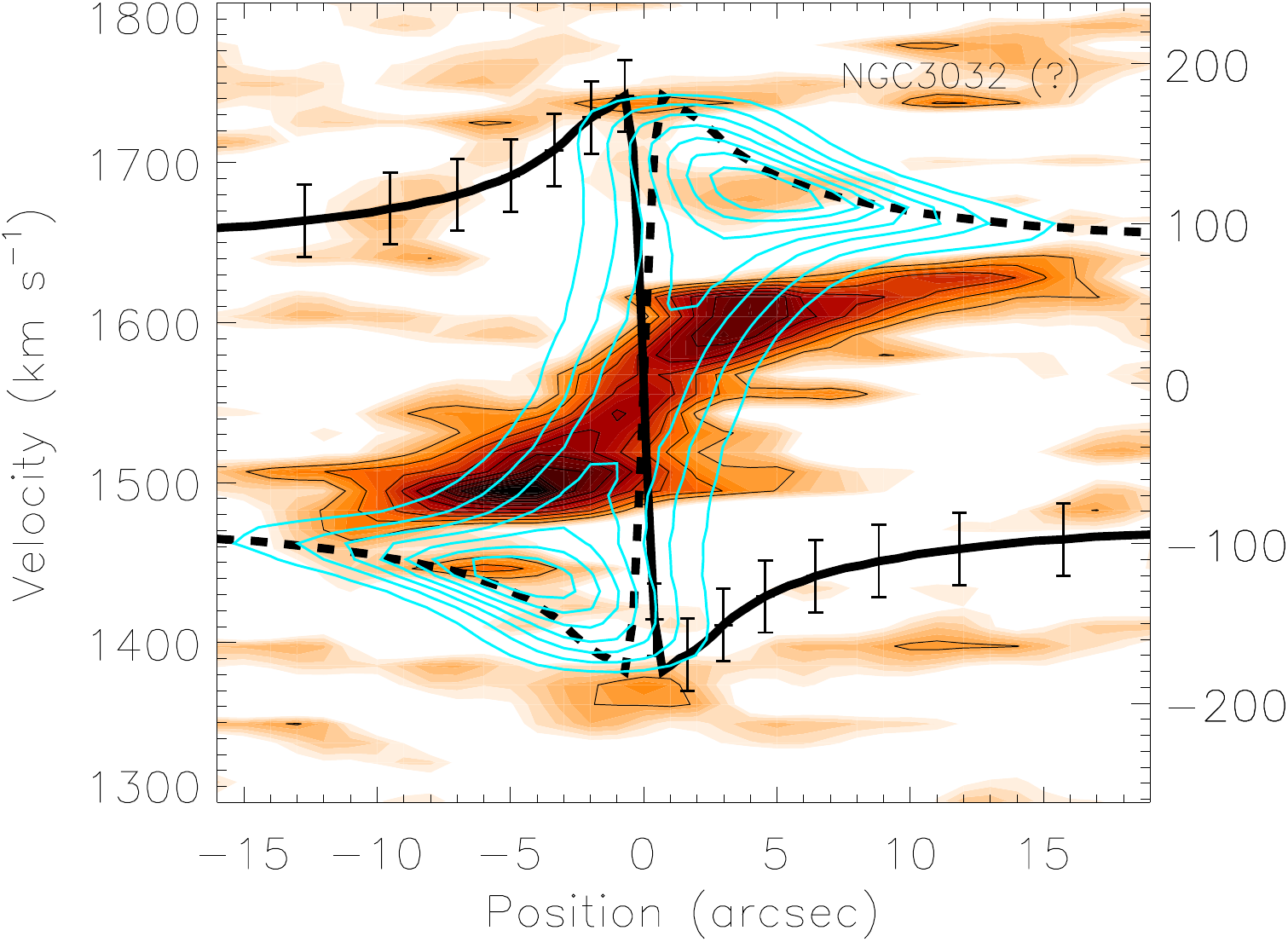}}
\subfigure{\includegraphics[scale=0.45]{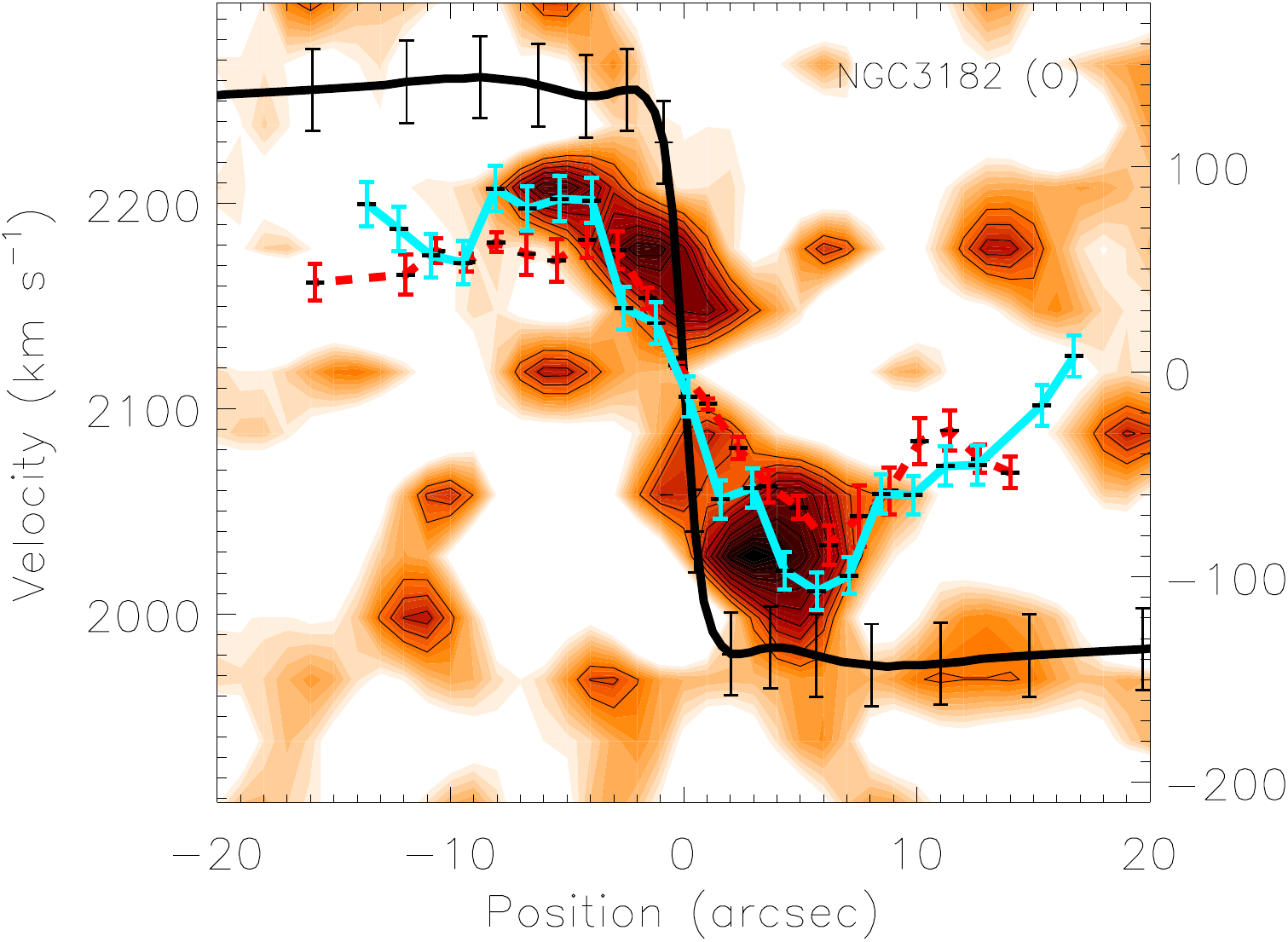}}
\subfigure{\includegraphics[scale=0.45]{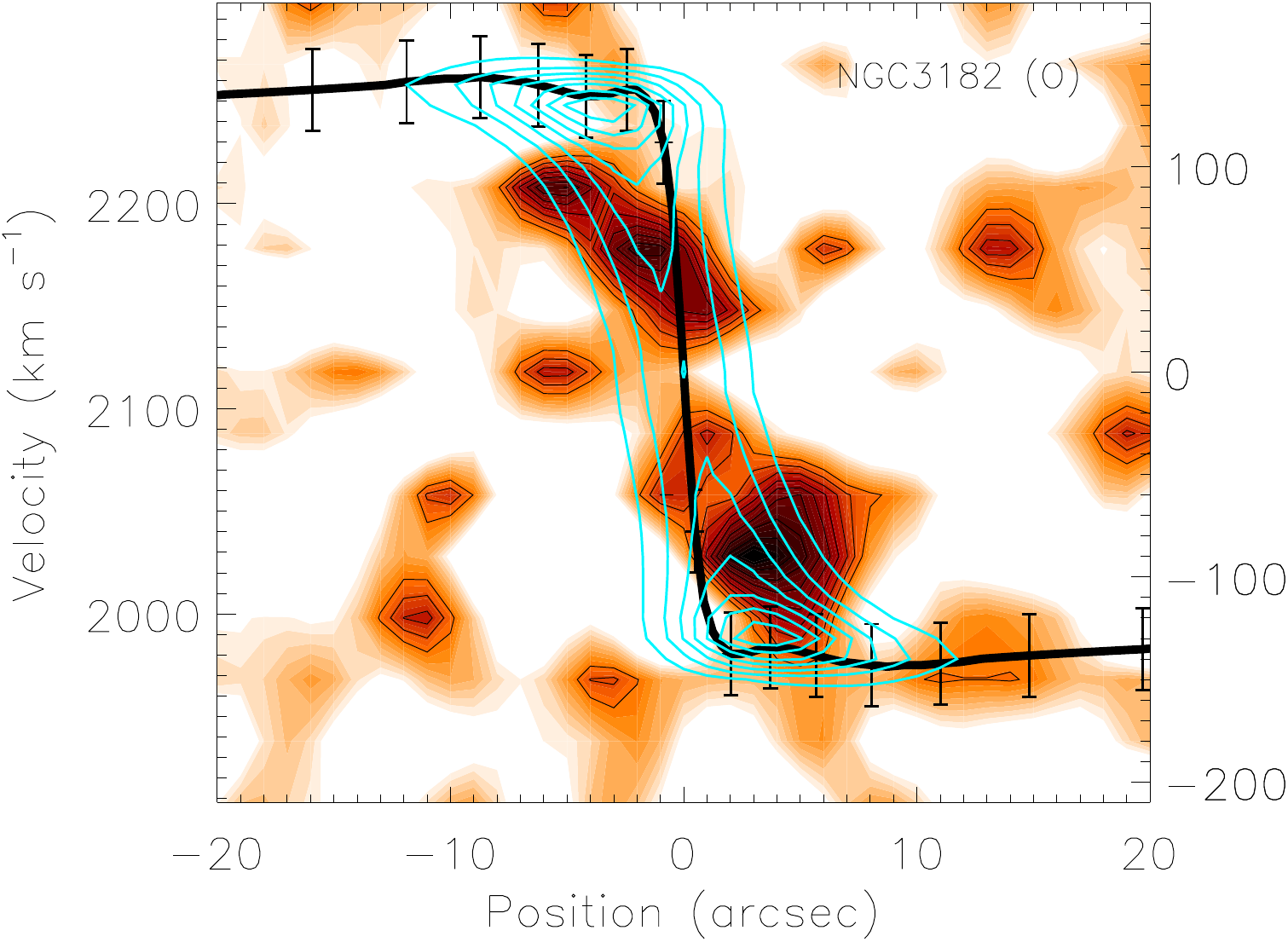}}
\subfigure{\includegraphics[scale=0.45]{plots/NGC3607pvdiag_comp_stars.pdf}}
\subfigure{\includegraphics[scale=0.45]{simplots/NGC3607pvsim_comp.pdf}}
\subfigure{\includegraphics[scale=0.45]{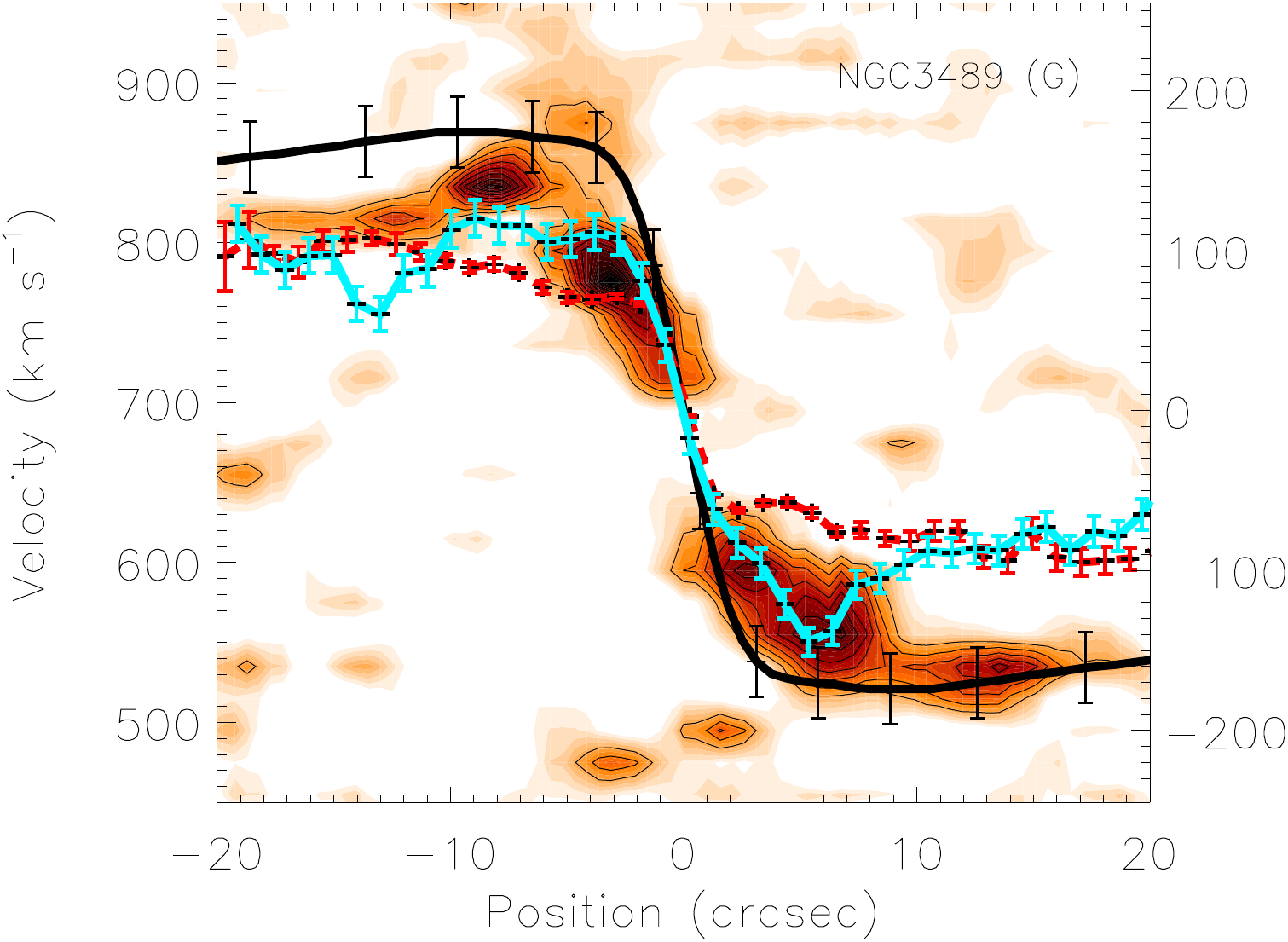}}
\subfigure{\includegraphics[scale=0.45]{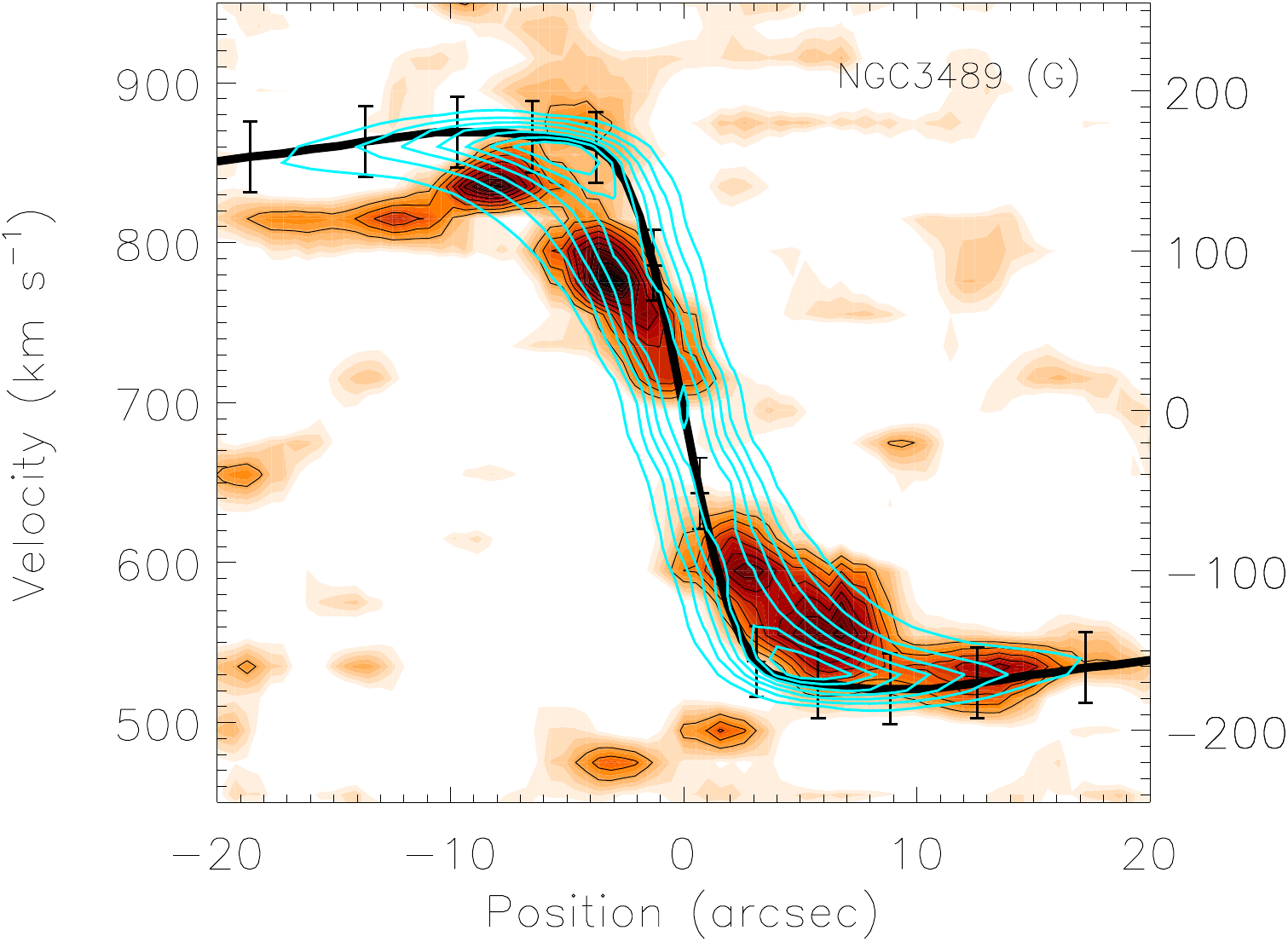}}
\parbox[t]{0.9 \textwidth}{ \textbf{Figure B1.} continued}
\end{figure*}
\begin{figure*}
\centering
\subfigure{\includegraphics[scale=0.45]{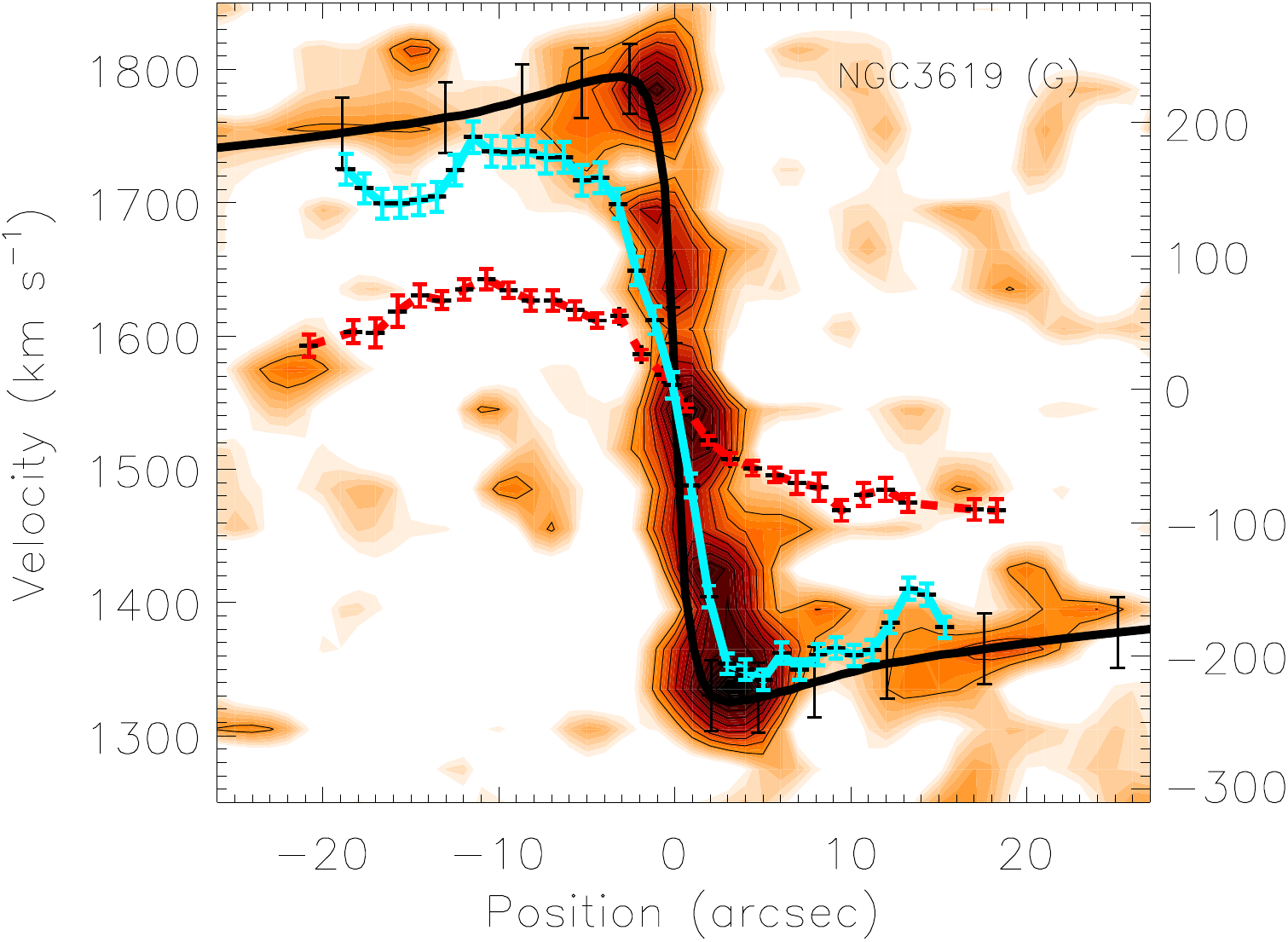}}
\subfigure{\includegraphics[scale=0.45]{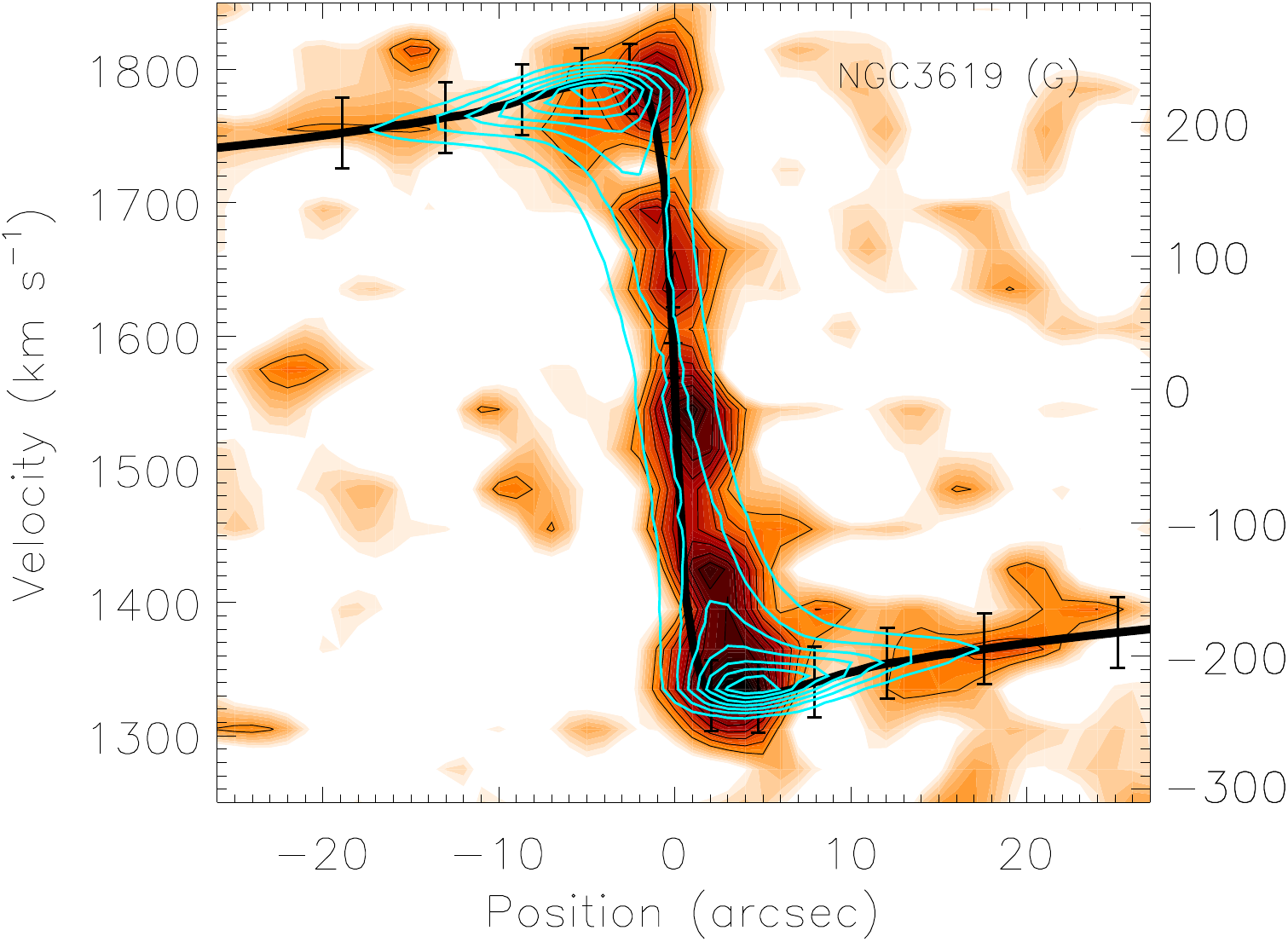}}
\subfigure{\includegraphics[scale=0.45]{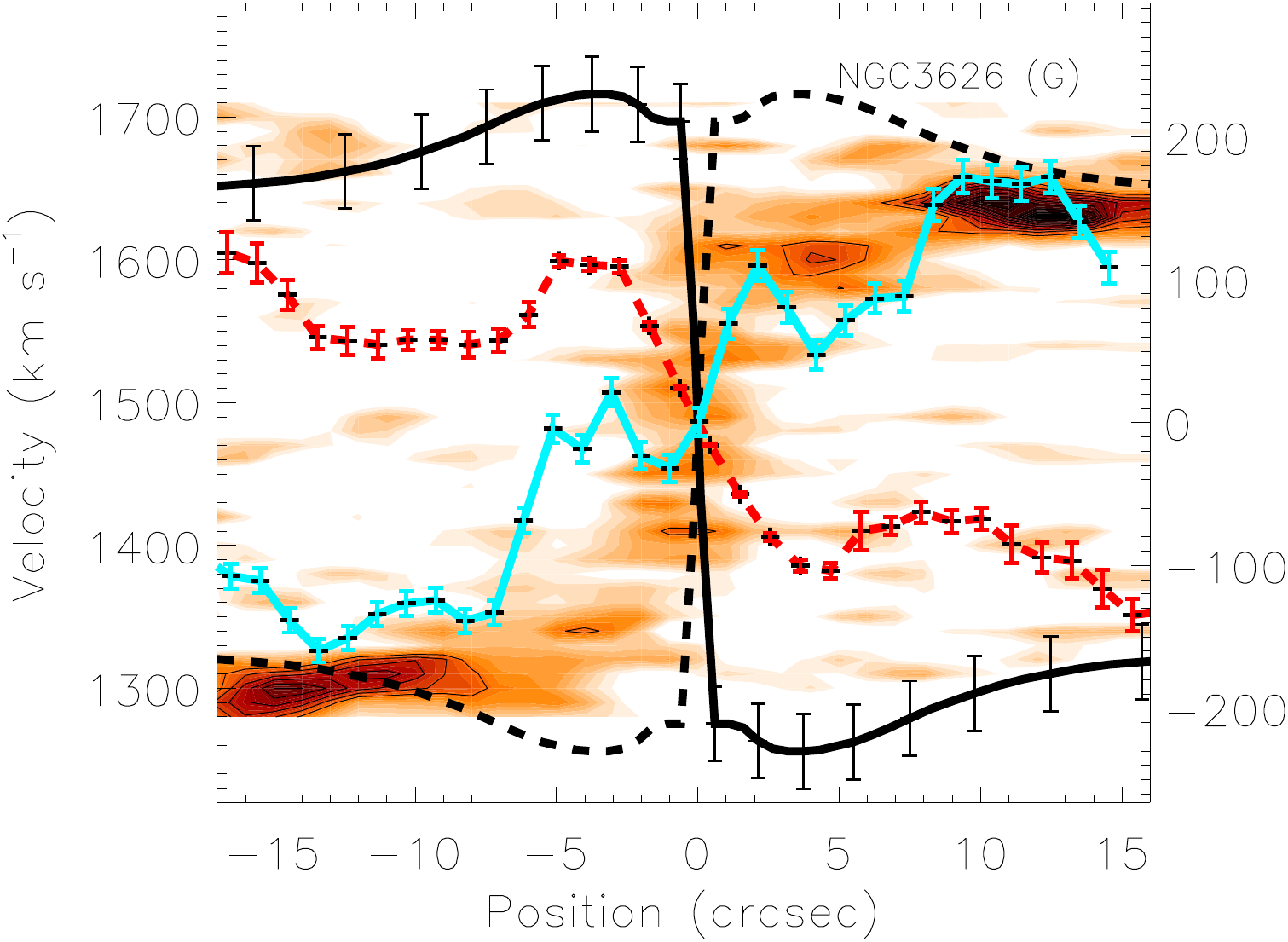}}
\subfigure{\includegraphics[scale=0.45]{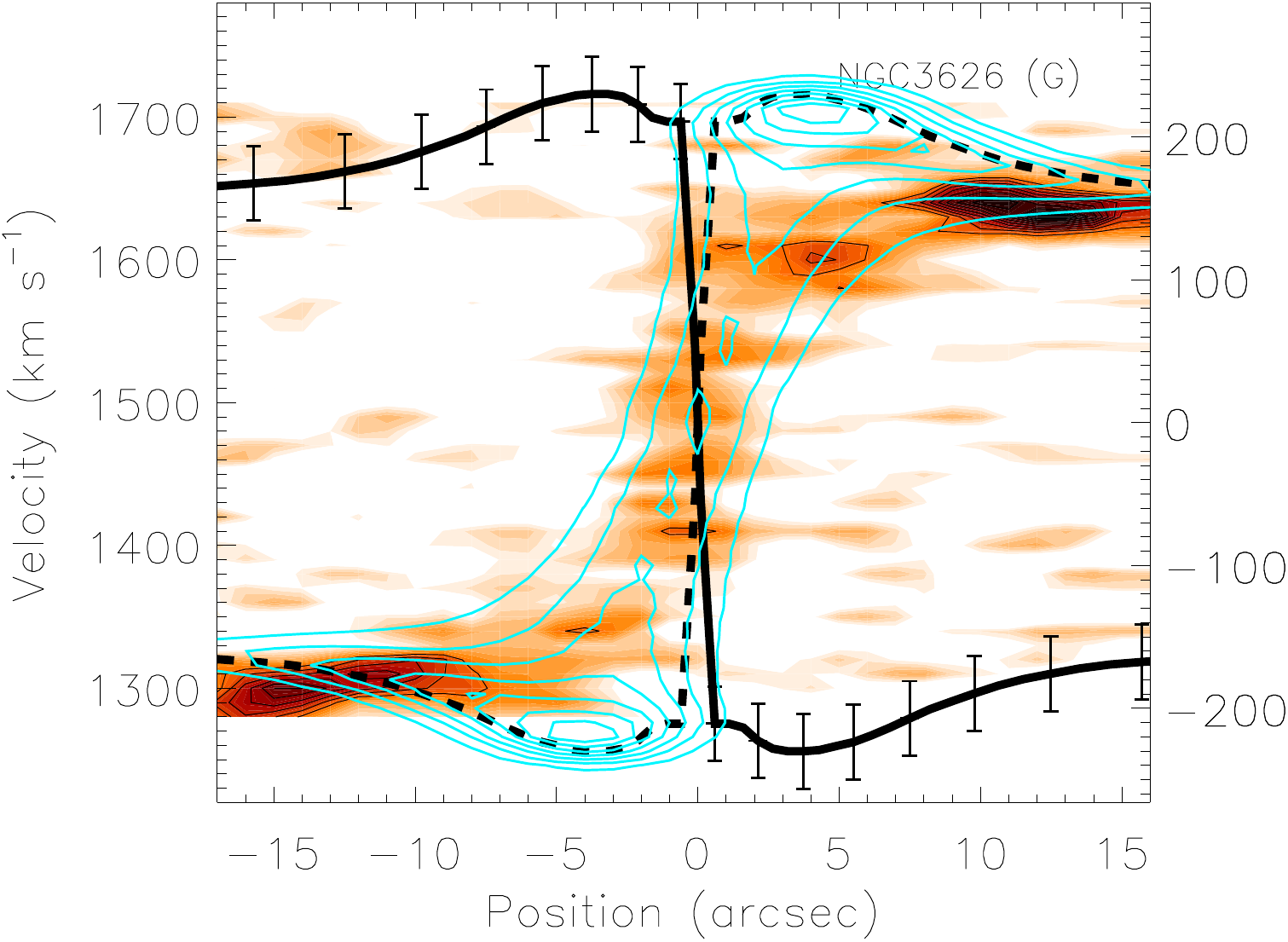}}
\subfigure{\includegraphics[scale=0.45]{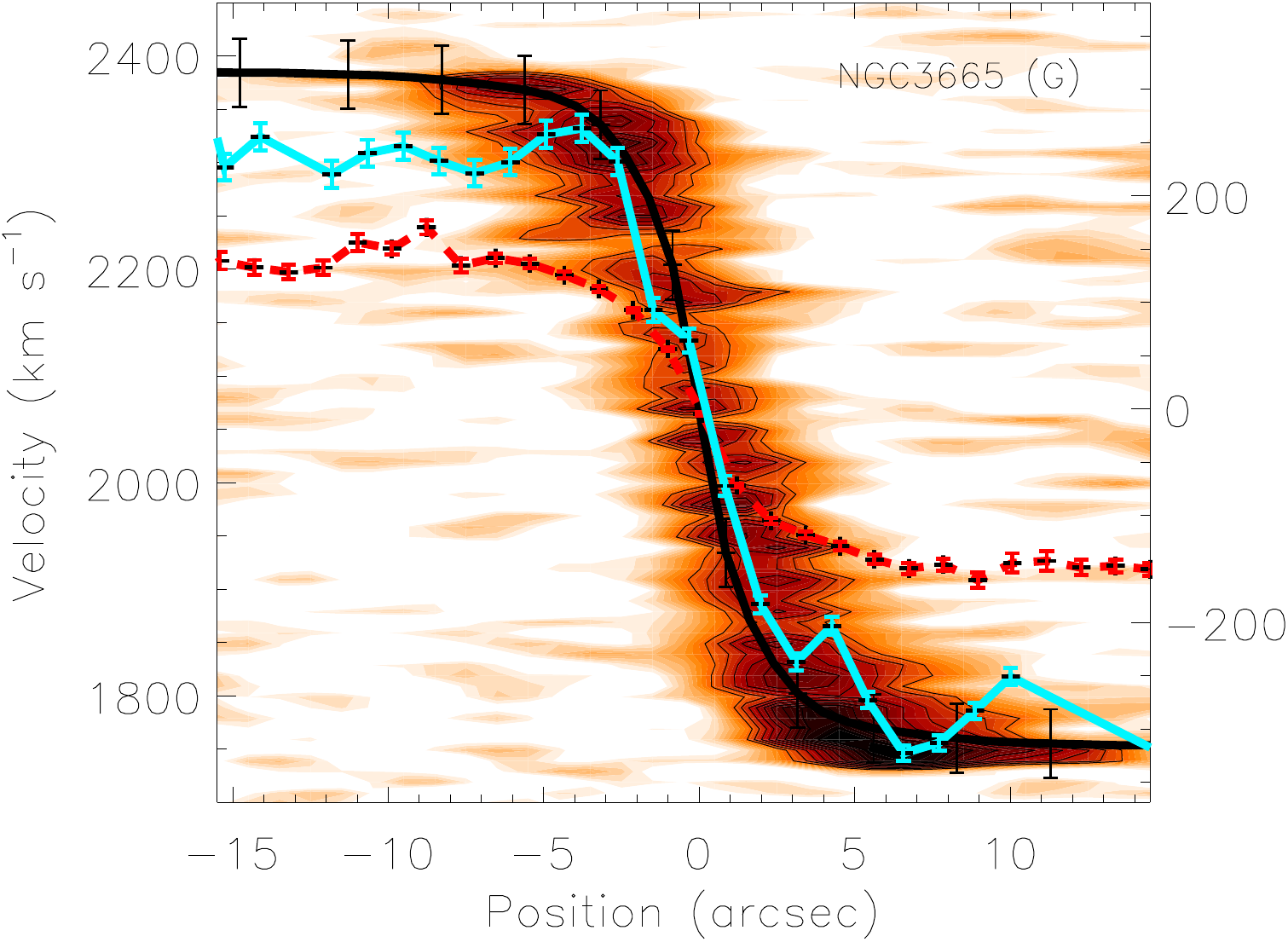}}
\subfigure{\includegraphics[scale=0.45]{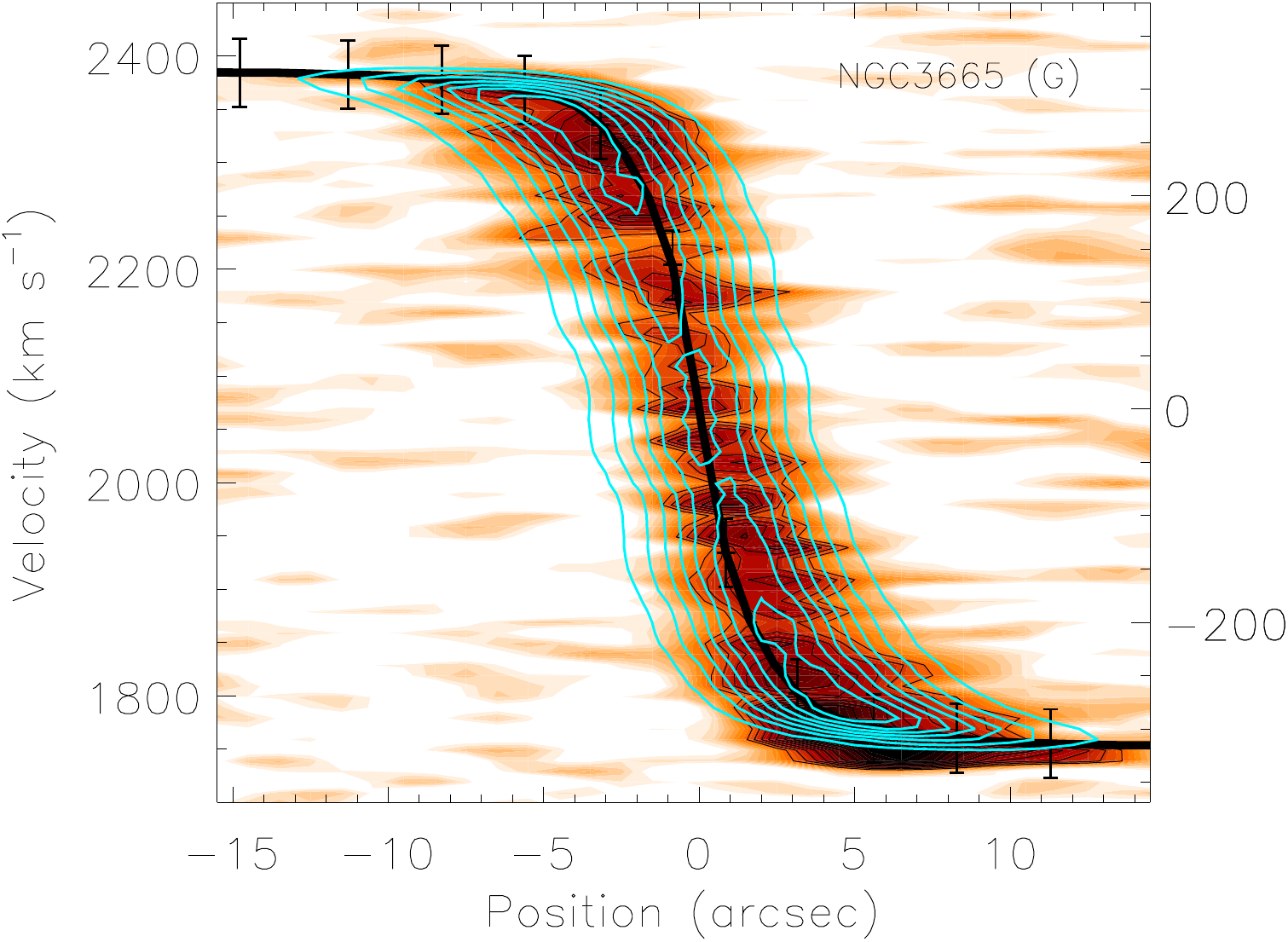}}
\subfigure{\includegraphics[scale=0.45]{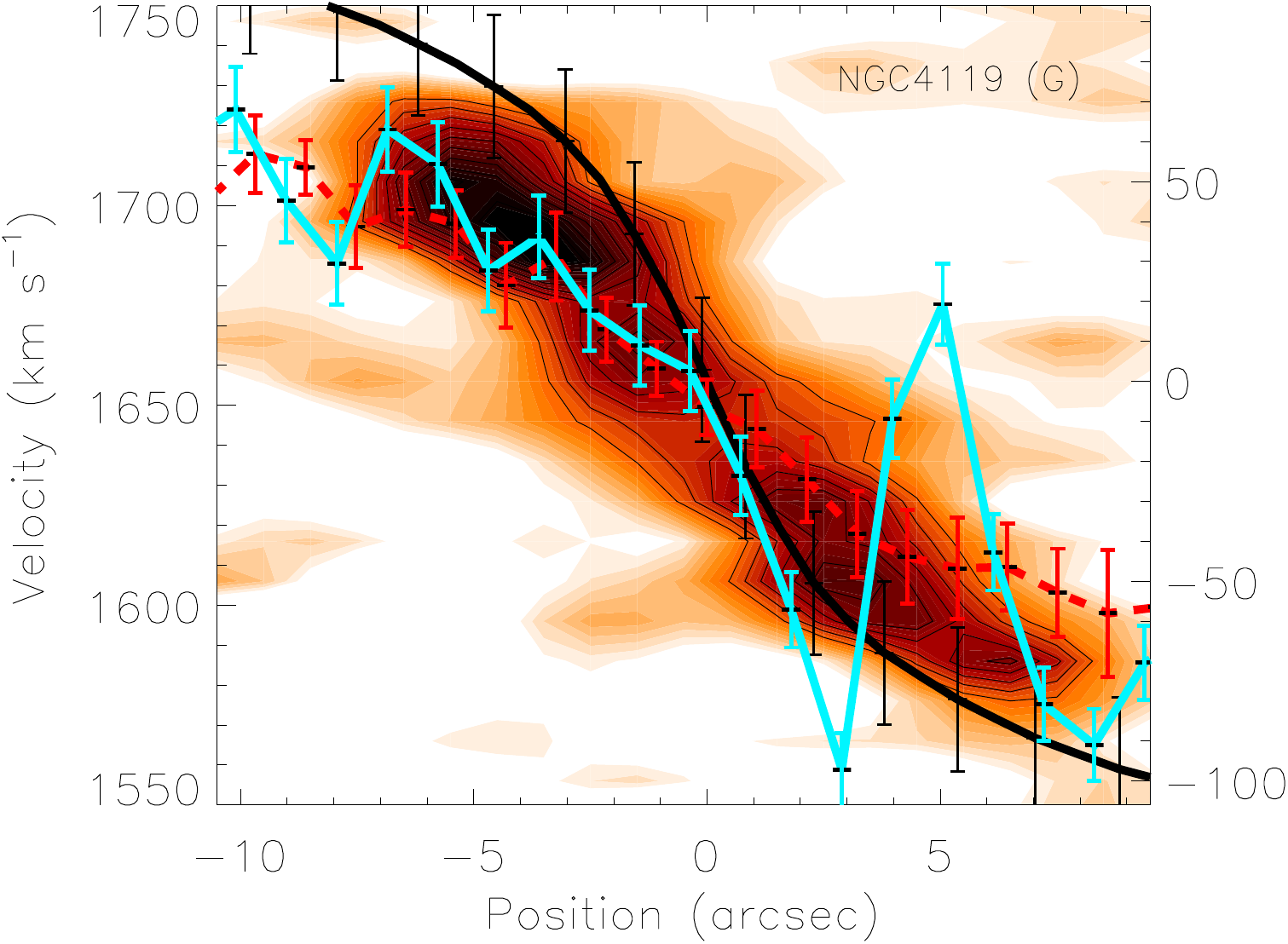}}
\subfigure{\includegraphics[scale=0.45]{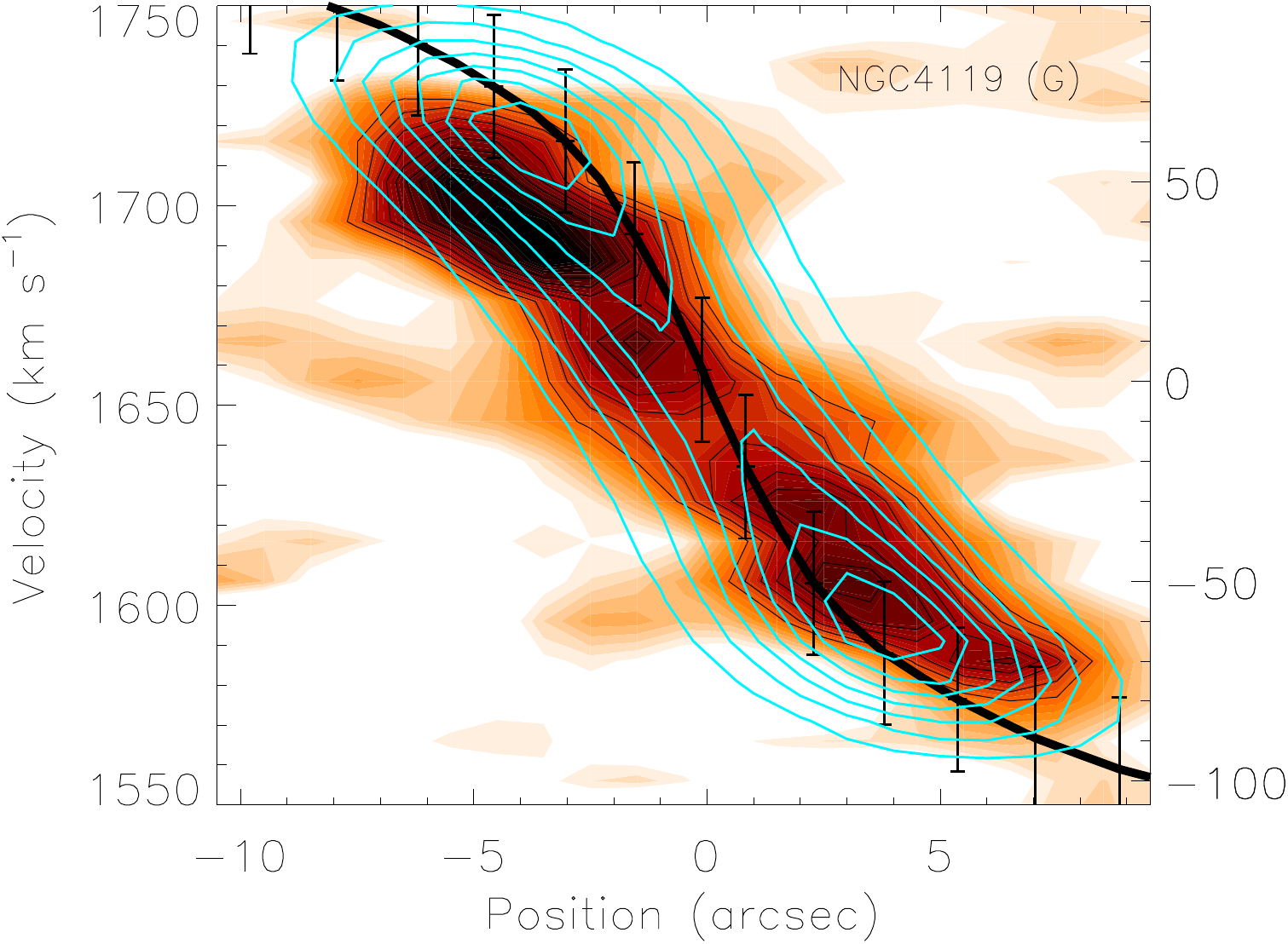}}
\parbox[t]{0.9 \textwidth}{ \textbf{Figure B1.} continued}
\end{figure*}
\begin{figure*}
\centering
\subfigure{\includegraphics[scale=0.45]{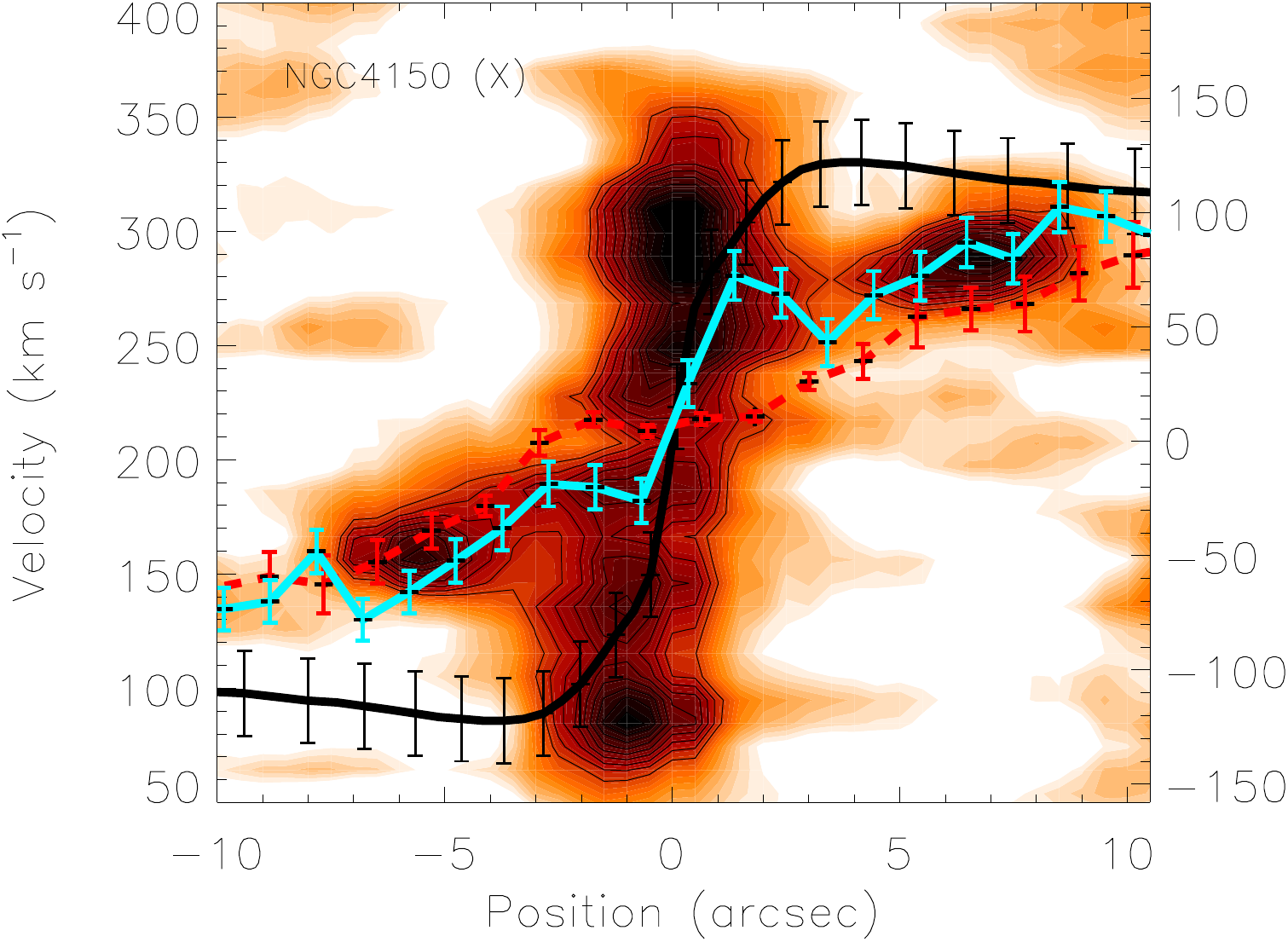}}
\subfigure{\includegraphics[scale=0.45]{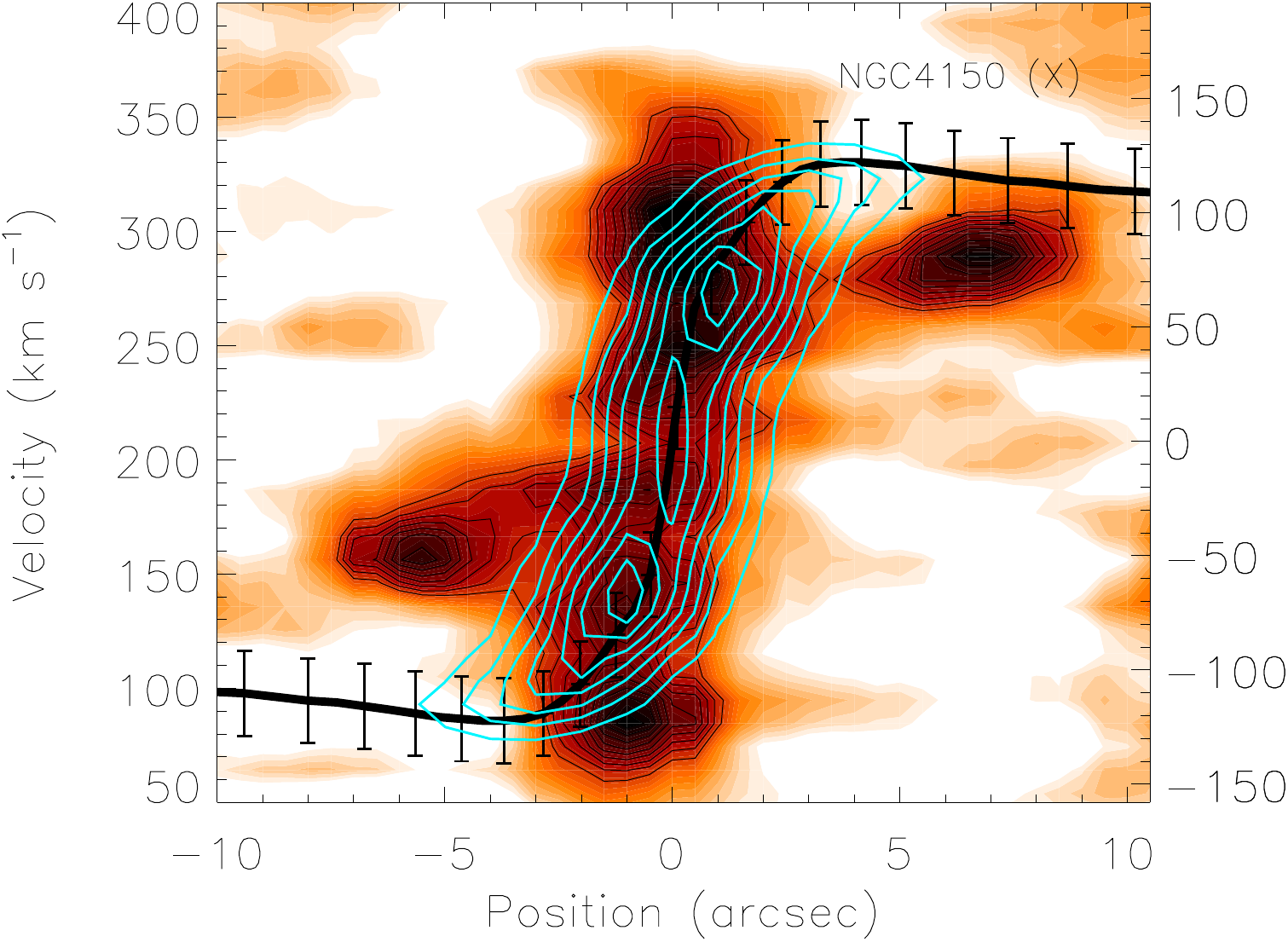}}
\subfigure{\includegraphics[scale=0.45]{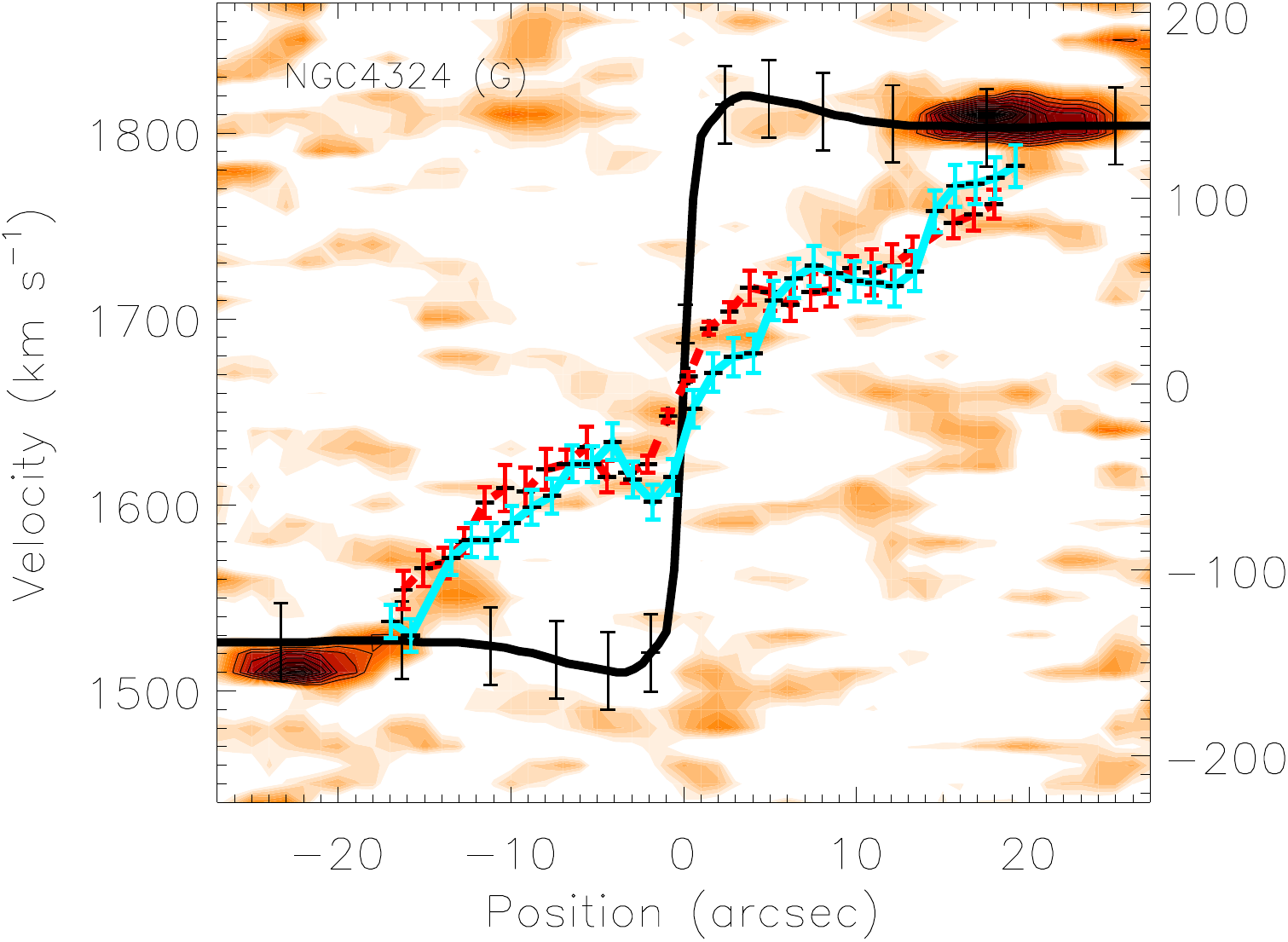}}
\subfigure{\includegraphics[scale=0.45]{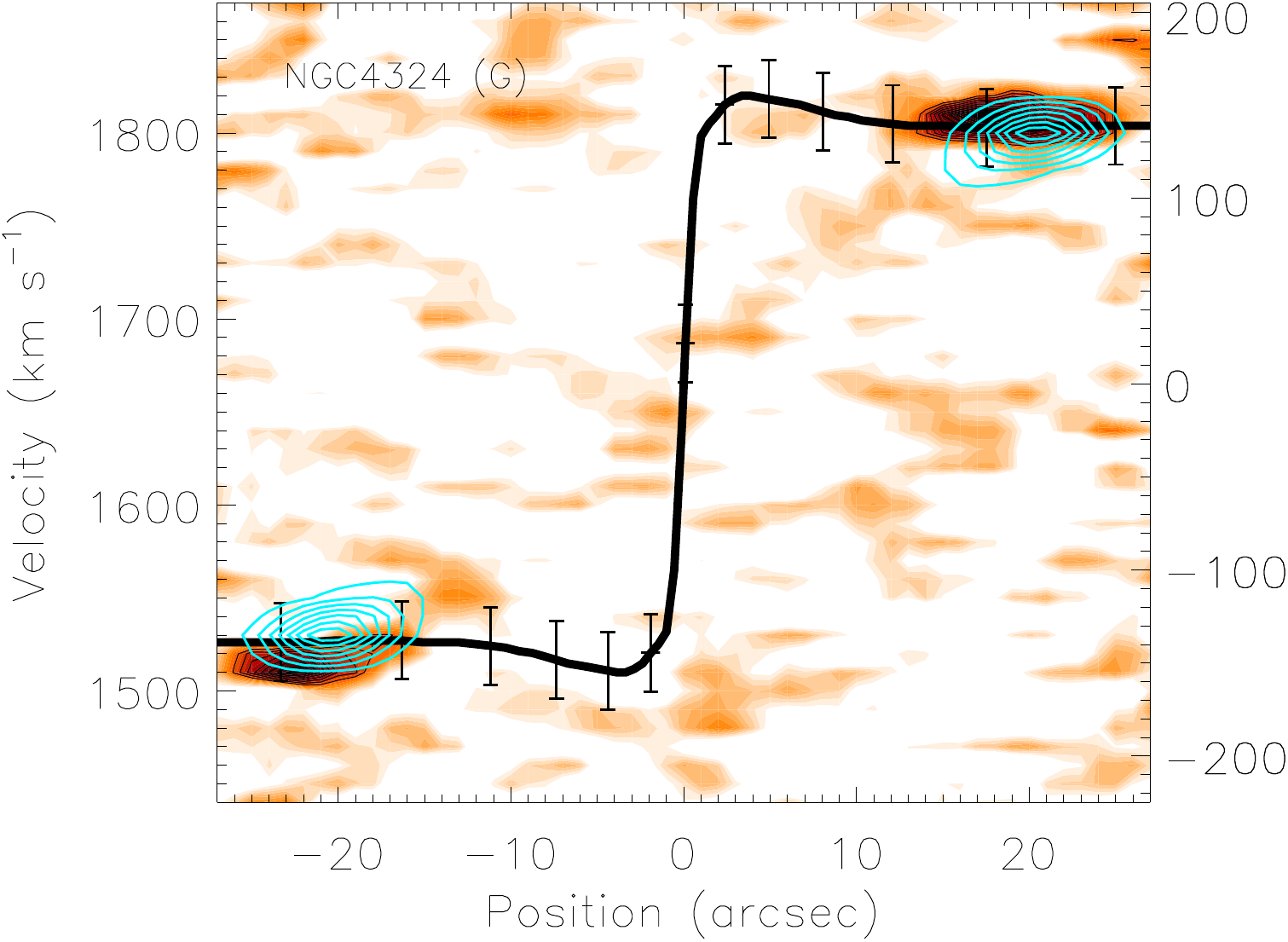}}
\subfigure{\includegraphics[scale=0.45]{plots/NGC4429pvdiag_comp_stars.pdf}}
\subfigure{\includegraphics[scale=0.45]{simplots/NGC4429pvsim_comp.pdf}}
\subfigure{\includegraphics[scale=0.45]{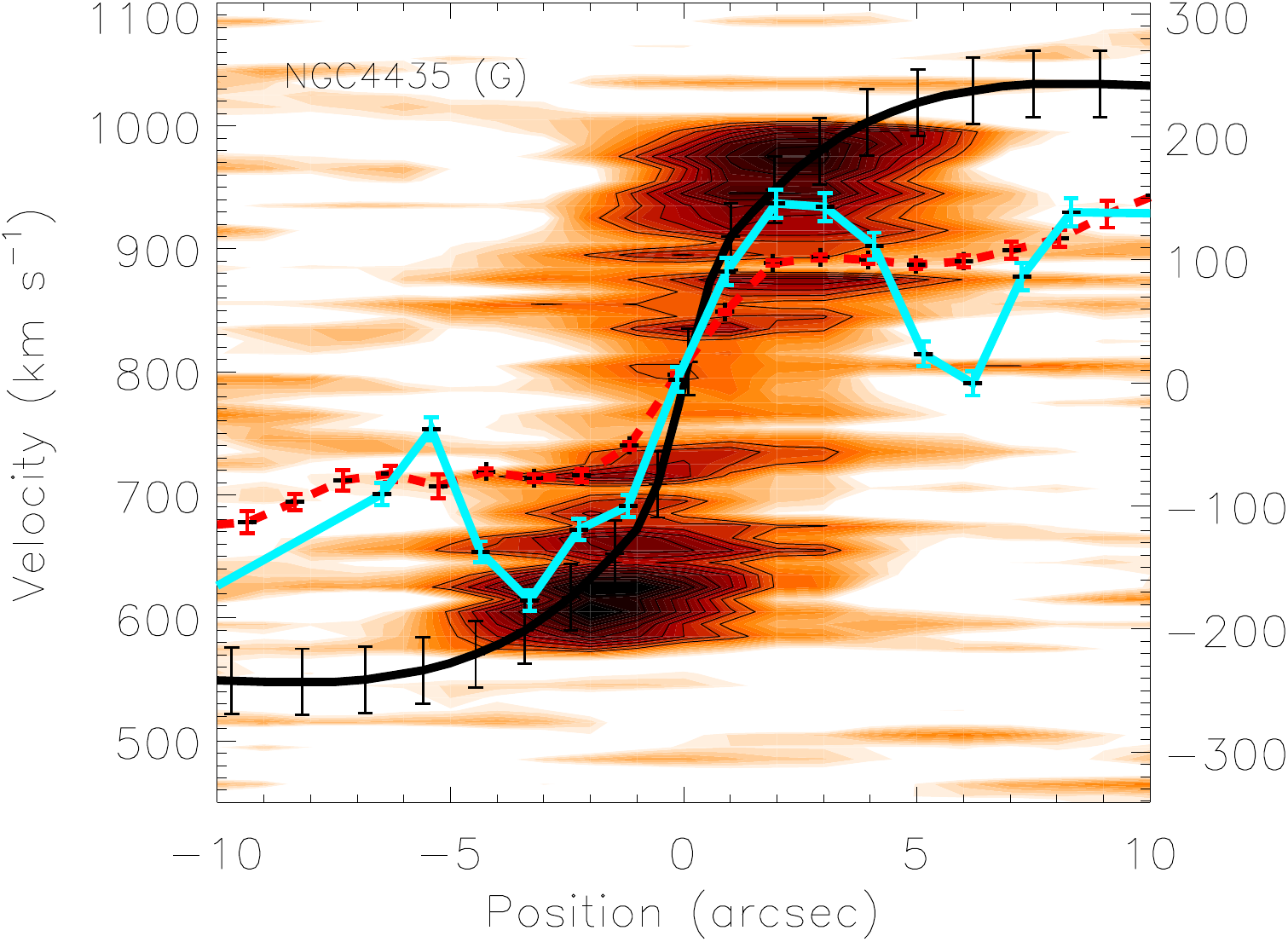}}
\subfigure{\includegraphics[scale=0.45]{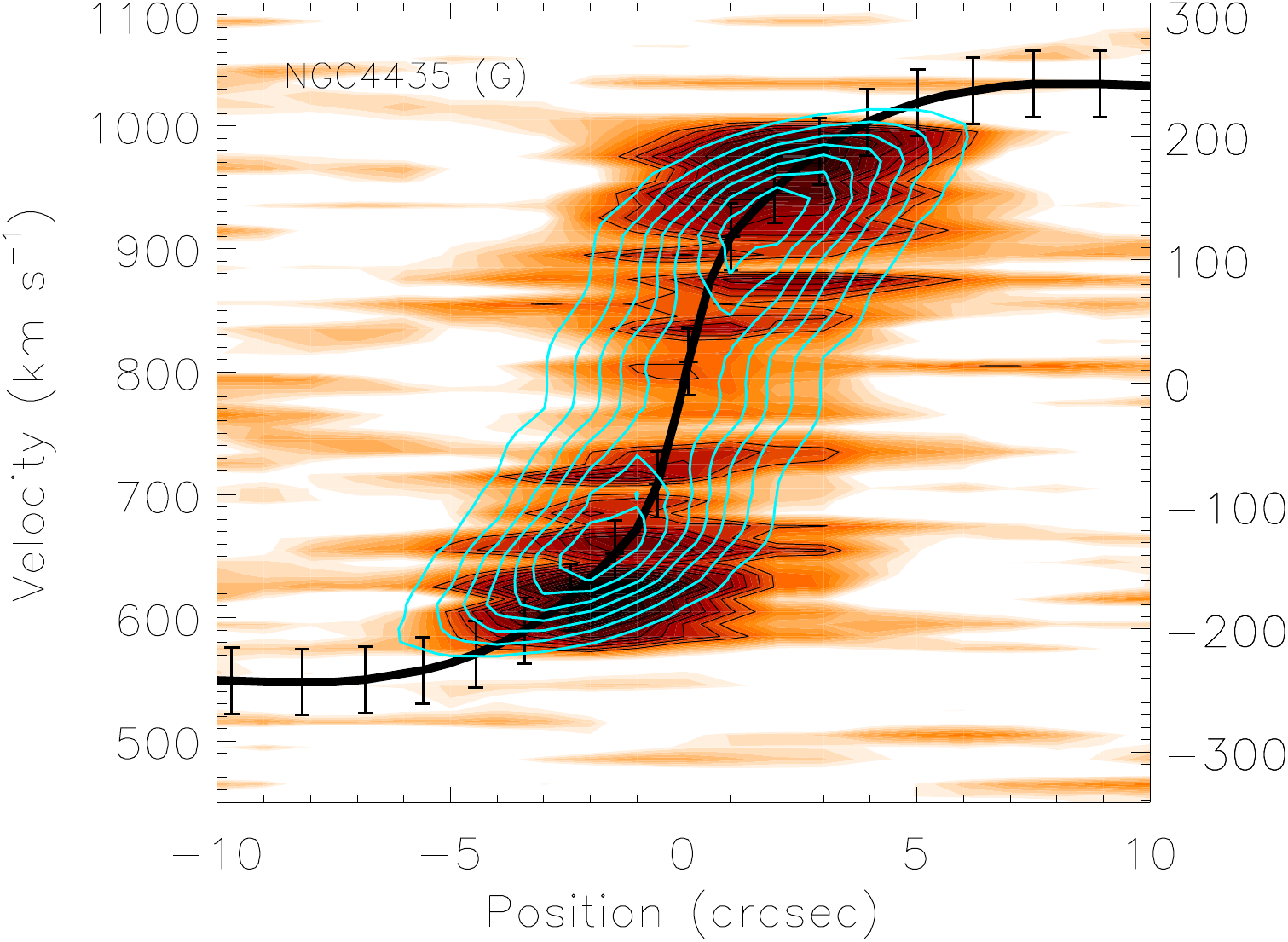}}
\parbox[t]{0.9 \textwidth}{ \textbf{Figure B1.} continued}
\end{figure*}
\begin{figure*}
\centering
\subfigure{\includegraphics[scale=0.45]{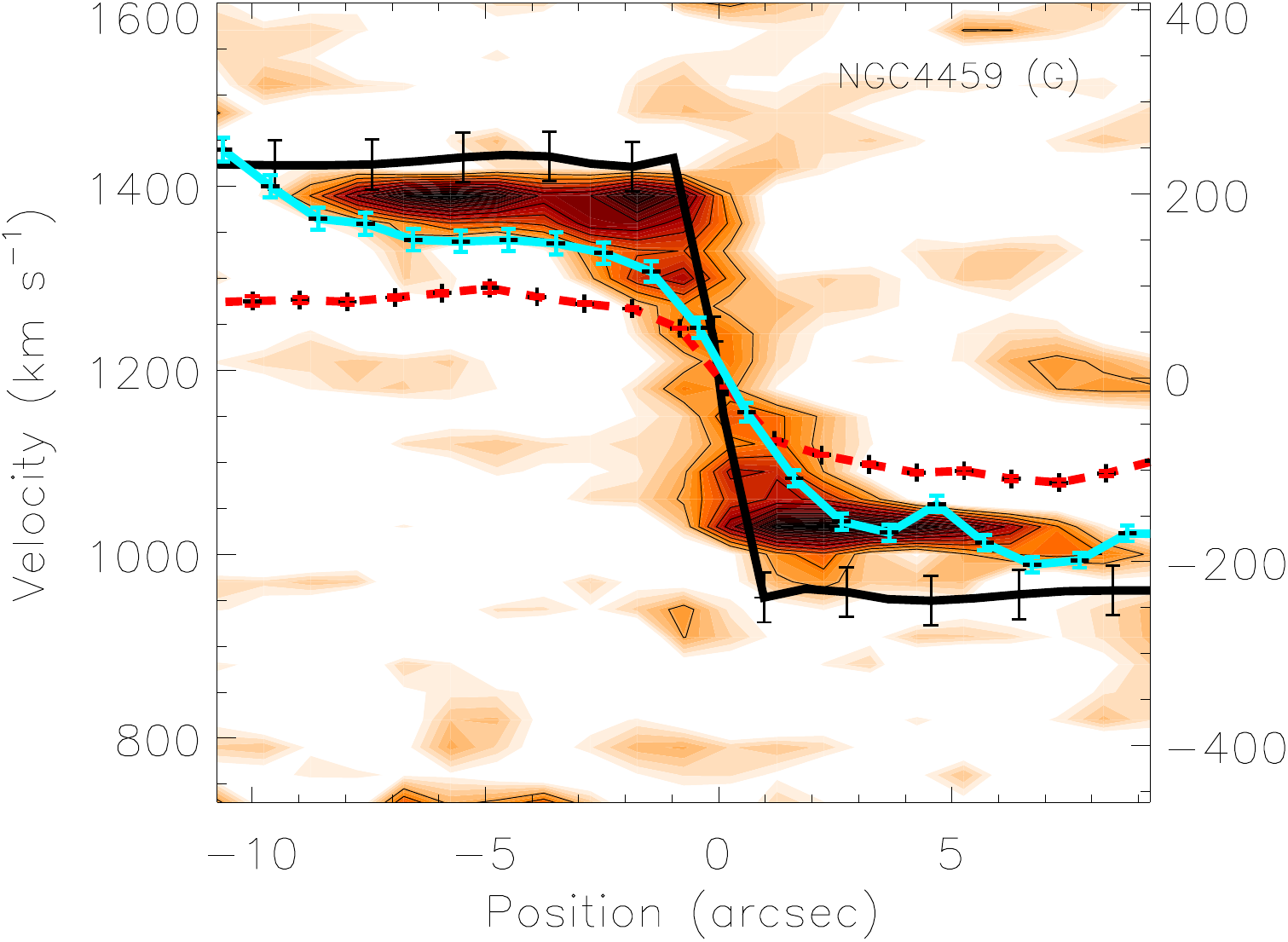}}
\subfigure{\includegraphics[scale=0.45]{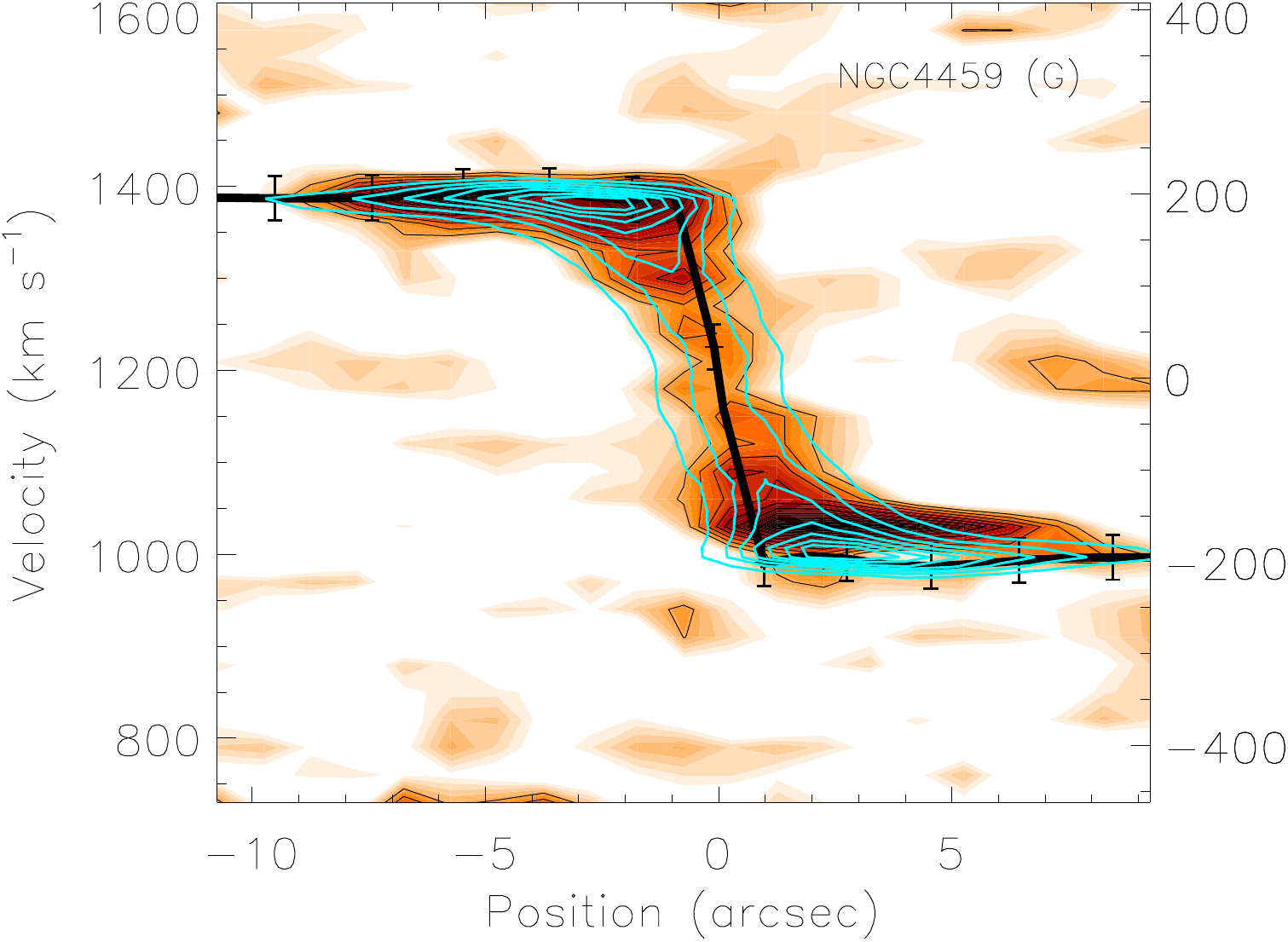}}
\subfigure{\includegraphics[scale=0.45]{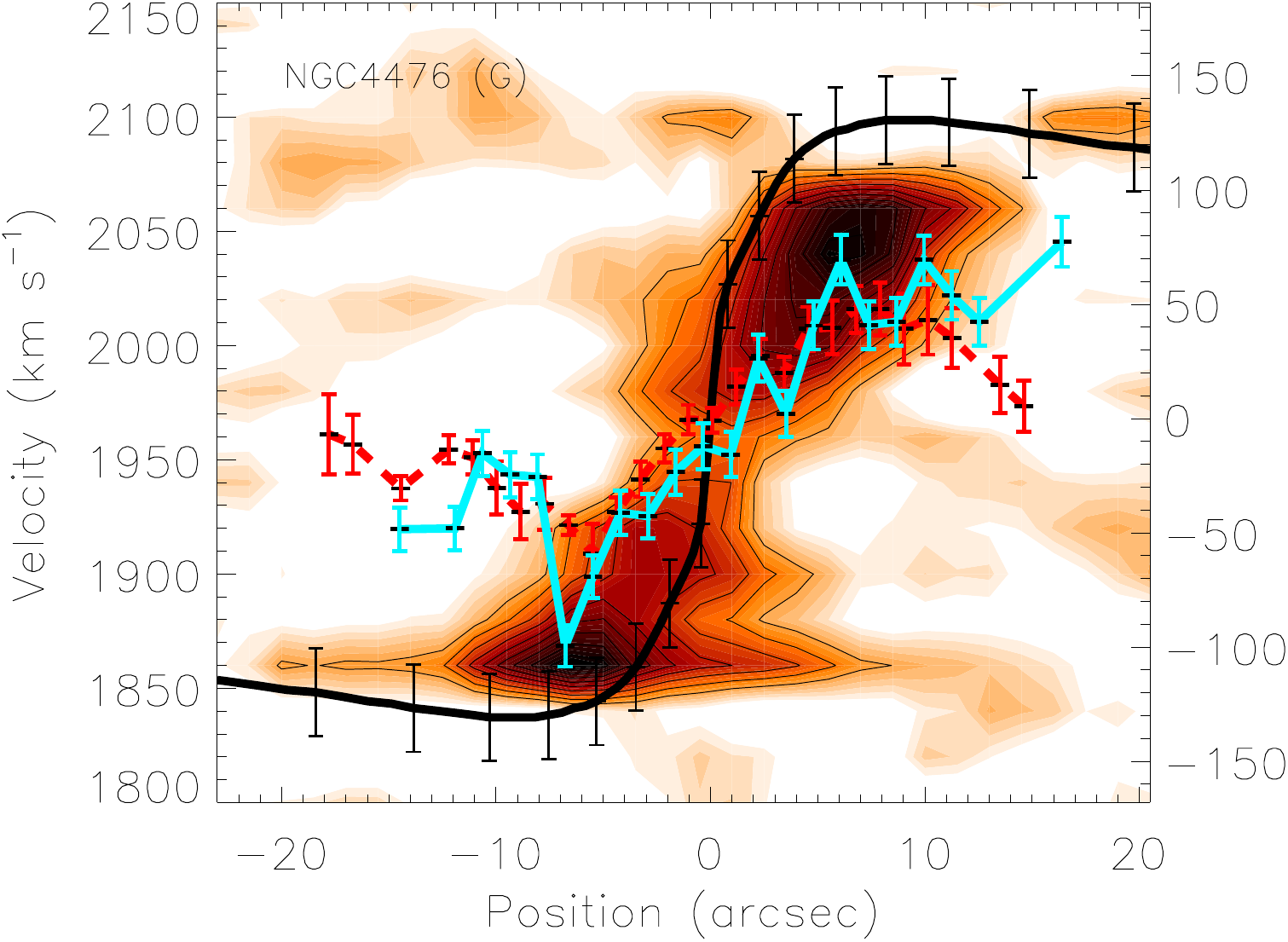}}
\subfigure{\includegraphics[scale=0.45]{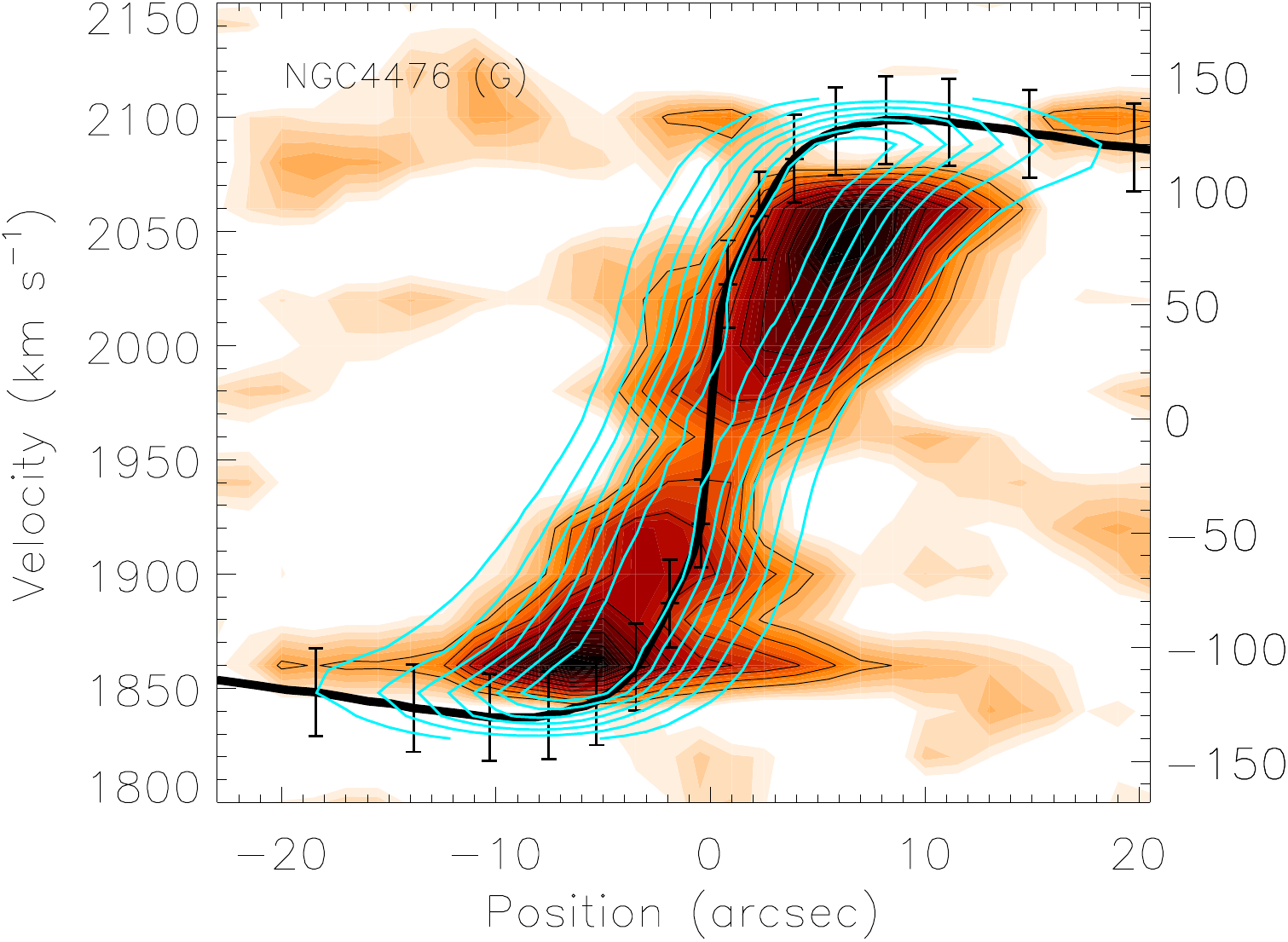}}
\subfigure{\includegraphics[scale=0.45]{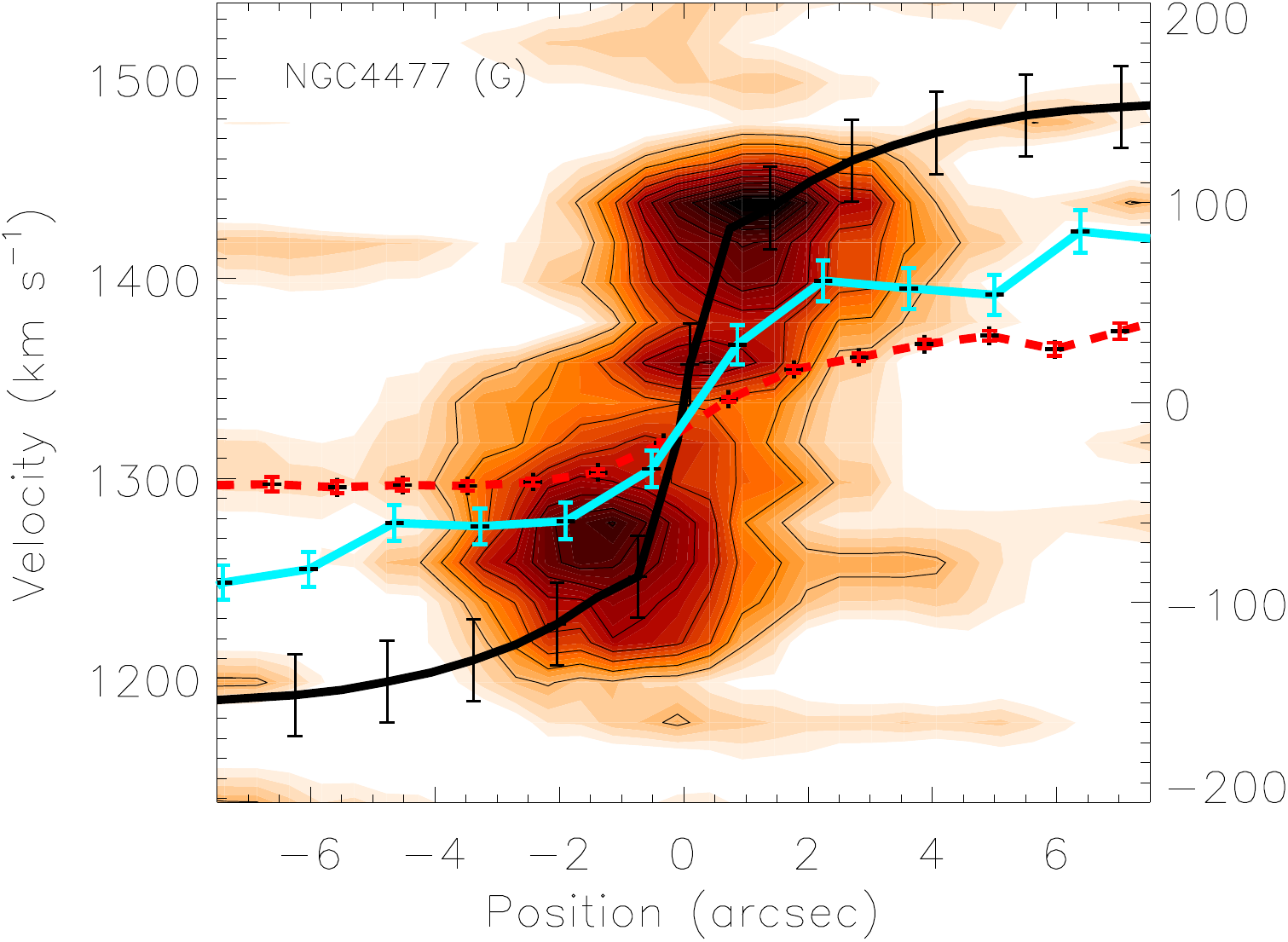}}
\subfigure{\includegraphics[scale=0.45]{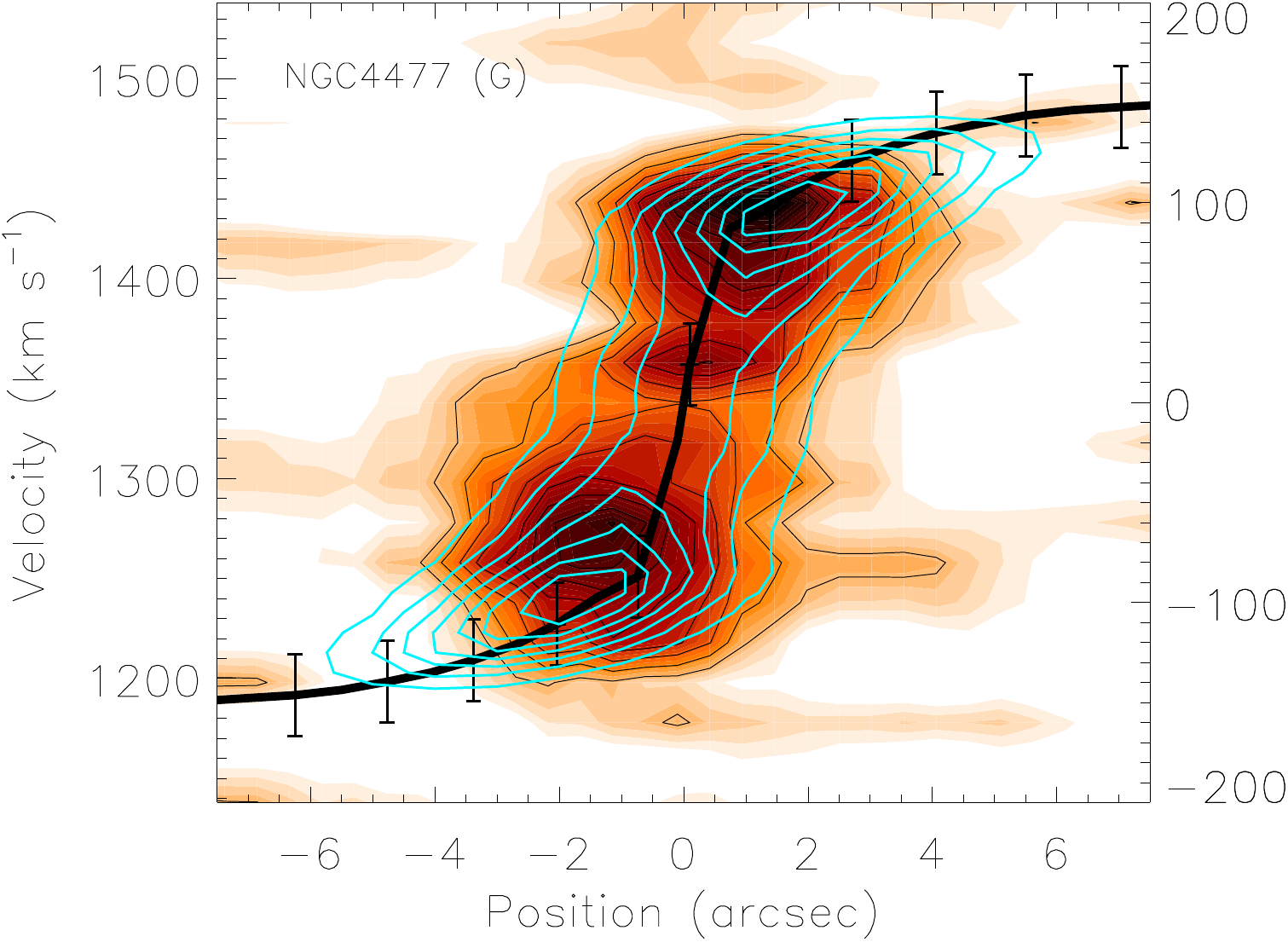}}
\subfigure{\includegraphics[scale=0.45]{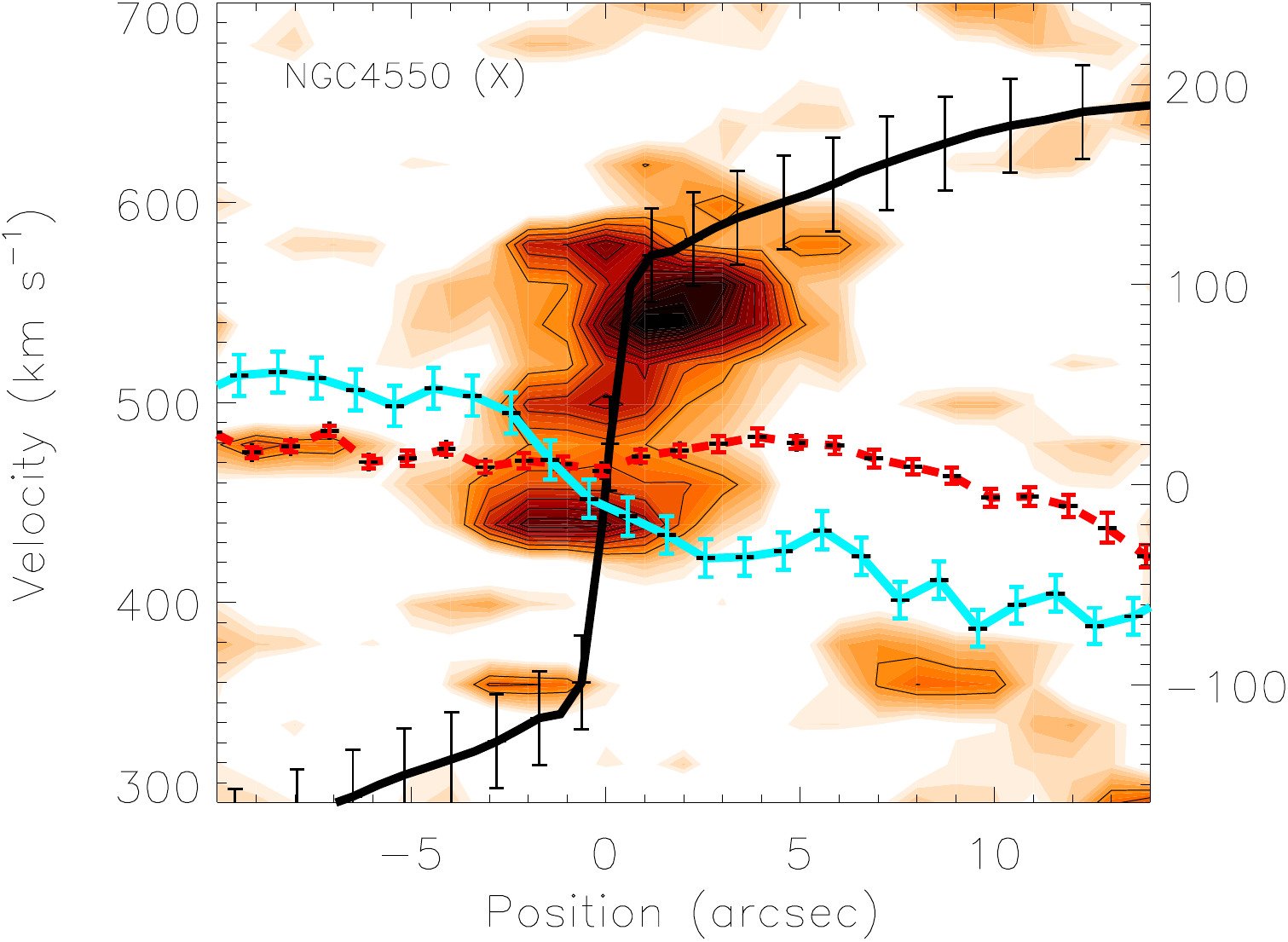}}
\subfigure{\includegraphics[scale=0.45]{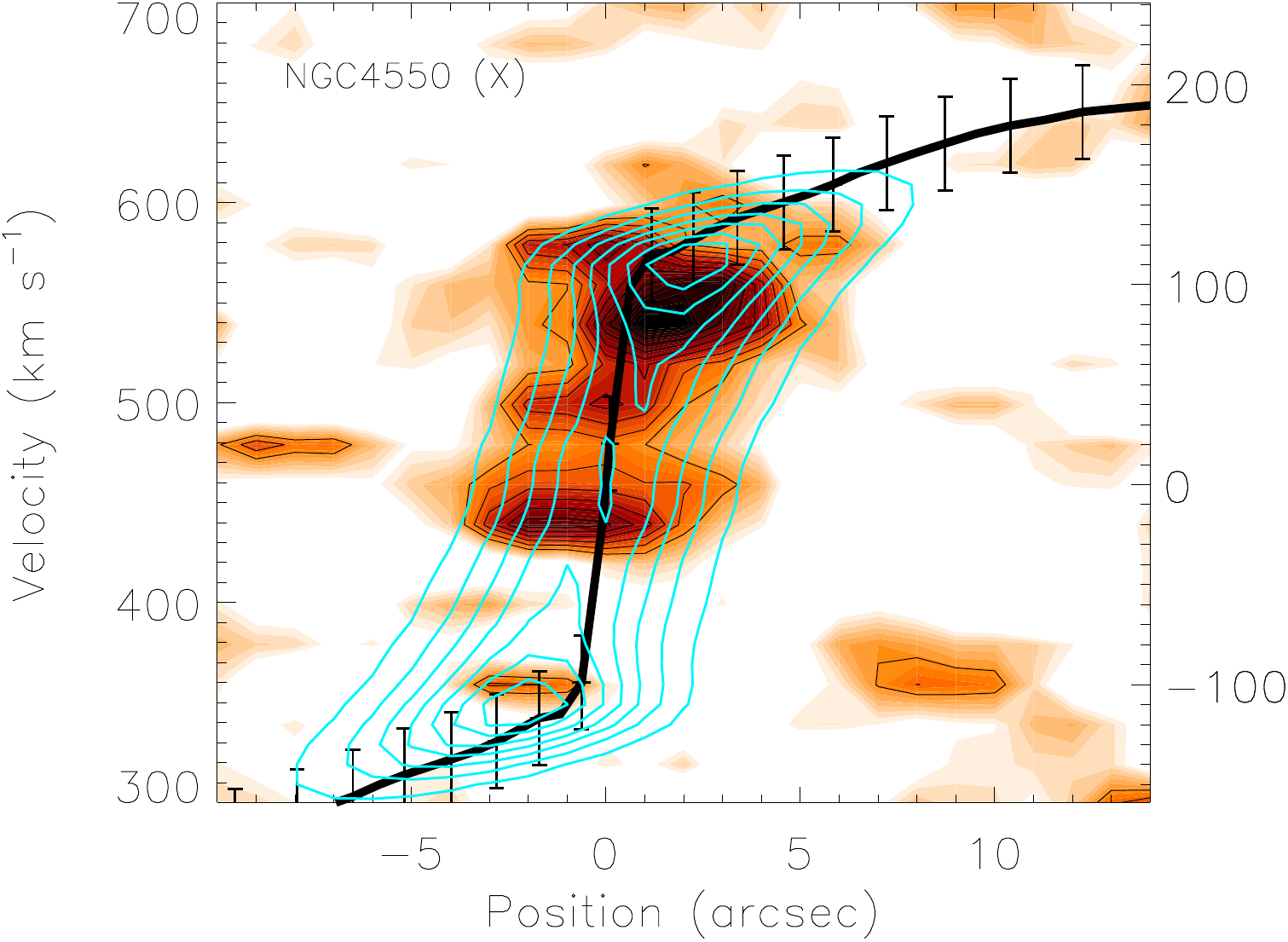}}
\parbox[t]{0.9 \textwidth}{ \textbf{Figure B1.} continued}
\end{figure*}
\begin{figure*}
\centering
\subfigure{\includegraphics[scale=0.45]{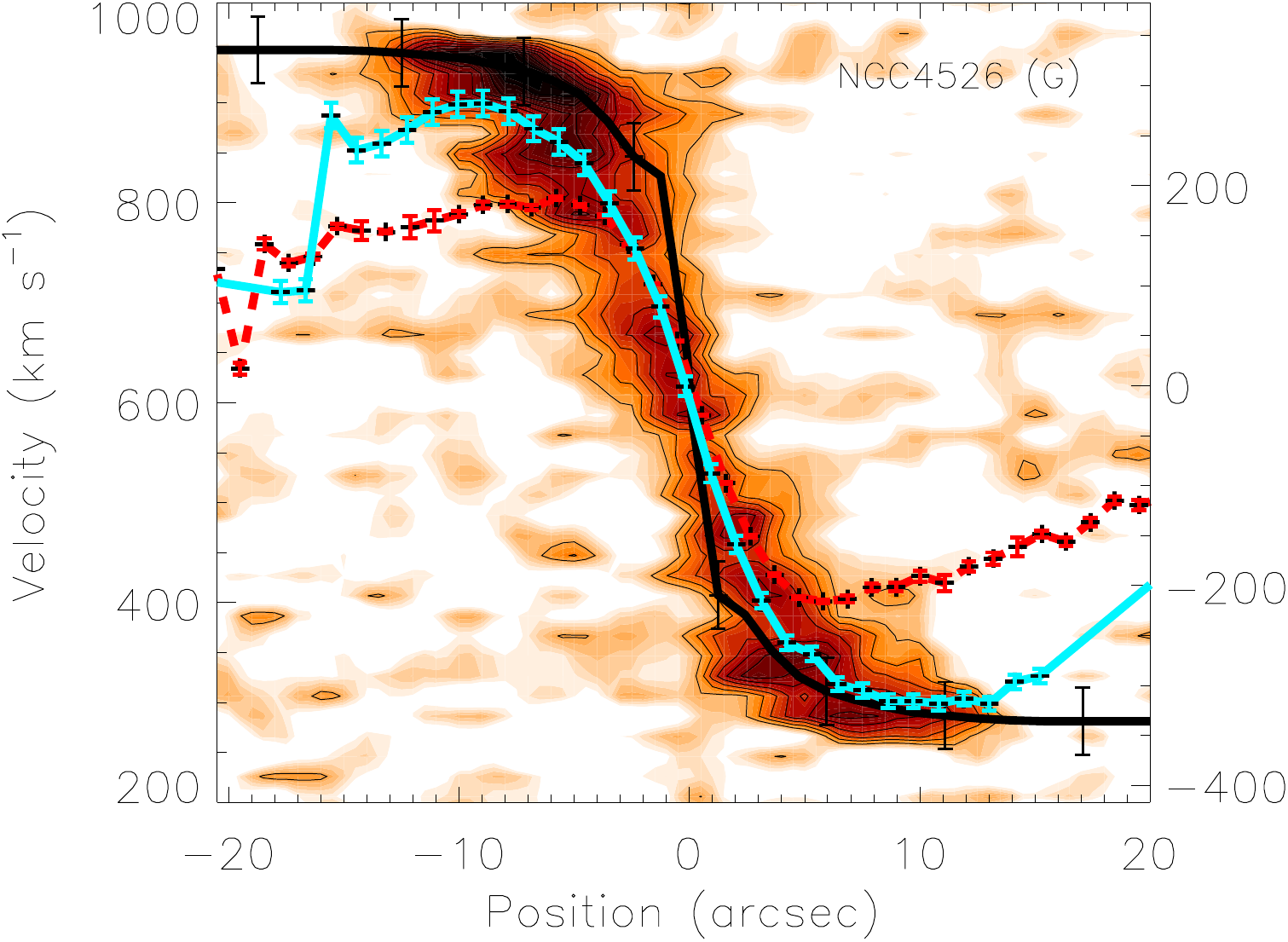}}
\subfigure{\includegraphics[scale=0.45]{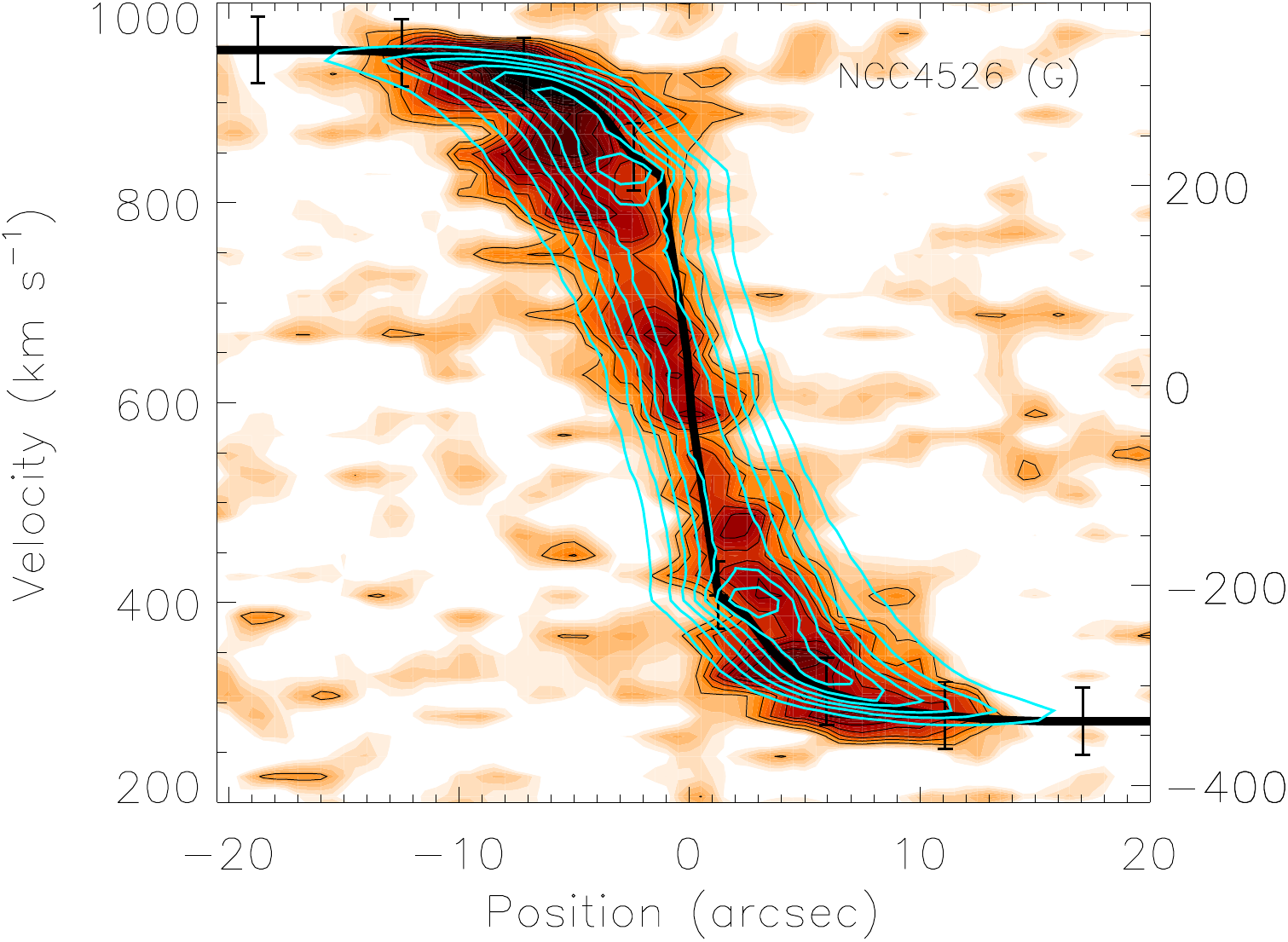}}
\subfigure{\includegraphics[scale=0.45]{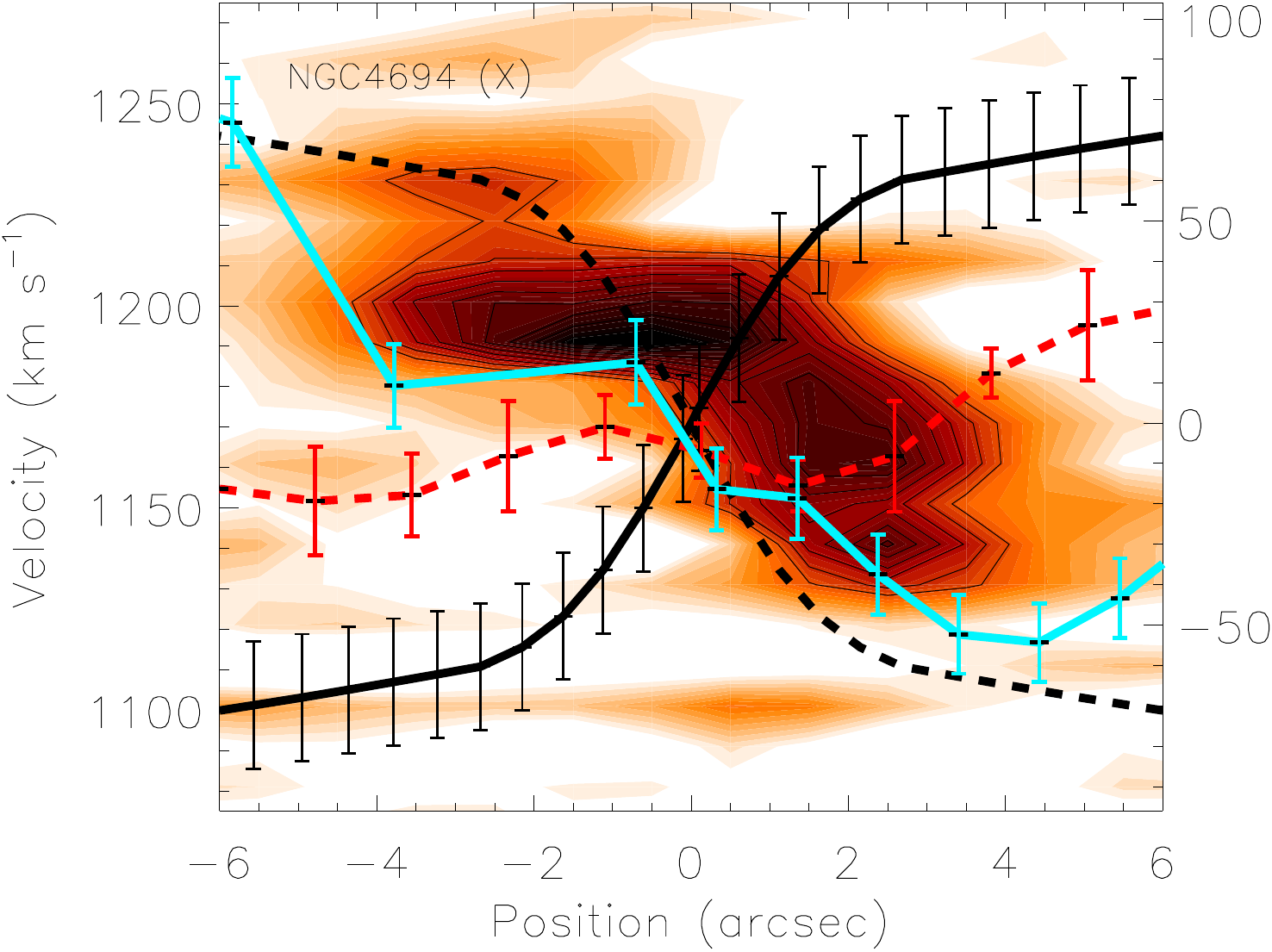}}
\subfigure{\includegraphics[scale=0.45]{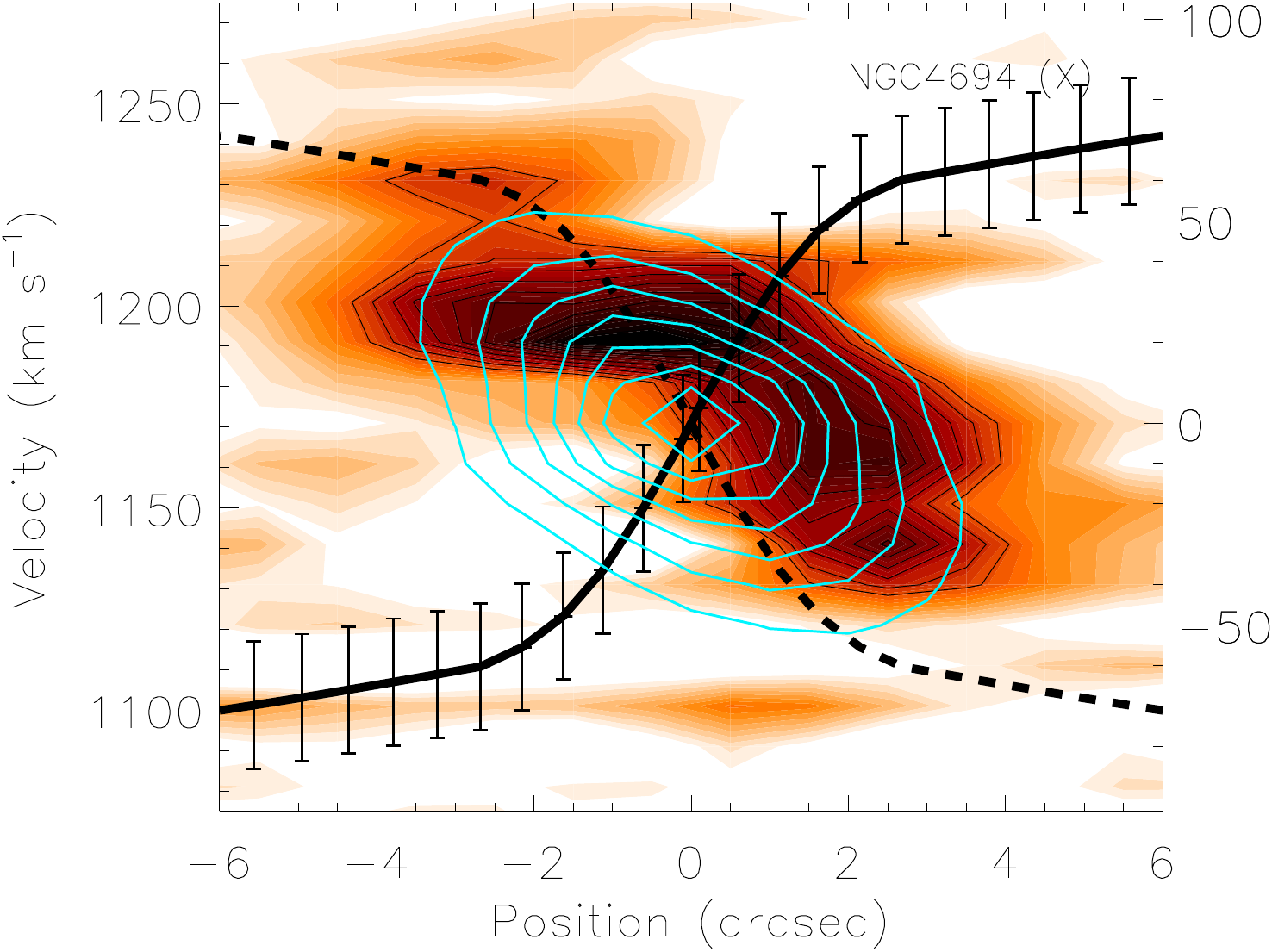}}
\subfigure{\includegraphics[scale=0.45]{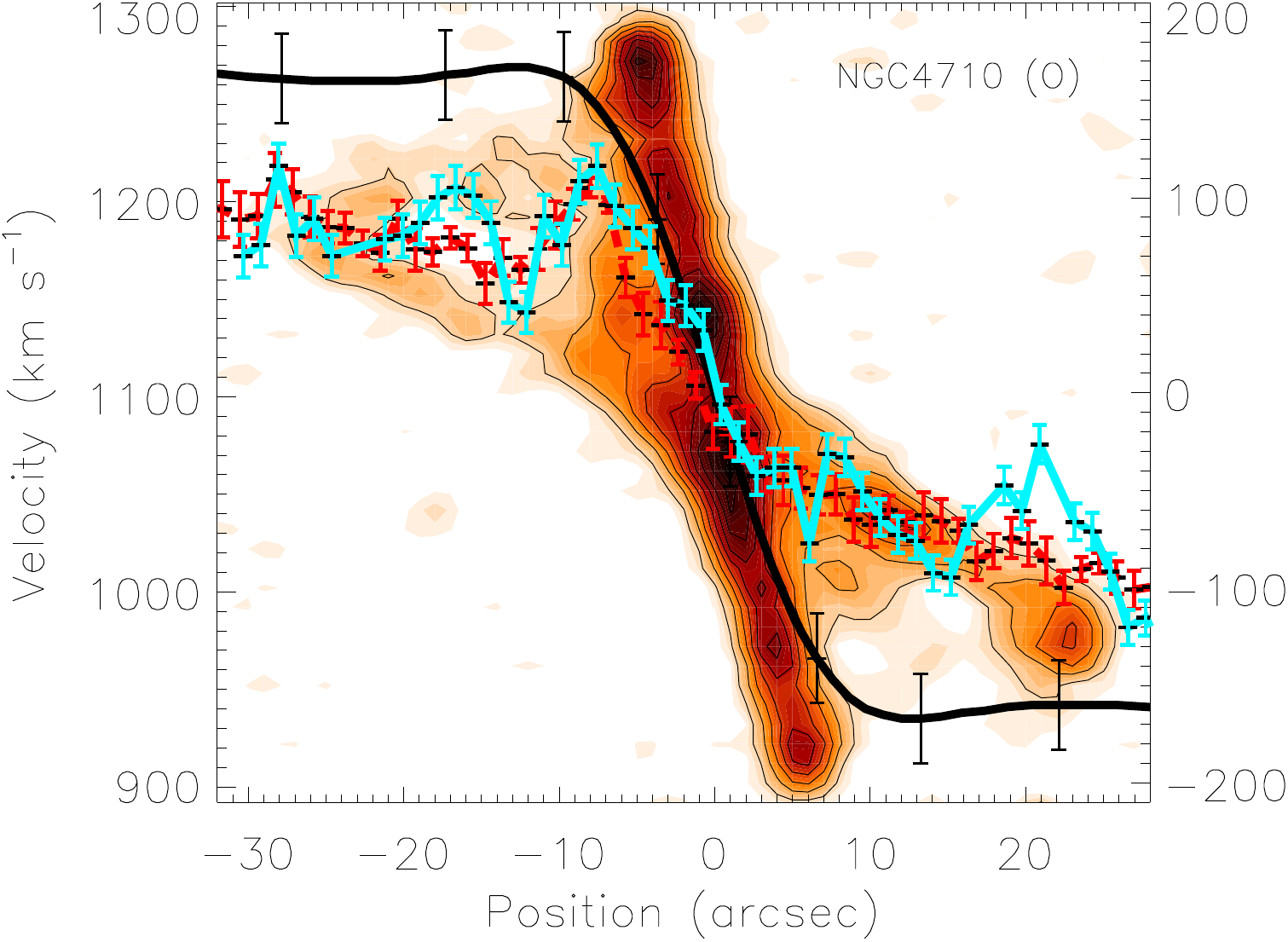}}
\subfigure{\includegraphics[scale=0.45]{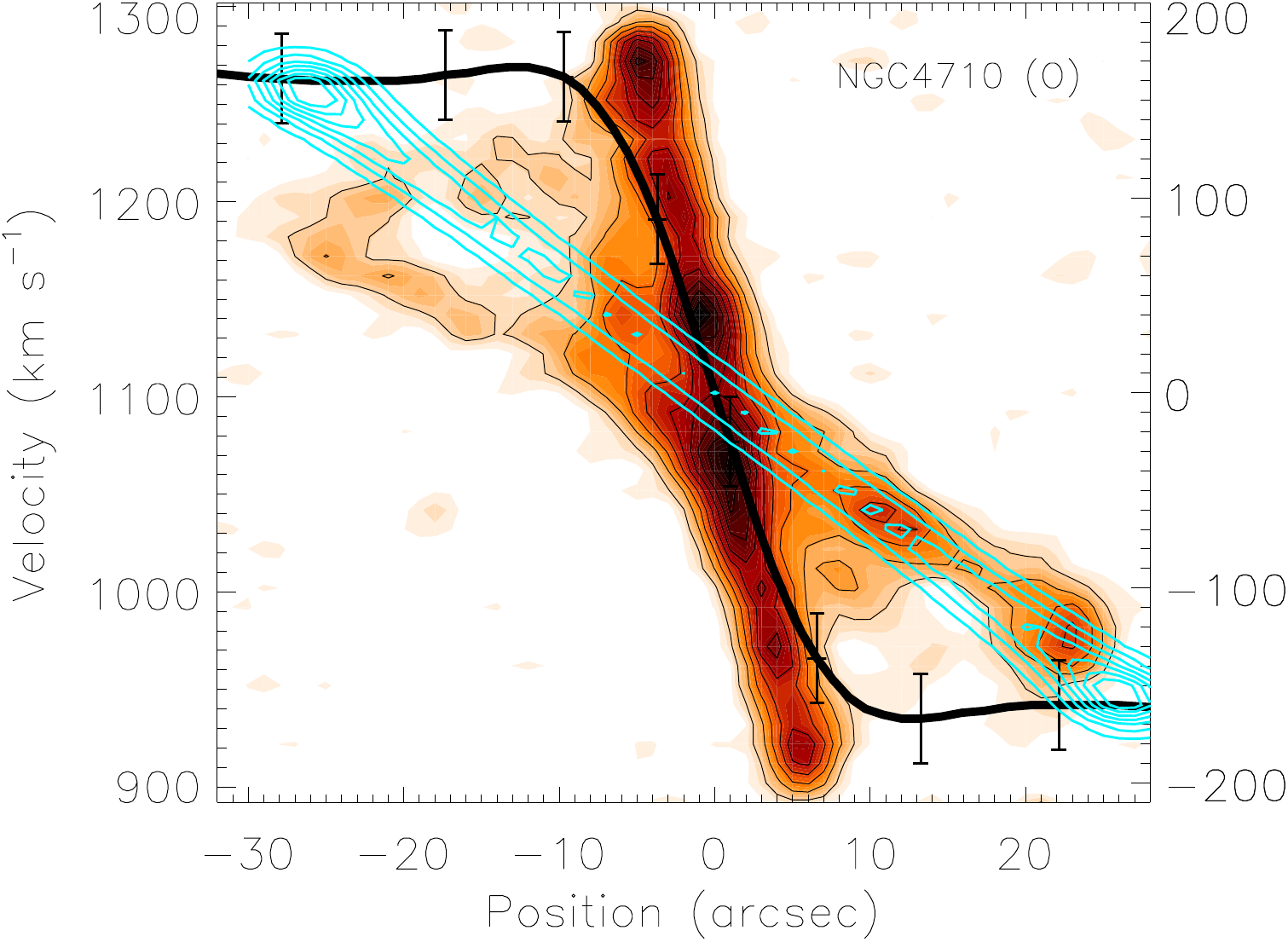}}
\subfigure{\includegraphics[scale=0.45]{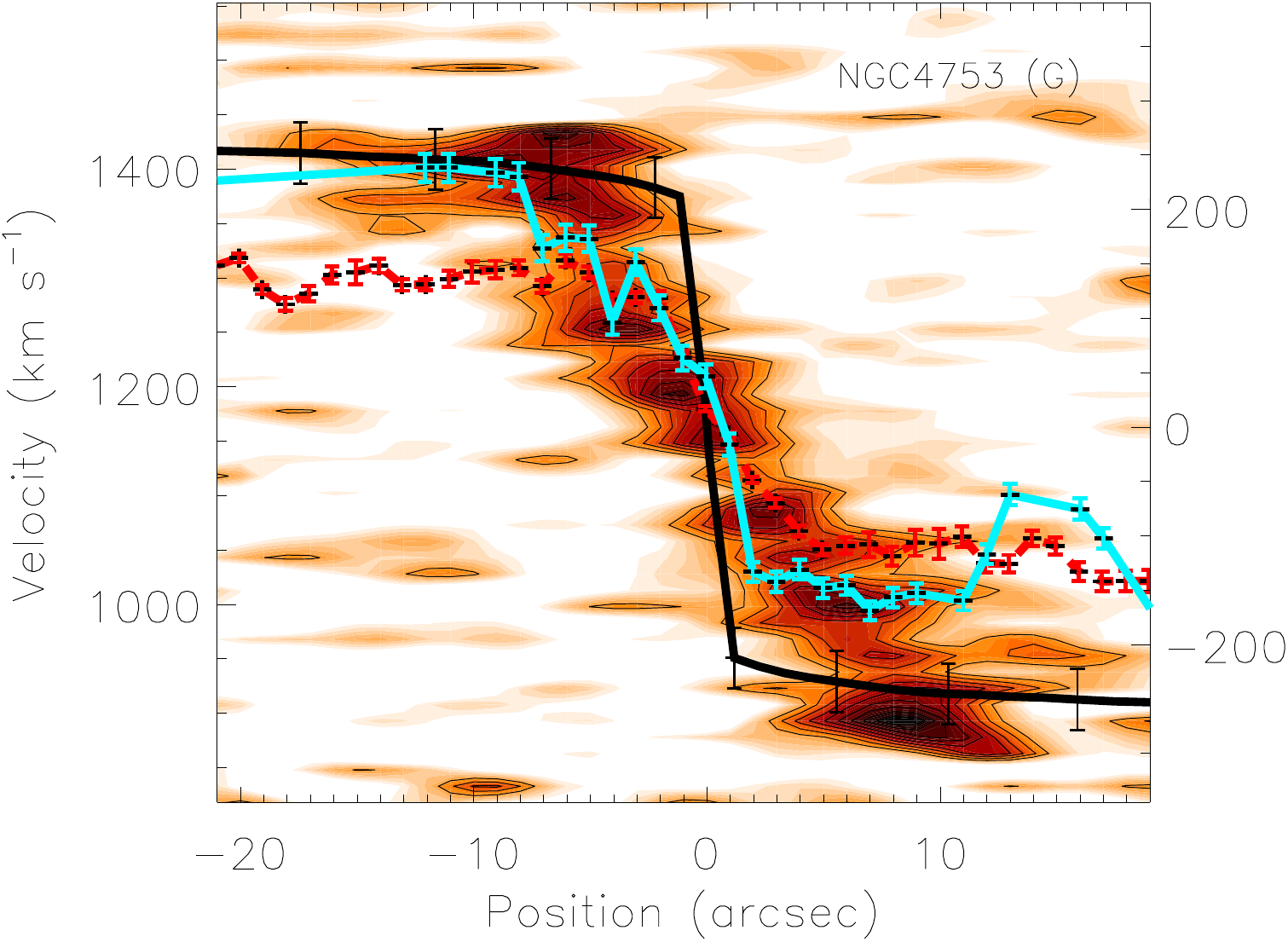}}
\subfigure{\includegraphics[scale=0.45]{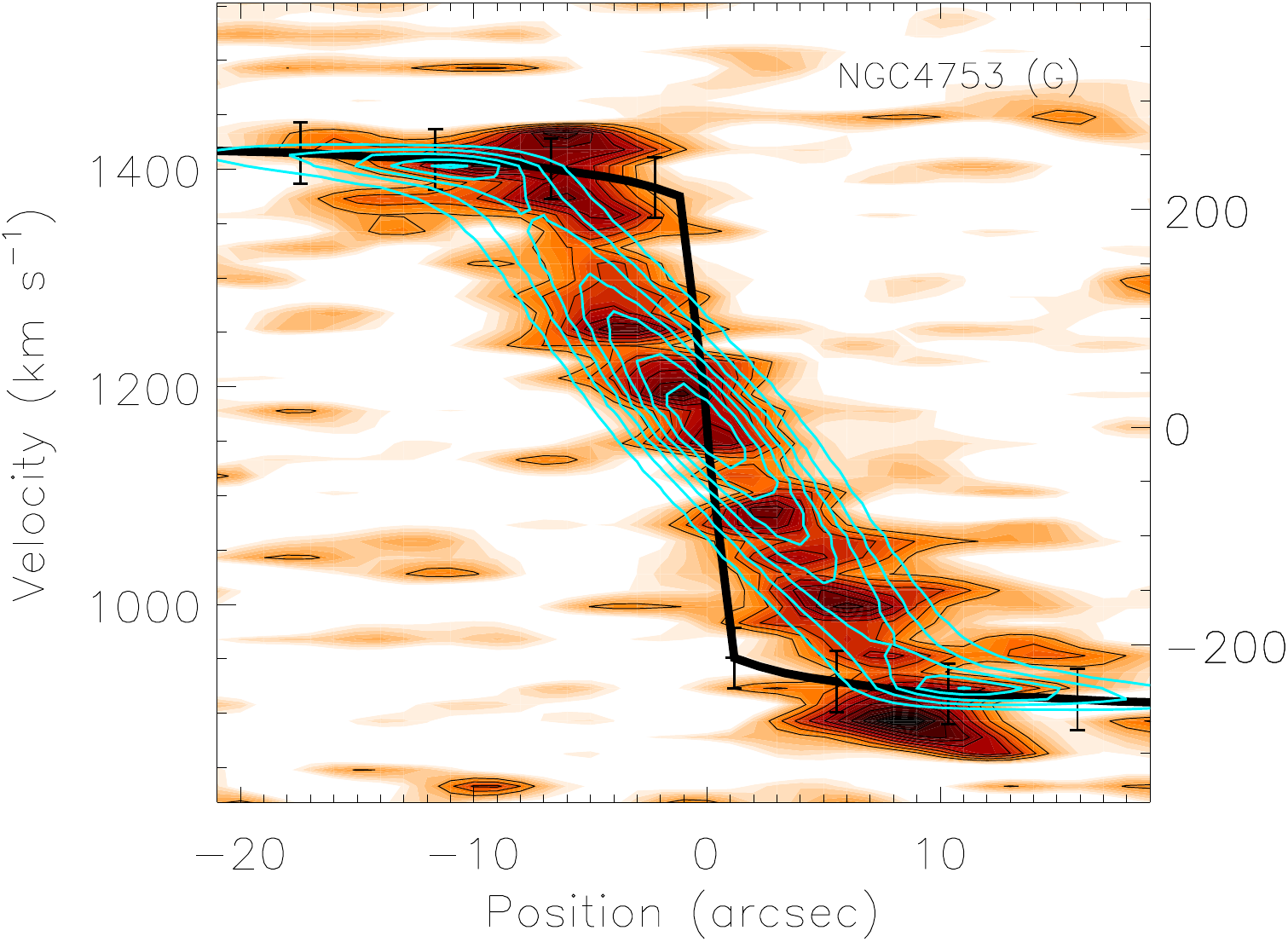}}
\parbox[t]{0.9 \textwidth}{ \textbf{Figure B1.} continued. NGC\,4710 has an MGE model heavily affected by dust, and bar, which together may cause the disagreement between model and observations (Scott et al., in prep).}
\end{figure*}
\begin{figure*}
\centering
\subfigure{\includegraphics[scale=0.45]{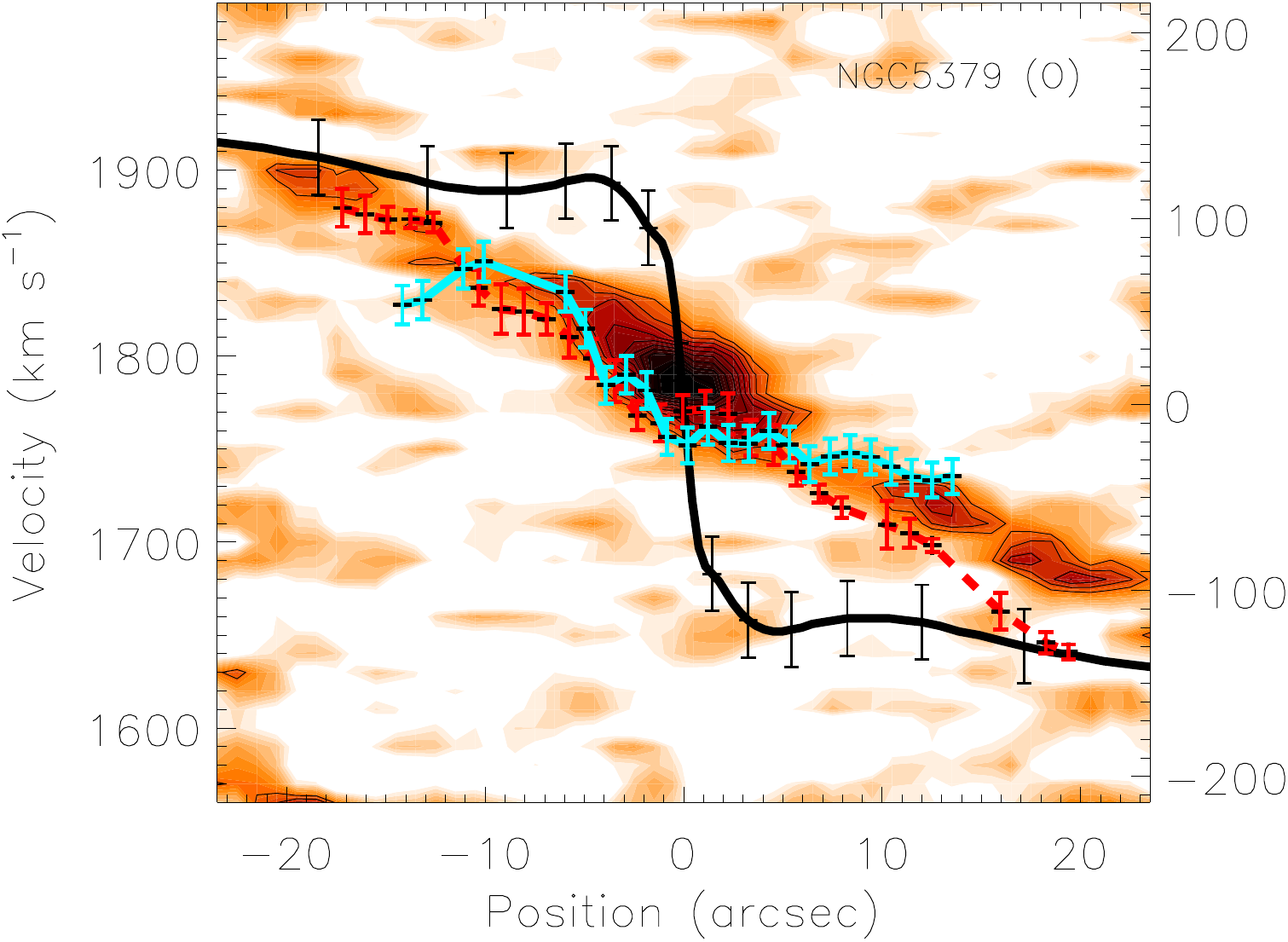}}
\subfigure{\includegraphics[scale=0.45]{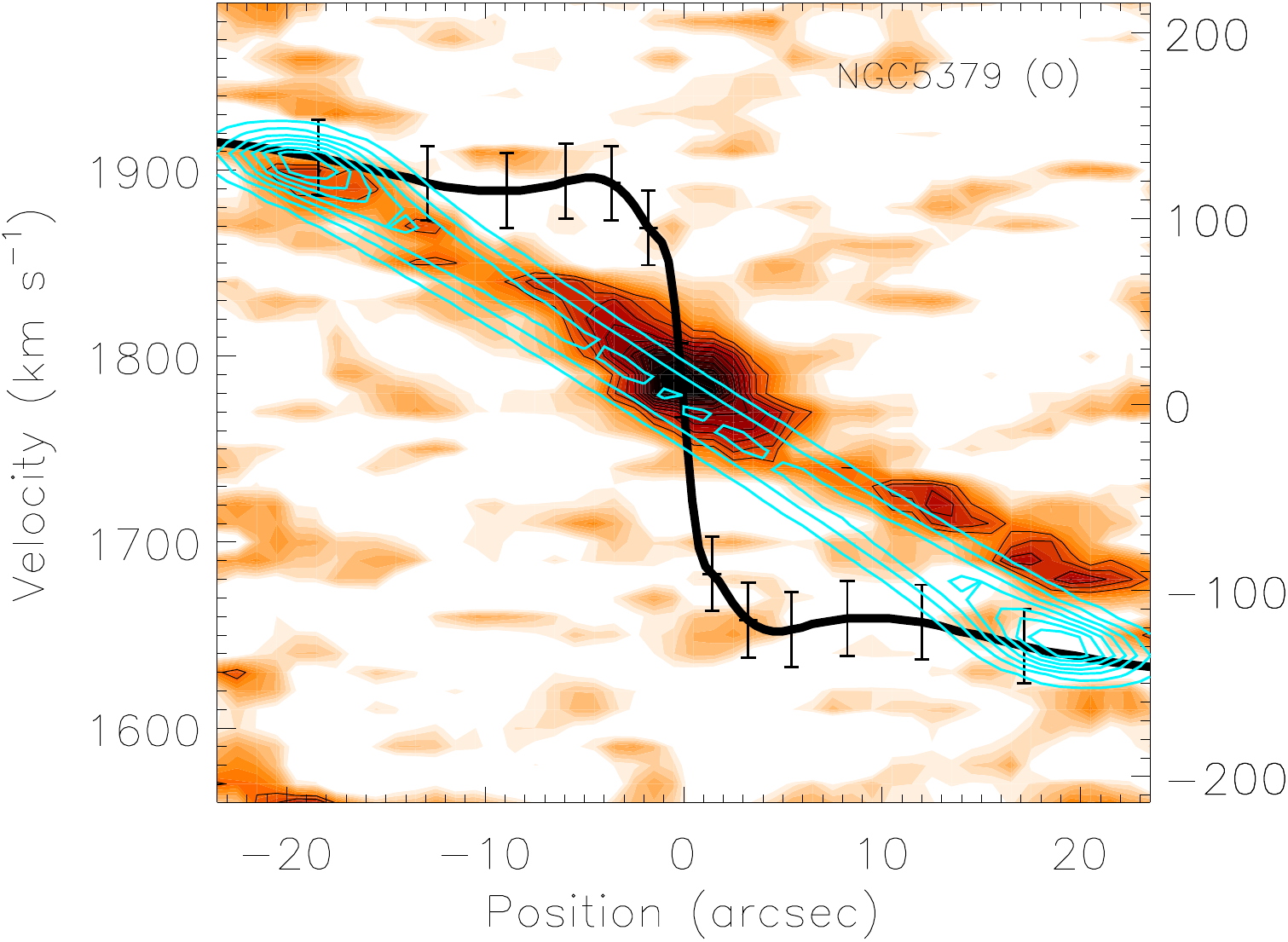}}
\subfigure{\includegraphics[scale=0.45]{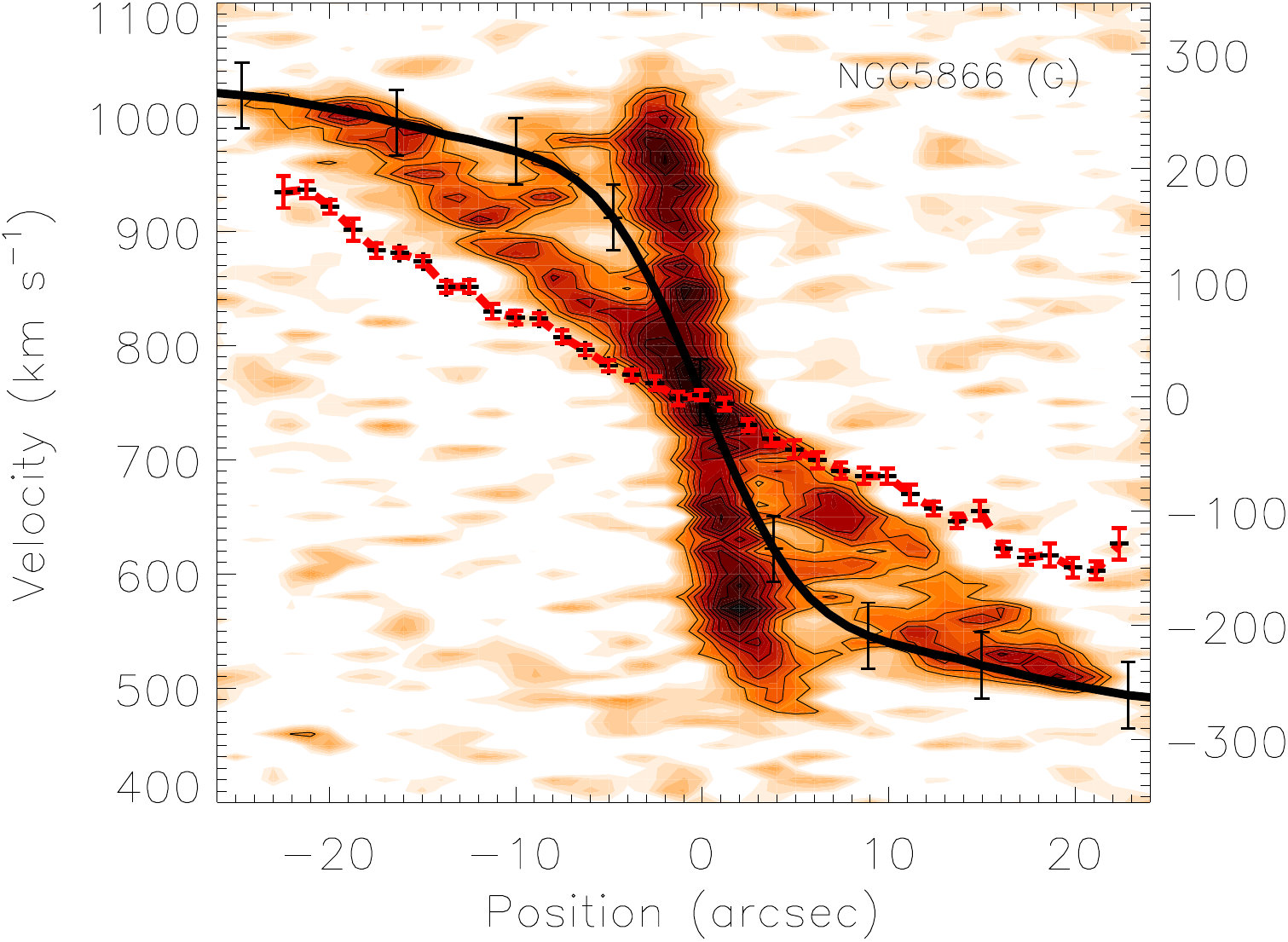}}
\subfigure{\includegraphics[scale=0.45]{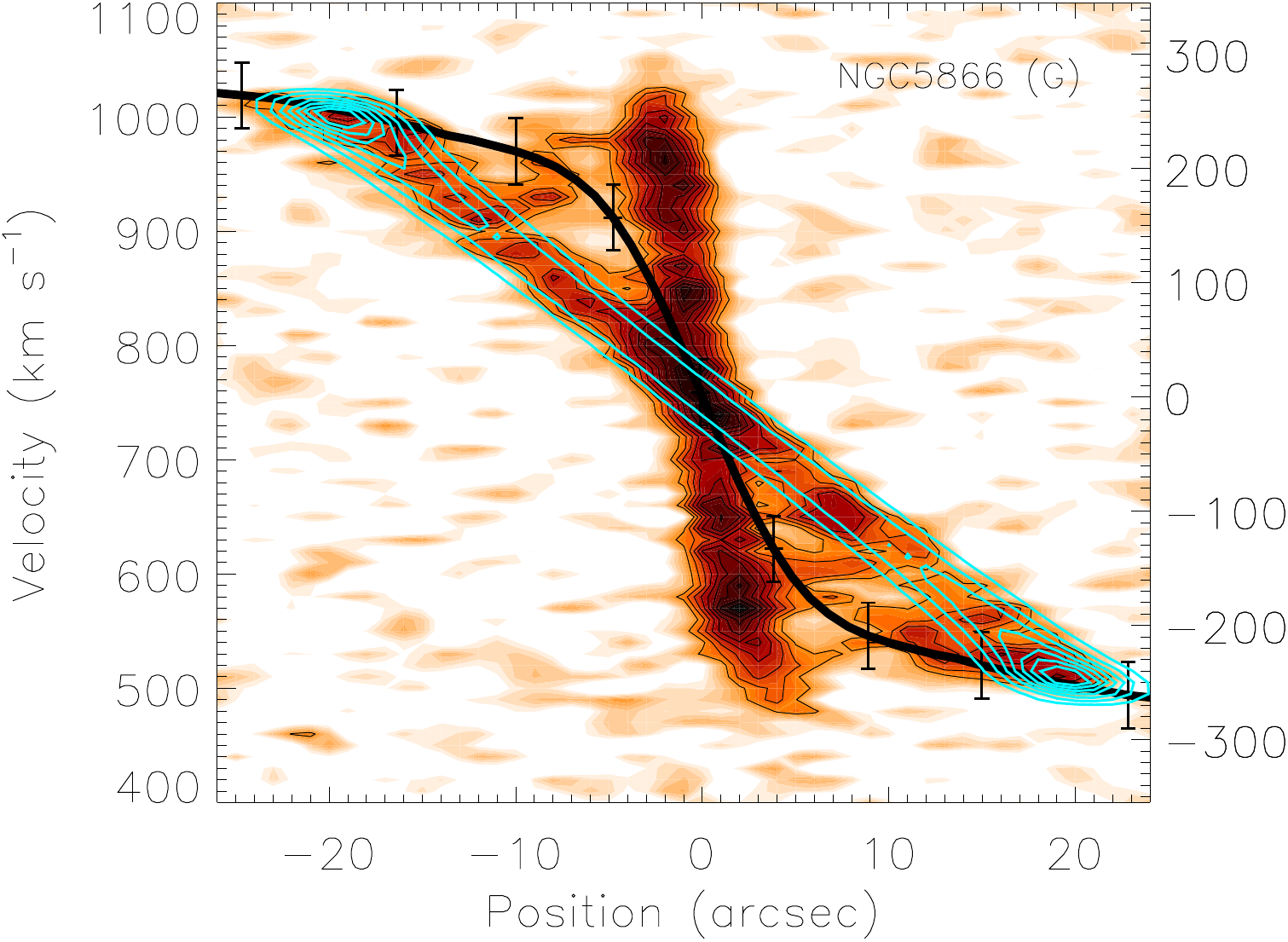}}
\subfigure{\includegraphics[scale=0.45]{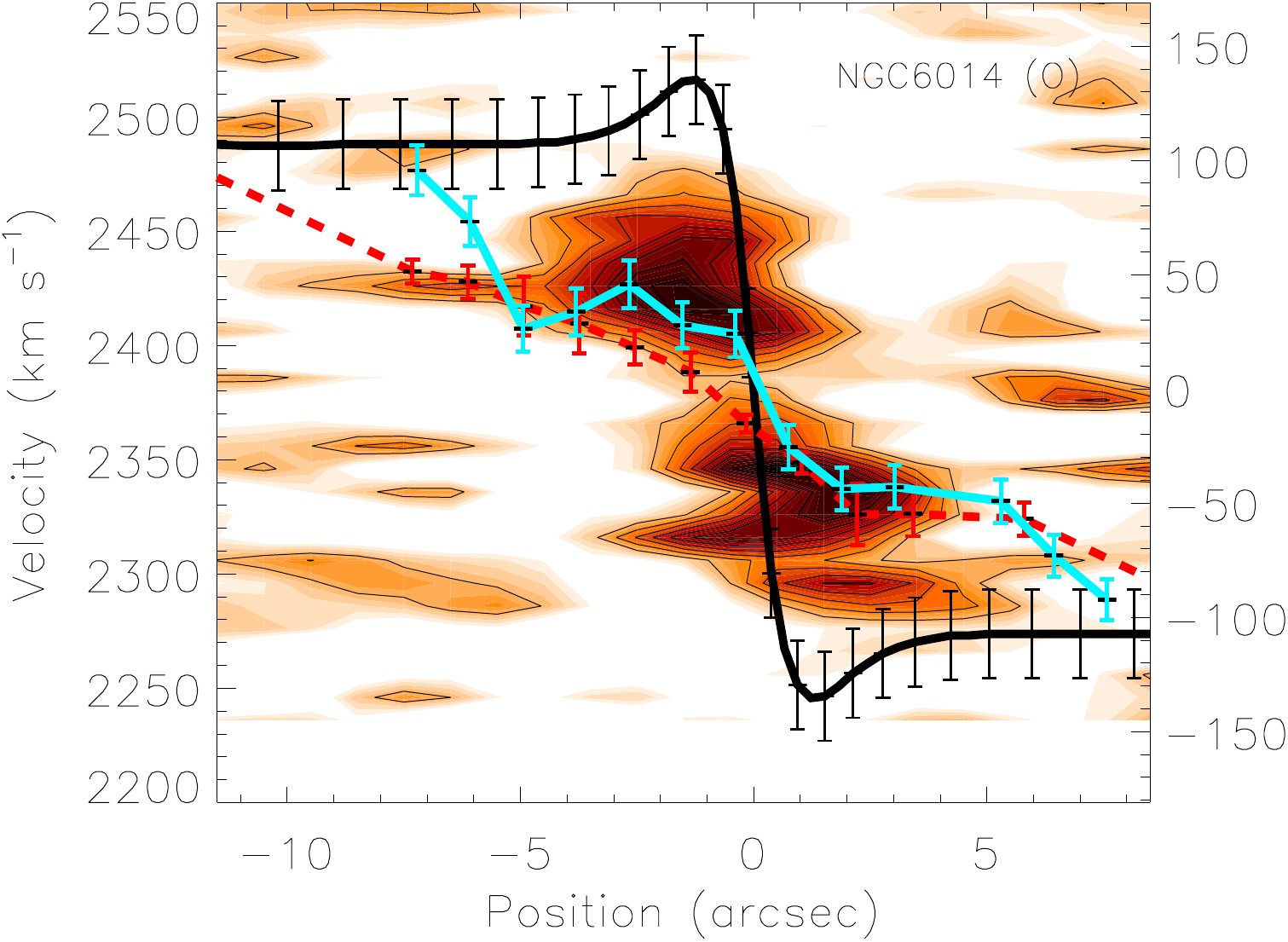}}
\subfigure{\includegraphics[scale=0.45]{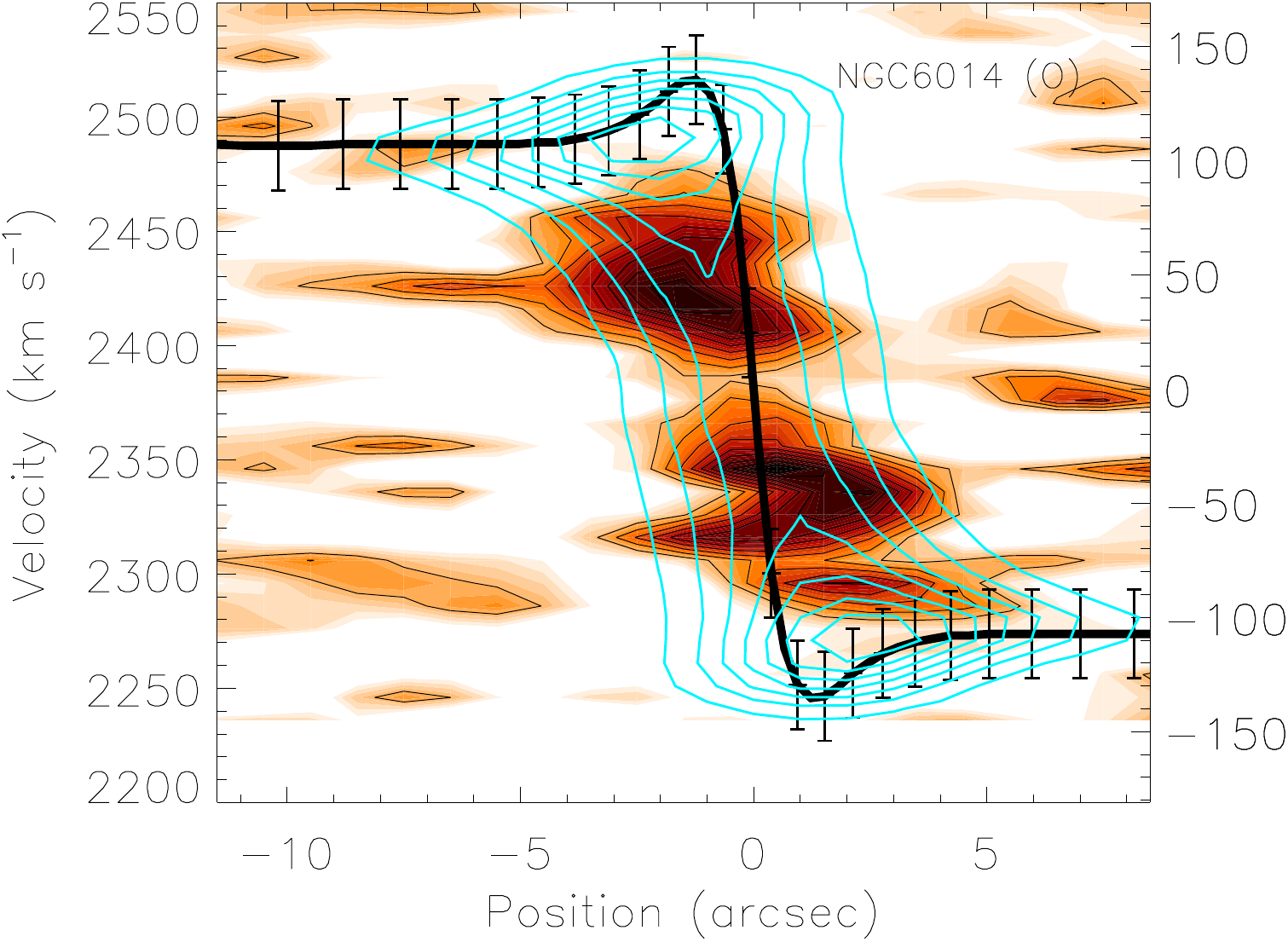}}
\subfigure{\includegraphics[scale=0.45]{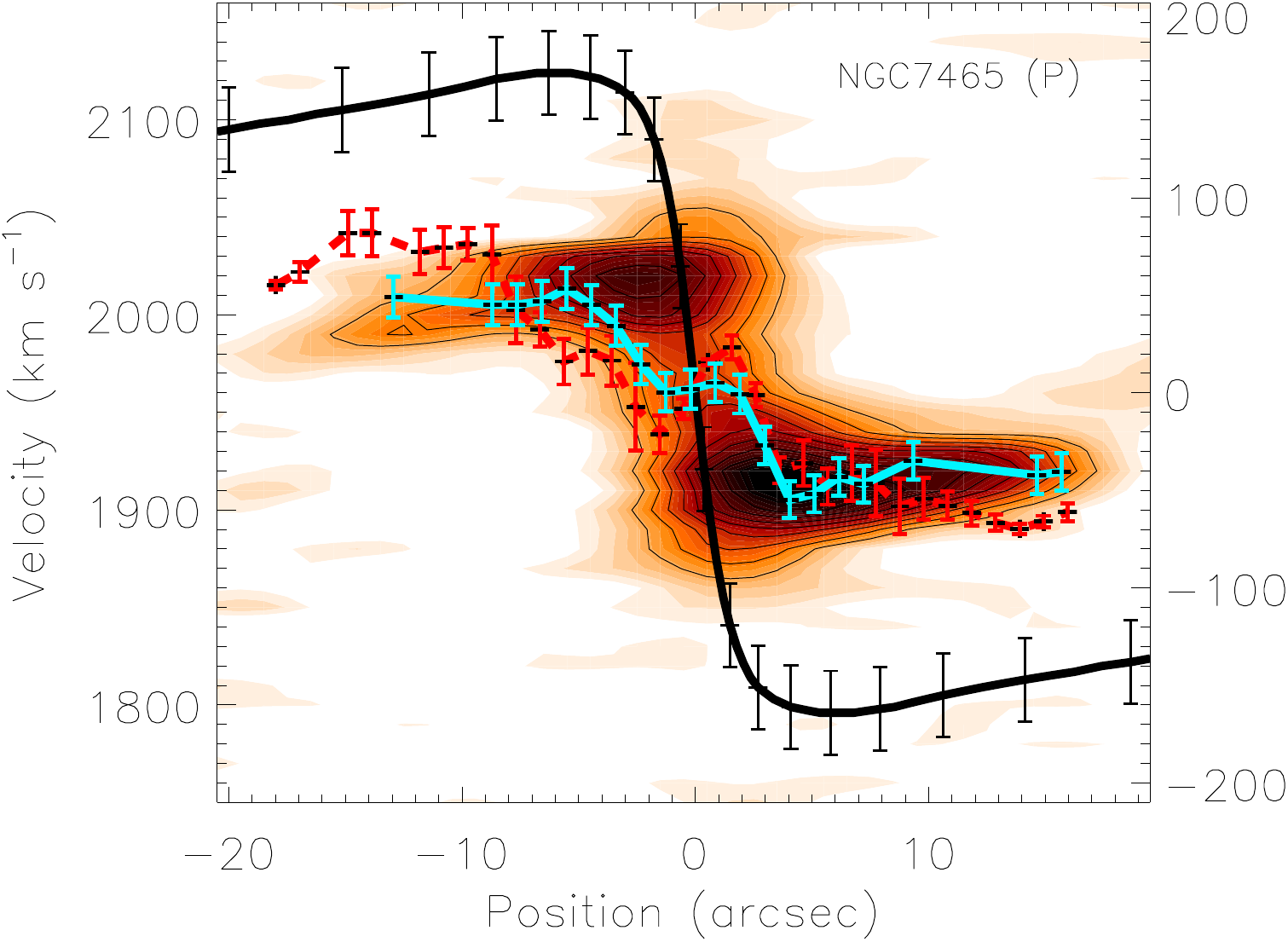}}
\subfigure{\includegraphics[scale=0.45]{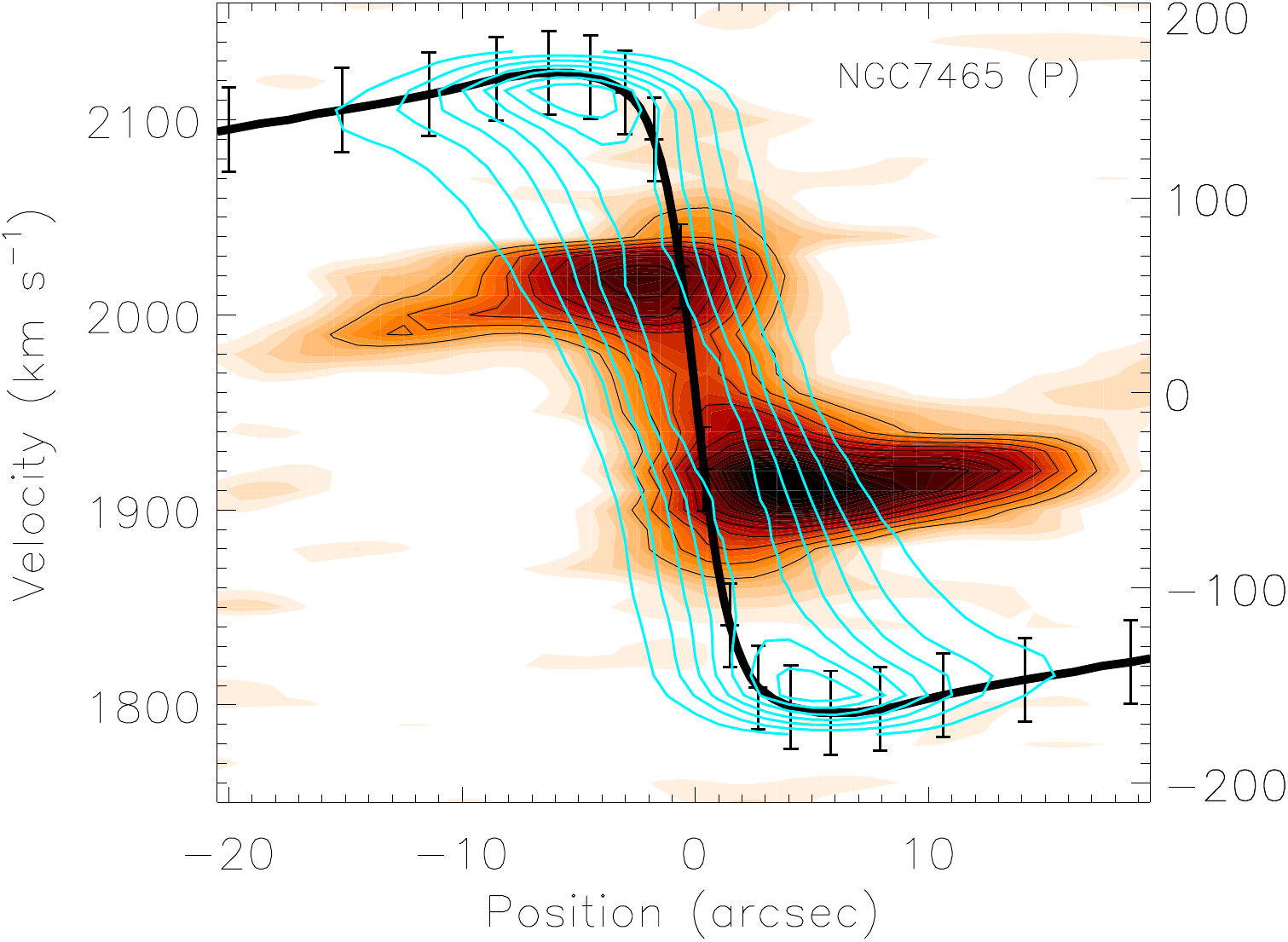}}
\parbox[t]{0.9 \textwidth}{ \textbf{Figure B1.} continued}
\end{figure*}
\begin{figure*}
\centering
\subfigure{\includegraphics[scale=0.45]{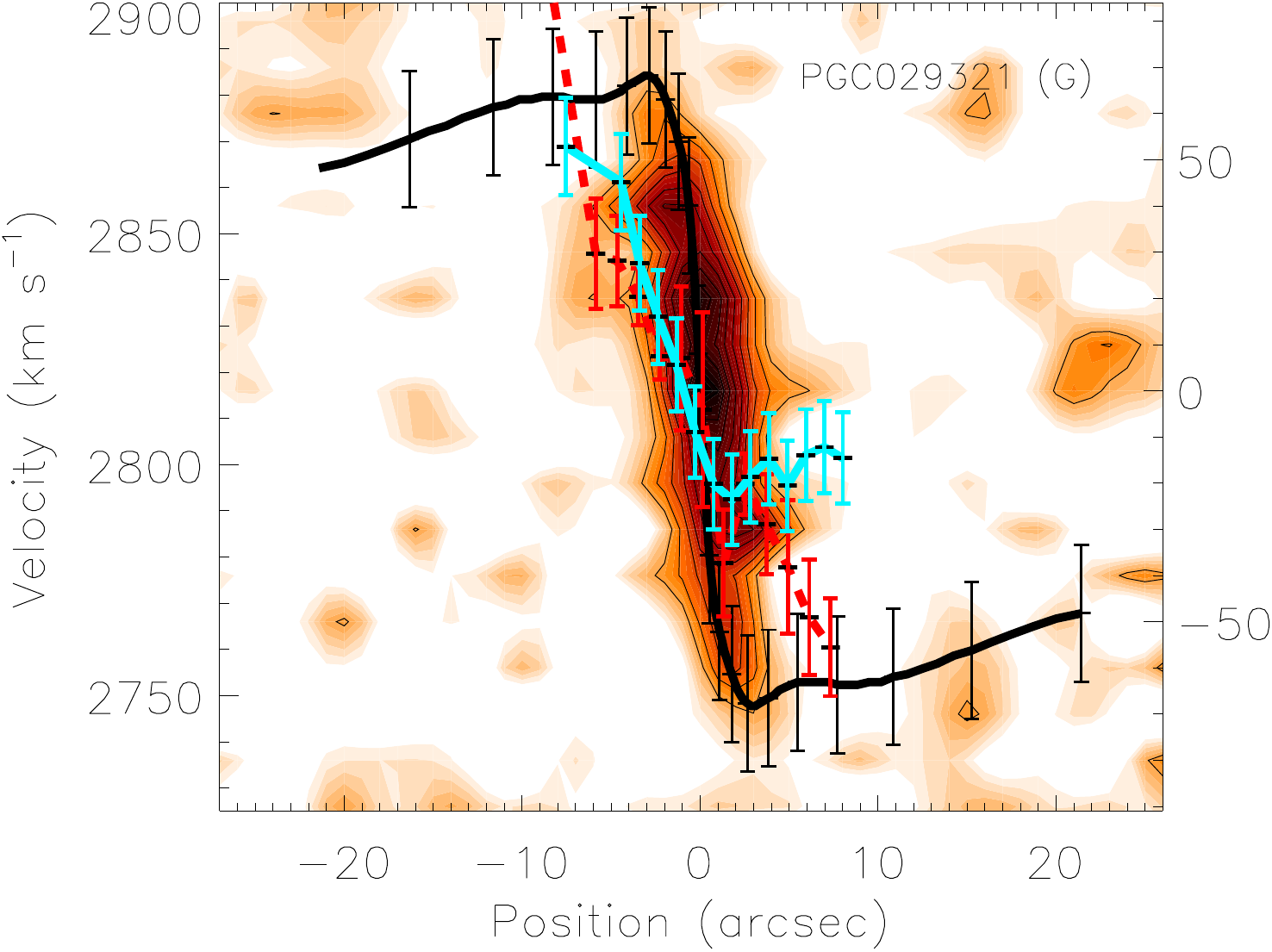}}
\subfigure{\includegraphics[scale=0.45]{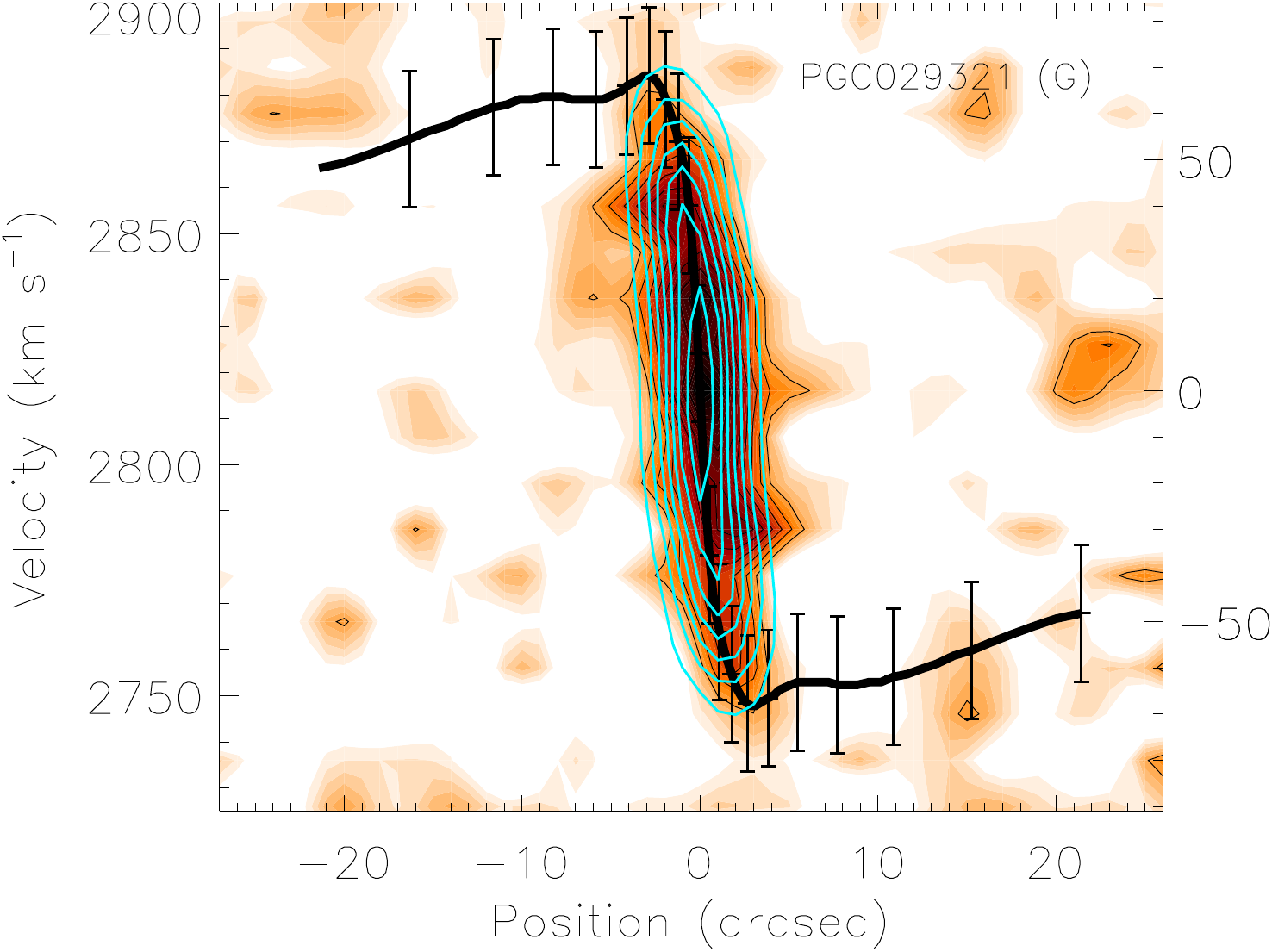}}
\subfigure{\includegraphics[scale=0.45]{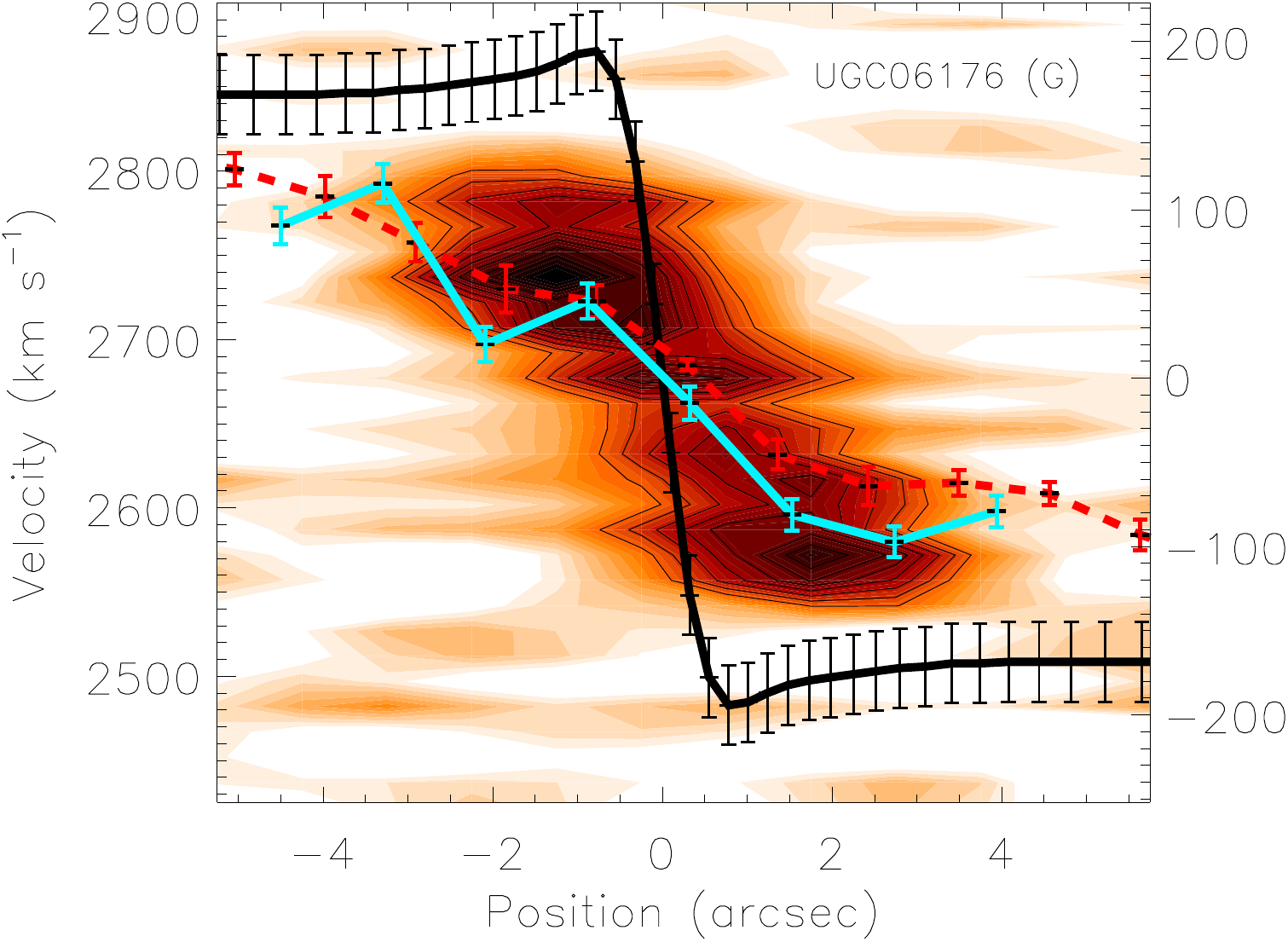}}
\subfigure{\includegraphics[scale=0.45]{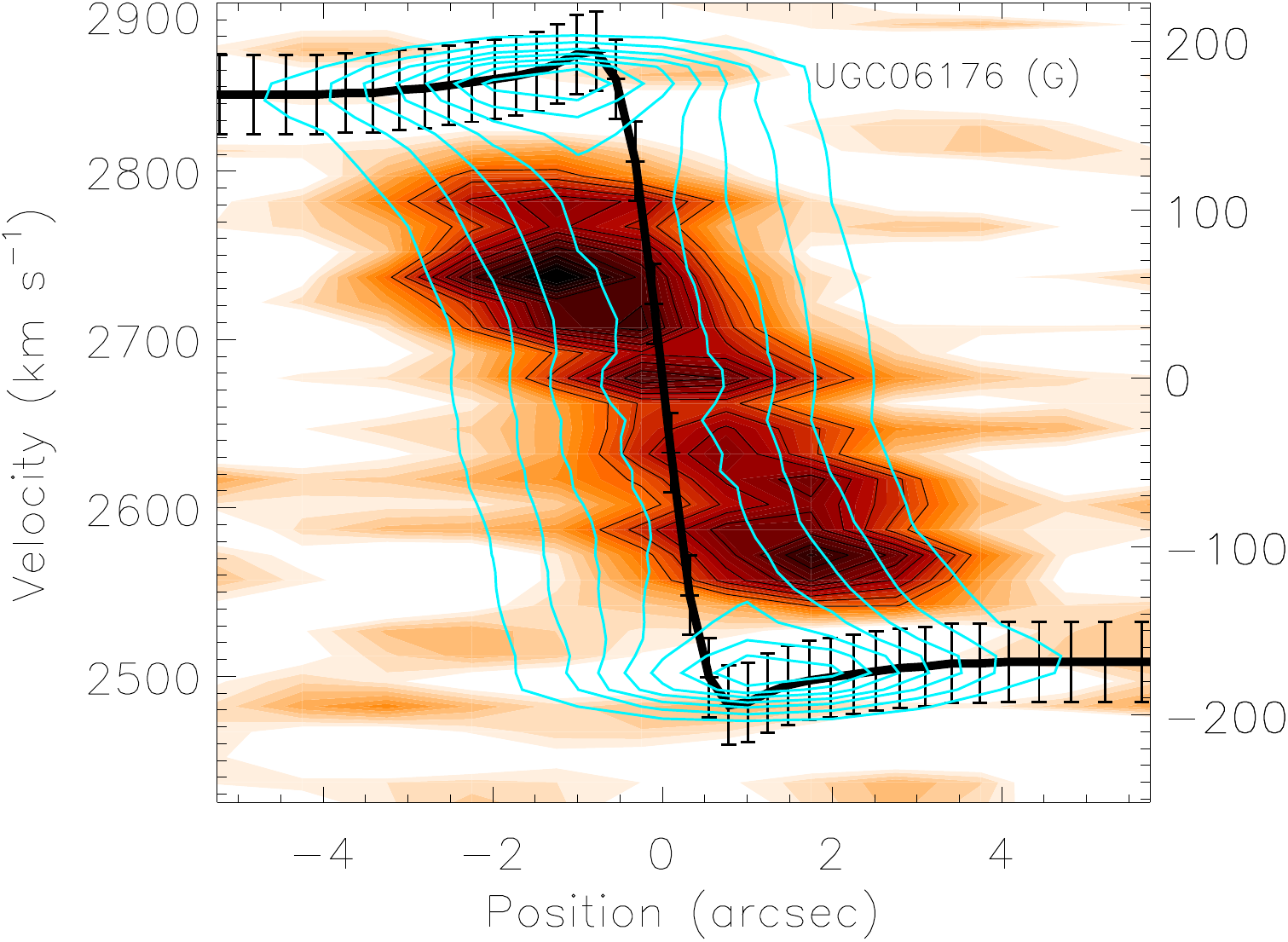}}
\subfigure{\includegraphics[scale=0.45]{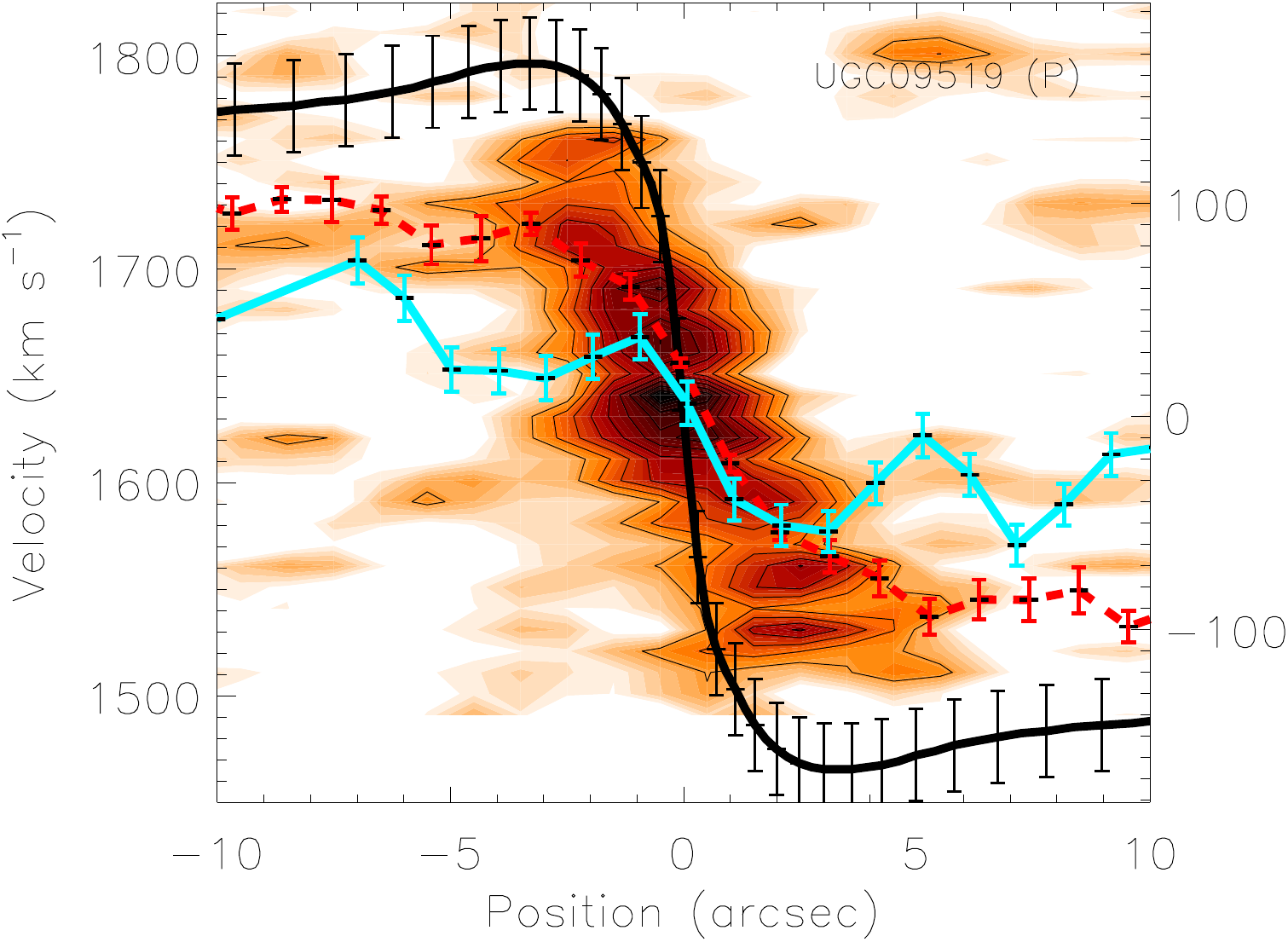}}
\subfigure{\includegraphics[scale=0.45]{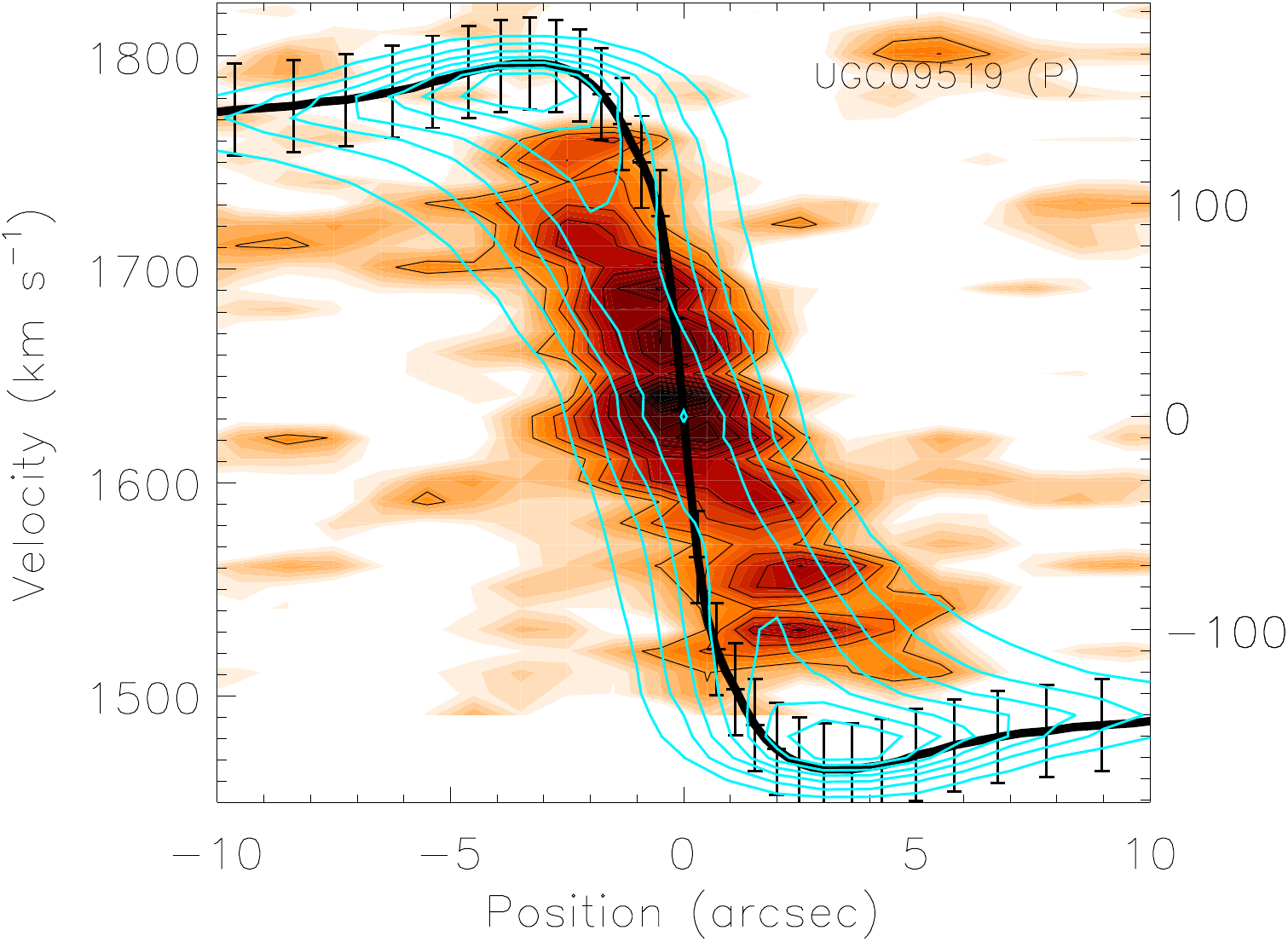}}
\parbox[t]{0.9 \textwidth}{ \textbf{Figure B1.} continued}
\end{figure*}